\PassOptionsToPackage{unicode}{hyperref}
\PassOptionsToPackage{hyphens}{url}
\PassOptionsToPackage{dvipsnames,svgnames,x11names}{xcolor}
\documentclass[12pt]{article}

\usepackage{romannum}
\usepackage{xcolor}
\usepackage{bm}

\newcommand{\blind}{1}
\newcommand{\Hawkes}[1]{}
\newcommand{\Ogata}[1]{}
\newcommand{\DV}[1]{}
\newcommand{\indsim}{\overset{\textit{\tiny{ind.}}}{\sim}}

\AtBeginDocument{\pagenumbering{arabic}}

\usepackage{amsmath,amssymb}
\usepackage{iftex}
\ifPDFTeX
  \usepackage[T1]{fontenc}
  \usepackage[utf8]{inputenc}
  \usepackage{textcomp} 
\else 
  \usepackage{unicode-math}
  \defaultfontfeatures{Scale=MatchLowercase}
  \defaultfontfeatures[\rmfamily]{Ligatures=TeX,Scale=1}
\fi
\usepackage{lmodern}
\ifPDFTeX\else  
\fi
\IfFileExists{upquote.sty}{\usepackage{upquote}}{}
\IfFileExists{microtype.sty}{
  \usepackage[]{microtype}
  \UseMicrotypeSet[protrusion]{basicmath} 
}{}
\makeatletter
\@ifundefined{KOMAClassName}{
  \IfFileExists{parskip.sty}{%
    \usepackage{parskip}
  }{
    \setlength{\parindent}{0pt}
    \setlength{\parskip}{6pt plus 2pt minus 1pt}}
}{
  \KOMAoptions{parskip=half}}
\makeatother
\usepackage{xcolor}
\makeatletter
\ifx\paragraph\undefined\else
  \let\oldparagraph\paragraph
  \renewcommand{\paragraph}{
    \@ifstar
      \xxxParagraphStar
      \xxxParagraphNoStar
  }
  \newcommand{\xxxParagraphStar}[1]{\oldparagraph*{#1}\mbox{}}
  \newcommand{\xxxParagraphNoStar}[1]{\oldparagraph{#1}\mbox{}}
\fi
\ifx\subparagraph\undefined\else
  \let\oldsubparagraph\subparagraph
  \renewcommand{\subparagraph}{
    \@ifstar
      \xxxSubParagraphStar
      \xxxSubParagraphNoStar
  }
  \newcommand{\xxxSubParagraphStar}[1]{\oldsubparagraph*{#1}\mbox{}}
  \newcommand{\xxxSubParagraphNoStar}[1]{\oldsubparagraph{#1}\mbox{}}
\fi
\makeatother

\usepackage{longtable,booktabs,array}
\usepackage{calc} 
\usepackage{etoolbox}
\makeatletter
\patchcmd\longtable{\par}{\if@noskipsec\mbox{}\fi\par}{}{}
\makeatother
\IfFileExists{footnotehyper.sty}{\usepackage{footnotehyper}}{\usepackage{footnote}}
\makesavenoteenv{longtable}
\usepackage{graphicx}
\makeatletter
\def\maxwidth{\ifdim\Gin@nat@width>\linewidth\linewidth\else\Gin@nat@width\fi}
\def\maxheight{\ifdim\Gin@nat@height>\textheight\textheight\else\Gin@nat@height\fi}
\makeatother
\setkeys{Gin}{width=\maxwidth,height=\maxheight,keepaspectratio}
\makeatletter
\def\fps@figure{htbp}
\makeatother

\makeatletter
\@ifpackageloaded{caption}{}{\usepackage{caption}}
\AtBeginDocument{%
\ifdefined\contentsname
  \renewcommand*\contentsname{Table of contents}
\else
  \newcommand\contentsname{Table of contents}
\fi
\ifdefined\listfigurename
  \renewcommand*\listfigurename{List of Figures}
\else
  \newcommand\listfigurename{List of Figures}
\fi
\ifdefined\listtablename
  \renewcommand*\listtablename{List of Tables}
\else
  \newcommand\listtablename{List of Tables}
\fi
\ifdefined\figurename
  \renewcommand*\figurename{Figure}
\else
  \newcommand\figurename{Figure}
\fi
\ifdefined\tablename
  \renewcommand*\tablename{Table}
\else
  \newcommand\tablename{Table}
\fi
}
\@ifpackageloaded{float}{}{\usepackage{float}}
\floatstyle{ruled}
\@ifundefined{c@chapter}{\newfloat{codelisting}{h}{lop}}{\newfloat{codelisting}{h}{lop}[chapter]}
\floatname{codelisting}{Listing}

\makeatother
\makeatletter
\@ifpackageloaded{caption}{}{\usepackage{caption}}
\@ifpackageloaded{subcaption}{}{\usepackage{subcaption}}
\makeatother

\ifLuaTeX
  \usepackage{selnolig}  
\fi
\usepackage[]{natbib}
\usepackage{bookmark}

\IfFileExists{xurl.sty}{\usepackage{xurl}}{} 
\hypersetup{
  pdftitle={Title},
  pdfauthor={Author 1; Author 2},
  pdfkeywords={3 to 6 keywords, that do not appear in the title},
  colorlinks=true,
  linkcolor={blue},
  filecolor={Maroon},
  citecolor={Blue},
  urlcolor={Blue},
  pdfcreator={LaTeX via pandoc}}

\begin{document}

\def\spacingset#1{\renewcommand{\baselinestretch}%
{#1}\small\normalsize} \spacingset{1}


\if1\blind
{
\title{\LARGE\bf Bayesian Nonparametric Marked Hawkes Processes for Earthquake Modeling}
\author{Hyotae Kim and Athanasios Kottas \thanks{
Hyotae Kim (hyotae.kim@duke.edu) is Postdoctoral Researcher, Department of Biostatistics 
\& Bioinformatics, Duke University, and Athanasios Kottas (thanos@soe.ucsc.edu) 
is Professor, Department of Statistics, University of California, Santa Cruz.} \\
}
    \maketitle
} \fi

\if0\blind
{
  \bigskip
  \bigskip
  \bigskip
  \begin{center}
    {\LARGE\bf Bayesian Nonparametric Marked Hawkes Processes for Earthquake Modeling}
\end{center}
  \medskip
} \fi


\bigskip
\begin{abstract}
The Hawkes process is a versatile stochastic model for point patterns that exhibit 
self-excitation, that is, the property that an event occurrence increases the rate of 
occurrence for some period of time in the future.
We present a Bayesian nonparametric modeling approach for temporal marked Hawkes 
processes. Our focus is on point process modeling of earthquake occurrences, where 
the mark variable is given by earthquake magnitude. 
We develop a nonparametric prior model for the marked Hawkes process excitation function, 
using a representation with basis components for the time lag and the mark, and basis
weights defined through a gamma process prior. We elaborate the model with a nonparametric 
prior for time-dependent background intensity functions, thus enabling a fully nonparametric 
approach to modeling the ground process intensity of marked Hawkes processes.
The model construction balances computationally tractable inference with flexible forms for 
marked Hawkes process functionals, including mark-dependent offspring densities. The posterior 
simulation method provides full inference, without any approximations to the Hawkes process 
likelihood. In the context of the application, the modeling approach enables estimation of 
aftershock densities that vary with the magnitude of the main shock, thus significantly 
expanding the inferential scope of existing self-exciting point process models for 
earthquake occurrences. We investigate different aspects of the methodology through study 
of model properties, and with inference results based on synthetic marked point patterns. 
The practical utility of modeling magnitude-dependent aftershock dynamics is demonstrated 
with analysis of earthquakes that occurred in Japan from 1885 through 1980.
\end{abstract}

\noindent%
\vfill

\newpage
\spacingset{1.8} 

\section{Introduction}
\label{sec:intro}
The Hawkes process (HP) \cite[][]{H1971a,H1971b} is a point process characterized by the 
property that each event increases the chance of subsequent events occurring within a certain 
time period. This \textit{self-exciting} HP property causes an event to generate descendant 
events, each of which can, in turn, trigger its own group of descendant events, resulting 
in event clusters akin to a family tree. 
A key example involves earthquake occurrences, where each earthquake acts as a 
catalyst for subsequent events, often forming clusters consisting of a main shock followed 
by multiple aftershocks. Indeed, HPs have been applied in seismology 
\cite[e.g.,][]{O1988, O1998, ZOV2002, VS2008, FSG2016, GPETAS2022}, as well as 
finance \cite[e.g.,][]{H2018}, 
biology/epidemiology \cite[e.g.,][]{HJS2022}, 
and criminology \cite[e.g.,][]{MSBST2011}.
\cite{Daley_Vere-Jones:2003,DV2007} provide theoretical background for (marked) HPs.
Reviews of HP models and their applications include \cite{LLT2021}, \cite{R2018}, 
and \cite{JABESreview2025}.

The marked Hawkes process (MHP) is an extension of the HP that incorporates marks, i.e., 
(observable) variables associated with each time event. 
The earthquake magnitude is the key mark variable in the analysis of earthquake occurrences.
Other application areas include criminology, where the type of crime can be used as 
a mark to study the dynamics of specific crime types, 
and epidemiology, where the disease strain can be treated as a mark for forecasting 
strain-specific epidemics of infectious diseases. 

We develop a flexible modeling approach for temporal MHPs. Our focus is on MHP modeling 
of earthquake occurrences, which involves a continuous mark (earthquake magnitude).
Here, the Epidemic Type Aftershock Sequences (ETAS) model provides the most widely 
used approach. A temporal earthquake point pattern comprises the time ($t_i$) and 
magnitude ($\kappa_i$) for earthquakes that exceed a magnitude cutoff, $\kappa_0 > 0$.
Then, the main component of the ETAS model conditional intensity (the ground process 
intensity $\lambda^\ast_g(t)$) is 
\begin{equation}
\lambda_g^\ast(t) \, = \, \mu \, + \, 
\sum_{t_i<t} a \, \exp\{ b \, (\kappa_i - \kappa_0) \} \, \text{Lomax}(t-t_i | p,c)
\label{eq:etas_model}
\end{equation}
where $\mu > 0$ is the background intensity, $\text{Lomax}(x|p,c)$ is the Lomax 
density, $p \, c^{p}/(c+x)^{p+1}$, for $x > 0$, with $p > 0$ and $c > 0$ the shape 
and scale parameter, respectively, and $a > 0$, $b > 0$.
The form in (\ref{eq:etas_model}) is a special case of the general MHP formulation, 
$\lambda^\ast_g(t) =$ $\mu(t) \, + \, \sum_{t_i<t} h(t-t_i,\kappa_i)$,
which involves time-dependent background intensity $\mu(t)$, and excitation function
$h(t-t_i,\kappa_i)$ that allows for interaction between the time lag from the shock 
at time $t_i$ and its magnitude $\kappa_i$. 
The ETAS model factorizes the excitation function into components that depend only 
on the time lag and magnitude. As discussed in Section \ref{sec:mhp_background}
(where we review the relevant background for MHPs), this is a restrictive assumption, 
since it implies that the distribution of aftershock occurrence times (more specifically, 
of the time lag from the main shock) does not depend on the main shock's magnitude.

The temporal ETAS model was introduced by \cite{O1988}, and it has since been extended in 
various directions, the main being to models for space-time point processes; see, e.g., 
\cite{JABESreview2025}. The space-time ETAS model adds a multiplicative 
(parametric) component to the excitation function in (\ref{eq:etas_model}) to account for 
the distance between the locations of the main shock and the aftershock (in general, 
the space component depends on earthquake magnitude). Therefore, the space-time ETAS model 
retains the restrictive structure of aftershock time densities that are not allowed to 
change with the magnitude of the main shock. 
To our knowledge, the literature does not include model-based inference methods for 
general excitation functions, either for earthquake modeling or for general temporal MHPs.

Our primary contribution is a Bayesian nonparametric model for MHP excitation functions 
with one continuous mark variable (earthquake magnitude). The excitation 
function is modeled as a weighted combination of basis functions for the time lag and 
the mark, specified such that the implied MHP functionals have desirable properties. 
The weights associated with the basis components are constructed through a 
nonparametric prior, which ensures effective use of the basis to capture 
different excitation function shapes.  
In the context of modeling earthquake occurrences, a key feature of the approach 
is that it removes the restrictive separability assumption between aftershock occurrence 
times and main shock magnitudes. As an important consequence, we can estimate 
magnitude-dependent aftershock densities. We demonstrate that this model property provides 
new insights into seismic patterns and it also improves forecasting. 
As a further contribution, we elaborate the ground process intensity model to allow for 
a time-dependent background intensity function, modeled with a nonparametric prior based
again on a flexible basis representation.

As discussed in Section \ref{sec:mhp_background}, MHPs admit a cluster representation,
which provides useful interpretation for the conditional intensity components, and it 
also facilitates likelihood or Bayesian estimation. 
However, even when augmented with latent variables that pertain to the cluster
representation, the MHP likelihood contains a complex normalizing term that involves 
$n$ integrals of the excitation function (where $n$ is the size of the point pattern). 
This may explain the limited work on models that are more flexible than the form 
in (\ref{eq:etas_model}); in particular, we are not aware of Bayesian nonparametric 
modeling for MHP excitation functions. Hence, a key consideration for our model 
construction is to balance flexible MHP functionals with a computationally tractable 
posterior simulation method, which does not resort to approximations to the MHP likelihood 
or to the posterior distribution.

The practical utility of the proposed modeling approach is explored with a simulation study 
(based on different scenarios for the excitation function structure) and with analysis of 
earthquakes that occurred in Japan from 1885 through 1980. The empirical investigation 
includes results from comparison with the ETAS model, and with a (novel) semiparametric 
extension of the ETAS model, which replaces the Lomax density in (\ref{eq:etas_model}) with 
a nonparametric prior model that supports general decreasing densities.

The outline of the paper is as follows. Section \ref{sec:mhp_background} collects the 
relevant background for MHPs. Section \ref{sec:mhp_method} develops the modeling approach, 
including details on model implementation. The simulation study and the earthquake data 
analysis are presented in Section \ref{sec:mhp_simulation_study} and 
Section \ref{sec:mhp_real_analysis}, respectively. Finally, Section \ref{sec:mhp_discussion} 
concludes with a summary and discussion.

%
%

\section{Background on marked Hawkes processes}
\label{sec:mhp_background}

We consider point processes over time, $t \in \mathbb{R}^{+}$, recorded in the observation 
window $(0,T)$. Denote by $\kappa$ the mark variable(s) taking values in mark space 
$\mathcal{K}$. The conditional intensity of a marked point process, including a MHP, 
can be generally expressed as $\lambda^\ast_g(t) \, f^\ast(\kappa|t)$, where
$f^\ast(\kappa|t)$ is the mark density function, and $\lambda^\ast_g(t)$ the
ground process intensity function.
Here, the asterisk ($*$) indicates dependence of the functions on the process history
up to time $t$, $\mathcal{H}(t)=$ $\{(t_i,\kappa_i): t_i < t\}$; in particular, the 
ground process intensity is a function of marks. 
Although the mark density may depend on the history of times and marks, in 
practice, simple forms are used, such as $f^\ast(\kappa|t)=$ $f(\kappa)$.

Let $\{(t_i,\kappa_i): i=1,\ldots,n\}$ be the observed point pattern, where 
$0 < t_1 < \ldots < t_n < T$ are the time events, and $\kappa_1,...,\kappa_n$ the 
associated marks. Under the MHP with ground process intensity $\lambda^\ast_g(t)$
and mark density $f(\kappa)$, the likelihood is given by  
\begin{equation*}
\Big[\prod_{i=1}^n f(\kappa_i)\Big] \, 
\exp\Big\{-\int_0^T \lambda^\ast_g(u) \, \text{d}u \Big\} \,
\Big[\prod_{i=1}^n\lambda^\ast_g(t_i) \Big]
\end{equation*}
\cite[][Proposition 7.3.\Romannum{3}]{Daley_Vere-Jones:2003}.

Our modeling focus is on the MHP ground process intensity function, which extends 
the HP intensity to incorporate dependence on both time and mark history. 
More specifically,  
\begin{equation}
\label{ground_process_intensity}
\lambda^\ast_g(t) \, \equiv \, \lambda_g(t|\mathcal{H}(t)) \, = \, 
\mu(t) \, + \, \sum_{t_i<t} h(t-t_i,\kappa_i),
\end{equation}
where $\mu(t)>0$, for $t>0$, and $h(x_i,\kappa_i)>0$, for $x_i=t-t_i>0$ and 
$\kappa_i \in \mathcal{K}$, are the background (immigrant) intensity and 
excitation (offspring intensity) functions, respectively.

The {\it immigrant} and {\it offspring} terminology 
originates from the HP cluster representation \citep{HO1974}, which facilitates 
inference by augmenting the MHP likelihood. The branching structure for the point pattern 
involves latent variables $\bm{y}=$ $\{ y_{i}: i=1,...,n \}$, such that 
$y_{i}=0$ if $t_{i}$ is an immigrant point, and $y_{i} = j$ if point $t_{i}$ is the offspring 
of $t_{j}$. Hence, given $\bm{y}$, the point pattern is partitioned into the set of immigrants, 
$I =$ $\{ t_{i}: y_{i} = 0 \}$, and sets of offspring, $O_{j} =$ $\{ t_{i}: y_{i} = j \}$, 
where $O_{j}$ collects all offspring of $t_{j}$. Moreover, conditioning on 
$\bm{y}$, the MHP can be constructed as the superposition of independent marked
non-homogeneous Poisson processes (NHPPs): $I$ follows a marked NHPP with intensity 
$\mu(t)$, and each $t_i \in I$ has mark $\kappa_i$ with density $f$; and, each $O_j$ follows
a marked NHPP with intensity $h(t - t_{j},\kappa_j)$, and each $t_i \in O_j$ has mark 
$\kappa_i$ with density $f$. 
Hence, using the branching variables $\bm{y}$, the augmented MHP likelihood can be 
written as 
\begin{equation}
\begin{split}
&  \Big[ \prod_{i=1}^n f(\kappa_i) \Big] \,
\Big[\exp\Big\{-\int_0^T \mu(u) \, \text{d}u \Big\} \, \prod_{\{t_i \in I\}}\mu(t_i) \Big]
\\
&  \times \,
\Big[ \exp\Big\{-\sum_{j=1}^n  \int_0^{T-t_j} h(u,\kappa_{j}) \, \text{d}u \Big\} \,
\prod_{\{t_i \in O\}} h(t_i-t_{y_i},\kappa_{y_i}) \Big],\\
\end{split}    
\label{eq:mhp_likelihood}
\end{equation}
where $O =$ $\cup_{j=1}^{n} O_{j} =$ $\{t_i: y_i\neq 0 \}$ is the set of all offspring points.

Denote by $\alpha(\kappa) = \int_0^\infty h(u,\kappa) \, \text{d}u$ the total offspring 
intensity at $\kappa$. This function controls the number of offspring of a predecessor with 
mark $\kappa$, and it thus extends the branching ratio of the HP. For any 
$\kappa \in \mathcal{K}$, the total offspring intensity must be finite to ensure a stable 
process, i.e., one that does not explode over time \cite[][p. 203]{Daley_Vere-Jones:2003}.
More precisely, when $f^\ast(\kappa|t)=$ $f(\kappa)$, the stability conditions for MHPs are: 
%
%
\[
\text{(i)} \,\,
\alpha(\kappa) = \int_0^\infty h(u,\kappa) \, \text{d}u < \infty, \,\, \text{for any} \,\,
\kappa \in \mathcal{K}; \,\,\,\,
\text{(ii)} \,\,
\rho = \text{E}(\alpha(\kappa)) = 
\int_\mathcal{K}\alpha(\kappa)f(\kappa) \, \text{d}\kappa < 1 .
\]
The more general version of condition (ii) 
is $\rho < \infty$. However, the constraint $\rho \in (0,1)$ ensures that the average 
cluster size produced by an immigrant is finite.
%
%

Note that the excitation function can be defined through $\alpha(\kappa)$ and 
$g_{\kappa}(x) =$ $h(x,\kappa)/\alpha(\kappa)$, for $x > 0$. 
In light of condition (i), the latter is a density on $\mathbb{R}^{+}$ for any 
$\kappa \in \mathcal{K}$. We refer to the mark-dependent densities $g_{\kappa}(x)$ as 
the offspring density functions of the MHP.

In the context of analysis of earthquake occurrences with MHPs, the ETAS model forms 
the basis of the most widely used modeling approaches. Here, the mark 
corresponds to earthquake magnitude, and the mark space is taken to be 
$\mathcal{K} =$ $(\kappa_0,\infty)$ for point patterns that comprise times 
of earthquakes that exceed a magnitude cutoff, $\kappa_0 > 0$.
The ETAS model builds from a (rescaled) exponential mark density, 
$f^\ast(\kappa|t) \equiv$ $f(\kappa) =$ 
$\psi \exp\{ -\psi \, (\kappa - \kappa_0) \}$, and ground process intensity function
given in (\ref{eq:etas_model}).
The use of the Lomax offspring density and the exponential 
mark density can be motivated by the seismological literature \cite[see, e.g.,][]{O1988}. 
Note that $\rho =$ $a \psi / (\psi - b)$, and thus the parameter constraints $\psi > b$
and $a \psi < \psi - b$ are needed for the $\rho \in (0,1)$ stability condition.

The ETAS model involves the restrictive assumption of an excitation function which is 
factorized into the total offspring intensity, $\alpha(\kappa) =$ 
$a \exp\{ b \, (\kappa - \kappa_0) \}$, and the Lomax offspring density, 
$g(x) =$ $\text{Lomax}(x|p,c)$,  which does not change with earthquake magnitude. 
Building a modeling approach that allows for flexible inference of magnitude-dependent 
offspring densities $g_{\kappa}(x)$ is a key objective of our methodology. 
Indeed, the data analysis of Section \ref{sec:mhp_real_analysis} illustrates that
offspring (aftershock) dynamics vary according to earthquake magnitude.
We also seek a flexible model for the total offspring intensity. 
For earthquake modeling, an increasing $\alpha(\kappa)$ function is a natural choice, as 
it reflects the assumption that a shock of higher magnitude will generate more subsequent 
shocks. However, the exponential form of the ETAS model may not adequately explain 
aftershock counts; indeed, alternative (parametric) functions have been proposed 
\cite[e.g.,][]{OZ2006}.

\section{Methodology}
\label{sec:mhp_method}

To address the limitations of the ETAS model, we develop a prior probability model 
for general MHP excitation functions (Section \ref{subsec:mhp_model_excitation}).
In Section \ref{subsec:immi_insty_model}, we elaborate the model for the MHP ground 
process intensity with a nonparametric prior for time-dependent immigrant 
intensity functions. A primary consideration for our methodology is to balance 
model flexibility for key MHP functionals (Section \ref{subsubsec:mhp_model_property})
with a tractable computational method for full posterior inference
(model implementation is discussed in Section \ref{model_implementation}).

\subsection{The modeling approach}
\label{subsec:mhp_model}

We consider one continuous mark (earthquake magnitude) with support $(\kappa_0,\kappa_{max})$.
The bounded support is a natural assumption in the context of earthquake modeling; analysis 
of earthquake occurrences almost exclusively involves earthquakes that exceed a particular 
magnitude threshold $\kappa_0 > 0$, and it is also straightforward to set the upper 
bound, $\kappa_{max}$.

As aforementioned, we focus on modeling the MHP ground process intensity function. 
As with existing methods for earthquake modeling, we use the simple form, 
$f^\ast(\kappa|t)=$ $f(\kappa)$, for the mark density. However, instead of the (rescaled)
exponential mark density of the ETAS model, we adopt a (rescaled) beta density with 
support $(\kappa_0,\kappa_{max})$, which facilitates deriving the expression for 
the expected total offspring intensity; see Section \ref{subsubsec:mhp_model_property}.
%
%
%

%
%

\subsubsection{Model formulation for the MHP excitation function}
\label{subsec:mhp_model_excitation}

Denote by $\text{Ga}(a,b)$ the gamma distribution with mean $a/b$, and by
$\text{ga}(\cdot |a,b)$ the corresponding density. We model the excitation function 
$h(x, \kappa)$ as a weighted combination of basis functions, each formed by the product 
of an Erlang density and a structured power function
with respective arguments $x$ and $\kappa$. More specifically,  
\begin{equation}
\begin{split}
h(x,\kappa) &= \sum_{l=1}^L\sum_{m=1}^M \nu_{lm} \, \text{ga}(x|l,\theta^{-1}) 
\, b_m(\kappa; d), \quad x \in \mathbb{R}^{+} \enspace \text{and} \enspace 
\kappa \in \mathcal{K} = (\kappa_0,\kappa_{max}) 
\\
\nu_{lm} &= H(A_{lm}), \quad H \mid H_0,c_0 \sim \mathcal{G}(H_0,c_0),
\label{eq:nonpara_model}
\end{split}
\end{equation}
where $b_m(\kappa; d)$ is the basis component for the mark (defined below),
and $A_{lm}=$ $[(l-1)\theta,l\theta)\times[(m-1)/M,m/M)$, for $l = 1,\ldots,L$ 
and $m = 1,\ldots,M$.
The weights, $\nu_{lm}$, are defined through increments of a random measure $H$ 
on $\mathbb{R}^+ \times (0,1)$, which is assigned a gamma process prior, 
$\mathcal{G}(H_0,c_0)$, with centering measure $H_0$ and precision parameter 
$c_0 > 0$.

\begin{figure}[t]
    \centering
\includegraphics[width=0.325\textwidth]{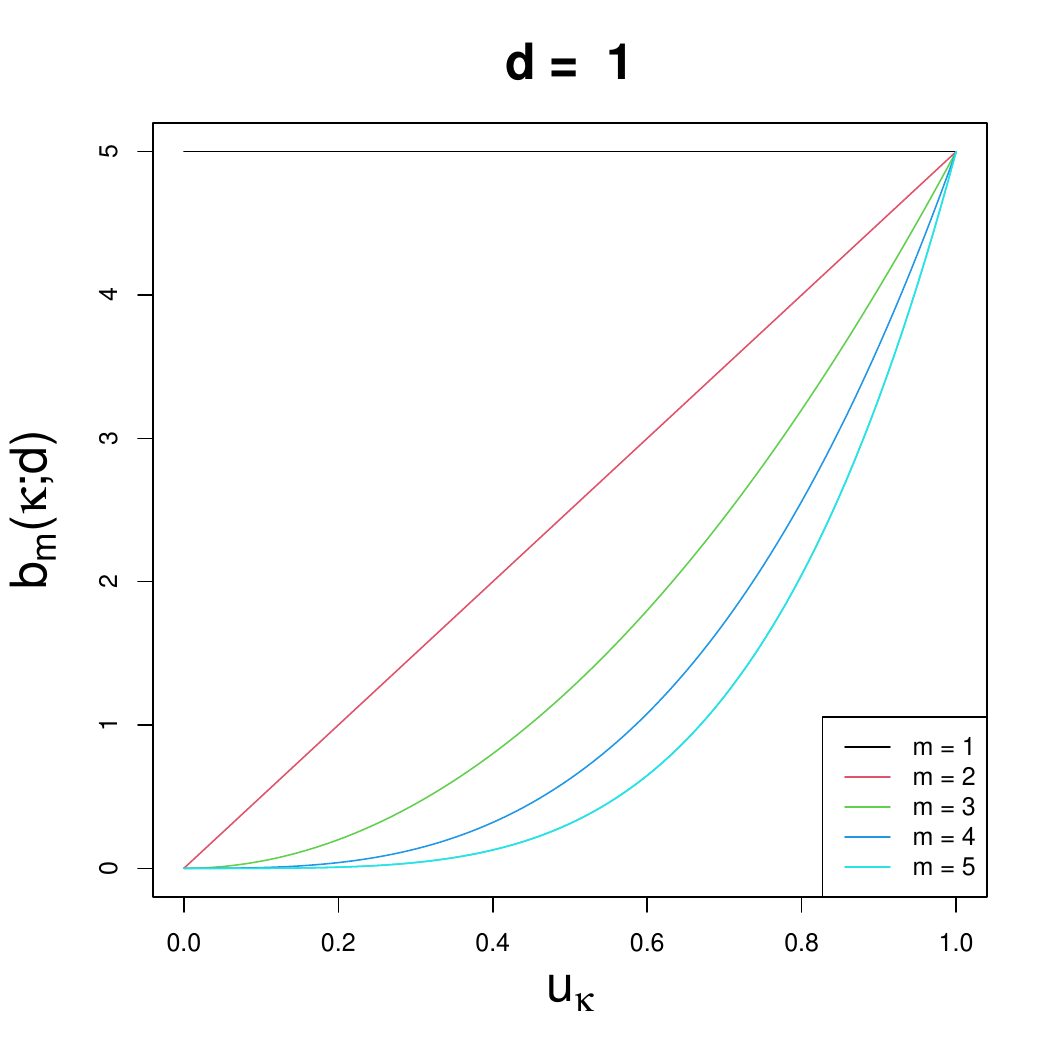}
\includegraphics[width=0.325\textwidth]{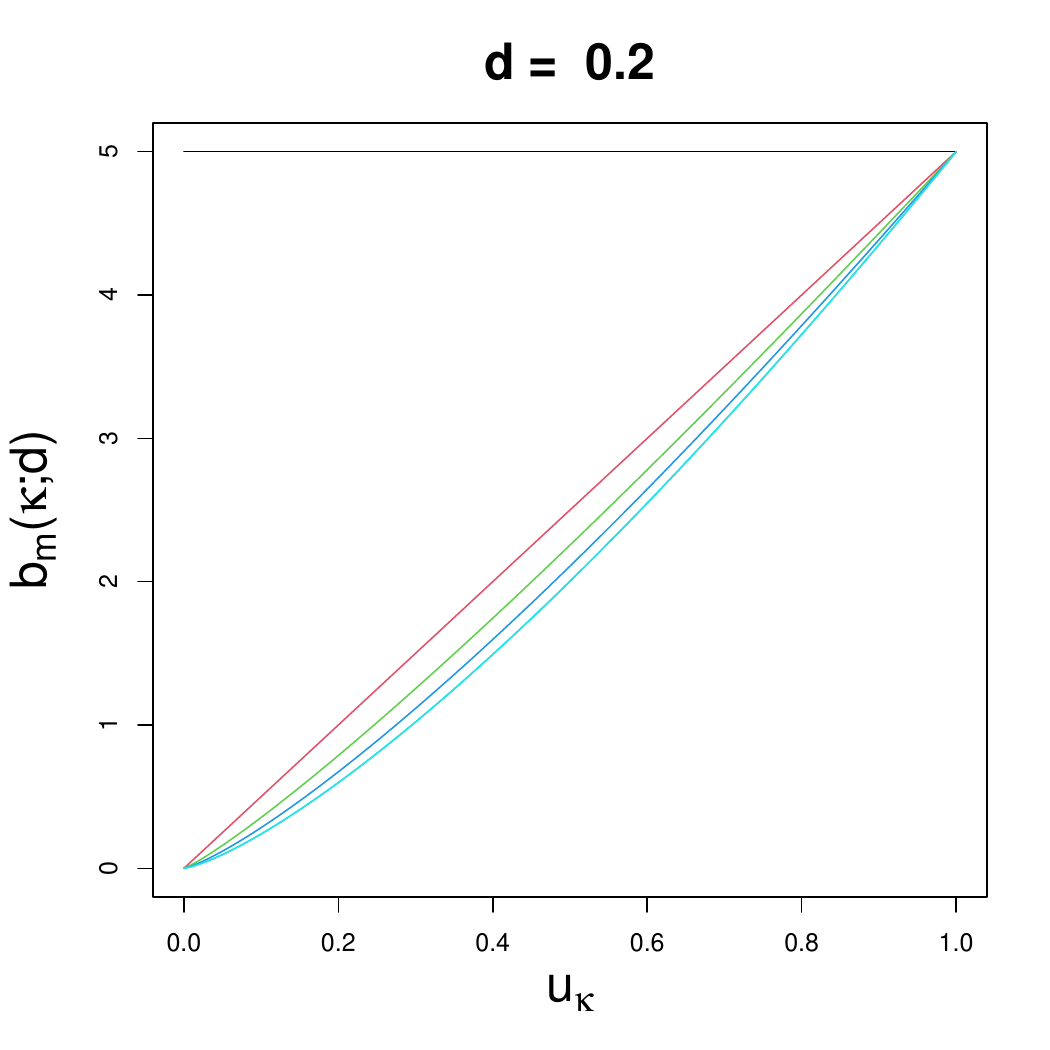}
\includegraphics[width=0.325\textwidth]{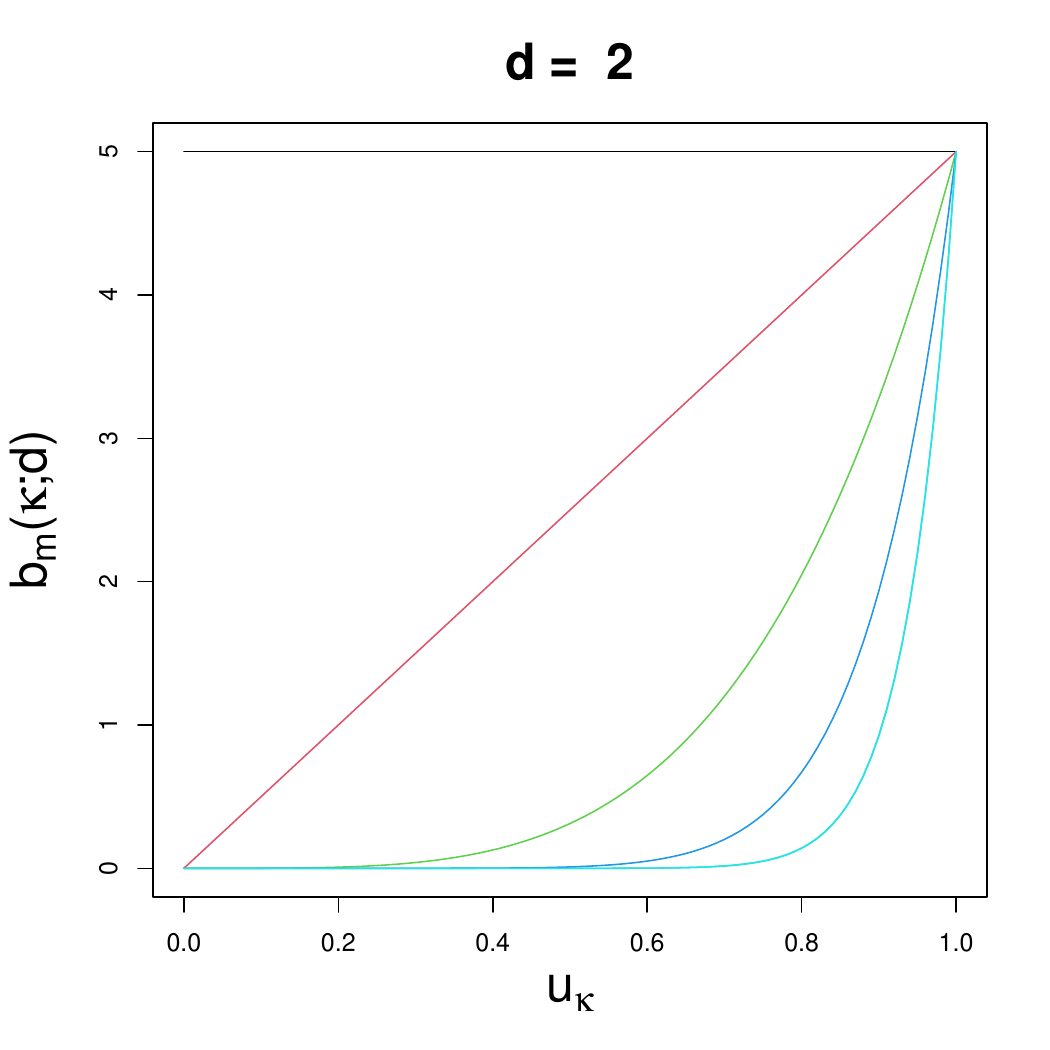}\\
\caption{
{\small Basis functions $b_m(\kappa; d)$, $m=1,\ldots,5$, for $d = 1$ (left), 
$0.2$ (middle), and $2$ (right).}}
    \label{fig:mhp_bmk}
\end{figure}

%
%
The basis representation uses the following component for earthquake magnitude:
\begin{equation*}
b_{m}(\kappa; d) 
\, = \, M \, u_\kappa^{(m-1)^d}; \,\,\,\,\,\,\,\,
u_\kappa \, = \, (\kappa-\kappa_0)/(\kappa_{max}-\kappa_0), 
\,\,\,\,\, \kappa \in (\kappa_0,\kappa_{max}) , 
\end{equation*}
where $d \in \mathbb{R}^+$ is a model parameter. The functions $b_{m}(\kappa; d)$ are
non-decreasing in $\kappa$. Taking $M > 2$, the basis includes a constant and a linear
function as its first two components. For given $d$, the slopes of the remaining 
basis components vary across $m$. For a specific basis component (with $m > 2$), 
parameter $d$ adjusts the slope, where small values ($d$ close to $0$) produce nearly 
linear increase, whereas larger values support a 
lower rate of increase. An illustration is provided in Figure \ref{fig:mhp_bmk}. 
Note that $b_{m}(\kappa; 1) \propto$ $\text{be}(u_\kappa |m,1)$, the beta density with 
shape parameters $m$ and $1$. The inclusion of parameter $d$ allows for a wider range 
of increasing shapes in the basis components for earthquake magnitude.
Viewing $d=1$ as a reference point, we use an exponential prior with mean $1$ 
for $d$. For all the data examples considered in Sections \ref{sec:mhp_simulation_study} 
and \ref{sec:mhp_real_analysis}, there is prior-to-posterior learning for $d$, with the 
posterior density concentrating on values $d < 1$ (see the Supplementary Material).

%
%

Under the gamma process prior for $H$ in (\ref{eq:nonpara_model}), the weights are
(conditionally) independent gamma distributed, with mean $H_0(A_{lm})$ and 
variance $H_0(A_{lm})/c_0$:
\begin{equation}
\nu_{lm} = H(A_{lm}) \mid H_0,c_0 \, \indsim \, 
\text{Ga}(c_{0} H_0(A_{lm}), c_{0}), 
\,\,\,\,\,\,\,  l=1,\ldots,L, \,\, m=1,\ldots,M .
\label{gamma_prior_for_weights}
\end{equation}
This prior structure applies for any $L$ and $M$, i.e., for any partition 
$\{ A_{lm} \}$ of $(0,L\theta) \times (0,1)$.

Let $A_{lm} = A_l\times A_m = [(l-1)\theta,l\theta)\times[(m-1)/M,m/M)$. We define 
the gamma process centering measure through the product of two components for $x$
and $\kappa$, $H_0(A_{lm}) =$ $H_{0x}(A_l) \, H_{0 \kappa}(A_m)$. The former is 
the Weibull cumulative hazard with shape parameter $b_1$ (and scale parameter 1), 
$H_{0x}(A_l) =$ $H_{0x}(l \theta) - H_{0x}( (l-1) \theta) )=$
$(l \theta)^{b_1} - \{ (l-1) \theta \}^{b_1}$, and the latter the exponential 
cumulative hazard with rate parameter $b_2$, $H_{0\kappa}(A_m) =$
$H_{0\kappa}(m/M) - H_{0\kappa}( (m-1)/M ) = b_2/M$.
The gamma process hyperparameters, $c_0$ and $(b_1,b_2)$, are included in 
the model parameters to be estimated given the observed point pattern. Their role 
in controlling model properties is discussed in the next section.

\subsubsection{MHP functionals}
\label{subsubsec:mhp_model_property}

The model formulation for the excitation function in (\ref{eq:nonpara_model}) yields
useful expressions for the key MHP functionals. First, the mark-dependent total 
offspring intensity is given by 
\begin{equation}
\alpha(\kappa) \, = \, \int_0^\infty h(x,\kappa) \, \text{d}x \, = \,
\sum_{m=1}^M V_m \, b_m(\kappa; d), \,\,\,\, \kappa \in \mathcal{K},
\label{eq:nonpara_function_total_offs}
\end{equation}
where $V_m = \sum_{l=1}^L \nu_{lm} \mid H_0, c_0
\indsim \text{Ga}\big(c_0 \sum_{l=1}^L H_0(A_{lm}), c_0\big)$, for $m=1,...,M$.
Hence, provided $M \geq 2$, the model implies the useful property of total 
offspring intensity which is increasing in earthquake magnitude. 
%
%

Under a (rescaled) beta mark density, $f(\kappa) =$ 
$\text{be}(u_\kappa|a_\beta,b_\beta) / (\kappa_{max} - \kappa_0)$, 
$\kappa \in (\kappa_0,\kappa_{max})$, we can evaluate the integral for the expected 
total offspring intensity:
\begin{equation}
\rho \, = \,
\int_{\kappa_0}^{\kappa_{max}} \alpha(\kappa) \, f(\kappa) \, \text{d}\kappa
\, = \, \frac{M}{\text{Beta}(a_\beta,b_\beta)}
\sum_{m=1}^{M} V_m \, \text{Beta}\big(a_\beta+(m-1)^d,b_\beta\big)
\label{eq:nonpara_function_branching_ratio}
\end{equation}
where $\text{Beta}(a,b)$ is the beta function. Although the stability condition 
$\rho < 1$ can not be expressed in terms of specific restrictions for model parameters,
their priors can be specified such that the implied prior distribution for $\rho$ places
(most of) its probability on $(0,1)$.

A key feature of the proposed model is that it yields mark-dependent offspring densities, 
$g_{\kappa}(x) =$ $h(x,\kappa)/\alpha(\kappa)$. Combining (\ref{eq:nonpara_model}) 
and (\ref{eq:nonpara_function_total_offs}), we obtain 
\begin{equation}
g_\kappa(x) \, = \, \sum_{l=1}^L W_l(\kappa) \, \text{ga}(x|l,\theta^{-1}), \,\,\,\,
x \in \mathbb{R}^{+};  \,\,\,\,\,\,\,\,\,\,
W_l(\kappa) = 
\frac{\sum_{m=1}^M \nu_{lm} \, b_m(\kappa; d)}{\sum_{l=1}^{L} \sum_{m=1}^M \nu_{lm} \, b_m(\kappa; d)}
\label{eq:nonpara_function_offs_dnsty}
\end{equation}
Hence, each offspring density can be expressed as a weighted combination 
of Erlang densities with local, mark-dependent weights. For earthquake modeling, 
this structure offers the capacity to estimate earthquake magnitude varying 
aftershock densities.

\begin{figure}[t]
\centering
\includegraphics[width=0.33\textwidth]
{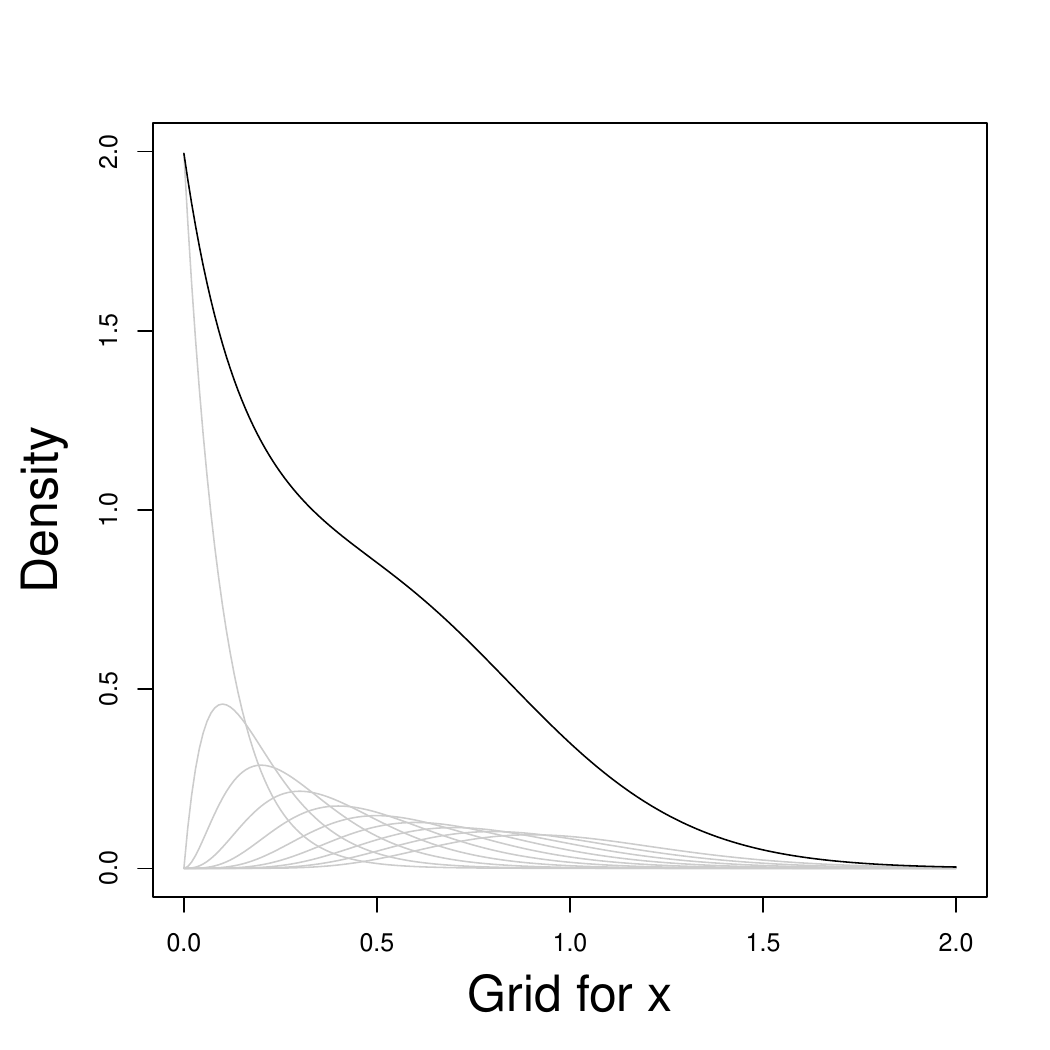}\includegraphics[width=0.33\textwidth]
{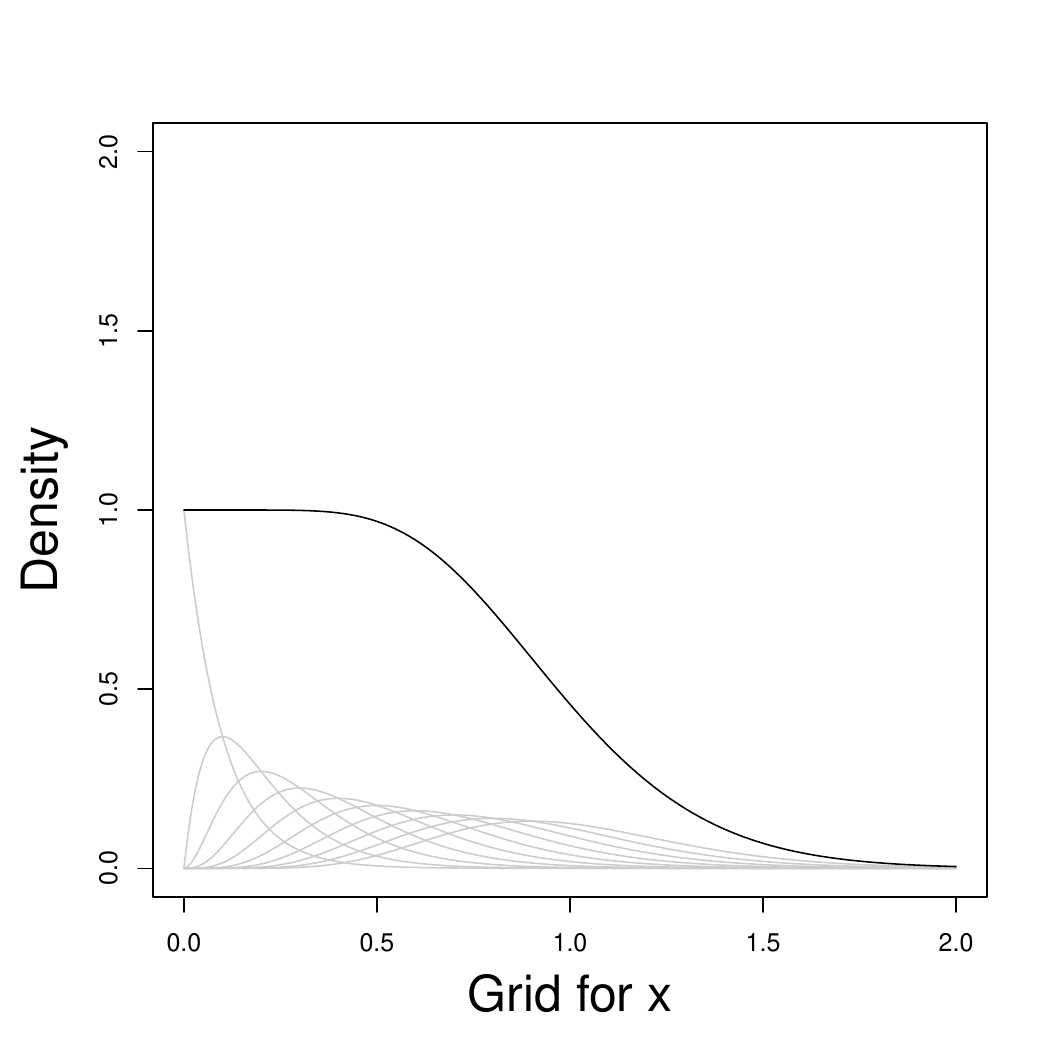}\includegraphics[width=0.33\textwidth]
{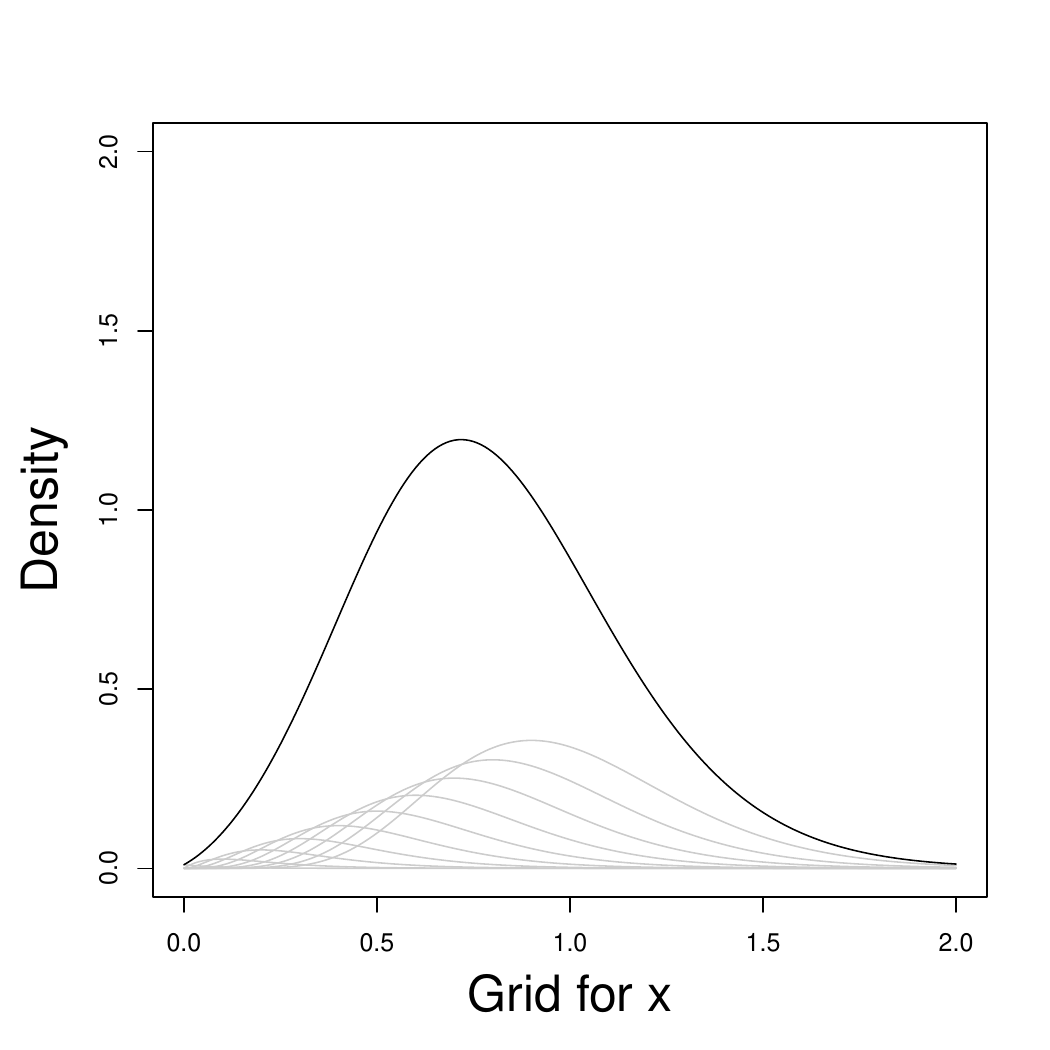}
\caption{
{\small  In each panel, the gray lines depict the weighted components 
$W_l(\kappa) \, \text{ga}(x|l,\theta^{-1})$, for $l = 1,\ldots,L=10$, and the black 
line the density $g_\kappa(x) =$ 
$\sum_{l=1}^L W_l(\kappa) \, \text{ga}(x|l,\theta^{-1})$, with $\kappa = 5.5$, 
where $\nu_{lm} = H_0(A_{lm})$. The left, middle, right panel corresponds to 
$b_1 = 0.7$, $1$, $3$, respectively.  
The other parameters are set to $b_2 = 0.2$, $\theta = 0.1$, $d = 1$, and $M = 5$.}}
\label{fig:mhp_offs_dnsty_b1}
\end{figure}

\begin{figure}[t!]
\centering
\includegraphics[width=0.244\textwidth]{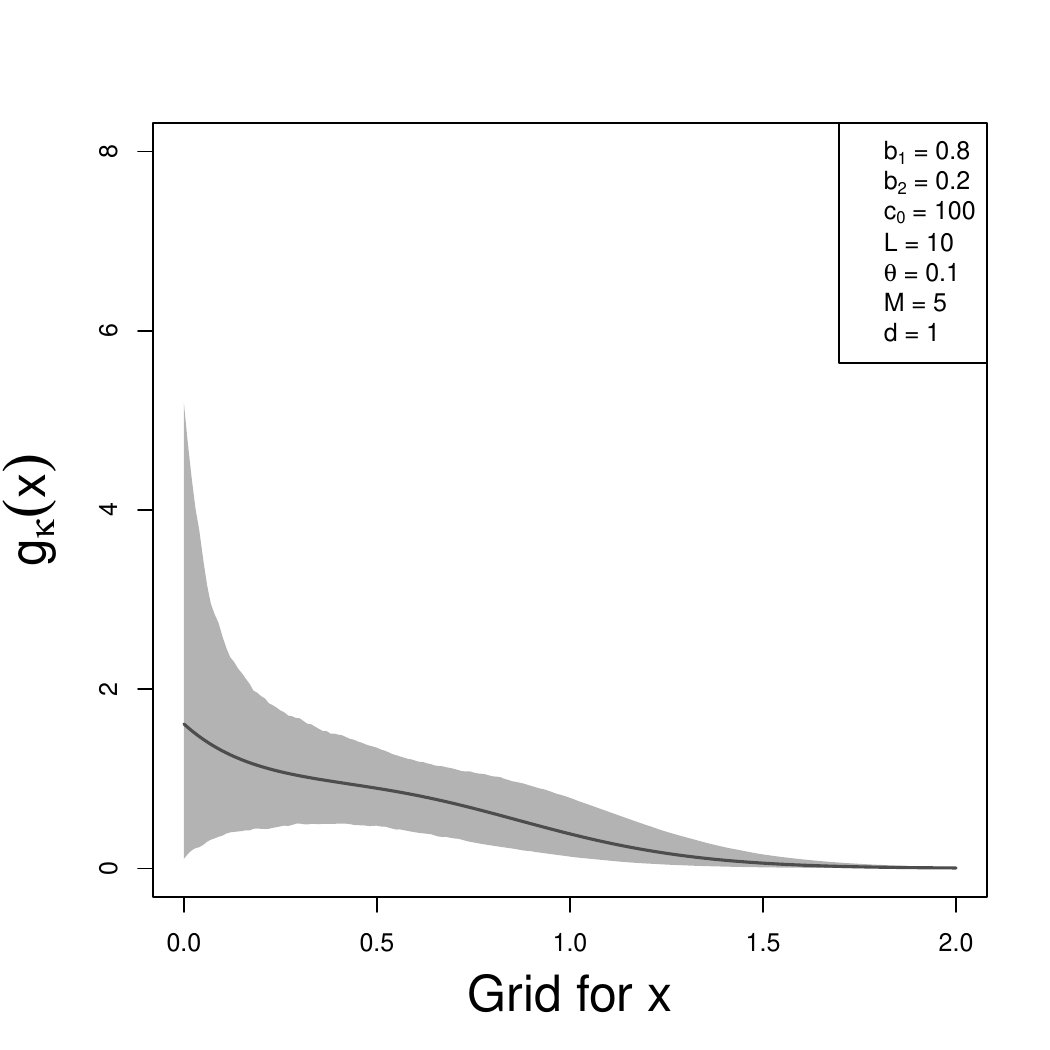}
\includegraphics[width=0.244\textwidth]{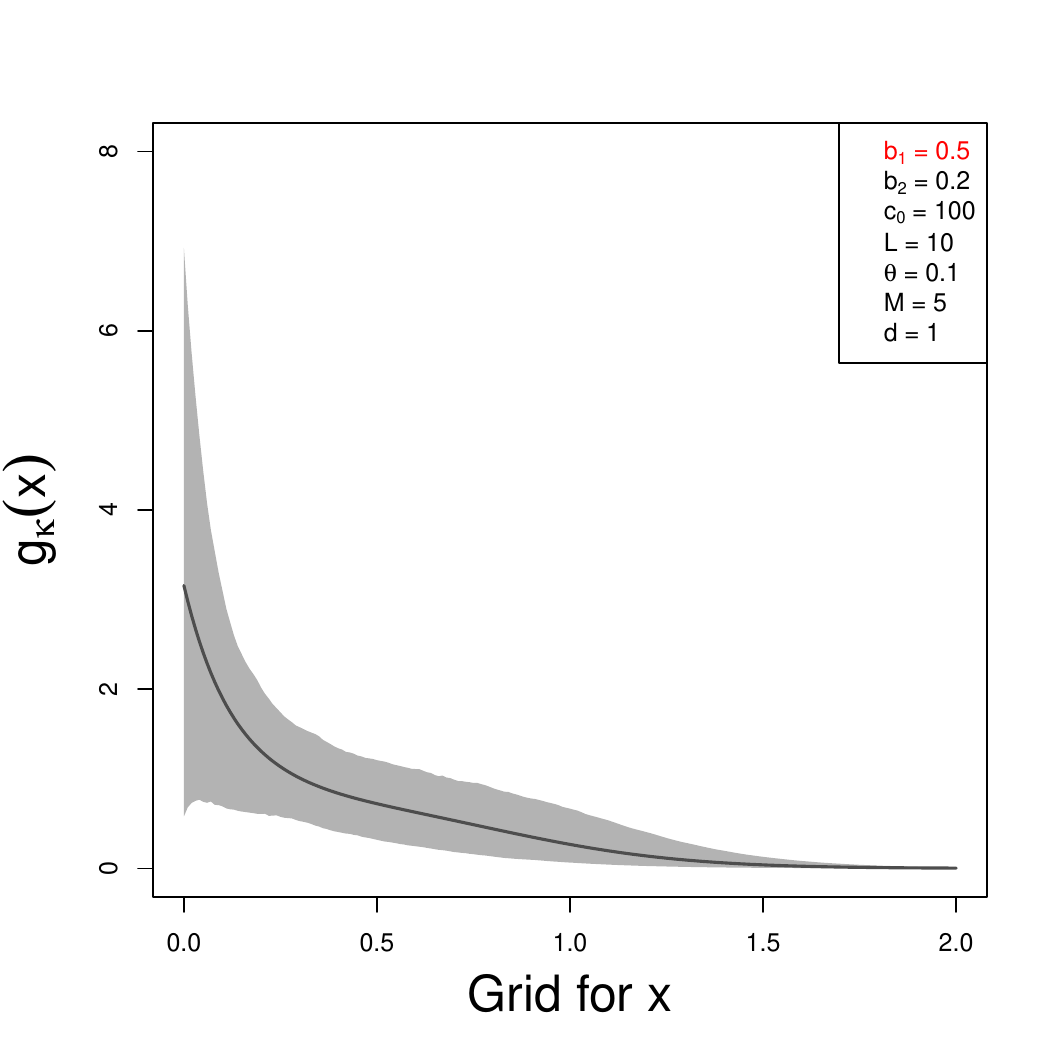}
\includegraphics[width=0.244\textwidth]{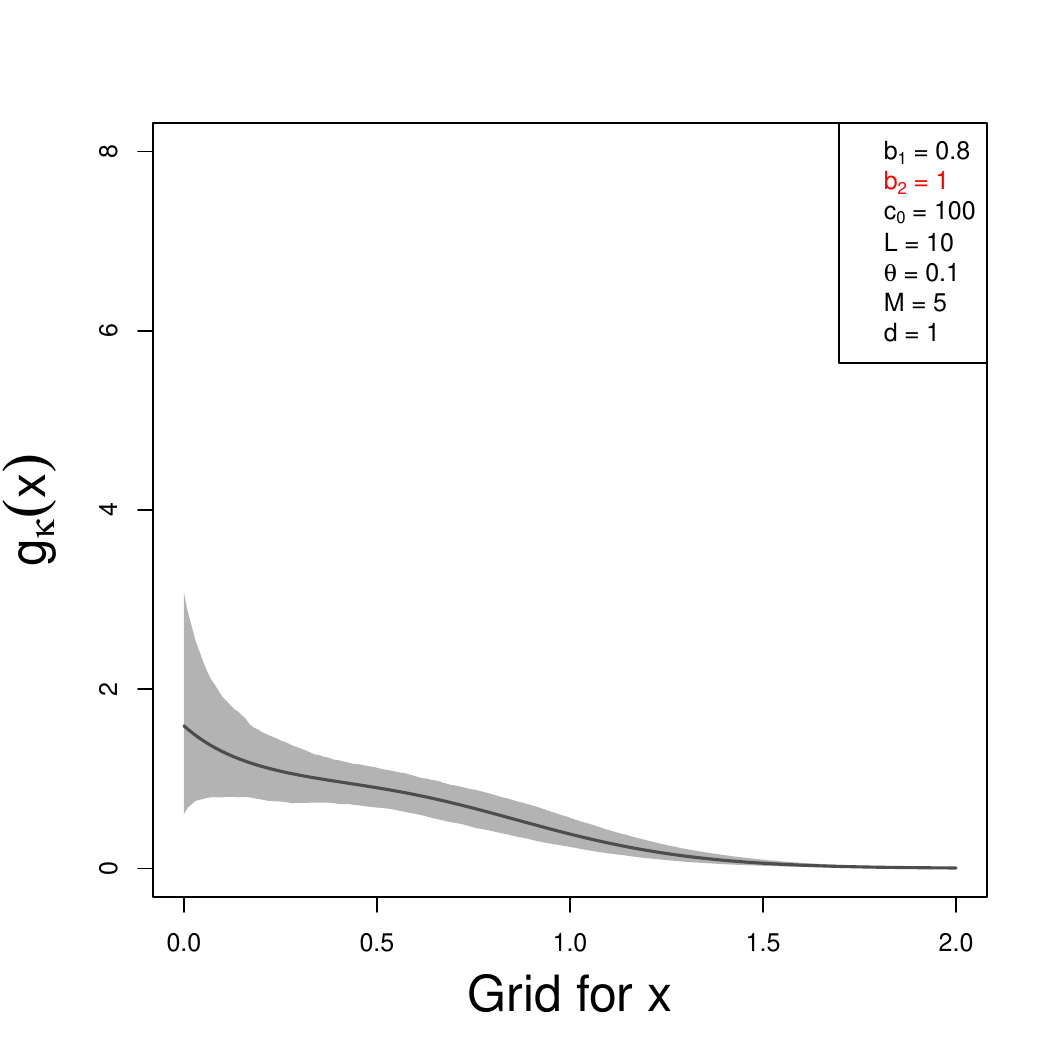}
\includegraphics[width=0.244\textwidth]{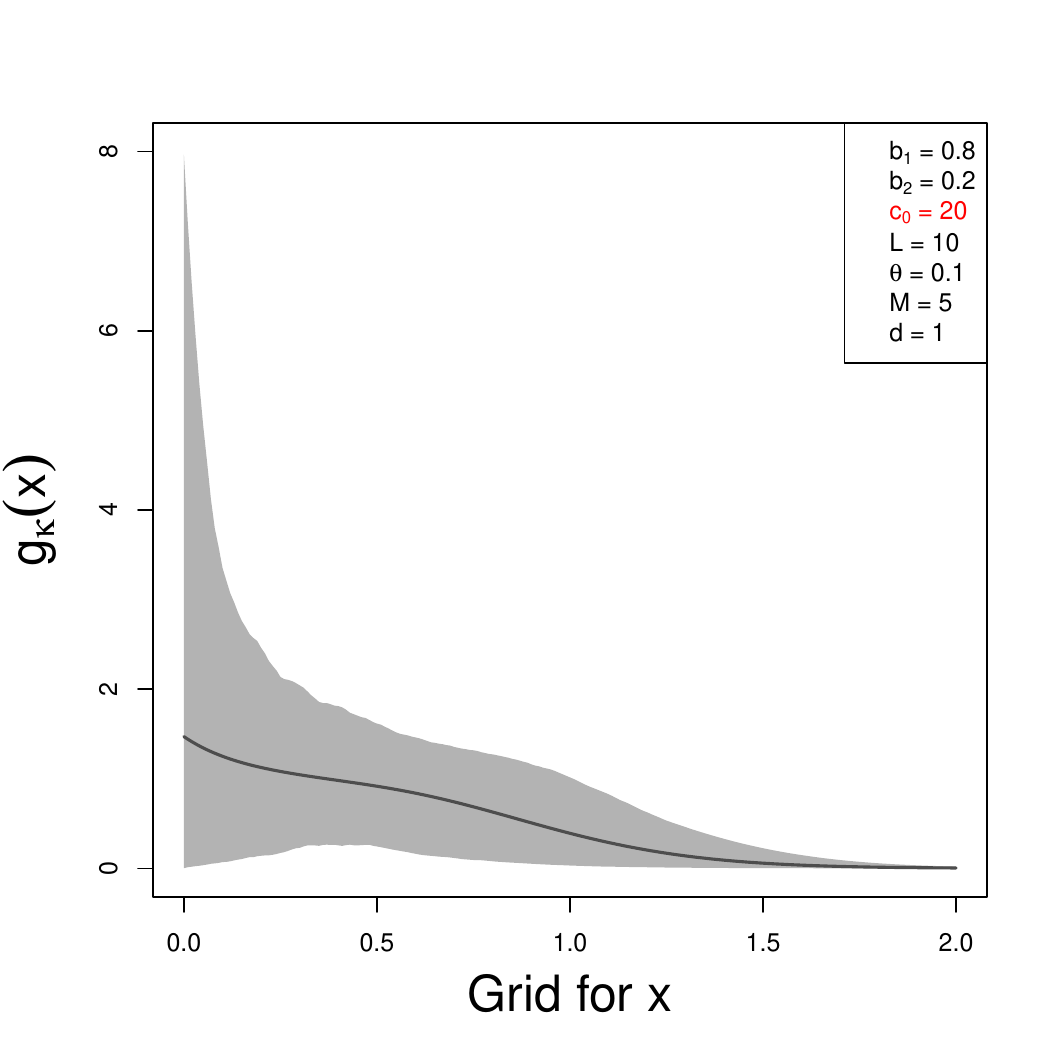}\\
\includegraphics[width=0.244\textwidth]{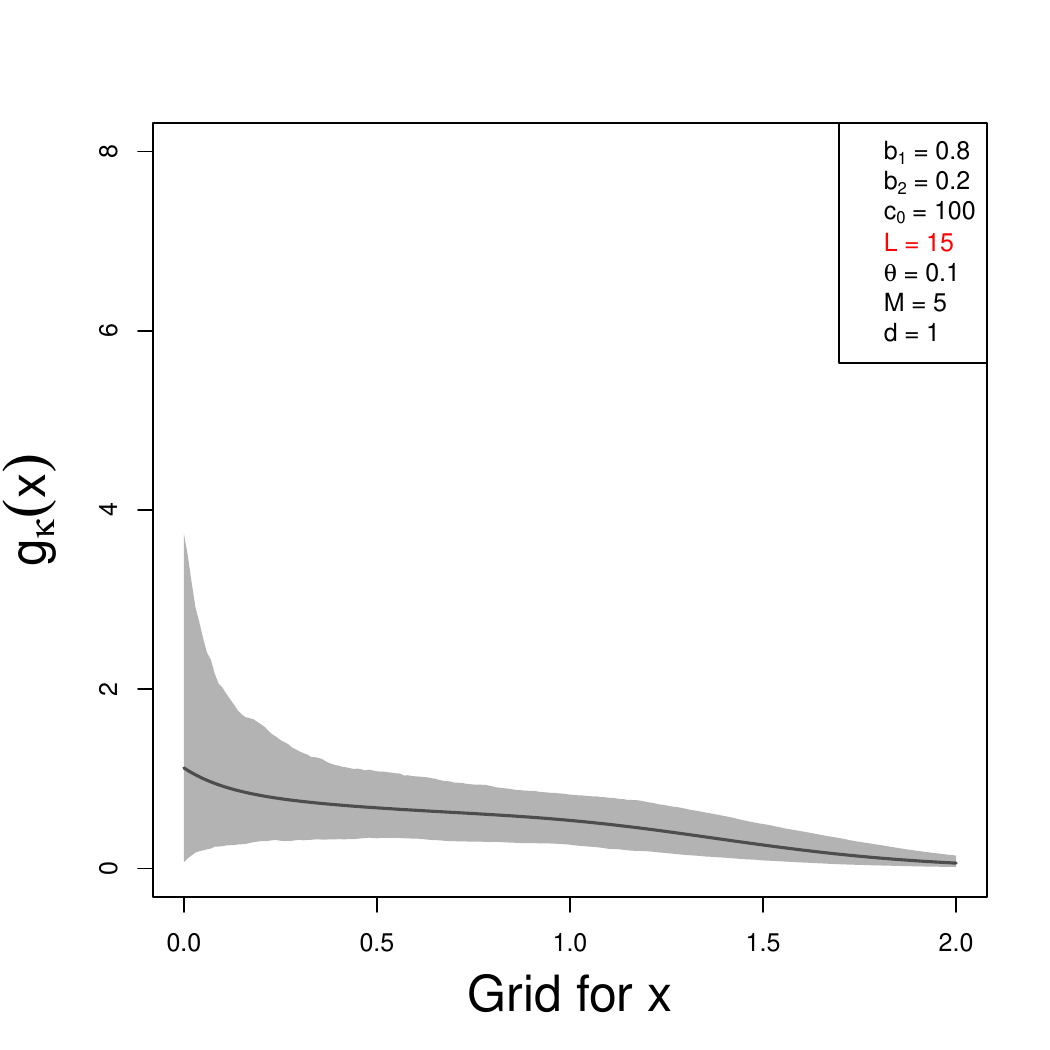}
\includegraphics[width=0.244\textwidth]{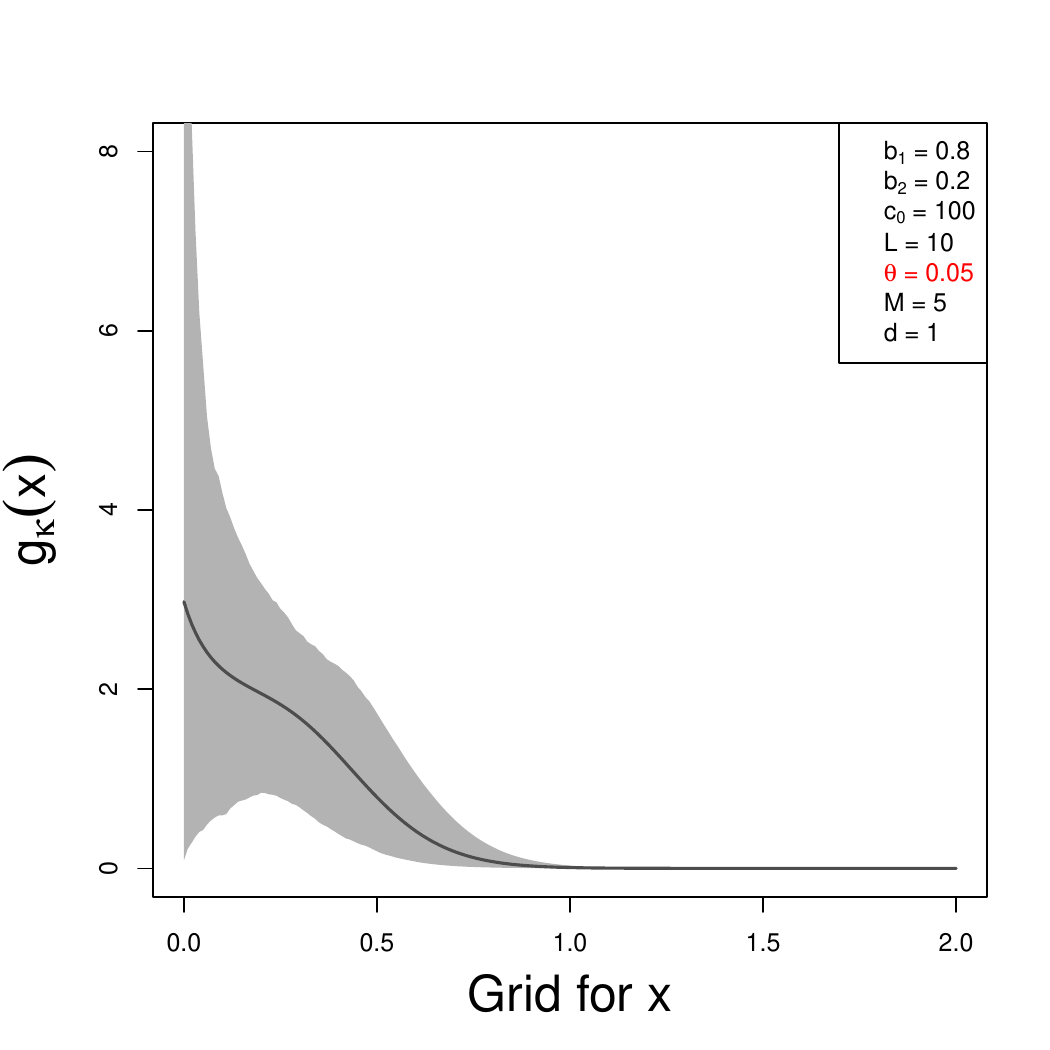}
\includegraphics[width=0.244\textwidth]{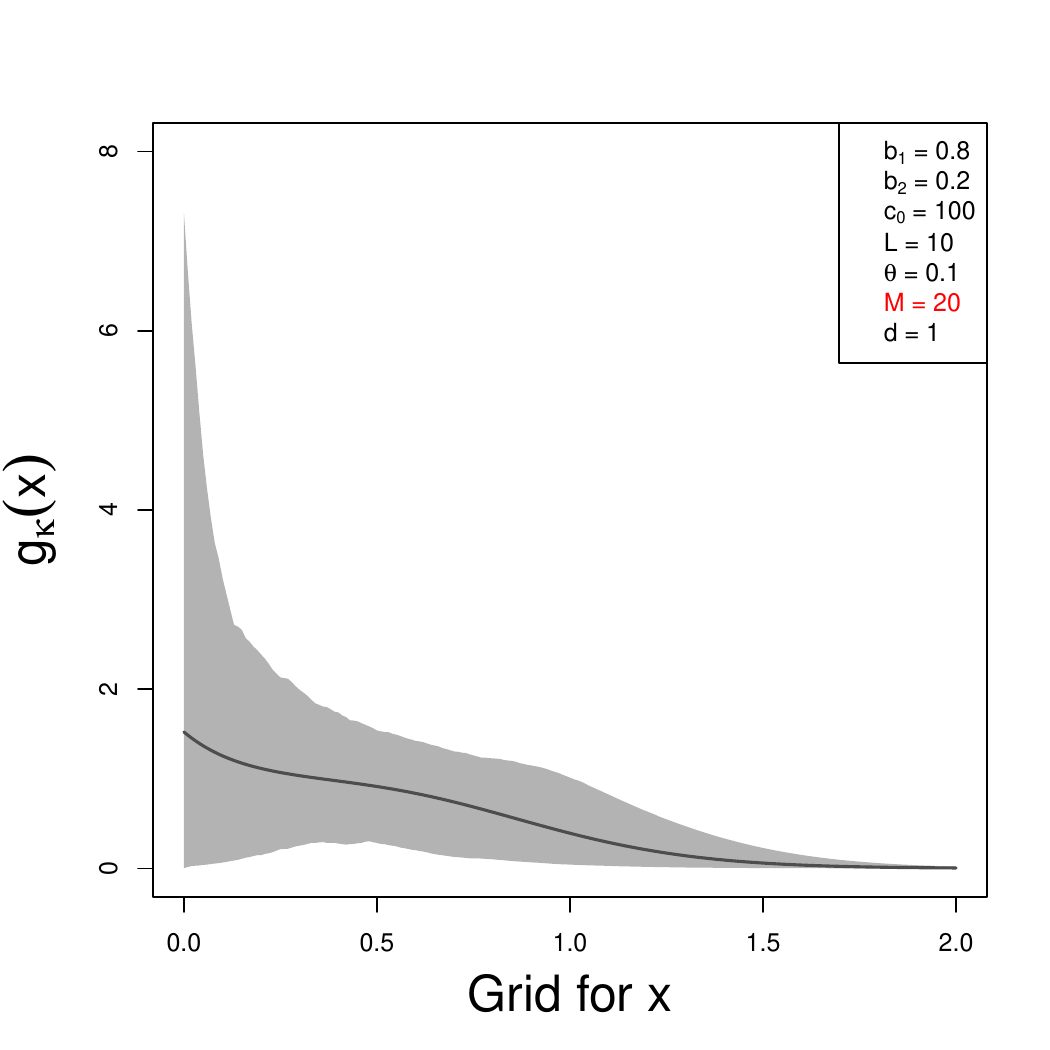}
\includegraphics[width=0.244\textwidth]{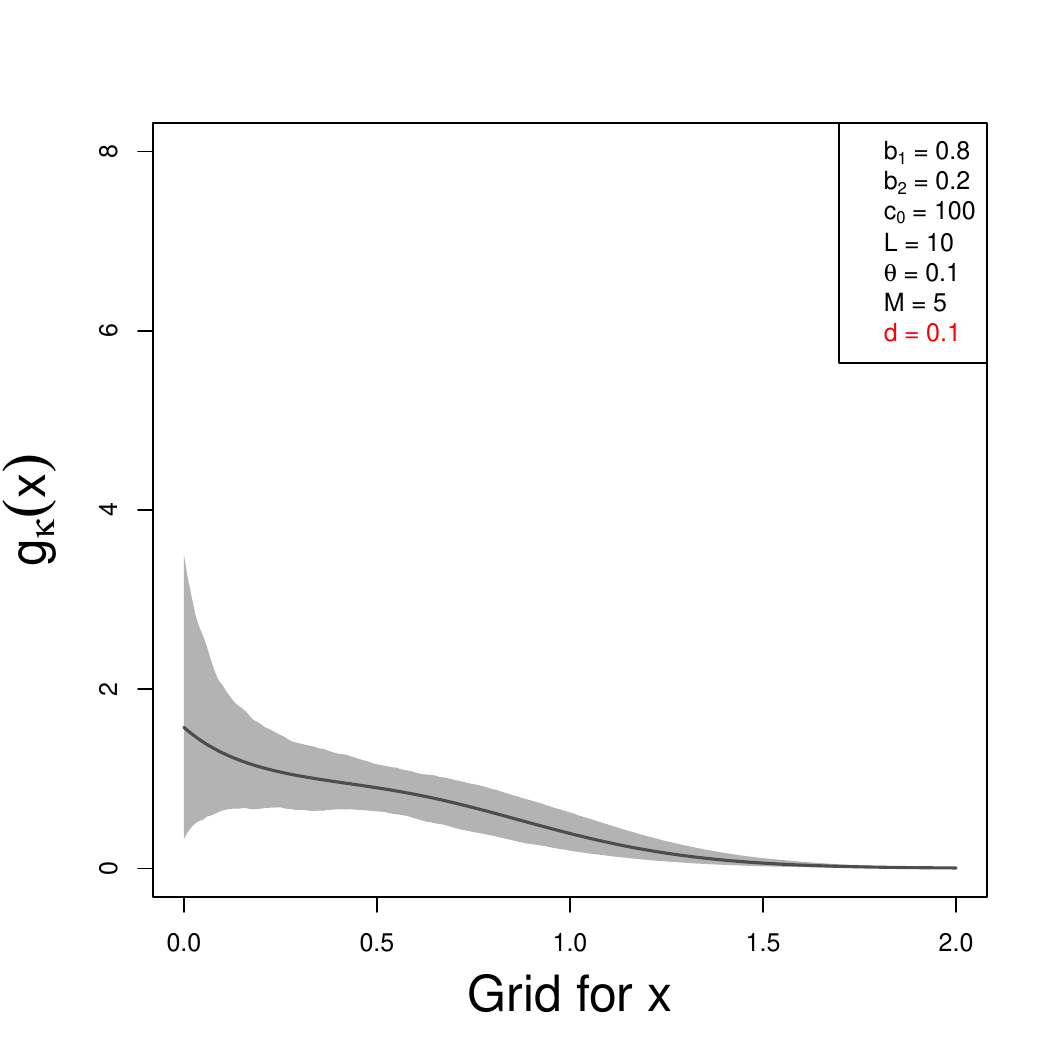}\\
\caption{
{\small  Prior mean (solid line) and 95\% interval estimates (gray bands) for 
$g_\kappa(x)$ (with $\kappa=5.5$) under different values for the model parameters. 
The prior point and interval estimates are based on 1,000 prior samples for the 
weights $\nu_{lm}$.}}
\label{fig:mhp_offs_dnsty}
\end{figure}

Existing models for earthquake occurrences do not allow for aftershock dynamics that change 
with earthquake magnitude.  
Parametric MHP models (including the ETAS model) are based on a decreasing offspring 
density, the decreasing shape suggested by seismology theory \cite[e.g.,][]{O1988}. 
The earthquake magnitude-dependent densities implied by our model are not necessarily 
decreasing. However, the decreasing shape in prior realizations for the $g_\kappa(x)$ 
densities can be favored through the gamma process hyperparameter $b_1$, in particular, 
for values $b_1 < 1$. This can be seen informally by examining the weights, $W_l(\kappa)$, 
in (\ref{eq:nonpara_function_offs_dnsty}). The weights are proportional to 
$\sum_{m=1}^M \nu_{lm} \, b_m(\kappa; d)$, the expectation of which over the prior for 
the $\nu_{lm}$ is $H_{0x}(A_l) \sum_{m=1}^{M} H_{0\kappa}(A_m) \, b_m(\kappa;d) =$
$C \, H_{0x}(A_l)$, where $C > 0$ does not depend on the index $l$ for the weights. 
Recall from Section \ref{subsec:mhp_model_excitation} that $H_{0x}(A_l) =$
$(l \theta )^{b_1} - \{ (l-1) \theta \}^{b_1}$, which decreases (increases) with $l$
when $b_1 < 1$ ($b_1 > 1$), thus favoring lower-index (higher-index) Erlang density 
components. A graphical demonstration is provided in Figure \ref{fig:mhp_offs_dnsty_b1}, 
where, for the purposes of this illustration, the weights $\nu_{lm}$ are fixed at 
their prior mean $H_0(A_{lm})$.
The model's capacity to favor decreasing offspring densities is also suggested by 
Figure \ref{fig:mhp_offs_dnsty}, where prior point and interval estimates for $g_{\kappa}(x)$
are obtained by sampling the weights $\nu_{lm}$ from their prior in (\ref{gamma_prior_for_weights}); 
note that $b_1 < 1$ in all Figure \ref{fig:mhp_offs_dnsty} panels.

In general, the role of the gamma process prior hyperparameters ($b_1$, $b_2$ and $c_0$), 
the basis component parameters ($\theta$ and $d$), as well as the (fixed) number of 
basis components ($L$ and $M$) can be studied through analytical results for prior 
expectations or by prior simulation. 
Figure \ref{fig:mhp_offs_dnsty} is an illustration of the latter. As an example 
of the former, taking expectations in (\ref{eq:nonpara_function_total_offs})
with respect to the distribution of the $V_m$, we obtain the (conditional) prior 
mean and variance of the total offspring intensity:
\begin{equation}
\begin{split}
\text{E}\{ \alpha(\kappa) | H_0, c_0, \theta, d \} & = 
\sum_{m=1}^M \text{E}(V_m | H_0, c_0) \, M u_\kappa^{(m-1)^d} = 
b_2 \, (L\theta)^{b_1} \sum_{m=1}^M u_\kappa^{(m-1)^d}
\\
\text{Var}\{ \alpha(\kappa) | H_0, c_0, \theta, d \} & = 
\sum_{m=1}^M \text{Var}(V_m | H_0, c_0) \, \Big\{M u_\kappa^{(m-1)^d}\Big\}^2 
= \frac{b_2 \,  (L\theta)^{b_1} \, M}{c_0} \sum_{m=1}^M u_\kappa^{2(m-1)^d}
\end{split}
\label{eq:analytical_alpha}
\end{equation}
since $\sum_{l=1}^L H_0(A_{lm}) =$ $b_2 \, (L\theta)^{b_1} / M$ under the choice 
for $H_0$ given in Section \ref{subsec:mhp_model_excitation}.
Note, for instance, that the precision parameter $c_0$ of the gamma process prior 
is inversely associated with the prior variability for the total offspring intensity.
%

Turning to Figure \ref{fig:mhp_offs_dnsty}, each panel shows the prior mean and 
95\% interval estimates for the offspring density $g_{\kappa}(x)$ (with $\kappa=5.5$), 
based on 1,000 prior samples for the weights $\nu_{lm}$. The top left panel involves 
a particular specification for the model parameters, which is then modified in the 
other panels by changing the value of one parameter at a time. 
Even though some of the  $g_{\kappa}(x)$ prior realizations used to obtain the point 
and interval estimates are not decreasing, a decreasing offspring 
density shape is favored in all cases. This is compatible with the earlier discussion 
regarding hyperparameter $b_1$, and the fact that $b_1 < 1$ in all panels. Note that 
reducing the value of $b_1$ results in higher concentration of probability mass at smaller 
values of $x$ (which is also consistent with the remarks following 
Eq. (\ref{eq:nonpara_function_offs_dnsty})). 
Since $g_{\kappa}(x)$ is built from Erlang basis densities with means $l \theta$, we can 
use $(L \theta) + 2 \theta \sqrt{L}$ (two standard deviations above the mean of the last
Erlang basis density) as a proxy for the upper bound of the effective support for
the offspring densities. The combined effect of $L$ and $\theta$ can be observed in 
the first two bottom left panels. Regarding the other model parameters, note that prior 
uncertainty decreases with increasing values of $b_2$ and $c_0$, and it increases 
with $M$ and $d$.

\subsubsection{Nonparametric prior for the immigrant intensity}
\label{subsec:immi_insty_model}

Here, we develop the general version of the nonparametric prior model for the 
MHP ground process intensity, by relaxing the assumption of constant immigrant intensity. 
As can be seen from the augmented MHP likelihood in (\ref{eq:mhp_likelihood}), the 
immigrant intensity can be viewed as the intensity of a NHPP. We thus adopt the 
nonparametric prior for NHPP intensities from \cite{KK2022}, which builds from a basis 
representation, in the same spirit with the model for the excitation function.

More specifically, the immigrant intensity function is modeled as a weighted 
combination of Erlang basis densities:
\begin{equation}
\begin{split}
\mu(t) &= \sum_{j=1}^{J} \omega_{j} \, \text{ga}(t \mid j,\phi^{-1}), 
\quad t \in \mathbb{R}^{+} \\
\omega_j &= G\big(j\phi\big) - G\big((j-1)\phi\big), \enspace j=1,\ldots,J, 
\quad G \mid G_{0},e_{0} \sim \mathcal{G}(G_0,e_0).
\end{split}
\label{eq:erlmix_immi}
\end{equation}
Here, the weights are increments of a (random) cumulative intensity 
$G$ on $\mathbb{R}^{+}$. Under the gamma process prior for $G$, we have 
$\text{E}(G(t) \mid G_0)=$ $G_0(t)$ and $\text{Var}(G(t) \mid G_0, e_0)=$ $G_0(t)/e_0$.
We take $G_0(t)=$ $t/b_{G_0}$, i.e., the cumulative hazard of the exponential 
distribution with scale parameter $b_{G_0}$.
This model extends Bayesian nonparametric Erlang mixtures for density estimation 
\citep{YLJLAK2024} to the problem of intensity estimation. 
Theoretical support for the model structure in (\ref{eq:erlmix_immi}) is provided by 
a convergence result \citep{KK2022}, which is analogous to the pointwise 
convergence result for the Erlang mixture density estimation model \cite[e.g.,][]{LinLee2010}. 
In particular, as $\phi \rightarrow 0$ and $J \rightarrow \infty$, 
$\sum_{j=1}^{J} \{ G(j\phi) - G((j-1)\phi) \} \, \text{ga}(t \mid j,\phi^{-1})$
converges pointwise to the intensity function associated with cumulative intensity $G$
(provided $G(t)=$ $O(t^{q})$, as $t \rightarrow \infty$, for some $q > 0$).

Similar to the remark regarding the role of $L$ and $\theta$ in (\ref{eq:nonpara_function_offs_dnsty}), 
the effective support of $\mu(t)$ in (\ref{eq:erlmix_immi}) is controlled by $J$ and $\phi$. 
Also useful for prior specification 
is the fact that 
the prior mean for the immigrant intensity is essentially constant at $1/b_{G_0}$ over the 
effective support of $\mu(t)$. As with the model for the excitation function, we work with a 
fixed number of basis components, $J$, and assign priors to $\phi$, and to the gamma 
process hyperparameters $b_{G_0}$ and $e_0$.

%
%

\subsection{Model implementation}
\label{model_implementation}

Here, we discuss posterior inference and prediction, with the technical details provided 
in the Supplementary Material.

\subsubsection{Posterior simulation}
\label{MCMC_simulation}

Of primary importance in model estimation are the weights, $\nu_{lm}$, which drive the basis
representation for the excitation function in (\ref{eq:nonpara_model}), as well as the 
weights, $\omega_{j}$, for the immigrant intensity prior in (\ref{eq:erlmix_immi}).
A key feature of the modeling approach is that we can achieve ready MCMC updates for 
these parameters, without resorting to any type of approximation to the MHP likelihood 
(or the posterior simulation method).

Denote by $\{0 < t_1 < \ldots < t_n < T\}$ the point pattern recorded in observation 
window $(0,T)$, and by $\kappa_i$ the observed mark associated with the $i$-th time event, 
for $i=1,...,n$. Recall from Section \ref{sec:mhp_background} that the introduction of 
the branching variables $\bm{y}=$ $(y_1,\ldots,y_n)$ facilitates interpretation of 
the MHP functionals. The branching structure is also key for likelibood-based estimation
procedures. In the context of MCMC posterior simulation, the augmented
MHP likelihood in (\ref{eq:mhp_likelihood}) highlights that the model parameters 
for the immigrant intensity and for the excitation function can be updated by focusing 
on distinct likelihood components that pertain to NHPPs. In both cases, the normalizing 
terms, which involve integrals of the immigrant intensity or the excitation function,
present the main challenge in working with flexible models for these MHP functionals.
Bayesian inference for the parametric ETAS model is discussed in, e.g., 
\cite{R2013} and \cite{R2021}.

The normalizing term associated with the set of all oﬀspring points is, in general, 
particularly challenging. However, under the model in (\ref{eq:nonpara_model}), we obtain 
$$
\exp \Big\{ -\sum_{j=1}^{n} \int_0^{T- t_{j}} h(u,\kappa_{j}) \, \text{d}u \Big\} \, = \,
\prod_{l=1}^L \prod_{m=1}^M \exp\{ - \nu_{lm} \, K_{lm}(\theta,d) \} ,
$$
where $K_{lm}(\theta,d) =$
$\sum_{j=1}^n b_m(\kappa_j; d) \{ \int_0^{T-t_j} \text{ga}(s|l,\theta^{-1}) \, \text{d}s \}$
depends only on parameters $\theta$ and $d$. 
This result, along with configuration variables that identify the excitation 
function basis component to which each offspring point is assigned, yields a gamma 
posterior full conditional distribution for each $\nu_{lm}$. And, importantly, given 
the other parameters (and the branching and basis configuration latent variables), 
the weights $\nu_{lm}$ are independent.

Similarly, based on the model in (\ref{eq:erlmix_immi}), 
$\exp \{ -\int_0^T \mu(u) \, \text{d}u \} =$ 
$\prod_{j=1}^{J} \exp\{ - \omega_{j} \, S_{j}(\phi) \}$,
where $S_{j}(\phi) =$ $\int_{0}^{T} \text{ga}(u \mid j,\phi^{-1}) \, \text{d}u$
depends only on parameter $\phi$. To update the immigrant intensity model parameters, 
we introduce configuration variables to identify the Erlang basis density for each 
immigrant time point. Again, the weights $\omega_j$ are conditionally independent, 
and each $\omega_j$ follows a gamma posterior full conditional distribution.
The Supplementary Material includes the full details for sampling the weight 
parameters, $\{ \nu_{lm} \}$ and $\{ \omega_j \}$, as well as for updating 
the other model parameters and the latent variables.

\subsubsection{Predictive inference}
\label{subsec:prediction}

For point process stochastic models implemented under the Bayesian framework, it is 
generally straightforward to obtain predictive inference for point process functionals. 
This can be done by computing the functional based on point patterns simulated from 
the stochastic model, with a different point pattern generated for each posterior sample 
for the model parameters \cite[e.g.,][]{LG_BA2017}.

For instance, we may be interested in the posterior predictive distribution for the 
event count in a specific time interval $B$, which may be within or outside the observation 
window for predictive model assessment or forecasting, respectively. For every MCMC 
posterior sample, we simulate a MHP realization, using the MHP functions defined by
the model parameters in the posterior sample. To simulate the point pattern, we
can use the MHP definition with the branching structure, which amounts to simulation
from independent marked NHPPs \cite[e.g.,][]{MR2006}. Then, recording from each simulated 
point pattern the count within interval $B$, $N_{pred}(B)$, yields a sample from the 
posterior predictive distribution for the number of events in interval $B$.

As a specific illustration in the context of the earthquake data analysis 
(Section \ref{sec:mhp_real_analysis}), we compare the performance of different models 
as it pertains to predicting the number of earthquakes within a {\it future} time window. 
The particular catalog of earthquakes spans years 1885 to 1980. We fit the models to the 
subset of the point pattern recorded up to the end of year 1949, and then forecast 
the earthquake count from 1950 onward. Hence, the posterior predictive distribution from each 
model can be contrasted with the actual earthquake count from 1950 to 1980, thus providing 
a means for comparison across models.

\section{Simulation study}
\label{sec:mhp_simulation_study}

We illustrate our modeling approach with three synthetic data examples, 
including comparison with the ETAS model and a semiparametric extension, presented
in Section \ref{models_for_comparison}.

For the simulation truth, we work with a MHP with ground process intensity:
\begin{equation}
\label{ground_intensity_simulation}
\lambda^\ast_g(t) \, = \, 
\mu \, + \, \sum_{t_i<t} \alpha(\kappa_i) \, g_{\kappa_i}(t-t_i),
\end{equation}
where $\alpha(\kappa)=$ $a \exp\{ b \, (\kappa - \kappa_0) \}$, with 
mark space, $\mathcal{K} =$ $(\kappa_0,\kappa_{max}) = (4, 10)$.
All three point patterns are generated from a MHP with immigrant intensity 
constant in time and an exponential function for the total oﬀspring 
intensity, that is, the ETAS model specification. Hence, the focus is on inference 
for the offspring density under different scenarios for its structure. 
We first test model performance when the point pattern is generated under the 
ETAS model Lomax offspring density, which does not change with earthquake magnitude 
(see the Supplementary Material). We then demonstrate the proposed
model's capacity to uncover mark-dependent offspring densities, under two 
parametric scenarios, a Lomax offspring density with mark-dependent shape parameter 
(Section \ref{subsec:mhp_simulation_dpowlaw}), and a two-component mixture of 
mark-dependent Lomax densities (Section \ref{subsec:mhp_simulation_dpowlawmix}).

For the simulation truth mark density, we take 
$$
f(\kappa) = 
\psi \exp(-\psi \, \kappa)/ \{ \exp(-\psi \, \kappa_0) - \exp(-\psi \, \kappa_{max}) \}, 
\,\,\,\, \kappa \in (\kappa_0,\kappa_{max})
$$
that is, an exponential mark density with rate parameter $\psi$, truncated over 
$(\kappa_0,\kappa_{max})$. Then, 
\[
\rho \, = \, \int_{\kappa_0}^{\kappa_{max}} \alpha(\kappa) f(\kappa) \, \text{d}\kappa 
\, = \, \frac{ a \, \psi}{\psi - b} \, 
\frac{ \exp(-b \kappa_0) \,
\{ \exp(-(\psi-b) \kappa_0) - \exp(-(\psi-b) \kappa_{max}) \} }
{ \{ \exp(-\psi \, \kappa_0) - \exp(-\psi \, \kappa_{max}) \} }
\]
where the second term arises due to the truncation applied to the exponential 
mark density. Therefore, we must have $\psi > b$ (to ensure $\rho > 0$), and 
$\rho < 1$. The latter constraint can not be easily expressed through a restriction 
on the model parameters, but it can be verified numerically for each simulation example.
In particular, the expected total offspring intensity under the data 
generating mechanism is $\rho =$ $0.8954$ for the example presented in the 
Supplementary Material, and $\rho = 0.8906$ for the examples in 
Sections \ref{subsec:mhp_simulation_dpowlaw} and \ref{subsec:mhp_simulation_dpowlawmix}.

\subsection{Parametric and semiparametric comparison models}
\label{models_for_comparison}

We apply the model for the excitation function (Section \ref{subsec:mhp_model_excitation}), 
with immigrant intensity $\mu$, which is constant in time. 
As previously discussed, existing modeling approaches for temporal MHPs do
not accommodate mark-dependent offspring densities. Hence, for comparison, we consider 
the parametric ETAS model in (\ref{eq:etas_model}). We also introduce here a semiparametric 
model which replaces the ETAS model Lomax offspring density with the nonparametric 
scale uniform mixture:
%
%
\begin{equation}
g(x) \, = \, \int_{\mathbb{R}^+} \theta^{-1}\bm{1}_{[0,\theta)}(x) \, \text{d}G(\theta), 
\,\,\,\, x \in \mathbb{R}^{+};  \,\,\,\,\,\,\,\,\,\,
G \sim \text{DP}(\alpha_0, F_0)        
\label{eq:semipara_model}
\end{equation}
where $\text{DP}(\alpha_0, F_0)$ denotes the Dirichlet process (DP) with precision 
parameter $\alpha_0$ and centering distribution $F_0$. We take $F_0 = \text{IG}(a_0,b_0)$, 
i.e., the inverse gamma distribution with mean $b_0/(a_0 - 1)$ (provided $a_0 > 1$).

This model builds from a probabilistic representation for non-increasing densities on 
$\mathbb{R}^+$; specifically, a density $g$ on $\mathbb{R}^{+}$ is non-increasing if and only 
if there exists a distribution $G$ on $\mathbb{R}^{+}$ such that Eq. (\ref{eq:semipara_model}) 
holds true. This representation has been used in the Bayesian nonparametrics literature 
for modeling of unimodal densities \cite[e.g.,][]{brunner1989,LM1995,KK2009}. 
As discussed earlier, the seismology literature supports a decreasing shape for the 
offspring density. Hence, the DP mixture model in (\ref{eq:semipara_model}) offers a 
practically relevant extension of parametric models for earthquake occurrences, such as 
the ETAS model. In particular, it supports both exponential and polynomial tails for the 
offspring density. The practical utility of the uniform mixture model for the offspring 
density was studied in \cite{K2021} for Hawkes processes without marks. 
However, although it extends the ETAS model, the semiparametric model does not overcome 
the main limitation of existing methods for MHPs, that is, it also does not allow for 
aftershock dynamics that change with earthquake magnitude.

For both the ETAS model and its semiparametric extension, the mark density is 
$f(\kappa) =$ $\psi \exp \{ -\psi \, (\kappa-\kappa_0) \}$, for 
$\kappa \in (\kappa_0, \infty)$, with $\kappa_0 = 4$, the value used for the data generation.
Hence, for both models, $\rho = a \psi/(\psi-b)$. We work with exponential priors for 
$a$, $b$ and $\psi$ (with rate parameters $\lambda_a = 2$, $\lambda_b = 0.2$, 
and $\lambda_\psi = 0.1$, respectively), with truncation applied to satisfy the stability 
condition $\rho \in (0,1)$ (with probability $1$ in the prior). 
%
%
%

Note that for the ETAS model (and, by extension, also for the semiparametric model), we 
use the standard version of the mark space that does not include an upper bound. 
However, there is no practical consequence, since the exponential mark densities used 
to generate the point patterns would place essentially all their probability mass in 
the interval $(4, 10)$, had they not been truncated above at $\kappa_{max} = 10$.
%
%

Regarding the priors for the offspring density, we assign $\text{Exp}(0.005)$ and 
$\text{Exp}(0.1)$ priors to the parameters $p$ and $c$, respectively, of the Lomax 
density for the ETAS model in (\ref{eq:etas_model}), where $\text{Exp}(\eta)$
denotes the exponential distribution with rate parameter $\eta$.
For the DP uniform mixture in (\ref{eq:semipara_model}) of the semiparametric model, 
we work with a $\text{Ga}(5,0.25)$ prior for $\alpha_0$, an 
$\text{Exp}(1)$ prior for $a_0$, and an $\text{Exp}(1.5)$ prior for $b_0$.

The priors for the nonparametric model in (\ref{eq:nonpara_model}) are given for each 
example in the corresponding subsection below (and in the Supplementary Material
for the first simulation example). 
The model is implemented with a fixed number of basis components. For all simulation examples, 
as well as for the earthquake data analysis in Section \ref{sec:mhp_real_analysis}, we use
sensitivity analysis to calibrate the values for $L$ and $M$. An illustration is provided 
with the simulation example presented in the Supplementary Material.
The Supplementary Material also includes more detailed results from sensitivity analysis with 
respect to the values of $L$ and $M$ for the earthquake data example.

For each example, we use the same prior for the immigrant intensity 
$\mu$ for the three models considered in the simulation study:
an $\text{Exp}(11)$, $\text{Exp}(22.3)$, and $\text{Exp}(22)$ prior for the example 
in the Supplementary Material, Section \ref{subsec:mhp_simulation_dpowlaw}, 
and Section \ref{subsec:mhp_simulation_dpowlawmix}, respectively.

Regarding inference results, the ensuing subsections focus on the total offspring intensity
and the offspring densities. Additional results are presented in the Supplementary Material, 
where we also discuss posterior simulation for the ETAS and semiparametric models.

\subsection{Mark-dependent Lomax density example}
\label{subsec:mhp_simulation_dpowlaw}

For this example, the point pattern is of size $449$ ($49$ immigrants, $400$ offspring
points), and it is recorded within interval $(0,T) =$ $(0,5000)$. 
The underlying point process is a MHP with ground intensity given by 
(\ref{ground_intensity_simulation}), with $\mu = 0.01$, $a = 0.32$, $b = 0.5$, 
and $g_{\kappa}(x) =$ $\text{Lomax}(x|5 + \kappa,1)$. The rate parameter of the 
truncated exponential mark density is $\psi = 0.6$. 
Note that, as $\kappa$ (i.e., earthquake magnitude) increases, $g_{\kappa}(x)$
concentrates more probability mass at smaller values. Hence, the offspring density  
is designed to reflect a  potential seismic pattern where aftershocks from parent 
earthquakes of higher magnitude emerge sooner.

\begin{figure}[!t]
\centering
\includegraphics[width=0.29\textwidth]{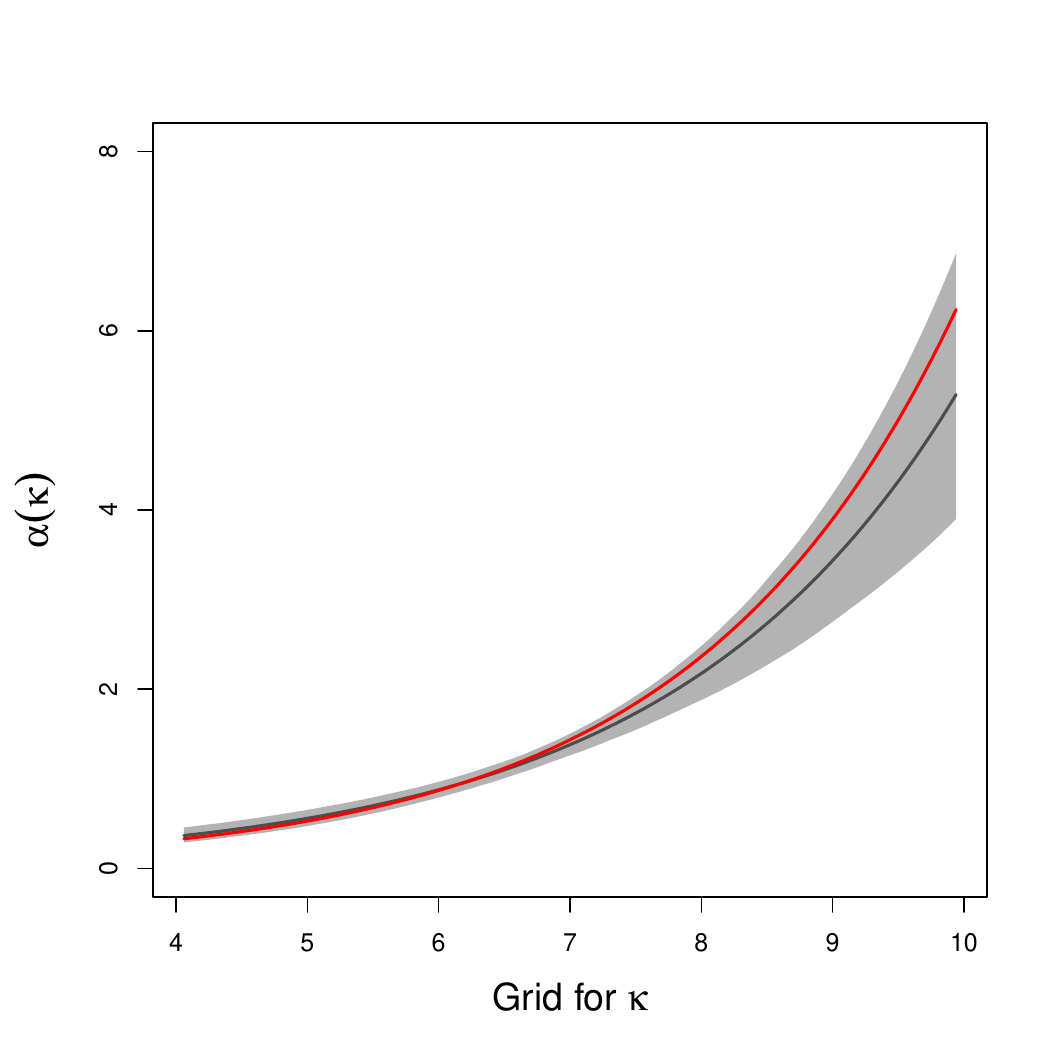}
\includegraphics[width=0.29\textwidth]{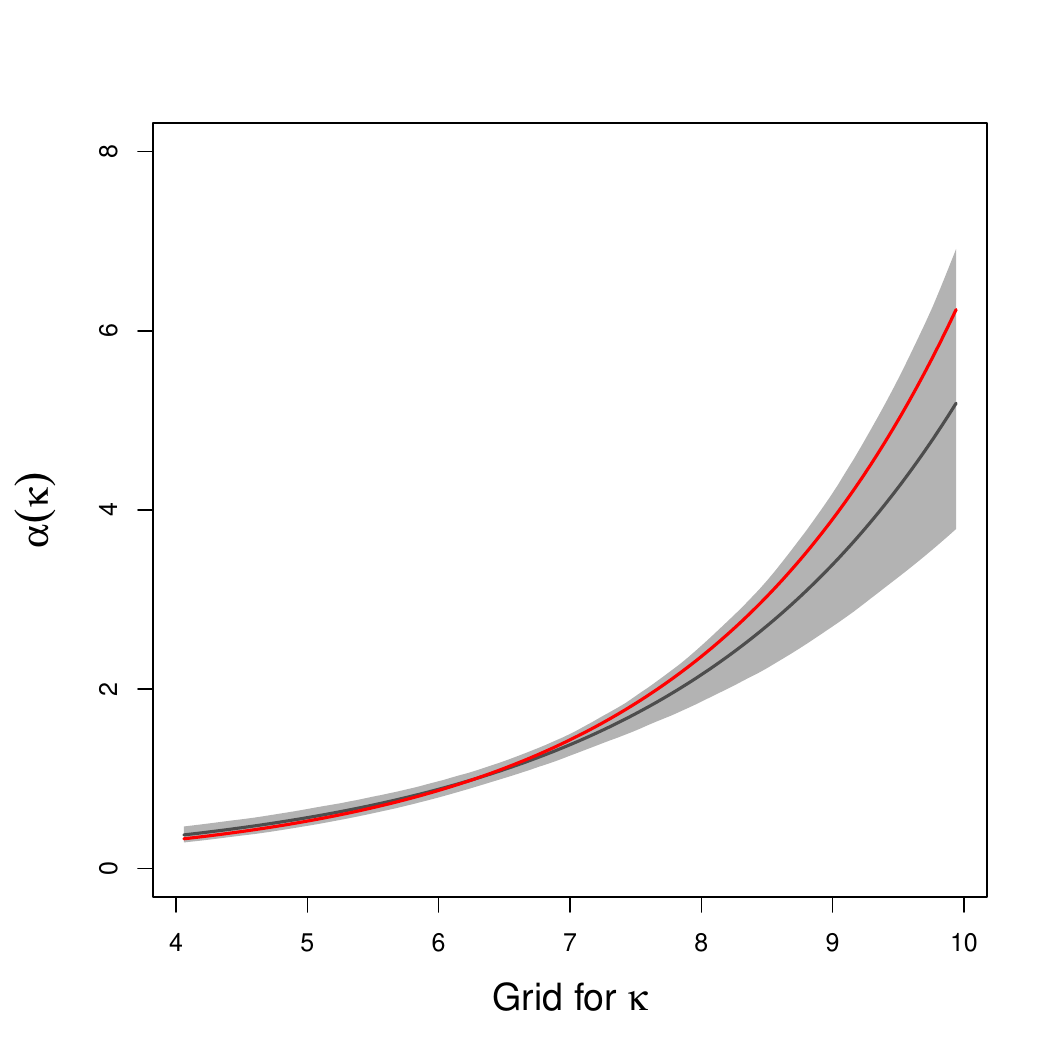}
\includegraphics[width=0.29\textwidth]{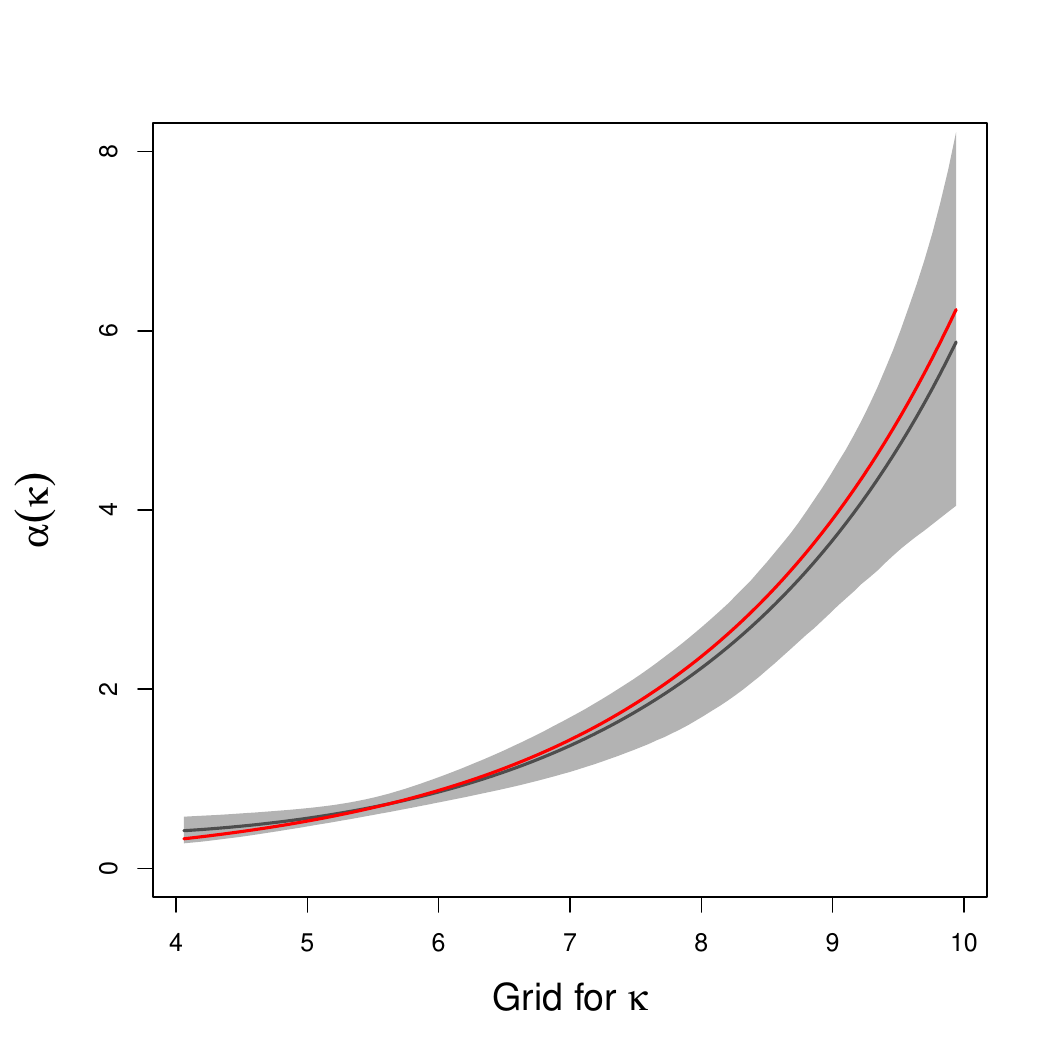} \\
\includegraphics[width=0.24\textwidth]{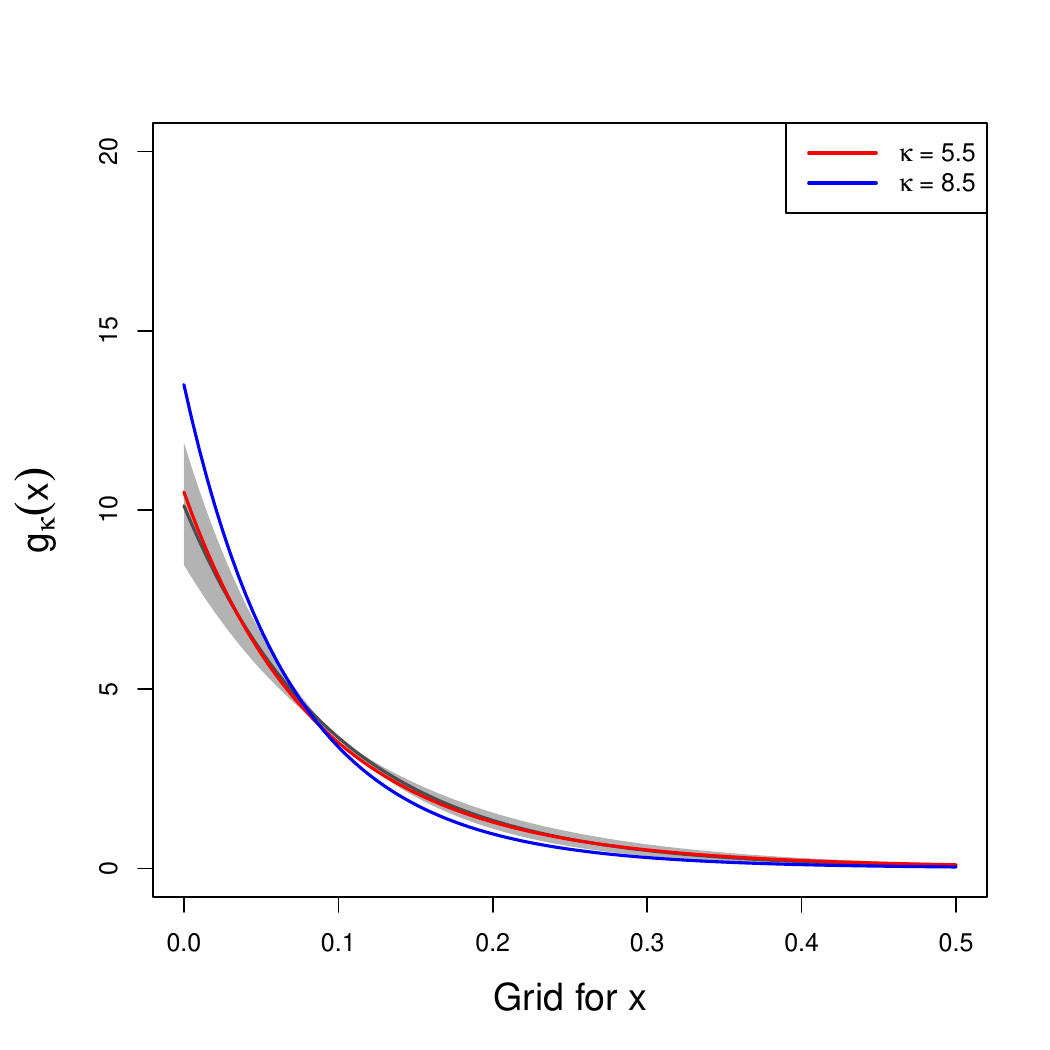}
\includegraphics[width=0.24\textwidth]{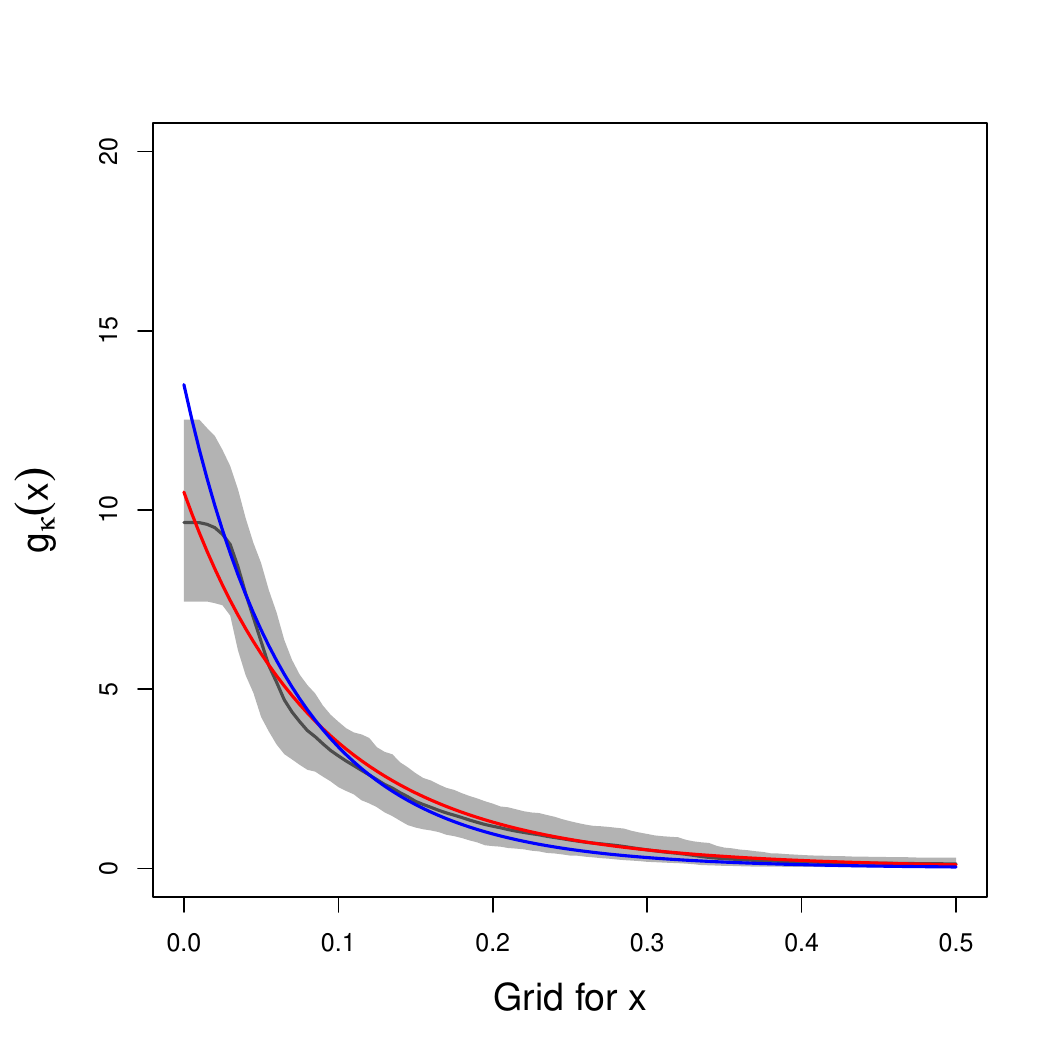}
\includegraphics[width=0.24\textwidth]{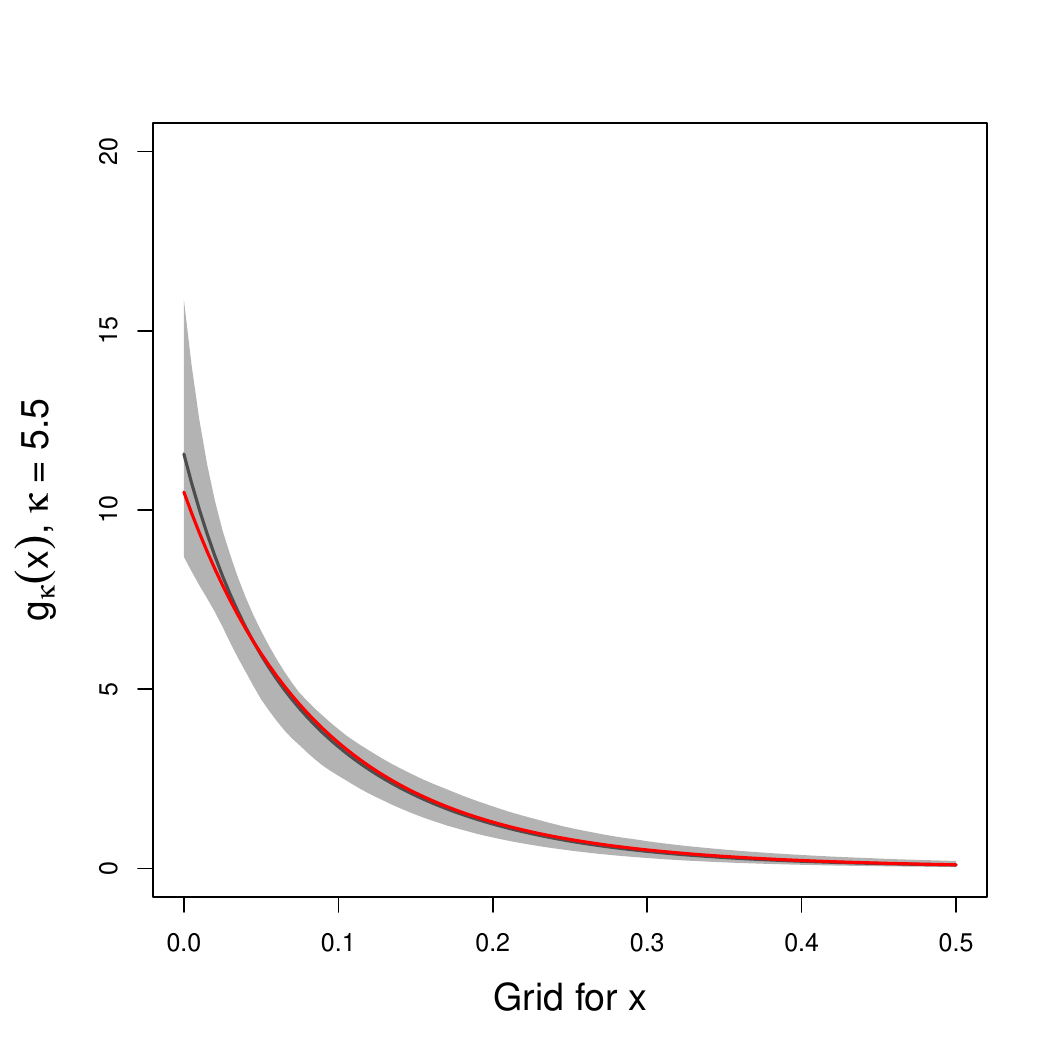}
\includegraphics[width=0.24\textwidth]{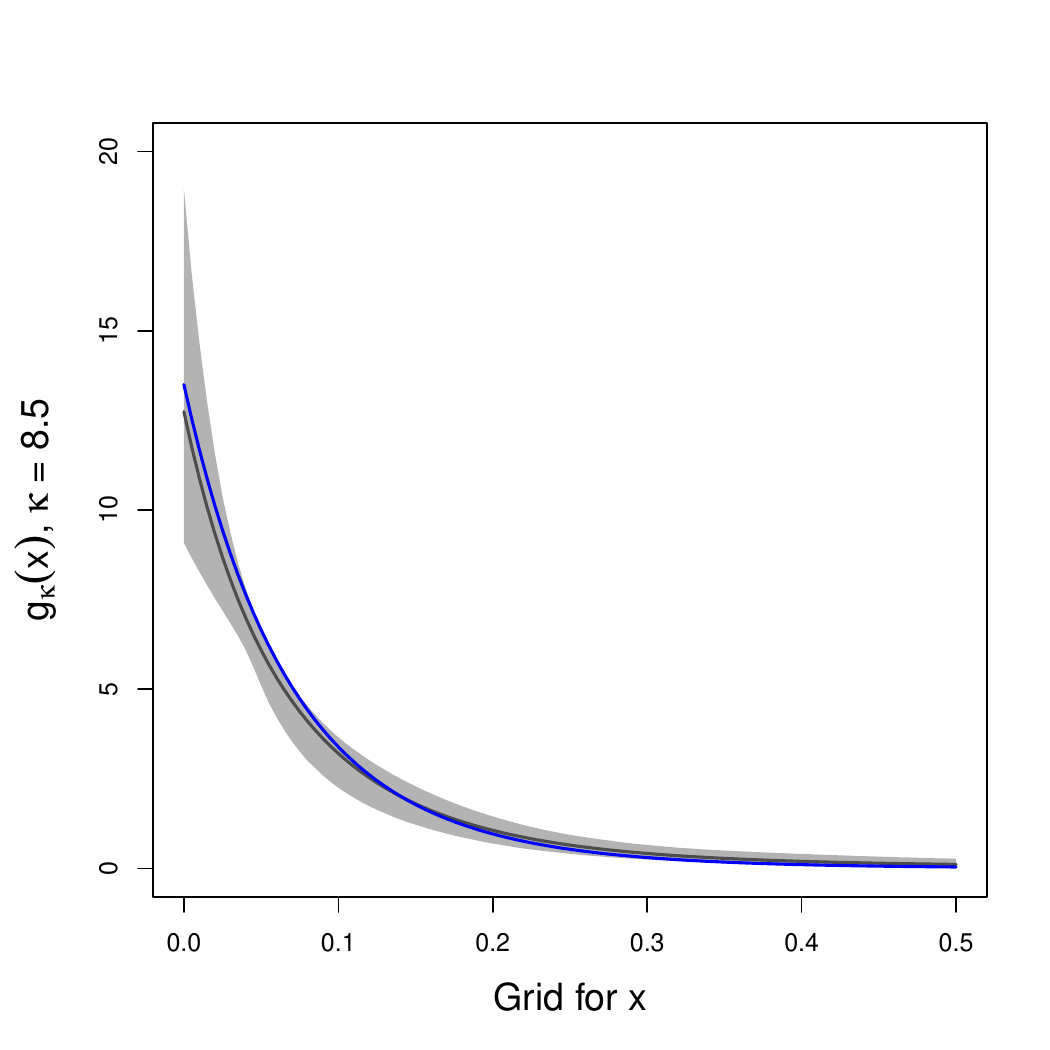} \\
\caption{
{\small Mark-dependent Lomax density simulation example. 
Top row: posterior mean (black line) and 95\% interval estimates for the total offspring 
intensity, under the ETAS, semiparametric, and nonparametric model (left, middle, right). 
Bottom row: posterior mean (black line) and 95\% interval estimates for $g(x)$ under the 
ETAS and semiparametric model (first and second column), and for $g_{5.5}(x)$ and $g_{8.5}(x)$ 
under the nonparametric model (last two columns). In each panel, the data generating function(s) 
are denoted by the red and/or blue line.}}
\label{fig:mhp_dpowlaw_excitation}
\end{figure}

The nonparametric model is implemented with $L=20$ and $M=15$, and the following priors: 
$\text{Lomax}(2,0.1)$ for $\theta$; $\text{Exp}(10)$ for both $b_1$ and $b_2$; 
$\text{Exp}(1)$ for $d$; $\text{Lomax}(2,2000)$ for $c_0$; and, $\text{Exp}(1)$ and 
$\text{Exp}(0.321)$ for $a_\beta$ and $b_\beta$.

Inference results for the offspring density at mark values $\kappa = 5.5$ and $\kappa = 8.5$
are provided in Figure \ref{fig:mhp_dpowlaw_excitation}. The nonparametric 
model recovers the varying shapes of the mark-dependent offspring densities, with the results 
somewhat more accurate for density $g_{5.5}(x)$. By their definition, the ETAS model and the 
semiparametric model can not estimate offspring densities that change with the mark.
Nonetheless, the posterior uncertainty bands produced by these models are of some interest. 
The ETAS model interval estimates miss the offspring density at the larger mark 
value, whereas the semiparametric model yields wider uncertainty bands that essentially 
include both densities $g_{5.5}(x)$ and $g_{8.5}(x)$.

As shown in Figure \ref{fig:mhp_dpowlaw_excitation}, all three models capture function 
$\alpha(\kappa)$ within their posterior uncertainty bands. However, the 
nonparametric model provides more accurate estimation, despite the fact that 
the ETAS and the semiparametric model are based on the same total offspring intensity 
parametric form with the data generating mechanism.

\subsection{Mixture of mark-dependent Lomax densities example}
\label{subsec:mhp_simulation_dpowlawmix}

The final simulation example involves a point pattern of size $454$
($50$ immigrants, $404$ offspring points), recorded in $(0,T) =$ $(0,5000)$. 
The data generating MHP is the same with the example of Section 
\ref{subsec:mhp_simulation_dpowlaw}, with the exception of the mark-dependent 
offspring density, which is given by $g_{\kappa}(x) =$
$0.6 \, \text{Lomax}(x | 10+\kappa,1) + 0.4 \, \text{Lomax}(x | 10,1+\kappa)$. 
This specification seeks to elaborate on the potential seismic pattern postulated 
in Section \ref{subsec:mhp_simulation_dpowlaw}. Here, the inclusion of the 
second mixture component (Lomax density with mark-dependent scale parameter)
yields offspring densities that, as earthquake magnitude increases, place more 
probability mass at smaller values and have a longer tail. That is, the specific 
choice for $g_{\kappa}(x)$ supports the premise that high-magnitude earthquakes 
trigger numerous immediate aftershocks, as well as further shocks over a prolonged period.

The nonparametric model is applied with $L=60$ and $M=15$, and priors: 
$\text{Lomax}(2,0.5)$ for $\theta$; $\text{Exp}(10)$ and $\text{Exp}(13)$ for $b_1$ and $b_2$;
$\text{Exp}(1)$ for $d$; $\text{Lomax}(2,2000)$ for $c_0$; and, $\text{Exp}(1)$ and 
$\text{Exp}(0.334)$ for $a_\beta$ and $b_\beta$.
As expected from the more complex offspring density shapes in this simulation scenario, 
a larger value for $L$ is needed for accurate estimation of functions $g_{\kappa}(x)$
(estimates stabilized at $L=50$, with $L=60$ used as a conservative choice).

\begin{figure}[!t]
\centering
\includegraphics[width=0.25\textwidth]{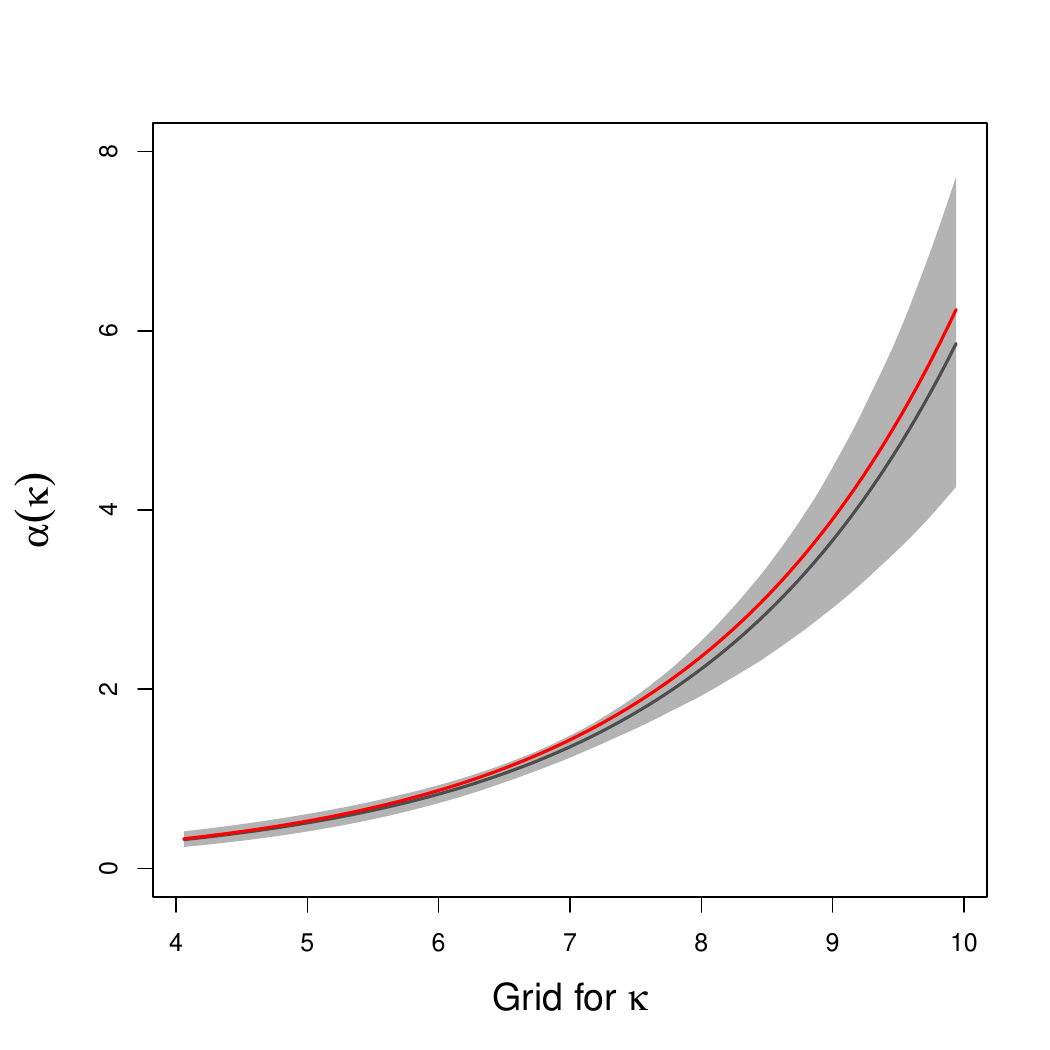}
\includegraphics[width=0.25\textwidth]{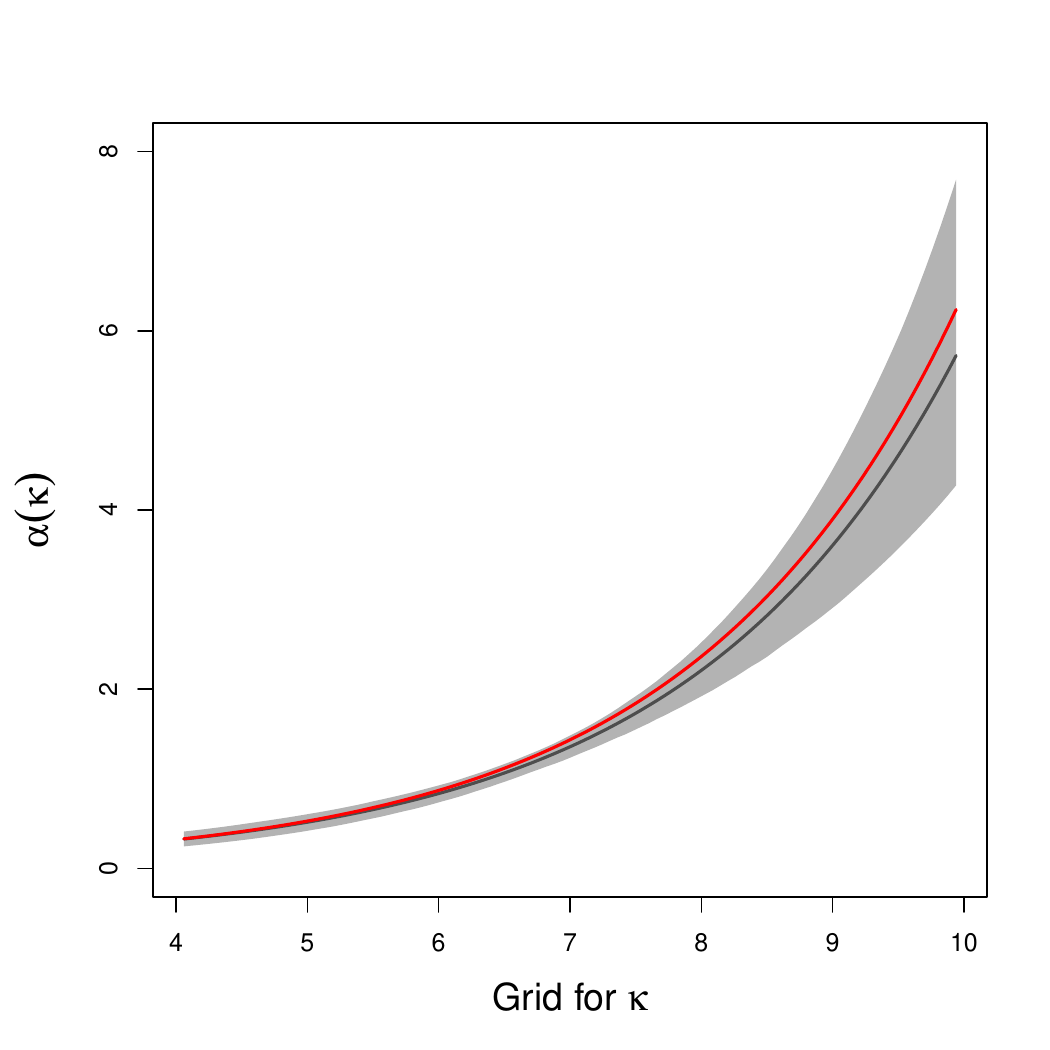}
\includegraphics[width=0.25\textwidth]{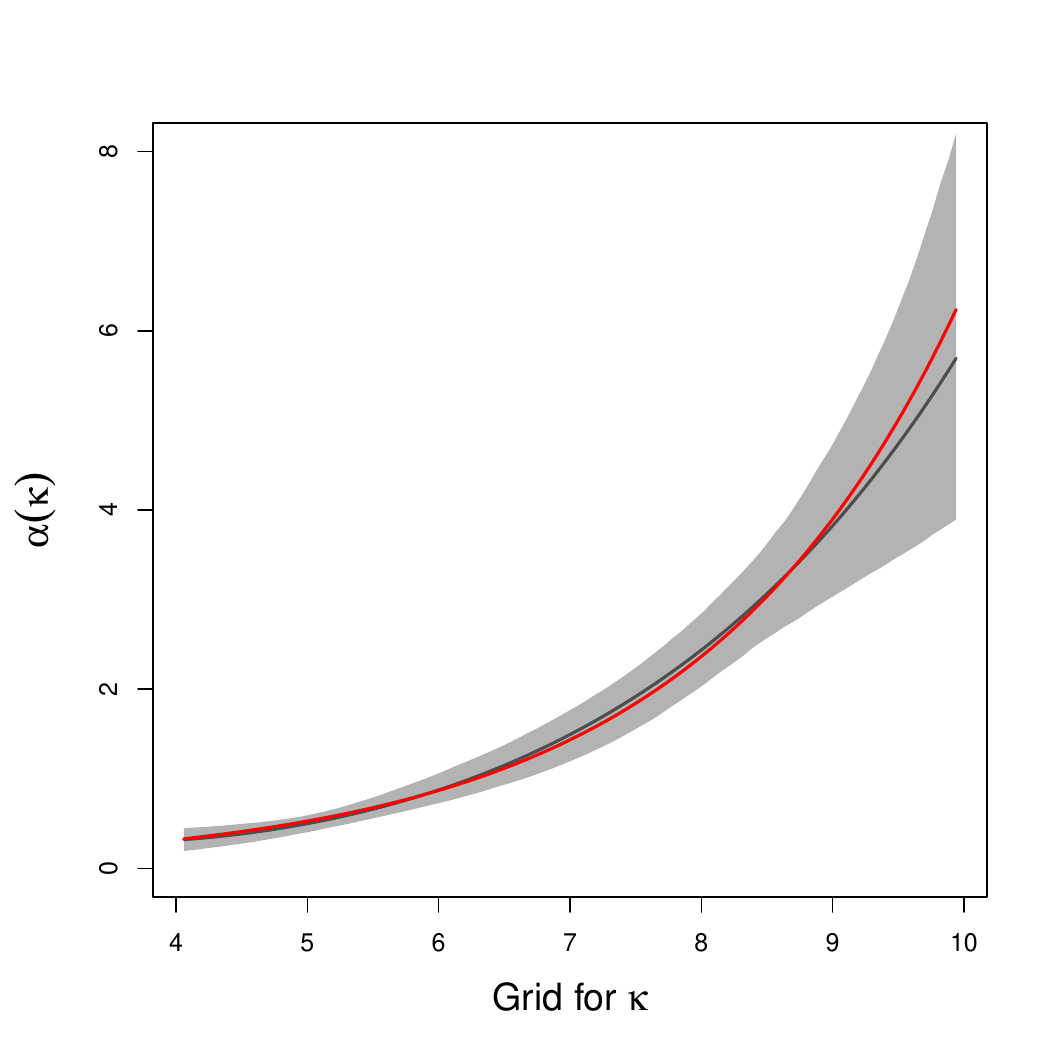} \\
\includegraphics[width=0.24\textwidth]{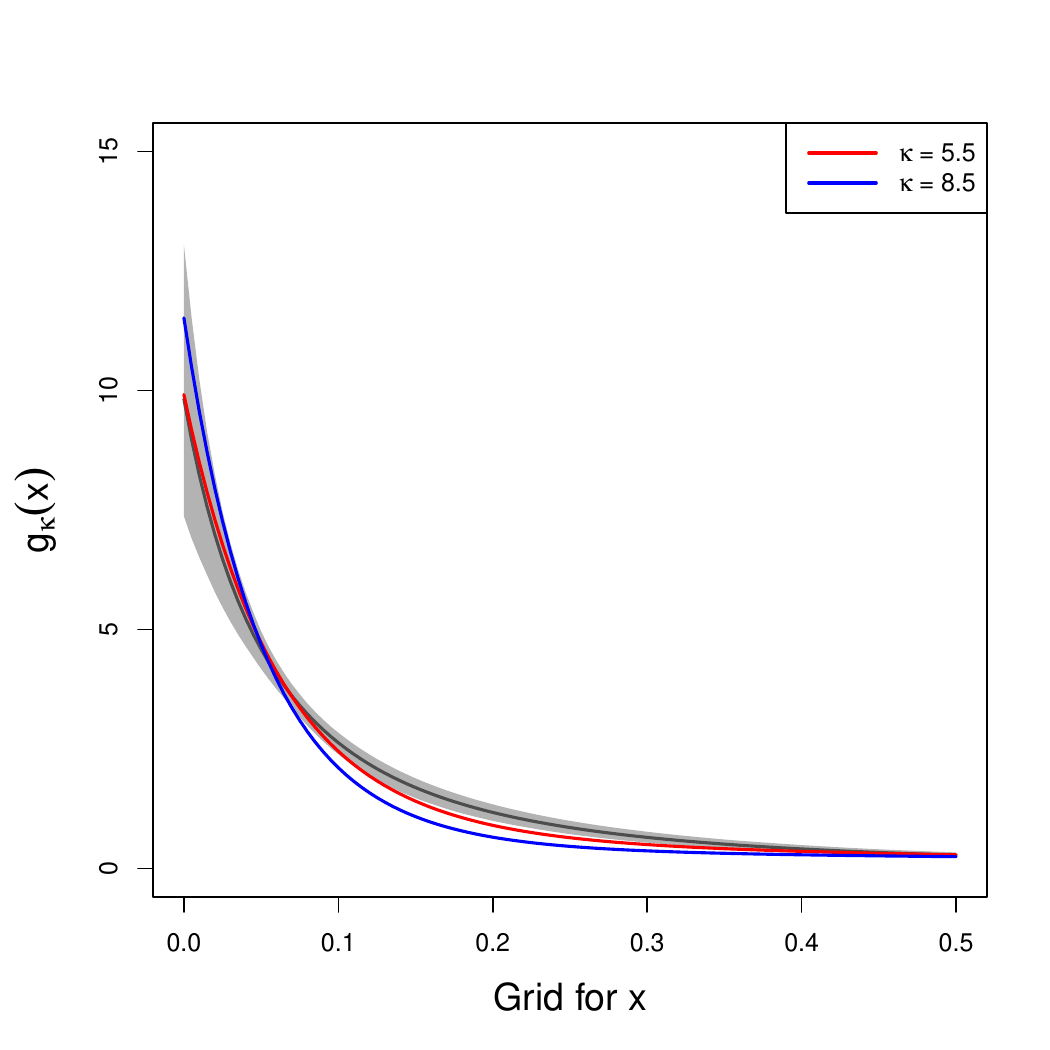}
\includegraphics[width=0.24\textwidth]{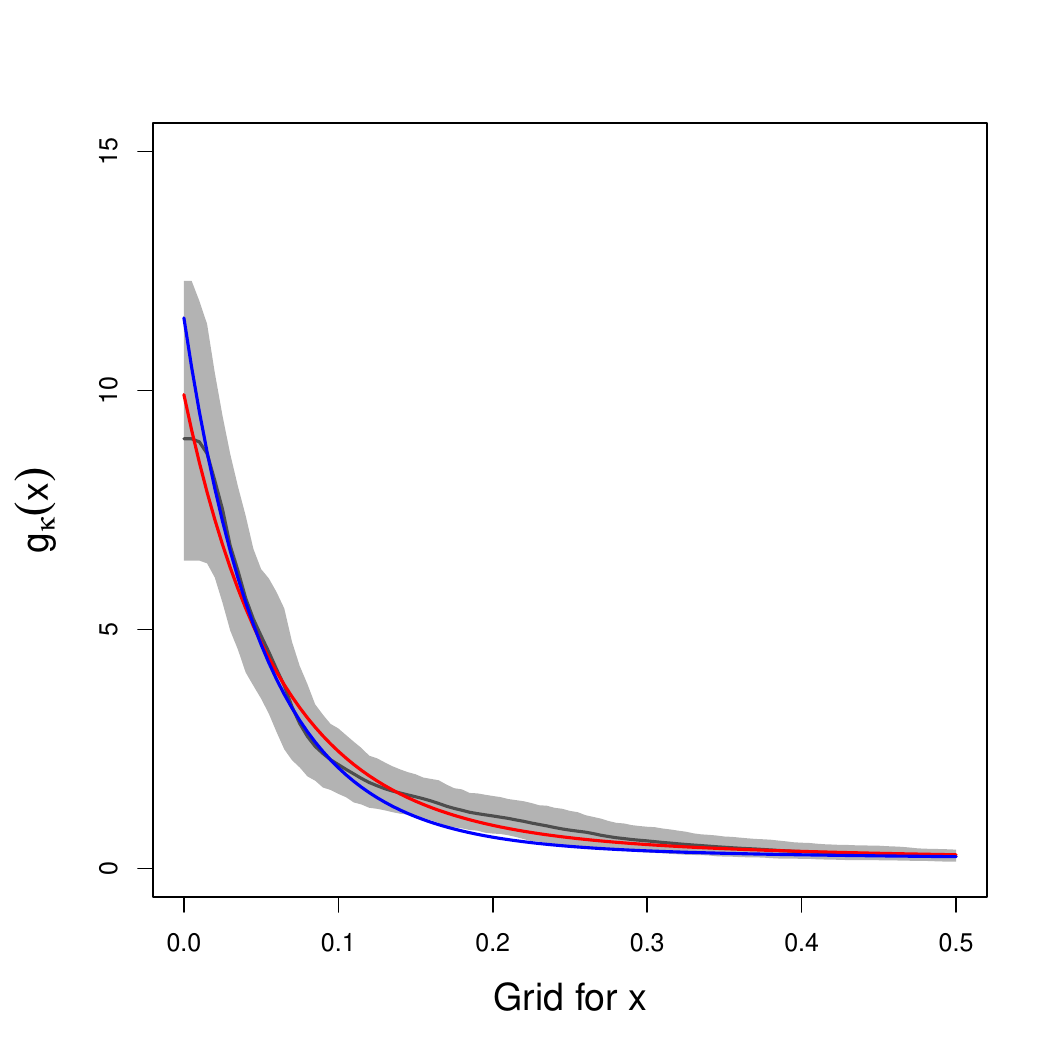}
\includegraphics[width=0.24\textwidth]{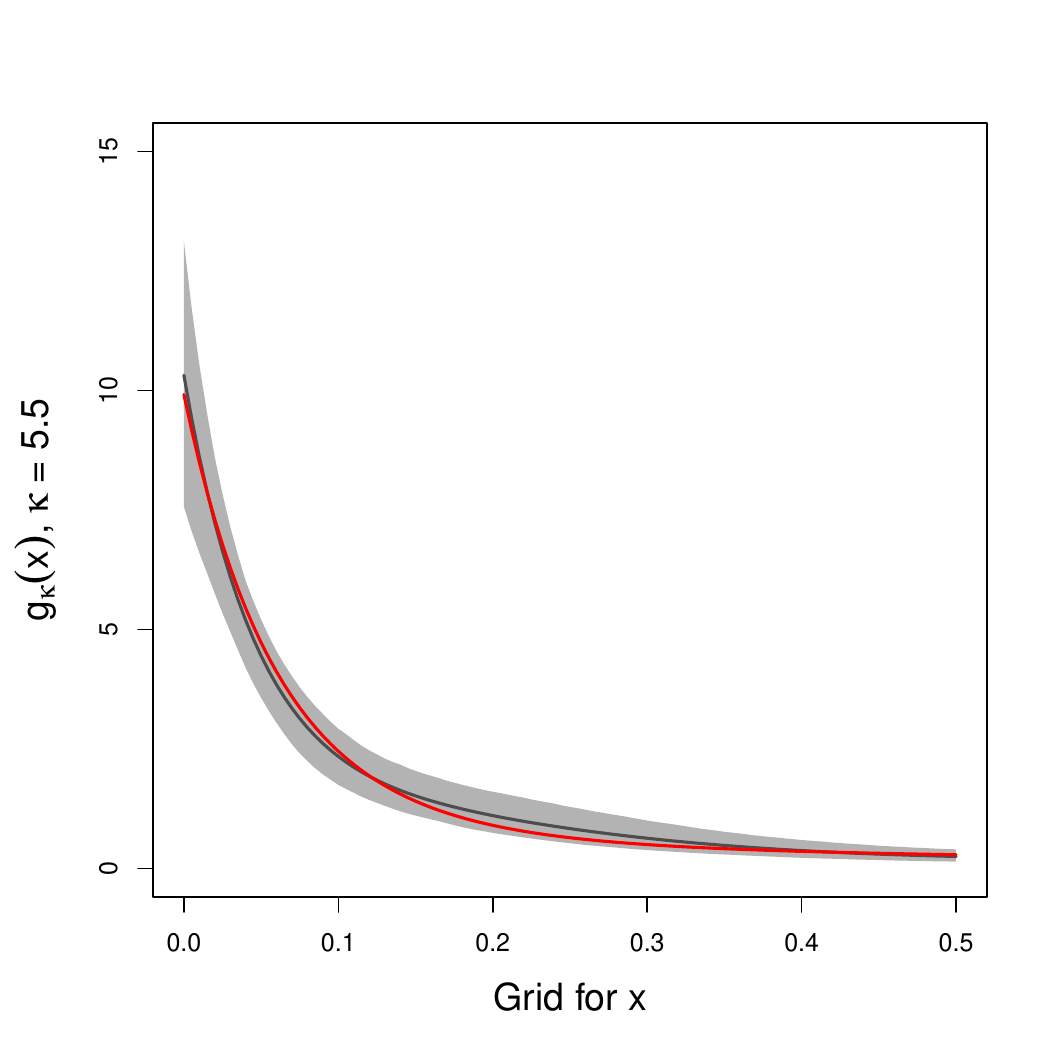}
\includegraphics[width=0.24\textwidth]{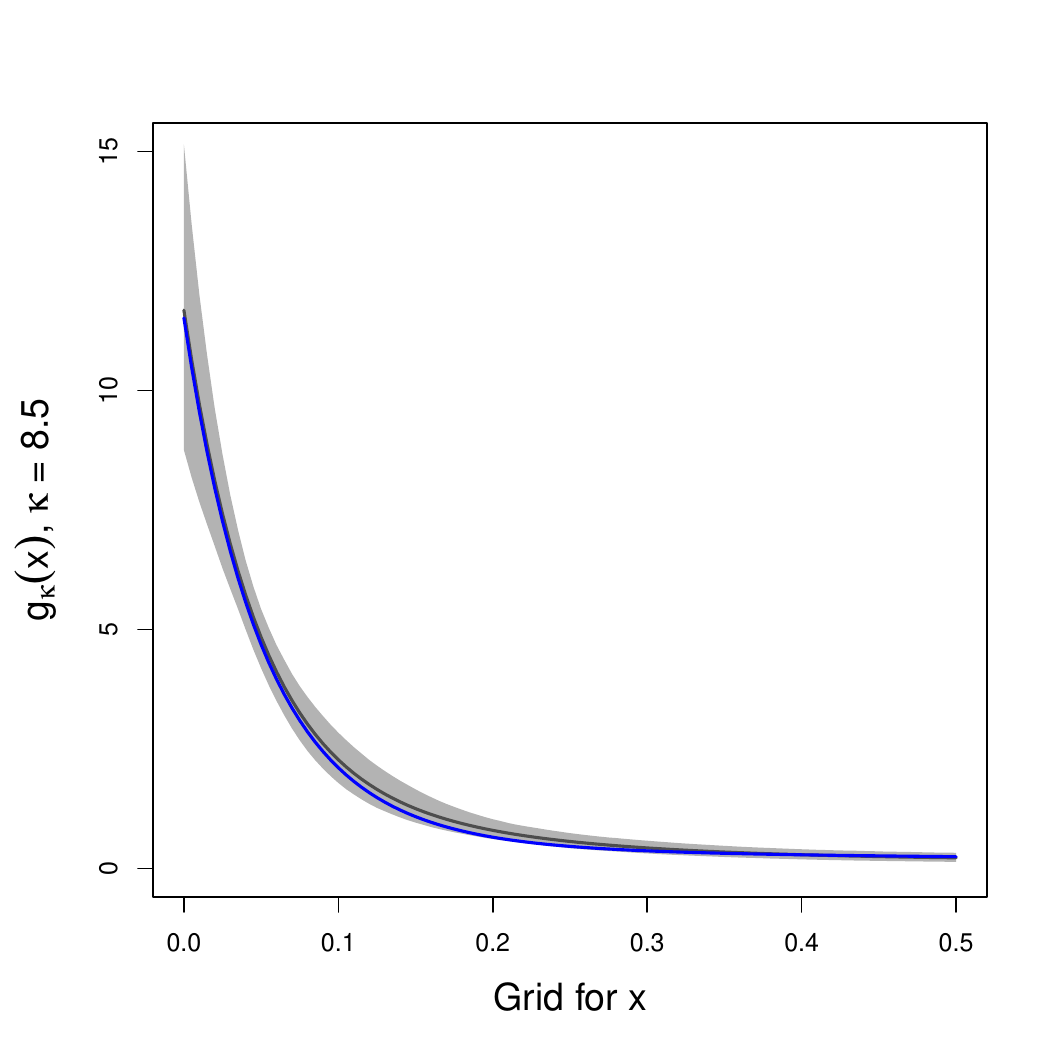} \\
\includegraphics[width=0.24\textwidth]{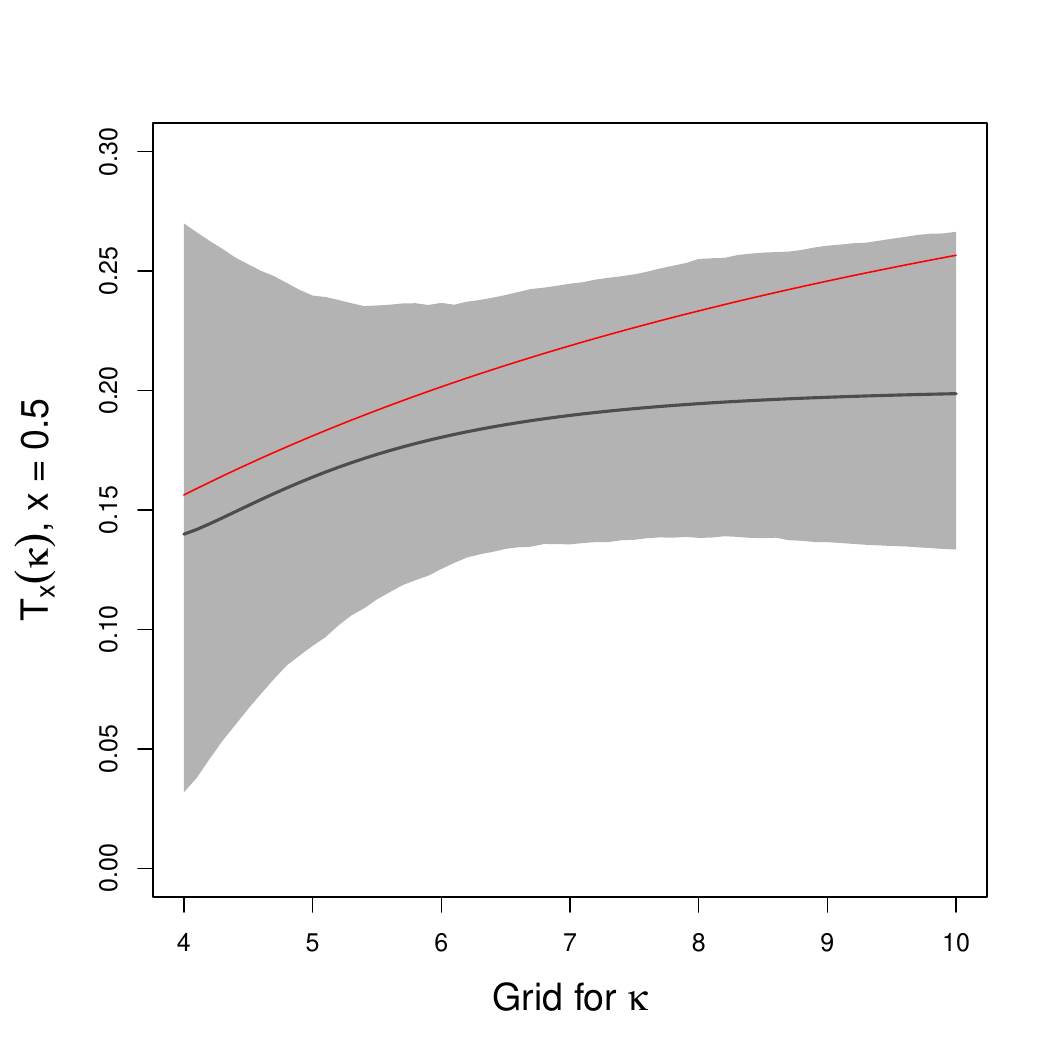}
\includegraphics[width=0.24\textwidth]{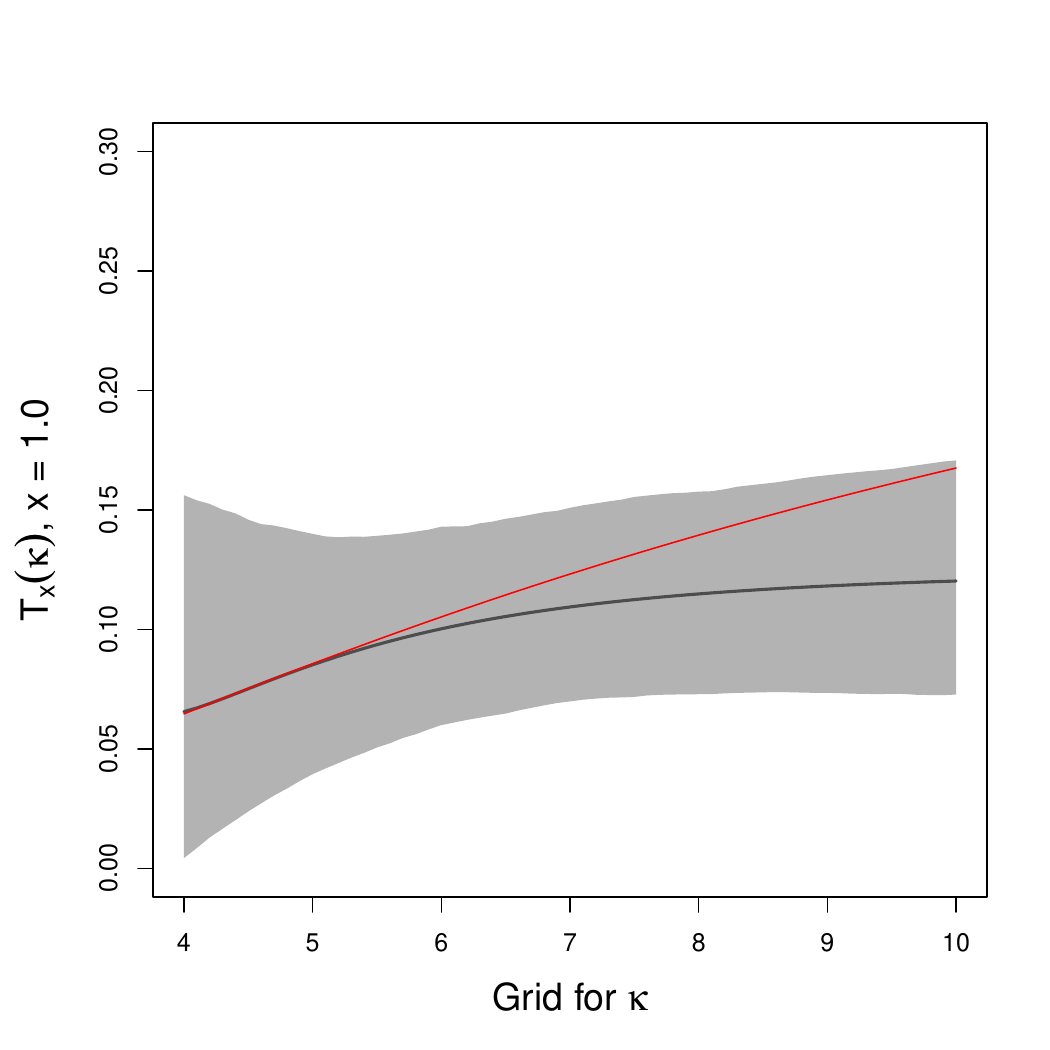}
\includegraphics[width=0.24\textwidth]{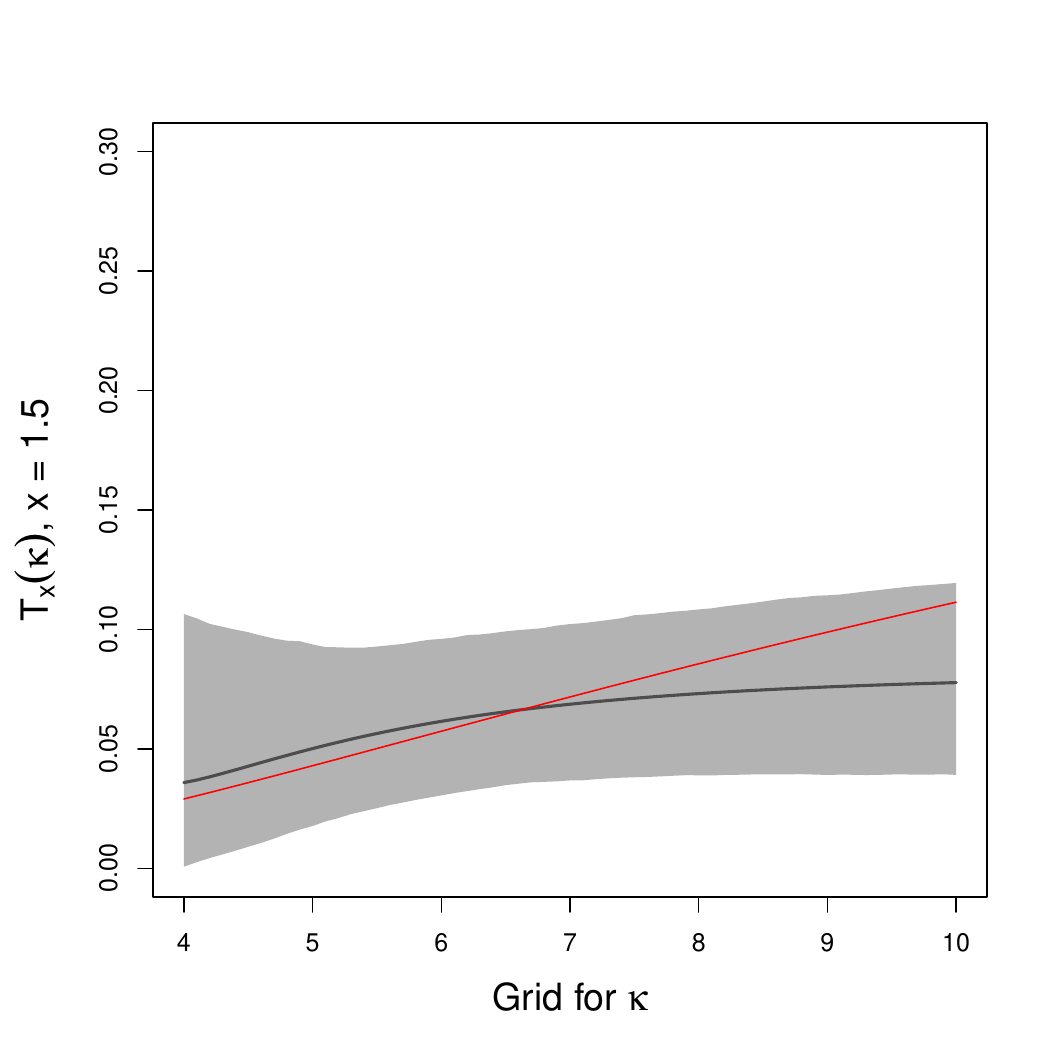}
\includegraphics[width=0.24\textwidth]{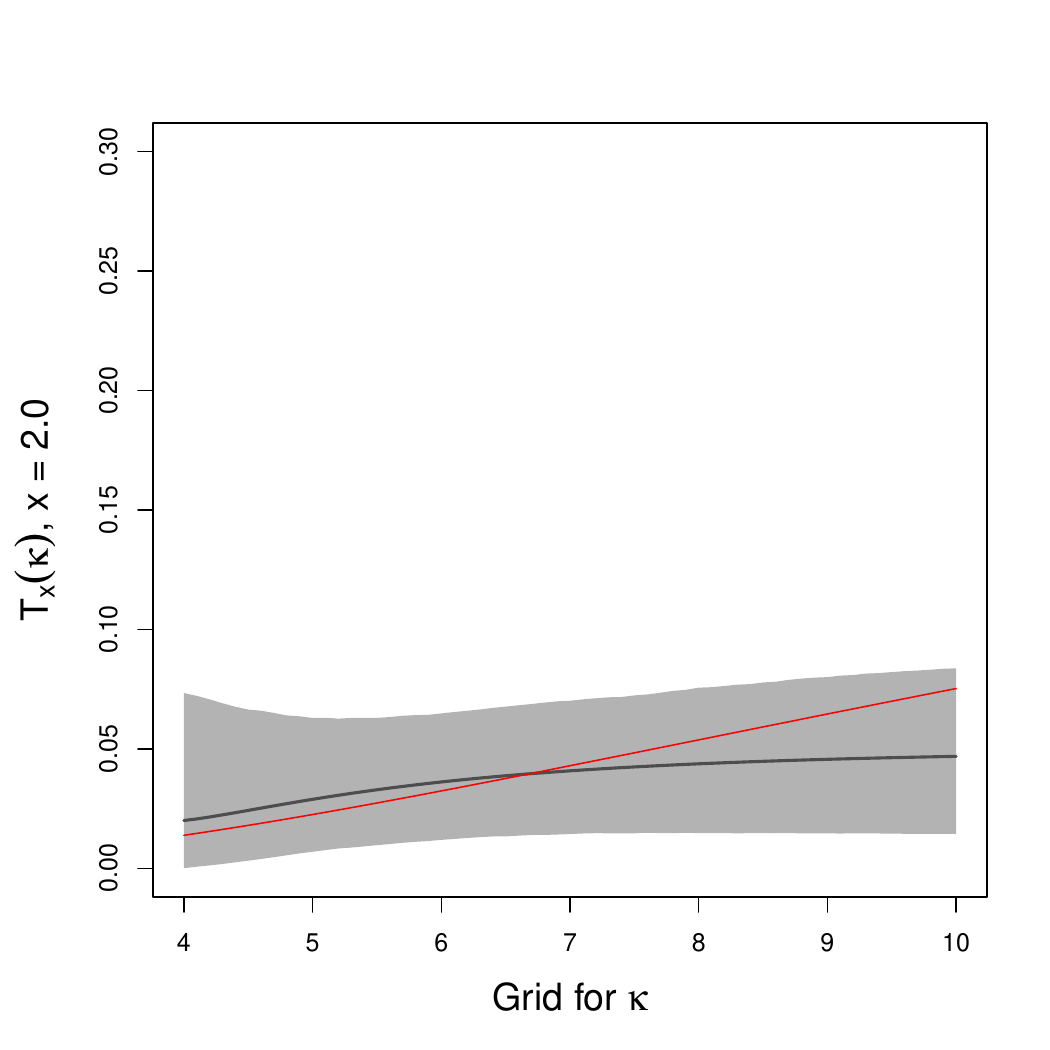} \\
\caption{
{\small Mixture of Lomax densities simulation example.
Top row: posterior mean (black line) and 95\% interval estimates for $\alpha(\kappa)$, 
under the ETAS, semiparametric, and nonparametric model (left, middle, right). 
Middle row: posterior mean (black line) and 95\% interval estimates for $g(x)$ under the 
ETAS and semiparametric model (first and second column), and for $g_{5.5}(x)$ and $g_{8.5}(x)$ 
under the nonparametric model (last two columns). 
Bottom row: nonparametric model posterior mean (black line) and 95\% interval estimates 
for the tail probability function, evaluated at four $x$ values.
In each panel, the data generating function(s) are denoted by the red and/or blue line.
}}
\label{fig:mhp_dpowlawmix_excitation}
\end{figure}

Inference results are reported in Figure \ref{fig:mhp_dpowlawmix_excitation}. 
Regarding the ETAS and the semiparametric model, the remarks are similar with the example 
of Section \ref{subsec:mhp_simulation_dpowlaw}. Even though the underlying mark-dependent 
offspring densities correspond to a structured, parametric setting, the nonparametric model 
captures well the different density shapes, including the steeper decline at larger 
mark values.

For a more detailed view at model performance, we study the more challenging task of 
inference for tail probabilities. More specifically, define the tail probability function 
at value $x$ as follows: $T_x(\kappa) =$ $1 - \int_{0}^{x} g_\kappa(u) \, \text{d}u$. 
Hence, $T_x(\kappa)$ represents the probability that an aftershock occurs more than $x$ 
time units after an earthquake of magnitude $\kappa$. We view this probability as 
a function of earthquake magnitude, for $x$ fixed at large values in the tails of the 
offspring densities. As expected given the structure of 
the data generating MHP, the true functions $T_x(\kappa)$ increase with $\kappa$; 
see the bottom row of Figure \ref{fig:mhp_dpowlawmix_excitation} for the underlying 
functions at $x = 0.5$, $1$, $1.5$, and $2$. In all four cases, the nonparametric model 
is able to contain the true tail probability function within its posterior interval 
estimates.

%
%

%
%

%
%

\section{Earthquake data analysis}
\label{sec:mhp_real_analysis}

\begin{figure}[!t]
\centering
\includegraphics[width=0.97\textwidth]{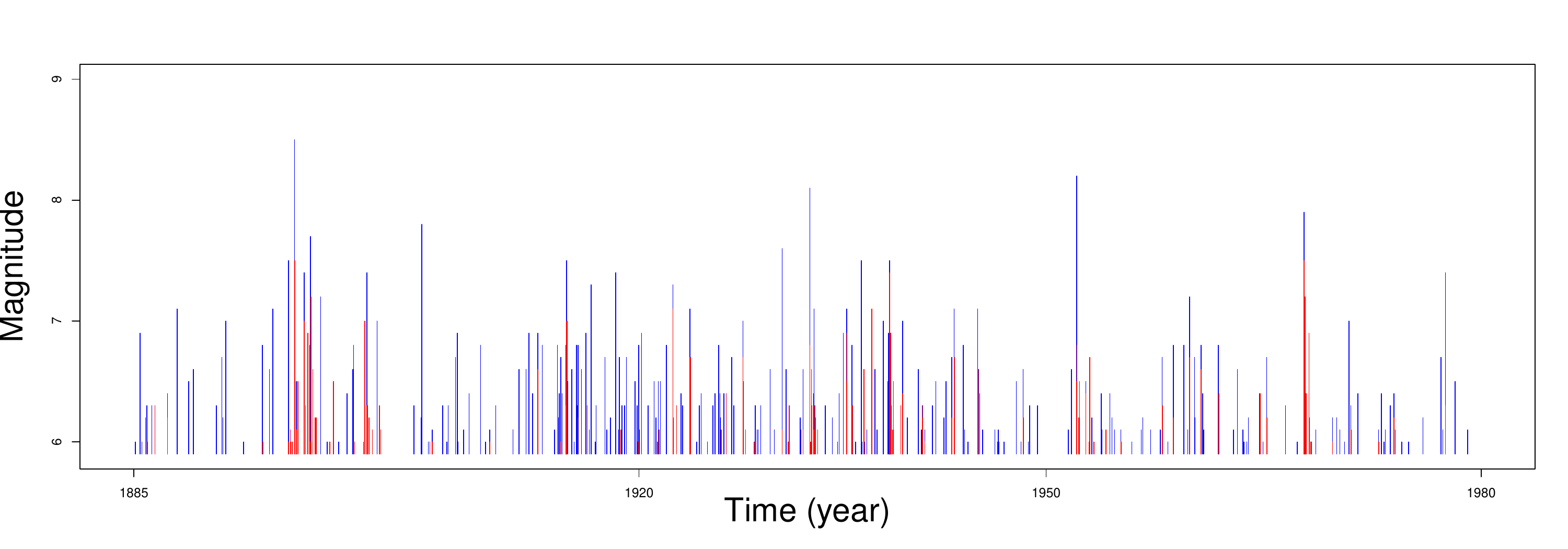}\\
\caption{{\small  
Earthquake occurrences (of magnitude $\geq 6$) in Japan and its vicinity from 1885 through 
1980. Main shocks and aftershocks are indicated by blue and red color, respectively.}}
\label{fig:mhp_real_japan_data}
\end{figure}

We consider a catalog of earthquakes with magnitude $\geq 6$ 
(and depth $< 100$ kilometers) that occurred in Japan and its vicinity over 
a time period of about $35000$ days, from 1885 through 1980 \citep{O1988}.  
A useful characteristic of this catalog is that each shock is classified as a 
main shock (258 points), aftershock (200 points) or foreshock (25 points). 
Figure \ref{fig:mhp_real_japan_data} displays the point pattern used
in our analysis, which excludes the 25 foreshocks.

Section \ref{real_data_constant_immigrant_intensity} compares the three 
models (with immigrant intensity constant in time) considered in the simulation study:
the ETAS model in (\ref{eq:etas_model}); the semiparametric model with 
the DP uniform mixture in (\ref{eq:semipara_model}) for the offspring density; and, 
the nonparametric model with excitation function in (\ref{eq:nonpara_model}). 
Section \ref{real_data_general_model} reports results under the general 
MHP model that combines the excitation function prior in (\ref{eq:nonpara_model})
with the immigrant intensity prior in (\ref{eq:erlmix_immi}).

The highest earthquake magnitude in the catalog is $8.5$. The mark space is set to 
$(6,8.6)$ for the nonparametric model, and $(6,\infty)$ for the ETAS and the 
semiparametric model.

All models are fitted to the point pattern up to the end of year 1949, 
which comprises $340$ earthquakes (192 main shocks, 148 aftershocks). Based on the approach
discussed in Section \ref{subsec:prediction}, the predictive performance of the models is 
compared using the remaining portion of the catalog, from the beginning of 
year 1950 to 1980, which involves $118$ earthquakes (66 main shocks, 52 aftershocks).
Although the main shock/aftershock classification available in this particular catalog
is not used in model fitting, it is useful as an additional means of model assessment
based on the estimation of the branching structure.

Here, we focus on inference for the components of the MHP ground process intensity function, 
including model comparison as discussed above. The Supplementary Material provides additional 
details for the nonparametric model, including prior specification, sensitivity analysis to 
calibrate the number of basis components ($L = 20$ and $M = 160$), prior sensitivity analysis, 
and inference results for the mark density and for model parameters.

\subsection{MHP models with immigrant intensity constant in time}
\label{real_data_constant_immigrant_intensity}

\begin{figure}[!t]
\centering
\includegraphics[width=0.44\textwidth]{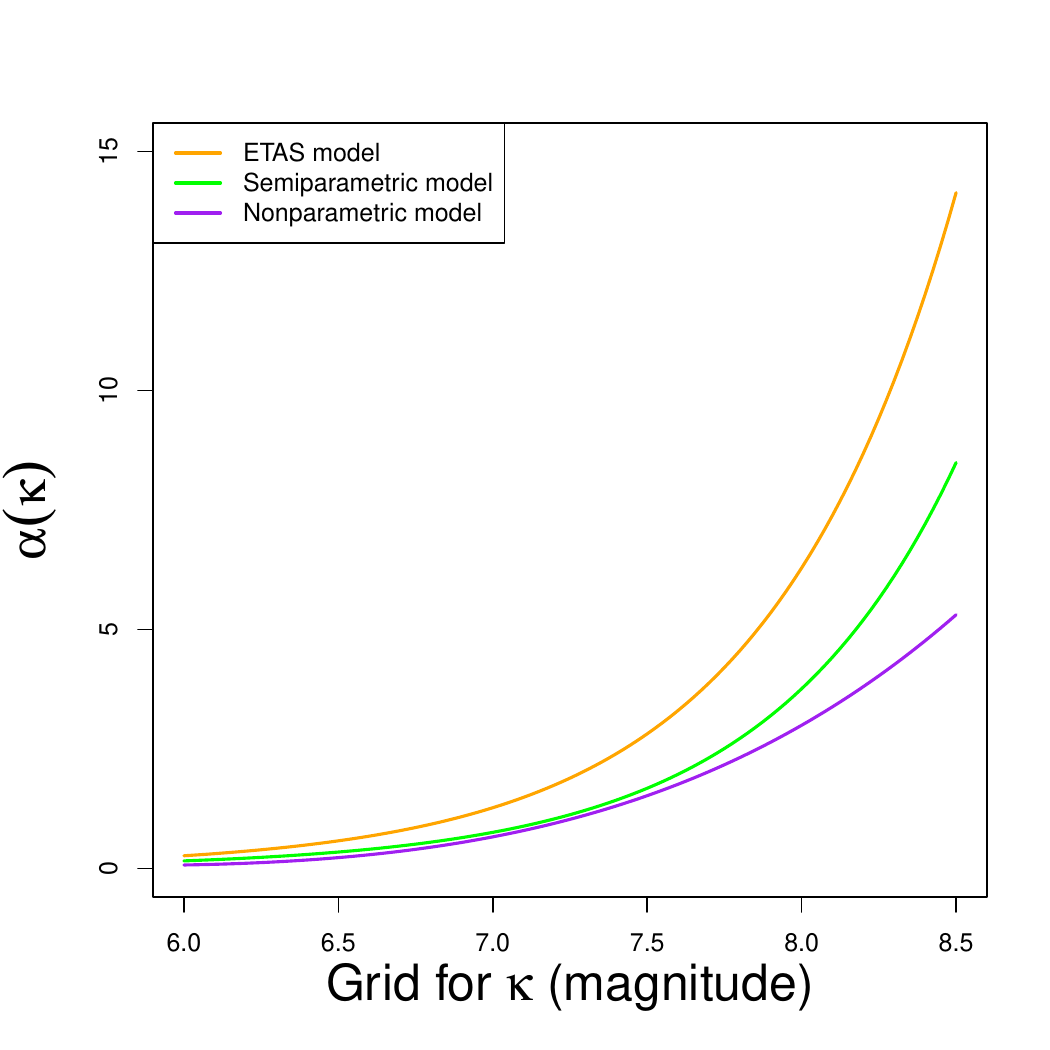}
\includegraphics[width=0.44\textwidth]{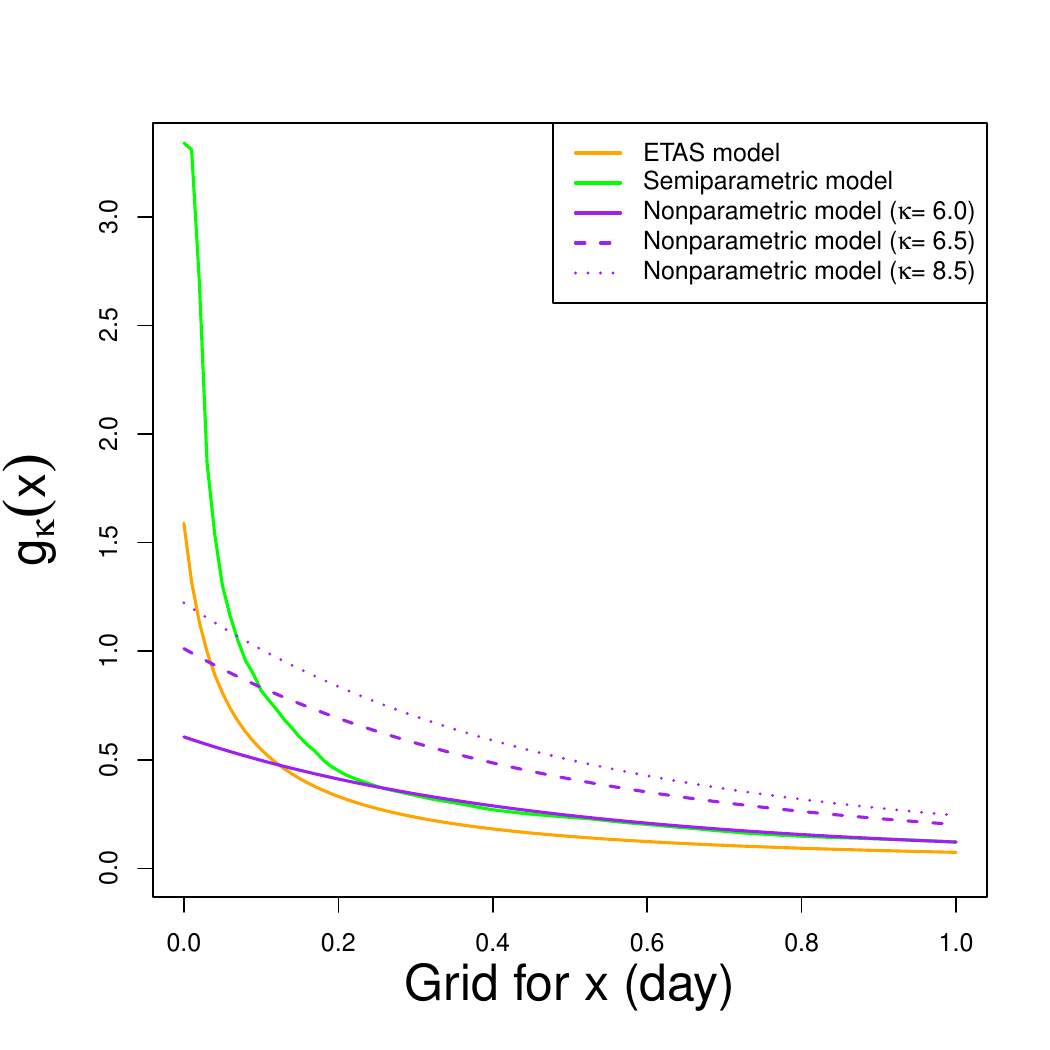} \\
\caption{{\small 
Earthquake data analysis (ETAS, semiparametric, and nonparametric model, with immigrant 
intensity constant in time). Posterior mean estimates of the total offspring intensity 
function (left) and offspring density (right). Nonparametric model offspring density 
estimates are shown at three magnitude values, $\kappa = 6$, $6.5$, $8.5$. }}    
\label{fig:mhp_real_japan_estm}
\end{figure}

Figure \ref{fig:mhp_real_japan_estm} plots the nonparametric model posterior mean estimates 
for the total offspring intensity, and for the offspring density at three magnitude values, 
$\kappa = 6, 6.5, 8.5$. Also plotted are the posterior mean estimates under the ETAS and the 
semiparametric model. Here, to facilitate graphical comparison, we show only point estimates; 
the Supplementary Material reports the posterior uncertainty bands for functions $\alpha(\kappa)$
and $g_{7.25}(x)$, under the nonparametric model. Although the tails of the offspring density 
estimates extend beyond the time interval of one day, we restrict the plot to that interval 
for better visualization.

The three models produce different estimates for the total offspring intensity, with the 
ETAS model estimating a substantially higher rate of increase with earthquake magnitude. 
The semiparametric and nonparametric model estimates are similar up to magnitude $7.5$, 
with the former model estimate increasing more sharply at higher magnitude values. 
The difference between the ETAS and the semiparametric model
is noteworthy, given that these two models are based on the same parametric 
function $\alpha(\kappa)$. However, recall that the semiparametric model 
is built from a prior that supports a larger range of decreasing offspring density shapes 
than the ETAS model Lomax density. The difference in the function $\alpha(\kappa)$ estimates
is reflected also in the estimation of the expected total oﬀspring intensity. 
The nonparametric model posterior mean and 95\% credible interval for $\rho$ are 
$0.288$ and $(0.218, 0.370)$; the corresponding estimates are $0.405$ and $(0.295, 0.540)$ 
under the semiparametric model, and $0.668$ and $(0.418, 0.890)$ under the ETAS model.

The nonparametric model offspring density estimates exhibit clear differences across
the three magnitude values, which span the range of observed earthquake magnitudes.
Offspring densities at higher magnitudes concentrate more probability mass at smaller 
values, thus indicating shorter waiting times for subsequent shocks. 
In Section \ref{real_data_general_model}, we report on more detailed inference for 
probabilities under the offspring densities, based on the general version of the 
nonparametric model with time-dependent immigrant intensity. 
Although the ETAS and the semiparametric model do not allow for magnitude-dependent 
offspring densities, it is useful to contrast their estimates with the nonparametric model. 
As expected from the difference in their function $\alpha(\kappa)$ estimates, the ETAS 
and semiparametric model also differ in their offspring density estimate.

%
%

Turning to model comparison, we consider estimation accuracy for the branching 
structure. Denote by $y_i^{\text{true}}$ the main shock ($y_i^{\text{true}} = 0$) or 
aftershock ($y_i^{\text{true}} \neq 0$) classification available in this earthquake 
catalog (again, this information is not used in model fitting). 
We use the posterior samples for the branching variables $y_i$ to compute
the misclassification rate, $R=$ $(M_I + M_O)/n$, where 
$M_I =$ $|\{i:y_i=0, \enspace y_i^{true} \neq 0, \enspace i=1,\ldots,n\}|$ and 
$M_O =$ $|\{i:y_i \neq 0, \enspace y_i^{true} = 0, \enspace i=1,\ldots,n\}|$ is the 
immigrant and offspring misclassification count, respectively, and $n=340$ is the 
size of the point pattern subset used for model fitting.
This criterion provides a ranking of the models. The nonparametric model yields
the smallest posterior mean and standard deviation for $R$ ($0.222$ and $0.011$), 
followed by the semiparametric model ($0.231$ and $0.024$), 
and the ETAS model ($0.267$ and $0.024$).

%
%

\begin{figure}[!t]
\centering
\includegraphics[width=0.32\textwidth]{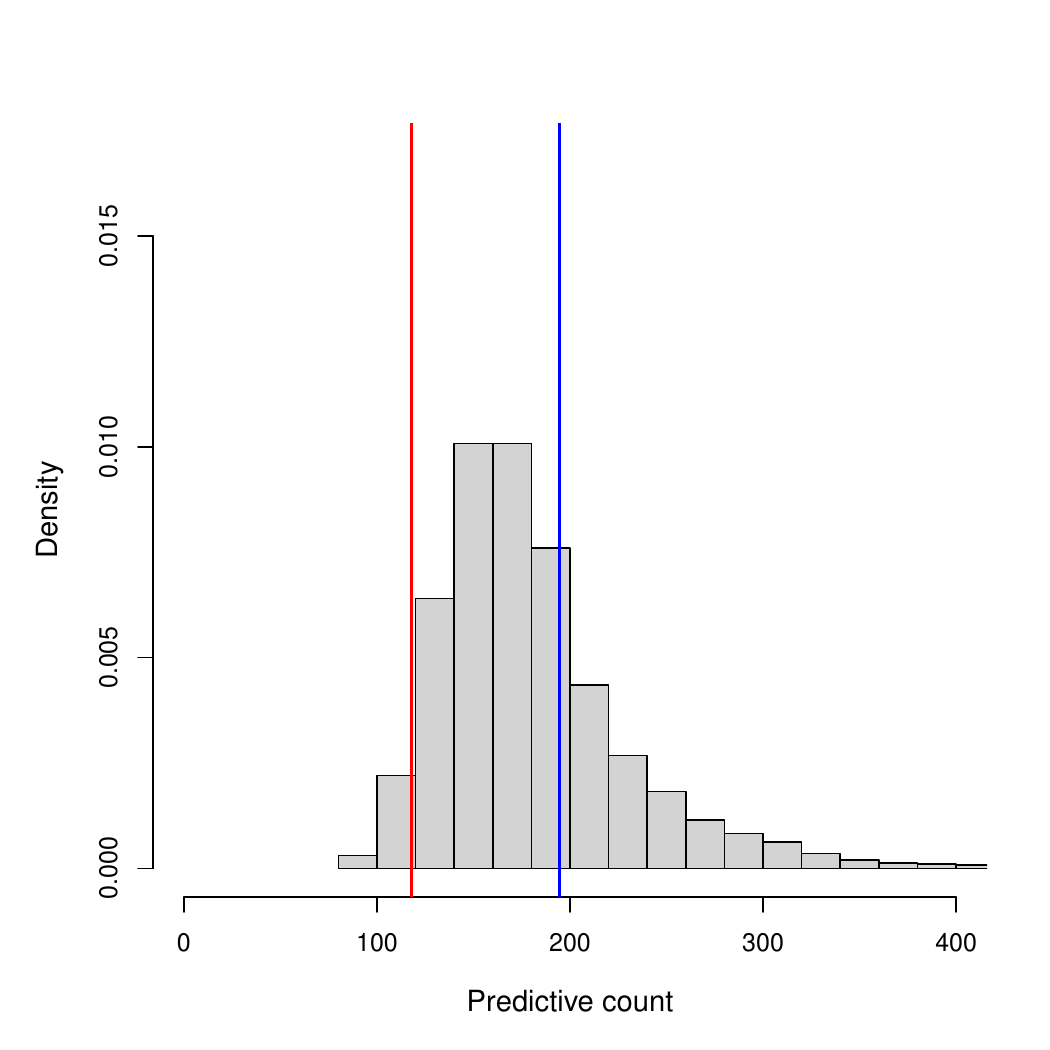}
\includegraphics[width=0.32\textwidth]{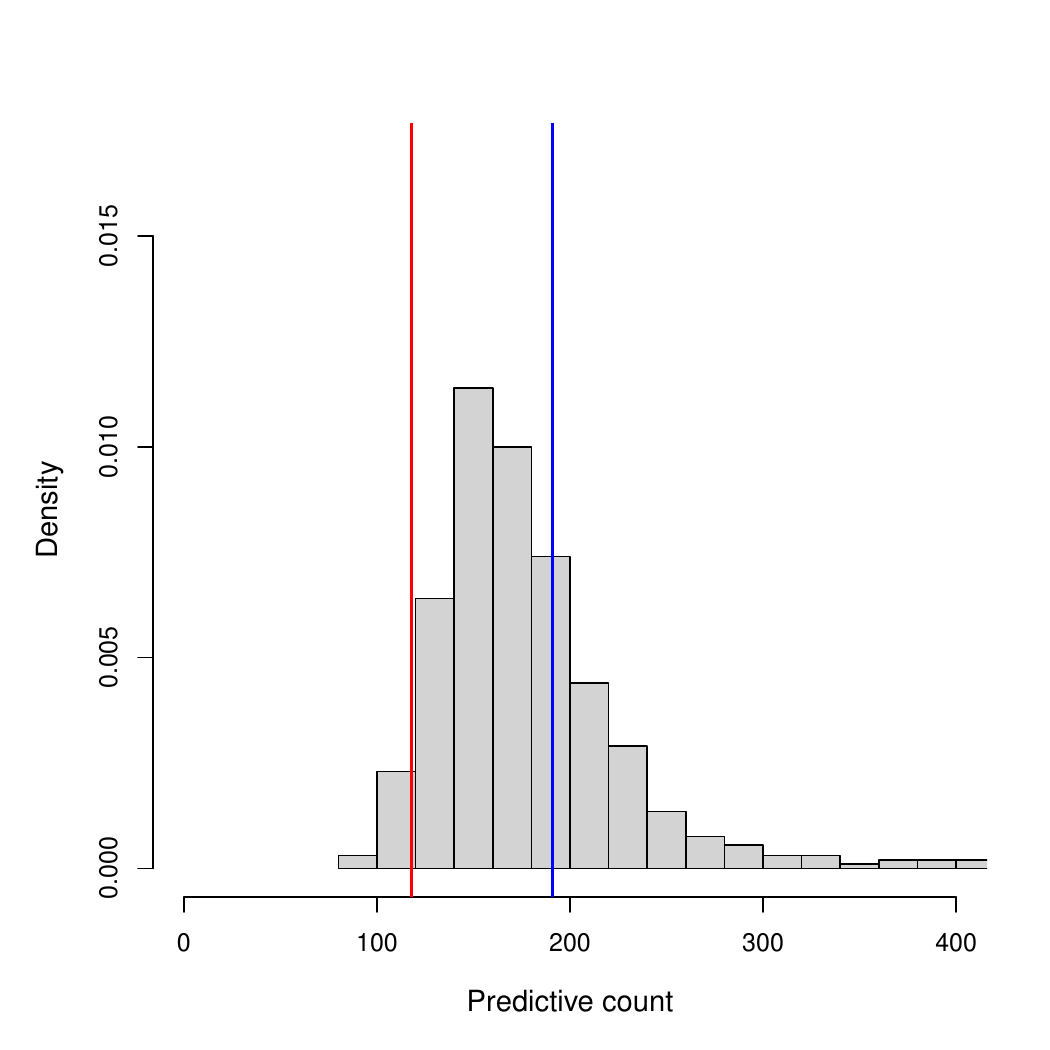}
\includegraphics[width=0.32\textwidth]{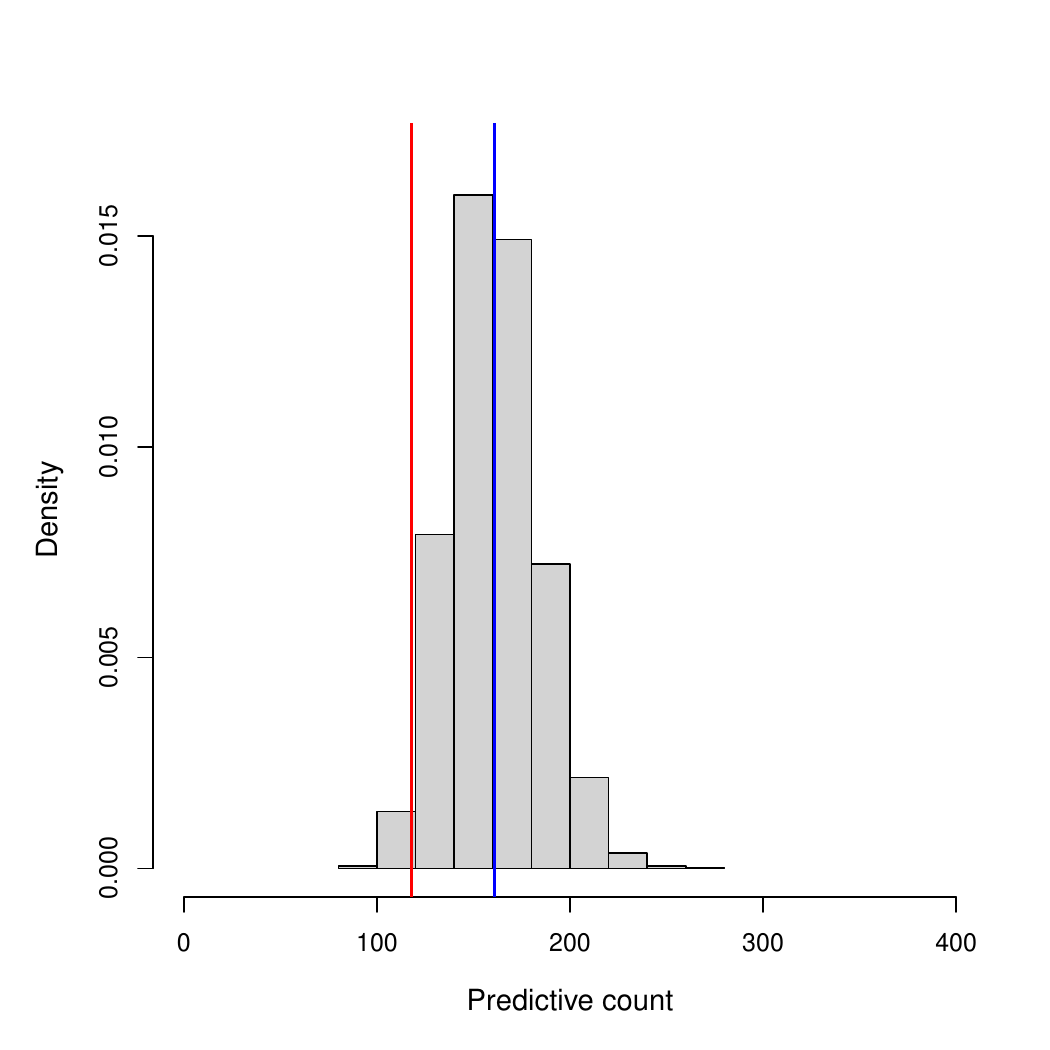} \\
\caption{{\small 
Earthquake data analysis. Posterior predictive distribution for the earthquake count 
from year 1950 to 1980, under the ETAS (left), semiparametric (middle), 
and nonparametric (right) model (all models have immigrant intensity which is constant 
in time). The blue line indicates the posterior predictive mean, and the red line the 
observed count of $118$ earthquakes.}}
\label{fig:mhp_real_japan_pred}
\end{figure}

Finally, we explore the forecast for the earthquake count from the beginning of year 1950 
to 1980, i.e., the part of the catalog not used in fitting the models. 
Figure \ref{fig:mhp_real_japan_pred} presents the posterior predictive distribution 
under each model, along with the count of $118$ earthquakes recorded in the forecast 
time interval. The ETAS and semiparametric models result in very similar posterior predictive 
distributions (these two models produced predictive draws above $400$, which are not
shown in the figure). Again, the nonparametric model outperforms 
the other two models, as it yields the smallest discrepancy between the observed count 
and the posterior predictive mean, as well as the least dispersed forecast distribution.

For numerical assessment of this (out-of-sample) prediction, we compute the 
interval score (IS) from \cite{gneiting2007}. Let $\tilde{y}$ $(= 118)$ be the count 
to be predicted, and $(l, u)$ the $(1-\alpha) 100 \%$ predictive interval.
The interval score, $\text{IS} =$
$(u - l) \, + \, (2 / \alpha) \, (l - \tilde{y}) \bm{1}(\tilde{y} < l)$ 
+ $(2 / \alpha) \, (\tilde{y} - u) \bm{1}(\tilde{y} > u)$, combines
a reward for a narrow predictive interval with a penalty when $\tilde{y}$ is 
not captured by the interval. 
The 95\% posterior predictive interval is: $(113, 377)$ (IS = 264) under the ETAS model;
$(115, 386)$ (IS = 271) under the semiparametric model; and, 
$(120, 211)$ (IS = 171) under the nonparametric model.

%
%

%
%

\subsection{General model for the MHP ground process intensity}
\label{real_data_general_model}

\begin{figure}[!t]
\centering
\includegraphics[width=6.44cm,height=6.44cm]{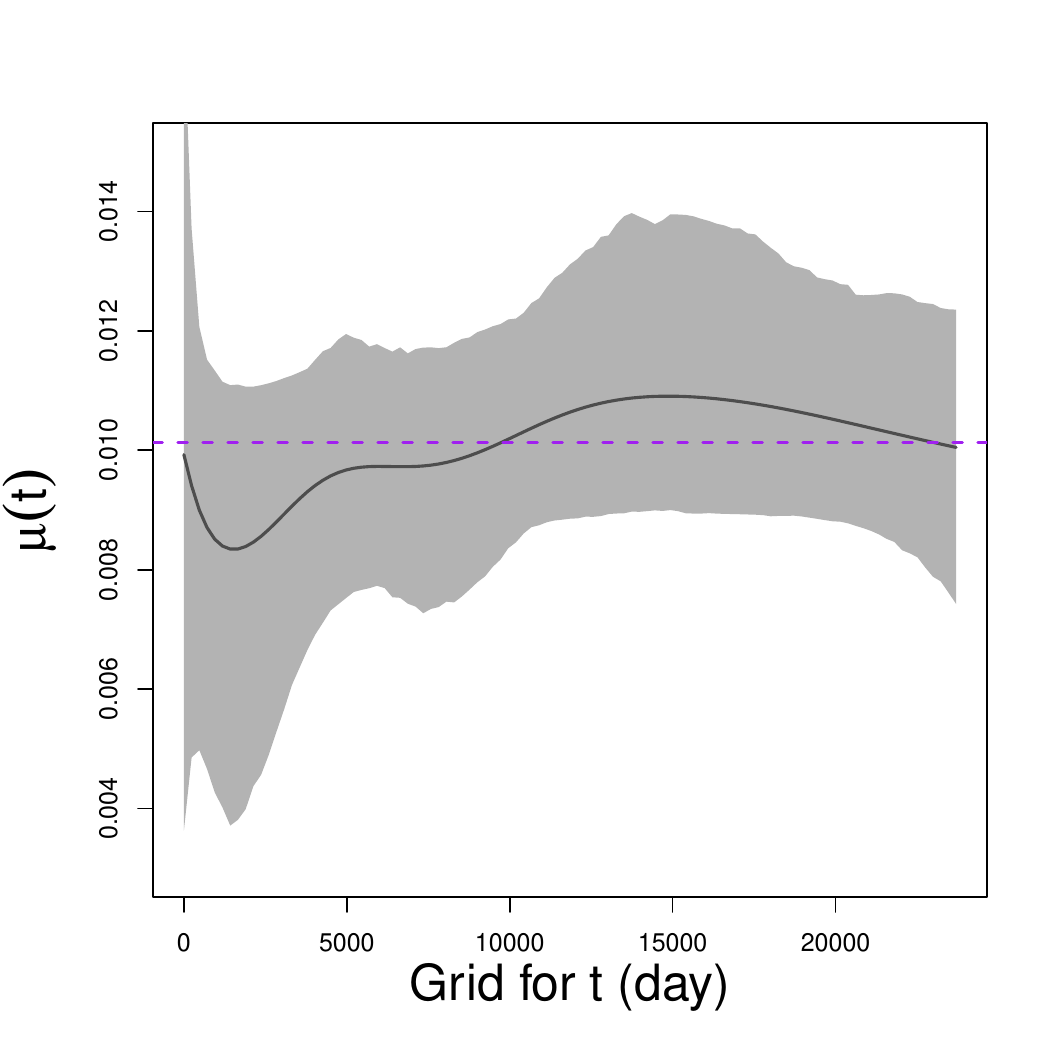}
\includegraphics[width=9.44cm,height=6.44cm]{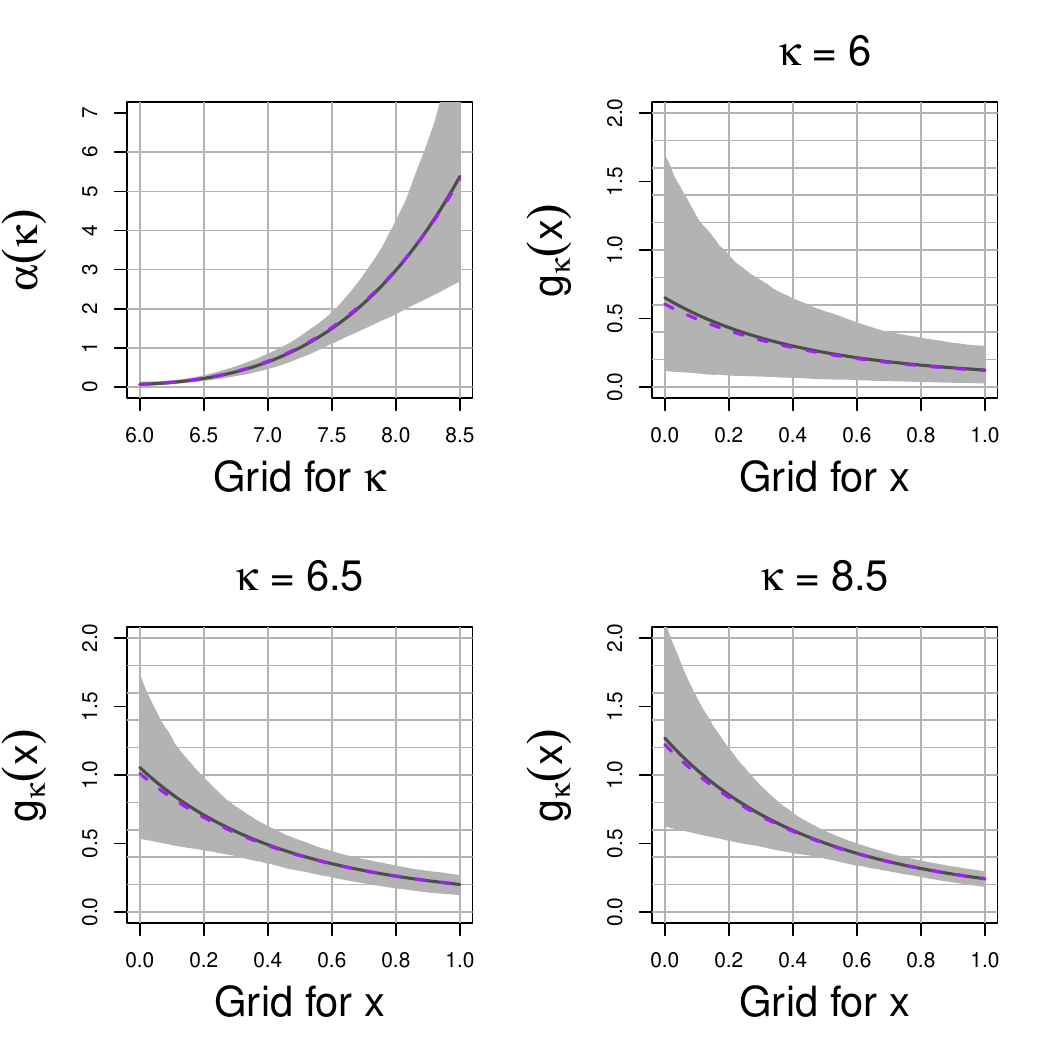}
\\
\includegraphics[width=5.14cm,height=4.44cm]{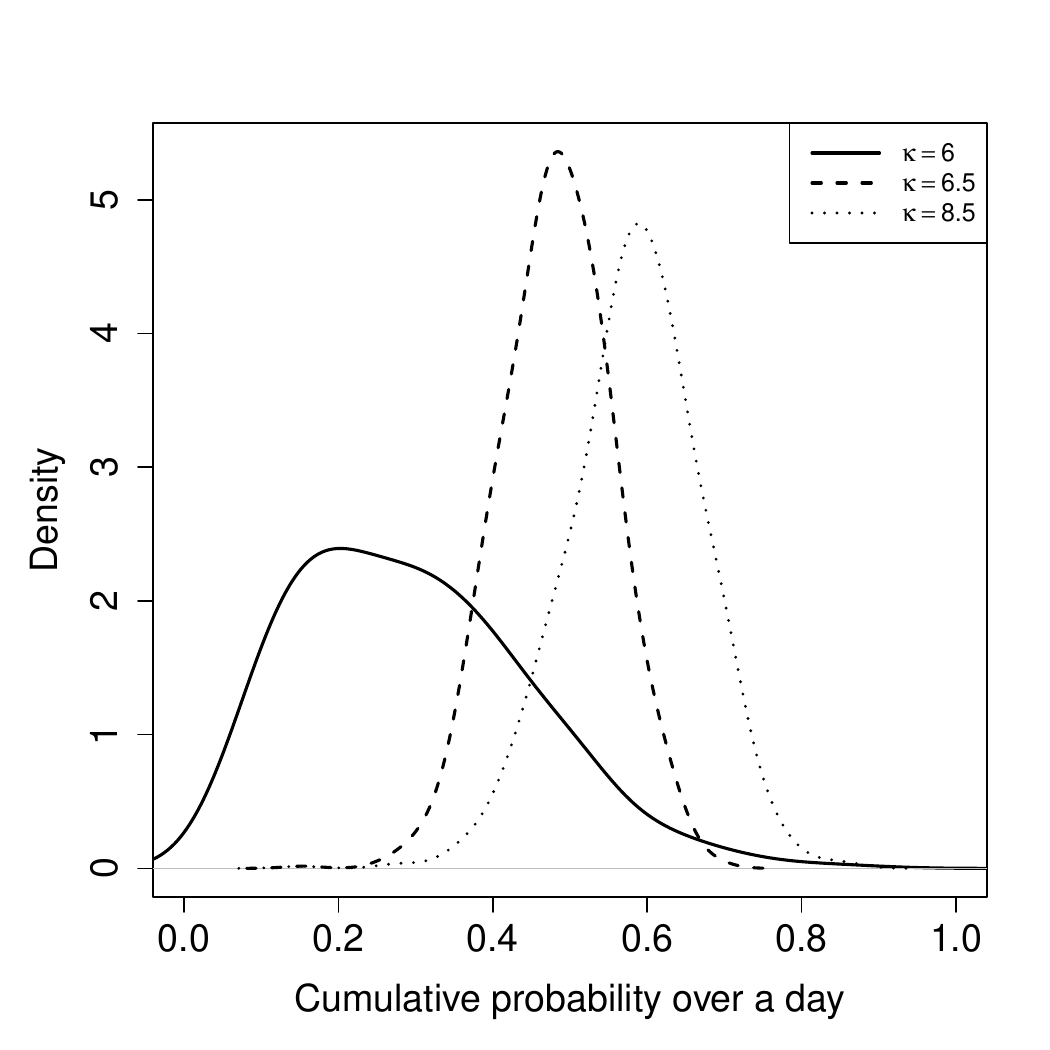}
\includegraphics[width=5.14cm,height=4.44cm]{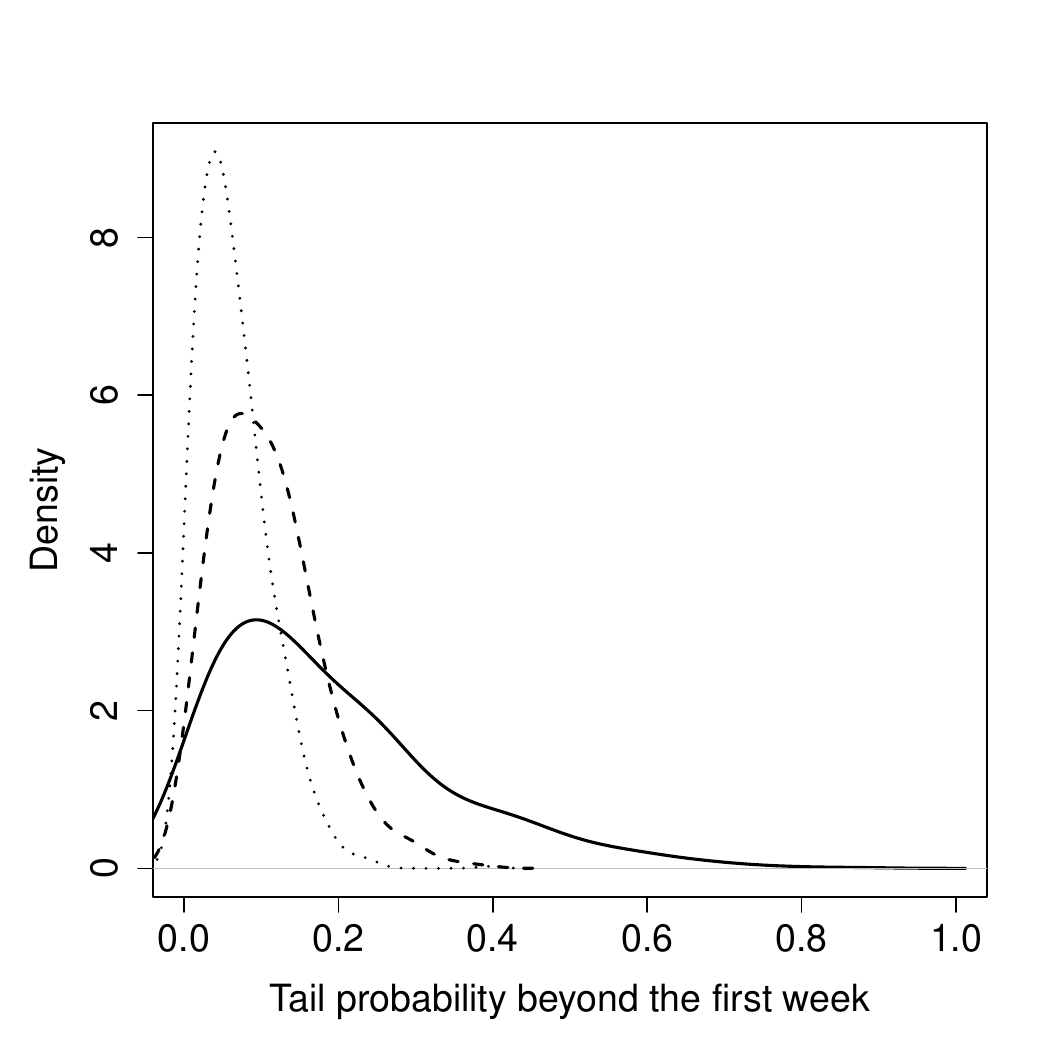} 
\includegraphics[width=5.14cm,height=4.44cm]{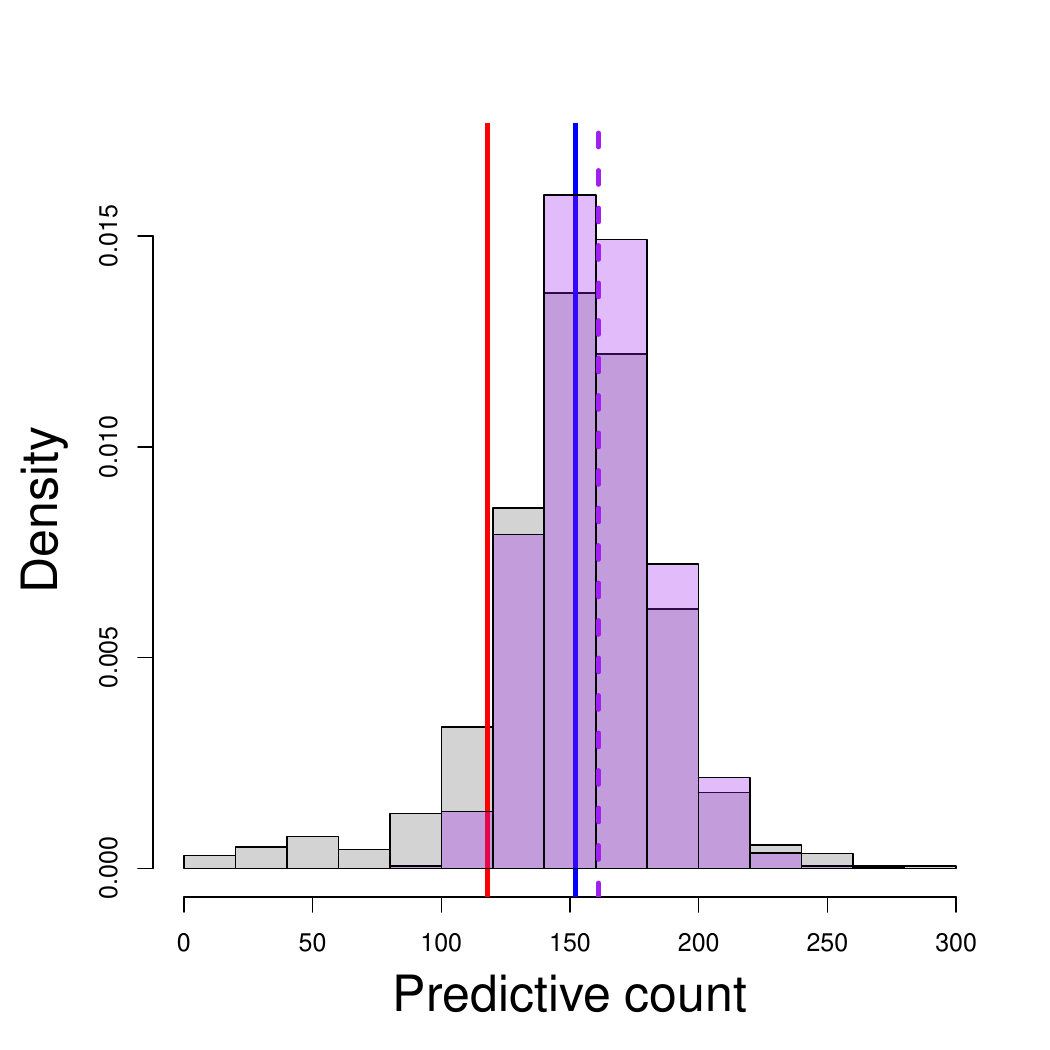} \\
\caption{{\small 
Earthquake data analysis (general nonparametric model).
Top row. Posterior mean and 95\% uncertainty bands for the immigrant intensity (left), 
and the total offspring intensity and offspring densities at magnitude $\kappa = 6$, 
$6.5$, $8.5$ (right). 
Bottom row. Posterior densities for the probability over the first day (left), and 
the tail probability beyond the first week (right), under the three offspring densities. 
The right panel plots the posterior predictive distribution for the earthquake count from 
year 1950 to 1980 (the blue/red line shows the mean/observed count). 
The purple dashed lines (purple histogram) indicate the posterior mean estimates 
(posterior predictive distribution) under the nonparametric model with immigrant 
intensity constant in time.}}
\label{fig:mhp_FBNP_real}
\end{figure}

Figure \ref{fig:mhp_FBNP_real} presents inference results using the extension of the 
nonparametric model that incorporates time-dependent immigrant intensity modeled with 
the prior in (\ref{eq:erlmix_immi}).

The estimated immigrant intensity exhibits variability over time, especially in the 
early part of the observation time window. Inference for the total offspring intensity 
and for the offspring densities (at magnitude values $\kappa = 6$, $6.5$, $8.5$) is very 
similar with the nonparametric model that has immigrant intensity constant in time. 
Figure \ref{fig:mhp_FBNP_real} explores probabilities under the three offspring densities. 
The bottom left panel plots the posterior densities for $\int_0^{1} g_\kappa(x) \, \text{d}x$, 
quantifying the increase in the probability accumulated within the first day after a shock 
of increasing magnitude. The bottom middle panel compares the posterior densities for 
$1 - \int_0^{7} g_\kappa(x) \, \text{d}x$, the tail probability beyond the first week.

Regarding the two model assessment metrics considered in 
Section \ref{real_data_constant_immigrant_intensity}, the 
posterior mean and standard deviation for the misclassification rate are 
$0.222$ and $0.012$, respectively, results which are almost identical to the ones
under the nonparametric model without time-dependent immigrant intensity.
The forecast distribution for the earthquake count from 1950 to 1980 
(Figure \ref{fig:mhp_FBNP_real}, bottom right panel) suggests improvement 
in predictive performance relative to the simpler nonparametric model. Indeed, the 
95\% posterior predictive interval under the general nonparametric model is 
$(53, 215)$, resulting in 162 for the interval score.

The analysis of the specific earthquake point pattern demonstrates the practical 
utility of the model for MHP excitation functions in (\ref{eq:nonpara_model}).
The key benefit is the capacity to estimate distinct offspring densities at different 
earthquake magnitude values for the parent shock. We quantified the improvement over 
existing methods that do not allow for magnitude-dependent offspring densities with a 
model comparison metric that is specific to MHPs (branching structure estimation) and 
a more generic one (count forecast). The general MHP model offers the additional benefit 
of investigating local temporal structure in the immigrant intensity; in this example, 
this yields further improvement in forecasting.

%
%

\section{Discussion}
\label{sec:mhp_discussion}

We have proposed a flexible Bayesian modeling method for marked Hawkes processes
(MHPs), the focus being on modeling of earthquake occurrences where the earthquake 
magnitude provides the (continuous) mark variable. We have developed a new prior 
probability model for the MHP excitation function, using a basis representation 
with components for the time lag and the mark. The weights for the basis functions 
are constructed as increments of a random measure, which is assigned a gamma 
process prior. The general version of the model includes a nonparametric prior for 
time-dependent background intensity functions, thus enabling a fully nonparametric 
approach to modeling the MHP ground process intensity. The model construction seeks 
to balance flexible forms for important MHP functionals (immigrant intensity, 
total offspring intensity, and offspring densities) with computationally tractable 
model implementation. The MCMC posterior simulation method yields full posterior 
inference (and prediction), without any approximations to the MHP likelihood or 
to the posterior distribution.

The literature on earthquake modeling is dominated by the Epidemic Type Aftershock 
Sequences (ETAS) model. The key motivation for our methodology is to enable flexible 
inference for aftershock densities that can change with earthquake magnitude. 
We therefore relax a restrictive assumption of the ETAS model, which uses the 
same (parametric) aftershock density across all magnitude values.

Analysis of a point pattern of earthquakes that occurred in Japan illustrated the 
benefits of estimating magnitude-dependent aftershock dynamics. In particular, 
earthquakes of larger magnitude tend to generate aftershocks that occur closer in 
time to their parent shock, resulting in aftershock densities that concentrate more 
probability mass at smaller values.
%
%
The nonparametric model outperformed the ETAS model in forecasting, as well as in 
estimation of the MHP branching structure. The model comparison study included
a semiparametric extension of the ETAS model, which replaces the Lomax aftershock 
density with a general nonparametric prior model, but retains the property of 
an aftershock density that does not change with magnitude. The ETAS model and 
its semiparametric extension performed similarly in forecasting, while the nonparametric 
model offered noticeable improvement. This result provides further empirical evidence 
regarding the practical utility of a model for magnitude-dependent aftershock densities.

%
%

Our modeling approach is based on the unpredictable mark assumption, where the mark 
density does not depend on the current time and the MHP (time and mark) history. 
To our knowledge, this assumption has been used in all applications of MHPs to 
earthquake modeling. A direction for future research involves incorporating more general 
mark densities that depend on the process history and/or the current time. 
Another promising direction is to develop models for space-time 
earthquake point patterns. This can be accomplished by expanding the basis representation 
for the MHP excitation function to include appropriate basis components for space, 
as well as by extending the prior model in (\ref{eq:erlmix_immi}) to accommodate 
background intensities that depend on space or on both time and space. 
We will report on the extension to modeling space-time MHPs in a future manuscript.
%
%

\section*{Acknowledgments}

This research was supported in part by the National Science Foundation 
under award SES 1950902.

\section*{Supplementary Material}

The Supplementary Material includes: details on MCMC posterior simulation for 
the nonparametric model; additional details and results for the simulation 
study (including the simulation example where the data generating mechanism is based 
on the ETAS model); and, additional results for the earthquake data analysis.

\bibliography{references}

\clearpage

%
%

\clearpage
\setcounter{page}{1}     
\appendix

\section*{Supplementary Material for ``Bayesian Nonparametric 
Marked Hawkes Processes for Earthquake Modeling''}
\begin{center}
\large
Hyotae Kim and Athanasios Kottas
\end{center}
\addcontentsline{toc}{section}{Supplementary Material} 
\setcounter{section}{0}
\renewcommand{\thesection}{S\arabic{section}}

\setcounter{figure}{0}
\renewcommand{\thefigure}{S\arabic{figure}}

\setcounter{table}{0}
\renewcommand{\thetable}{S\arabic{table}}

\setcounter{equation}{0}
\renewcommand{\theequation}{S\arabic{equation}}

\section{Posterior inference}
\label{sec:mhp_posterior_inference}

Here, we present the details for the MCMC algorithm used to implement the model 
developed in Section 3.1 of the main paper.

Using the notation for the branching structure from Section 2 of the main paper, 
the rescaled beta mark density, $f(\kappa | a_\beta,b_\beta) =$ 
$\text{be}(u_\kappa|a_\beta,b_\beta) / (\kappa_{max} - \kappa_0)$, 
$\kappa \in (\kappa_0,\kappa_{max})$, 
and the model for the MHP excitation function developed in Section 3.1.1 of the main paper, 
the augmented MHP likelihood can be written as 
\begin{equation}
\begin{split}
&\Big[\prod_{i=1}^n f(\kappa_i |a_\beta,b_\beta)\Big] \, 
\exp\Big\{-\int_0^T \mu(u) \, \text{d}u \Big\} \, \prod_{\{ t_i \in I \}}\mu(t_i) \\
&\times 
\exp\Big\{-\sum_{l=1}^L \sum_{m=1}^M \nu_{lm} \, K_{lm}(\theta,d)\Big\} \, 
\prod_{\{ t_i \in O \}}\Big\{\sum_{l=1}^L\sum_{m=1}^M \nu_{lm} \,
\text{ga}(t_i-t_{y_i}|l,\theta^{-1}) \, b_m(\kappa_{y_i}; d)\Big\},
\end{split}
\label{eq:nonpara_lkd}
\end{equation}
where $K_{lm}(\theta,d) =$
$\sum_{j=1}^n b_m(\kappa_j; d)\int_0^{T-t_j} \text{ga}(s|l,\theta^{-1}) \, \text{d}s$. 
The immigrant intensity function is modeled as either constant in time, 
$\mu(t) \equiv \mu$, or with the nonparametric Erlang mixture prior discussed in 
Section 3.1.3 of the main paper.

We further augment the likelihood in (\ref{eq:nonpara_lkd})
with auxiliary variables $\bm{\xi} =$
$ \{\bm{\xi_i}: i = 1,\ldots,N_O\}$, where $\bm{\xi_i} = (\xi_{i1},\xi_{i2})$ 
identifies the basis function to which a marked point $(t_i,\kappa_i) \in O$ is assigned. 
Let $N_O$ denote the number of offspring points categorized by the branching structure, 
i.e., $N_O =$ $|O| = |\{t_i: y_i \neq 0, i=1,\ldots,n\}|$, where $|A|$ is the cardinality 
of set $A$. Similarly, $N_I = |I| = |\{t_i: y_i = 0, i=1,\ldots,n\}|$ is the number of 
immigrant points.
Let $\bm{\Theta} = (\bm{y},\bm{\tau}_\mu,\theta,d,\bm{\xi},\bm{\nu},a_\beta,b_\beta)$, 
where $\bm{\tau}_\mu$ is a vector of parameter(s) in $\mu(t)$. 
The full hierarchical model for observed point patterns can then be written as:
\begin{equation}
    \begin{split}
        p\big((\bm{t},\bm{\kappa})|\bm{\Theta}\big) &\propto 
        \Big[\prod_{i=1}^n \text{be}(u_\kappa|a_\beta,b_\beta) \Big]\\
        &\times \exp\Big\{-\int_0^T \mu(u) du\Big\}\exp\Big\{-\sum_{l=1}^L\sum_{m=1}^M\nu_{lm}K_{lm}(\theta,d)\Big\}\\
        &\times \prod_{t_i \in I}\mu(t_i)\prod_{t_i \in O}\Big\{\Big(\sum_{r_1=1}^L\sum_{r_2=1}^M \nu_{r_1 r_2}\Big) \text{ga}(t_i-t_{y_i}|\xi_{i1},\theta^{-1})b_{\xi_{i2}}(\kappa_{y_i}; d)\Big\};\\
        p(\bm{\xi_i}|\bm{\nu}) &\propto \sum_{l=1}^L\sum_{m=1}^M \frac{\nu_{lm}}{\Big(\displaystyle\sum_{r_1=1}^L\sum_{r_2=1}^M \nu_{r_1 r_2}\Big)} \delta_{(l,m)}(\xi_{i1},\xi_{i2}), \quad i=1,\ldots,N_O;\\
        p(\bm{y}) &\propto \delta_0(y_1)\prod_{i=2}^n \text{unif}(y_i|0,1,\ldots,i-1);\\
        \nu_{lm}|c_0,b_1,b_2,\theta &\indsim \text{Ga}(c_0 H_0(A_{lm}),c_0), \quad l=1,\ldots,L, \quad m=1,\ldots,M;\\
        (\bm{\tau}_\mu,\theta,d) &\sim p(\bm{\tau}_\mu)\text{Lomax}(2,a_\theta)\text{Exp}(a_d);\\
        (c_0,b_1,b_2) &\sim \text{Lomax}(2,a_{c_0})\text{Exp}(a_{b_1})\text{Exp}(a_{b_2});\\
        (a_\beta,b_\beta) &\sim \text{Exp}(a_{a_\beta})\text{Exp}(a_{b_\beta}),\\
    \end{split}
\end{equation}
where $\delta_0(x)$ and $\delta_{(a,b)}(x,y)$ are delta functions defined as $\delta_0(x)=1$ if $x=1$ and $0$ otherwise, and $\delta_{(a,b)}(x,y)=1$ if $(x,y)=(a,b)$ and $0$ otherwise. $\text{unif}(x|0,1,\ldots,i-1)$ is the discrete uniform probability mass function. The prior distribution for $\bm{\tau}_\mu$, denoted by $p(\bm{\tau}_\mu)$, is specified as either an exponential distribution for the constant immigrant intensity $\mu$, or as the product of a Lomax distribution and two exponential distributions for the parameters $(\phi, e_0, b_{G_0})$ in the Erlang–mixture model for the immigrant intensity given in \textcolor{red}{(10)}. Note that the mixture weights $(\omega_j)$ in the immigrant intensity mixture model follow gamma distributions induced by the gamma process prior placed on $G$.

Posterior inference was carried out using the Gibbs sampler, in which simple and efficient updates of $\bm{\xi}$, $\bm{y}$, and $\bm{\nu}$ were made through their closed-form posterior full conditional distributions. We denote the observed data by $\bm{D} = \{\bm{t}, \bm{k}\}$ for use in the derivation of the full conditionals below.

The full conditional for the auxiliary variable $\bm{\xi}_i$ is an independent discrete distribution supported on $\{(l,m): l=1,\ldots,L, \enspace m=1,\ldots,M\}$, with
$\text{Pr}\big((\xi_{i1}=l,\xi_{i2}=m)|\bm{y},\theta,d,\bm{\nu},\bm{D}\big) \propto \nu_{lm}\text{ga}(t_i-t_{y_i}|l,\theta^{-1})b_m(\kappa_{y_i}; d)$.

Given the discrete uniform priors, the branching structures $y_i$ for $i=2,\ldots,n$ 
also have discrete distributions as their posterior full conditionals with probabilities
\begin{equation*}
    \begin{split}
    	\text{Pr}(y_i=k|\mu,\theta,\bm{\nu},d,\bm{D}) = 
    	\begin{cases}
    		\dfrac{\mu}{\mu+\displaystyle\sum^{i-1}_{r=1}\sum_{l=1}^L\sum_{m=1}^M \nu_{lm} \text{ga}(t_i-t_r|l,\theta^{-1})b_m(\kappa_r; d)}, \enspace k=0;\\
    		\\
    		\dfrac{\displaystyle\sum_{l=1}^L\sum_{m=1}^M \nu_{lm} \text{ga}(t_i-t_k|l,\theta^{-1})b_m(\kappa_k; d)}{\mu+\displaystyle\sum^{i-1}_{r=1}\sum_{l=1}^L\sum_{m=1}^M \nu_{lm} \text{ga}(t_i-t_r|l,\theta^{-1})b_m(\kappa_r; d)}, \enspace k=1,\ldots,i-1.
    	\end{cases}		
    \end{split}		
\end{equation*}

Let $n_{lm} = |{t_i : \xi_{i1} = l,, \xi_{i2} = m,, y_i \neq 0}|$ for $l = 1,\ldots,L$ and $m = 1,\ldots,M$. The full conditional distribution for $\bm{\nu}$ can be expressed as 
$$
p(\bm{\nu}|\bm{\xi},\theta,c_0,b_1,b_2,\bm{\kappa}) \propto 
\prod_{l=1}^L\prod_{m=1}^M \Big[\exp\Big\{-\nu_{lm}K_{lm}(\theta,d)\Big\} \,
\nu_{lm}^{n_{lm}} \, \text{ga}(\nu_{lm}|c_0 H_0(A_{lm}),c_0)\Big] .
$$
Hence, given the other model parameters and the data, the mixture weights are independently 
gamma distributed, with each $\nu_{lm}$ sampled from $\text{Ga}(c_0H_0(A_{lm})+n_{lm},c_0+K_{lm}(\theta,d))$. 

Other parameters $\theta$, $d$, $c_0$, $b_1$, $b_2$, $a_\beta$, and $b_\beta$ are updated with the Metropolis-Hastings (M-H) algorithm, in which log-normal distributions are used as proposal distributions.

For the constant immigrant intensity $\mu$, the $\text{Exp}(\lambda_\mu)$ prior results in a gamma posterior full conditional with shape $N_I+1$ and rate $T+\lambda_\mu$. For the Erlang mixture-based immigrant intensity, we first represent the likelihood of the immigrant process, $\exp\Big\{-\sum_{j=1}^{J} \omega_{j}\int_0^T \text{ga}(u \mid j,\phi^{-1}) du\Big\}\prod_{t_i \in I}\sum_{j=1}^{J} \omega_{j}\text{ga}(t_i \mid j,\phi^{-1})$, as a hierarchical model with auxiliary variables $\bm{\zeta}=\{\zeta_i:i=1,\ldots,N_I\}$, that is,
\begin{equation*}
    \begin{split}
       p(\bm{t}|\bm{y},\phi,\bm{\zeta},\bm{\omega}) &\propto \exp\Big\{-\sum_{j=1}^{J} \omega_{j}\int_0^T \text{ga}(u \mid j,\phi^{-1}) du\Big\}\prod_{t_i \in I}\Big\{\Big(\sum_{j=1}^{J} \omega_{j}\Big)\text{ga}(t_i \mid \zeta_i,\phi^{-1})\Big\};\\
       p(\zeta_i|\bm{\omega}) &\propto \sum_{j=1}^J \frac{\omega_j}{\Big(\sum_{r=1}^J \omega_r\Big)}\delta_j(\zeta_i), \quad i=1,\ldots,N_I.
    \end{split}
\end{equation*}
Each mixture weight $\omega_j$, under its 
$\text{Ga}(e_0\phi/b_{G_0},e_0)$ prior, is updated independently from the gamma full conditional: $\text{Ga}\big(e_0\phi/b_{G_0}+n_j,e_0+\int_0^T \text{ga}(u \mid j,\phi^{-1}) du \big)$, with $n_j=|\{t_i: \zeta_i = j, y_i = 0\}|$. 
The auxiliary variables $(\zeta_i)$ have independent discrete full conditional distributions 
with $\text{Pr}(\zeta_i=j|\phi,\bm{\omega},\bm{D}) \propto \omega_{j}\text{ga}(t_i|j,\phi^{-1})$. 
Updates of the other parameters $(\phi,e_0,b_{G_0})$ are accomplished via the M-H algorithm 
with log-normal proposal distributions.

%
%

\section{Simulation study: additional details and results}
\label{sec:additional_results_syntheticdata}

\subsection{MCMC posterior simulation for the comparison models}

The simulation study includes comparison with the ETAS model, and the semiparametric 
model extension that replaces the Lomax offspring density with the DP uniform mixture 
presented in Section 4.1 of the main paper.

Posterior simulation for the ETAS model is performed with a Gibbs sampler, which includes
Metropolis-Hastings (M-H) updates for some of the parameters. Each ETAS model parameter
is assigned an exponential prior distribution, with its rate parameter denoted by 
$\lambda_{d}$, where $d$ corresponds to the specific parameter. Recall that the exponential 
priors for parameters $a$, $b$ and $\psi$ are truncated subject to the constraints 
required to satisfy the stability condition $0 < a \psi/(\psi-b) < 1$
(with probability $1$ in the prior). Denote by $N_I =$ $| \{ t_i : y_i = 0 \} |$ 
and $N_O =$ $| \{t_i : y_i \neq 0 \} |$ the number of observations classified as immigrant 
points and offspring points, respectively,

The posterior full conditional for the immigrant intensity parameter, $\mu$, is a gamma 
distribution, $\text{Ga}(N_I + 1,\, T + \lambda_\mu)$, where $T$ is the upper bound of 
the observation window. 
The posterior full conditional for $a$ is a gamma distribution, with shape 
parameter $N_O + 1$, and rate parameter 
$\lambda_{a} + \sum_{j=1}^{n} \exp \{ b \, (\kappa_j - \kappa_0) \}
\int_0^{T - t_j} \text{Lomax}(u| p, c)\, \text{d}u$, truncated to $a < (\psi - b)/\psi$.
The posterior full conditional for $\psi$ is a gamma distribution, with 
shape parameter $n + 1$, and rate parameter 
$\lambda_{\psi} + \sum_{i=1}^{n} (\kappa_i - \kappa_0)$, truncated to $\psi > b / (1 - a)$. 
Finally, parameter $b$ and the Lomax offspring density parameters, $p$ and $c$, are updated
using M-H steps.

MCMC posterior simulation is more involved for the semiparametric model. However, using 
slice sampling auxiliary variables (in addition to the branching structure latent variables),
it is possible to employ established techniques for DP mixture models. Details are 
provided in \cite{K2021}, where the DP uniform mixture was used to build a nonparametric prior 
for the offspring density of Hawkes processes without marks.

\subsection{Lomax offspring density example}
\label{SM:mhp_simulation_powlaw}

Here, we present the first simulation example, where the point pattern is generated 
from the ETAS model. 

The synthetic point pattern comprises $906$ time points ($102$ immigrants, 
$804$ offspring points), recorded within interval $(0,T) =$ $(0,5000)$. 
The pattern is generated from a MHP with ground intensity given by Equation (11) 
of the main paper, with $\mu = 0.02$, $a = 0.47$, $b = 0.5$, 
and $g_{\kappa}(x) \equiv g(x) =$ $\text{Lomax}(x|20,2)$. The rate parameter of the 
truncated exponential mark density is $\psi = 1$. This is essentially the ETAS model 
adjusted for the bounded mark space.

The priors for the nonparametric model are: $\text{Lomax}(2,0.1)$ for $\theta$; 
$\text{Exp}(10)$ and $\text{Exp}(8)$ for $b_1$ and $b_2$; $\text{Exp}(1)$ for $d$; 
$\text{Lomax}(2,2000)$ for $c_0$; and, $\text{Exp}(1)$ and $\text{Exp}(0.204)$ for 
$a_\beta$ and $b_\beta$. Moreover, the numbers for the basis components were
set at $L=15$ and $M=20$.

\begin{figure}[!t]
\centering
\includegraphics[width=0.29\textwidth]{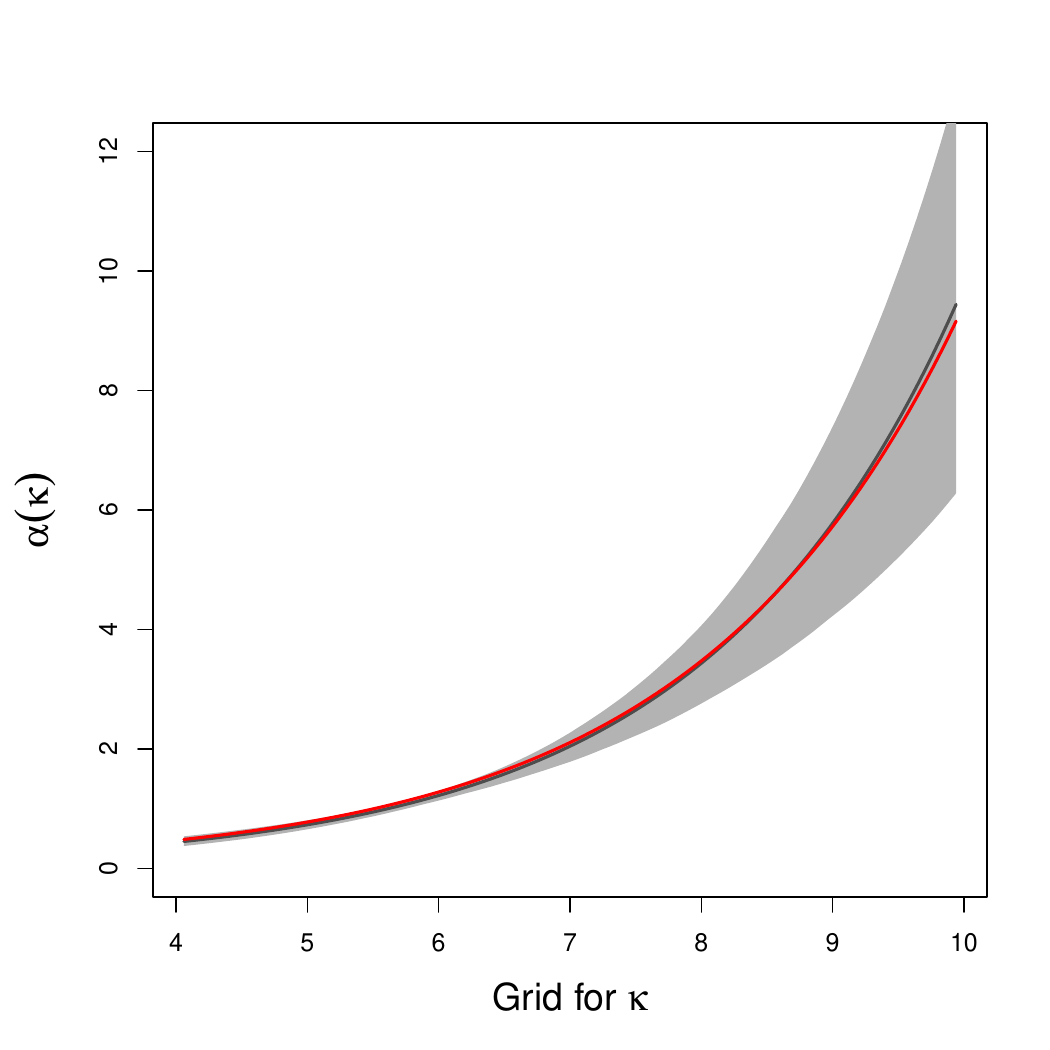}
\includegraphics[width=0.29\textwidth]{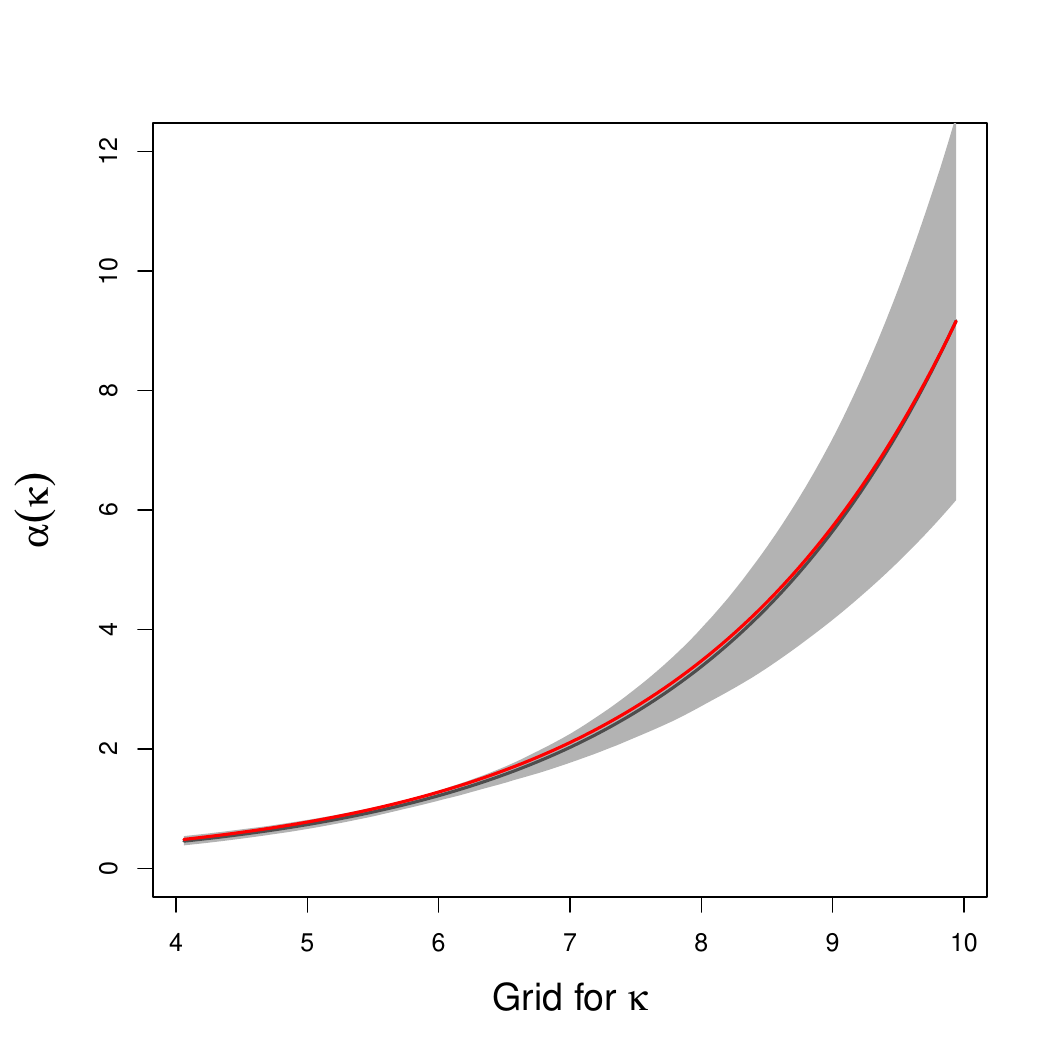}
\includegraphics[width=0.29\textwidth]{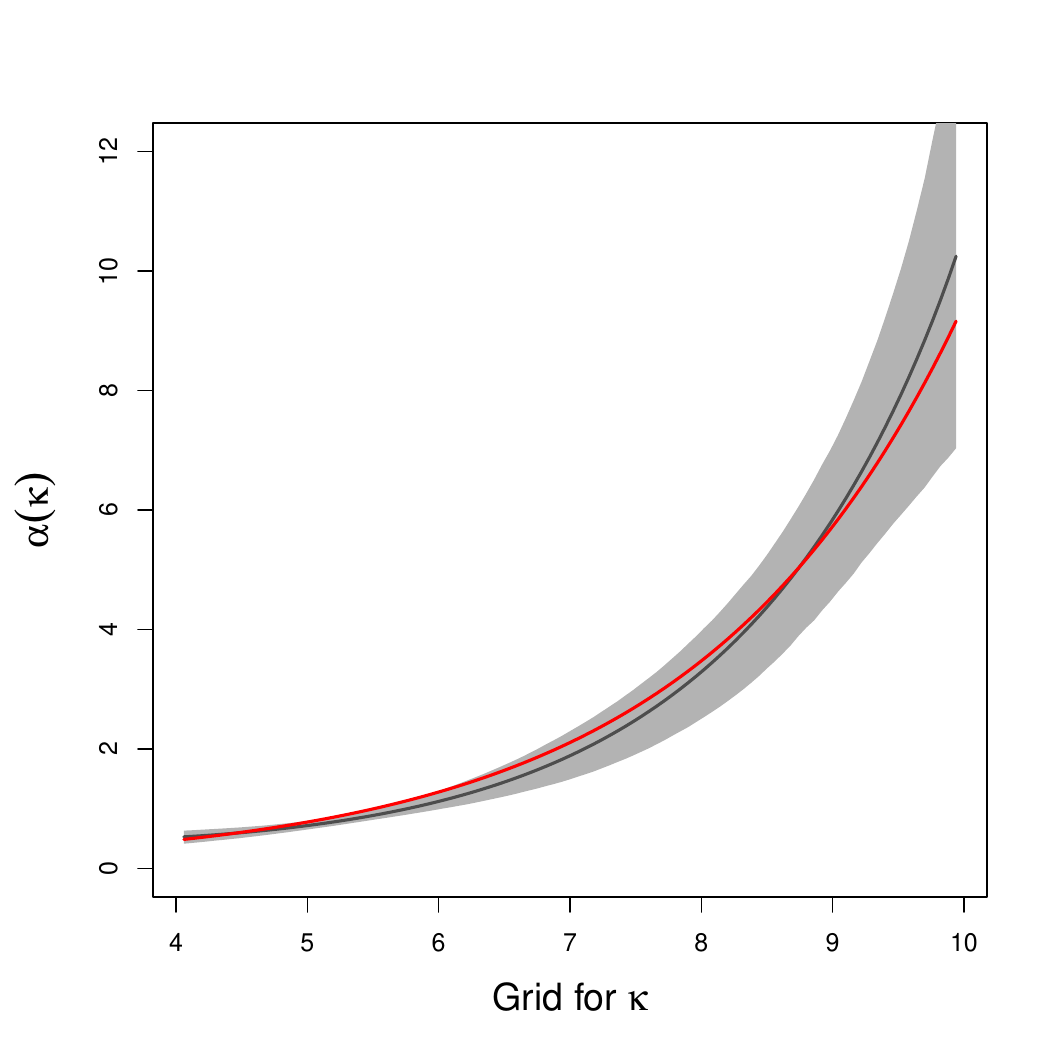} \\
\includegraphics[width=0.24\textwidth]{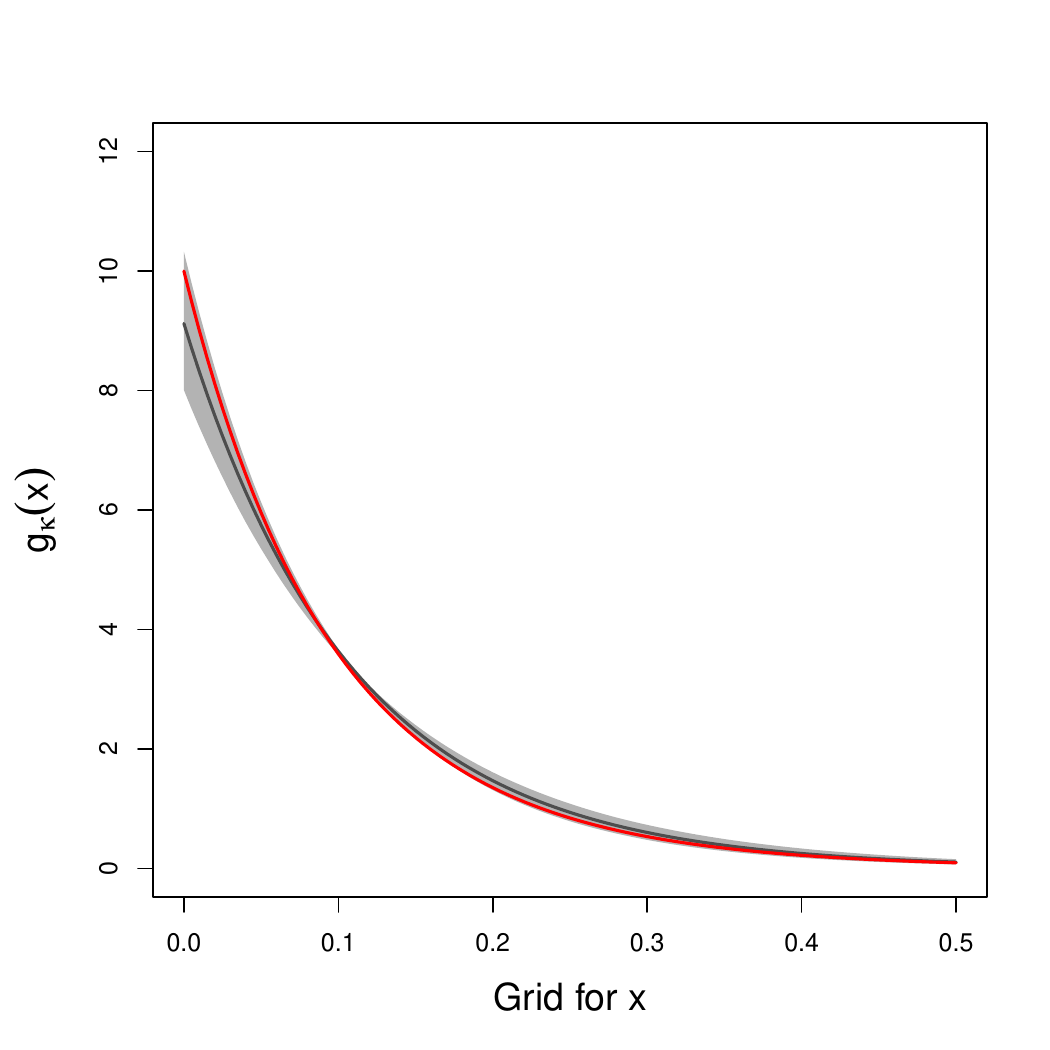}
\includegraphics[width=0.24\textwidth]{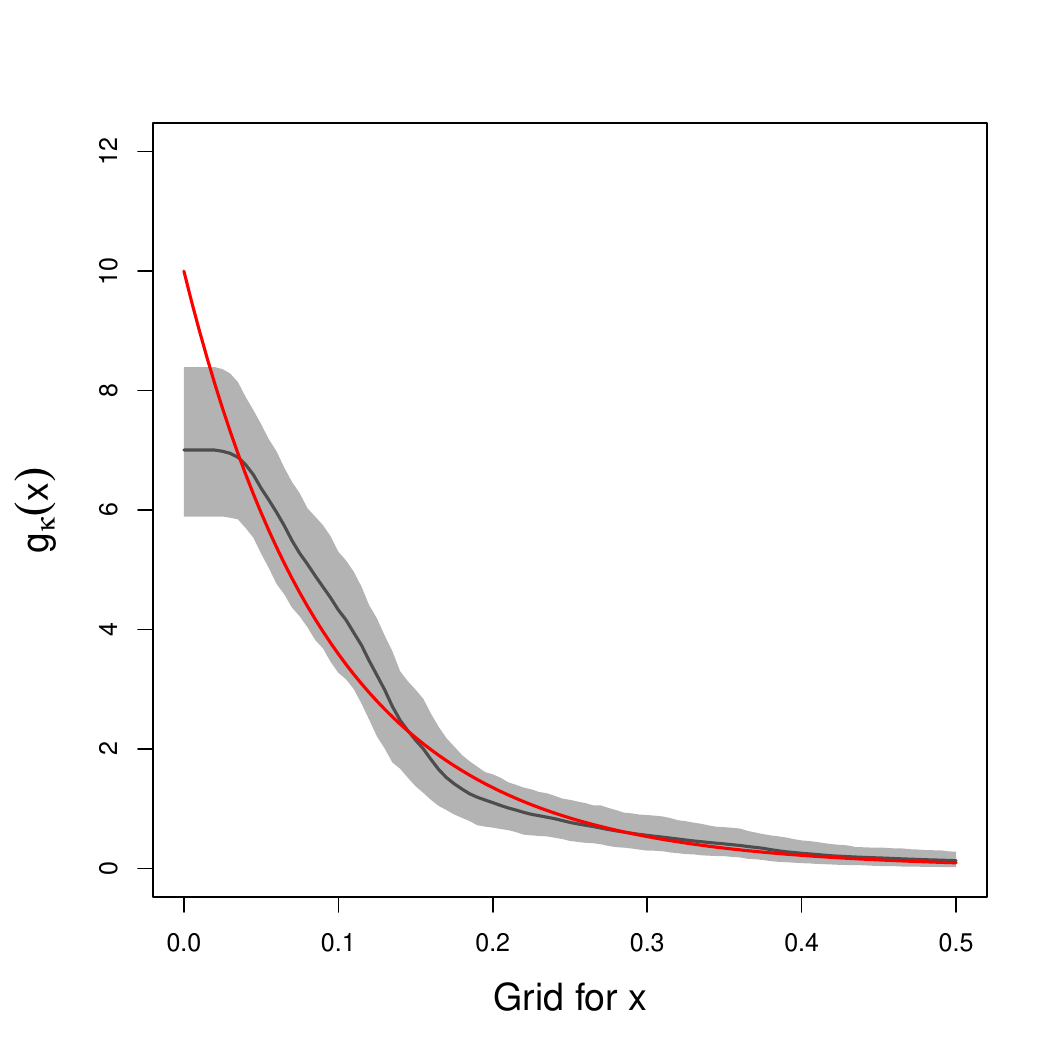}
\includegraphics[width=0.24\textwidth]{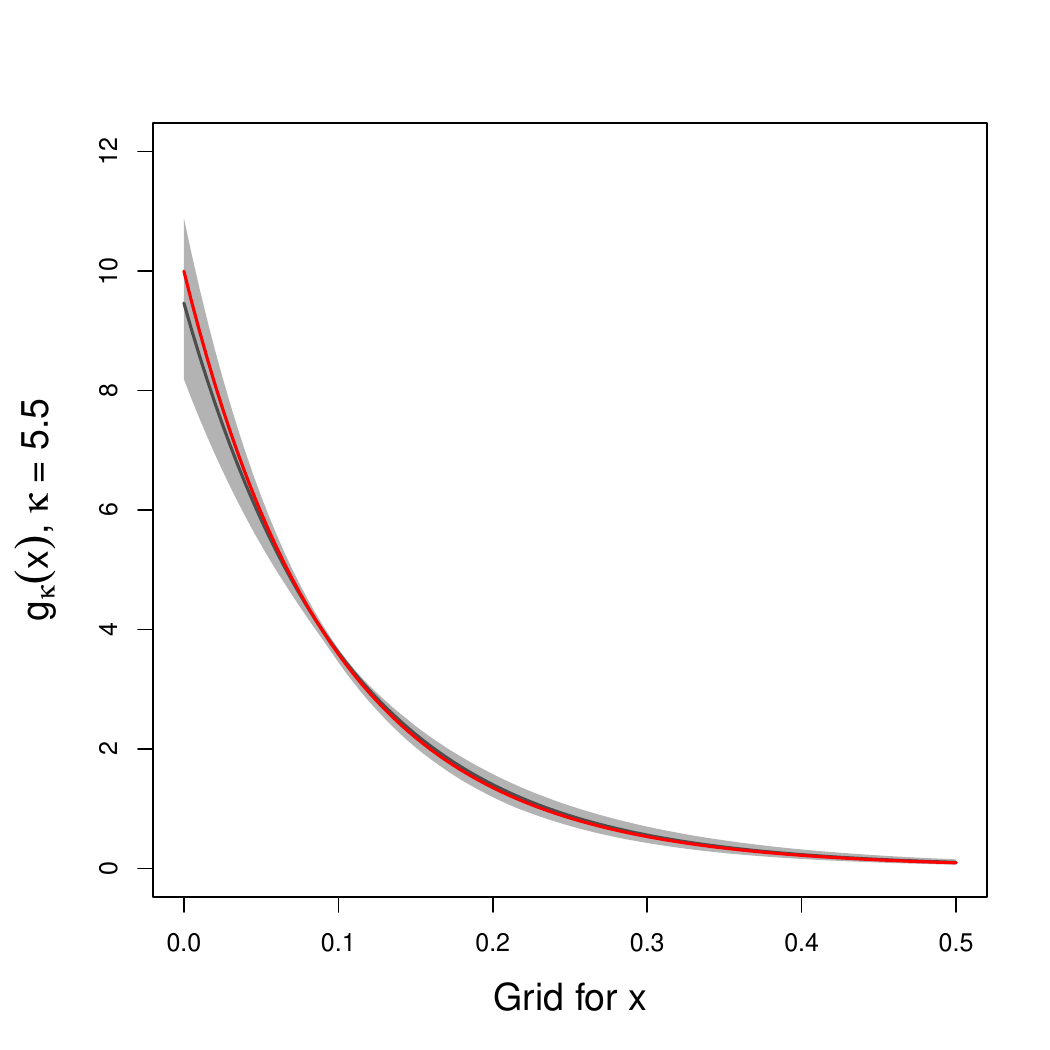}
\includegraphics[width=0.24\textwidth]{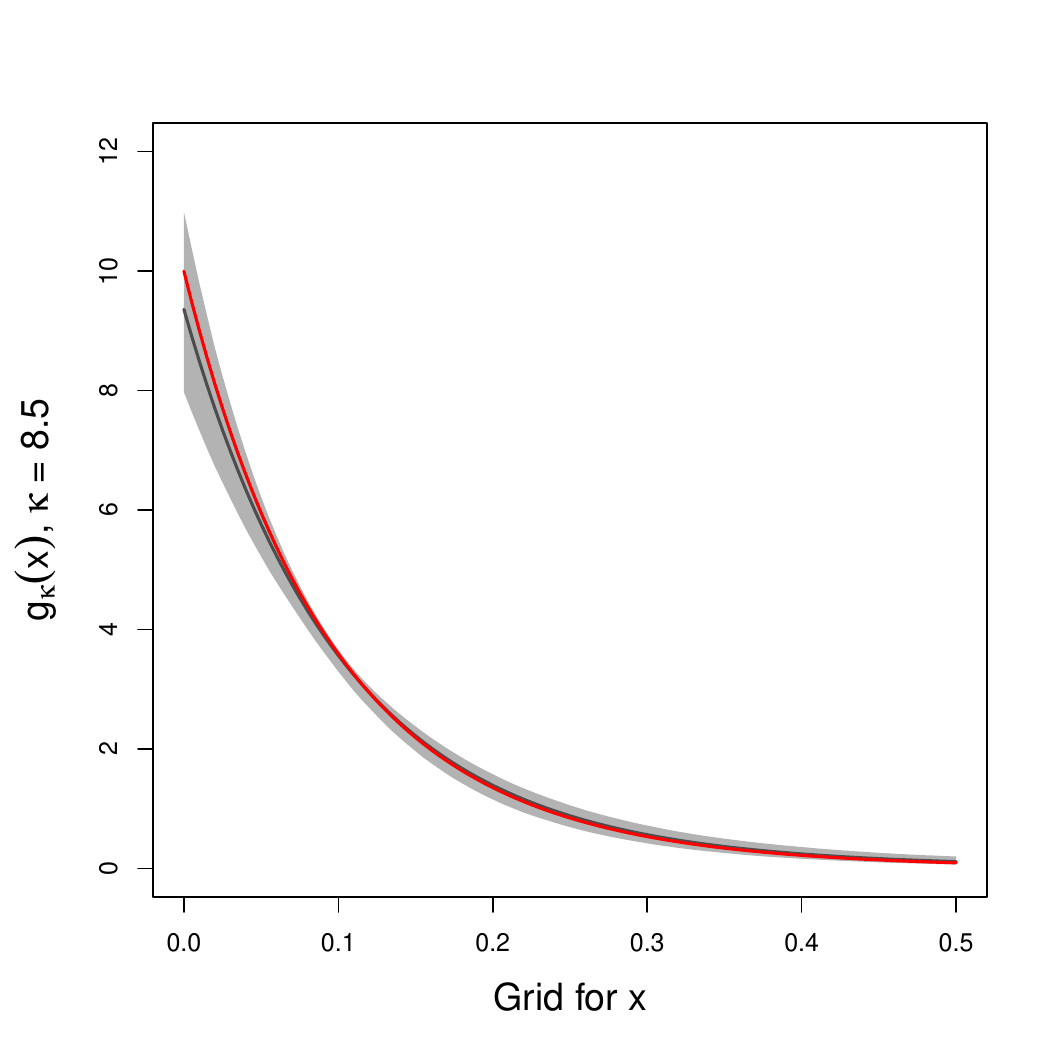} \\
\caption{
{\small Lomax density simulation example.
Top row: posterior mean (black line) and 95\% uncertainty bands for the total offspring 
intensity, under the ETAS, semiparametric, and nonparametric model (left, middle, right). 
Bottom row: posterior mean (black line) and 95\% uncertainty bands for the offspring density,  
under the ETAS model (first column), semiparametric model (second column), and nonparametric 
model for two mark values, $\kappa = 5.5, 8.5$ (last two columns). In each panel, the 
function used to generate the data is denoted by the red line.}}
\label{fig:mhp_powlaw_excitation}
\end{figure}

For this example, the ETAS and semiparametric models use the same (exponential) total offspring 
intensity function with the simulation truth; furthermore, the ETAS model is based on the 
offspring density used to generate the point pattern. This is reflected in the corresponding 
inference results reported in Figure \ref{fig:mhp_powlaw_excitation}, where the ETAS model 
point and interval estimates can be viewed as the gold standard. 
The nonparametric model yields a somewhat less accurate posterior mean estimate for the total 
offspring intensity than the ETAS model and its semiparametric extension. Again, those two 
models use the same parametric form for $\alpha(\kappa)$ with the data generating mechanism.

Evidently, mark-dependent offspring densities are not required for this simulation setting. 
Nonetheless, it is encouraging that the nonparametric model produces essentially the same 
point and interval estimates for the offspring density $g_{\kappa}(x)$ across the mark space; 
Figure \ref{fig:mhp_powlaw_excitation} plots the estimates for $\kappa = 5.5$ and $8.5$. 
It is also noteworthy that the posterior uncertainty bands are only slightly wider than under 
the ETAS model. This is in contrast with the semiparametric model for which the increase in 
posterior uncertainty is noticeable.

\subsection{Additional results for the nonparametric model}
\label{subsec:additional_figures_simulation}

\begin{figure}[!t]
\centering
\includegraphics[width=0.19\textwidth]{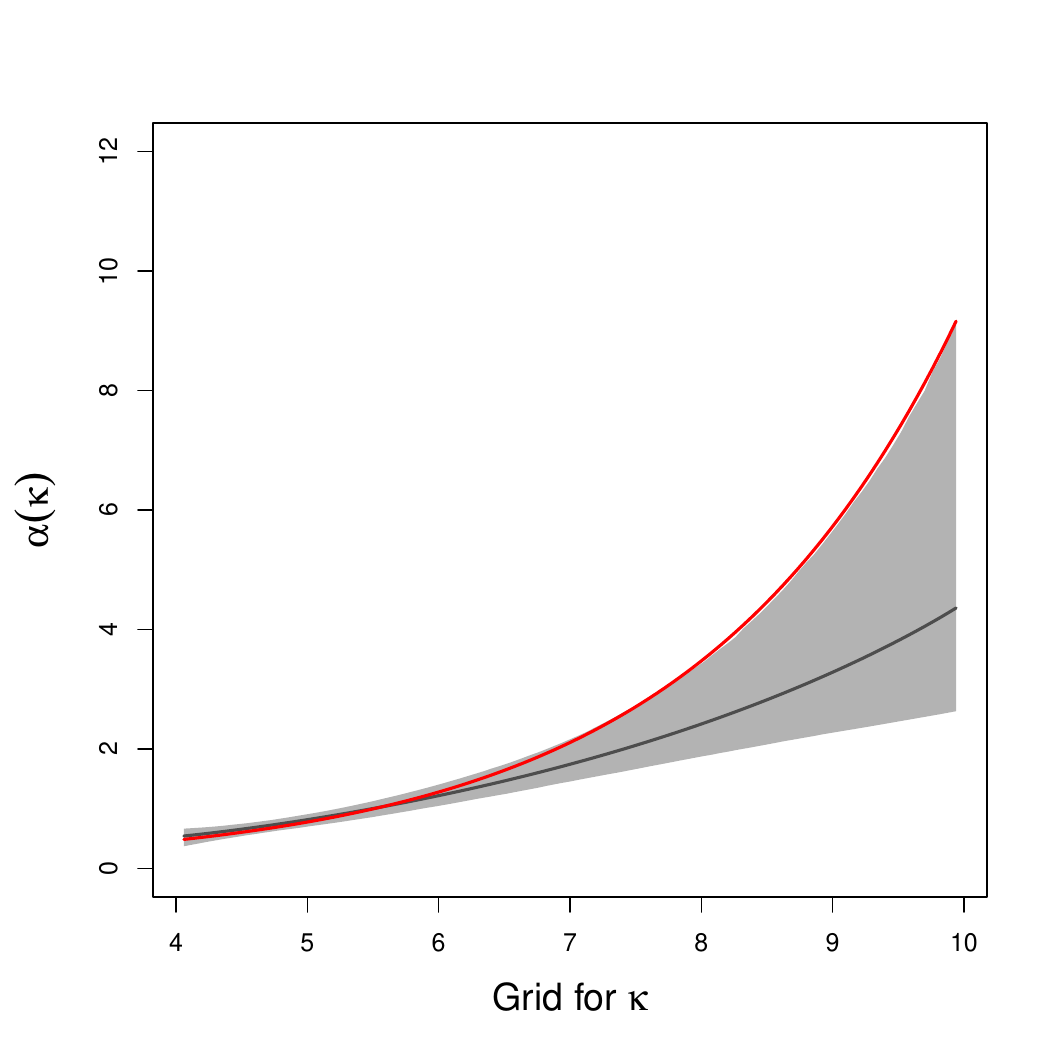}
\includegraphics[width=0.19\textwidth]{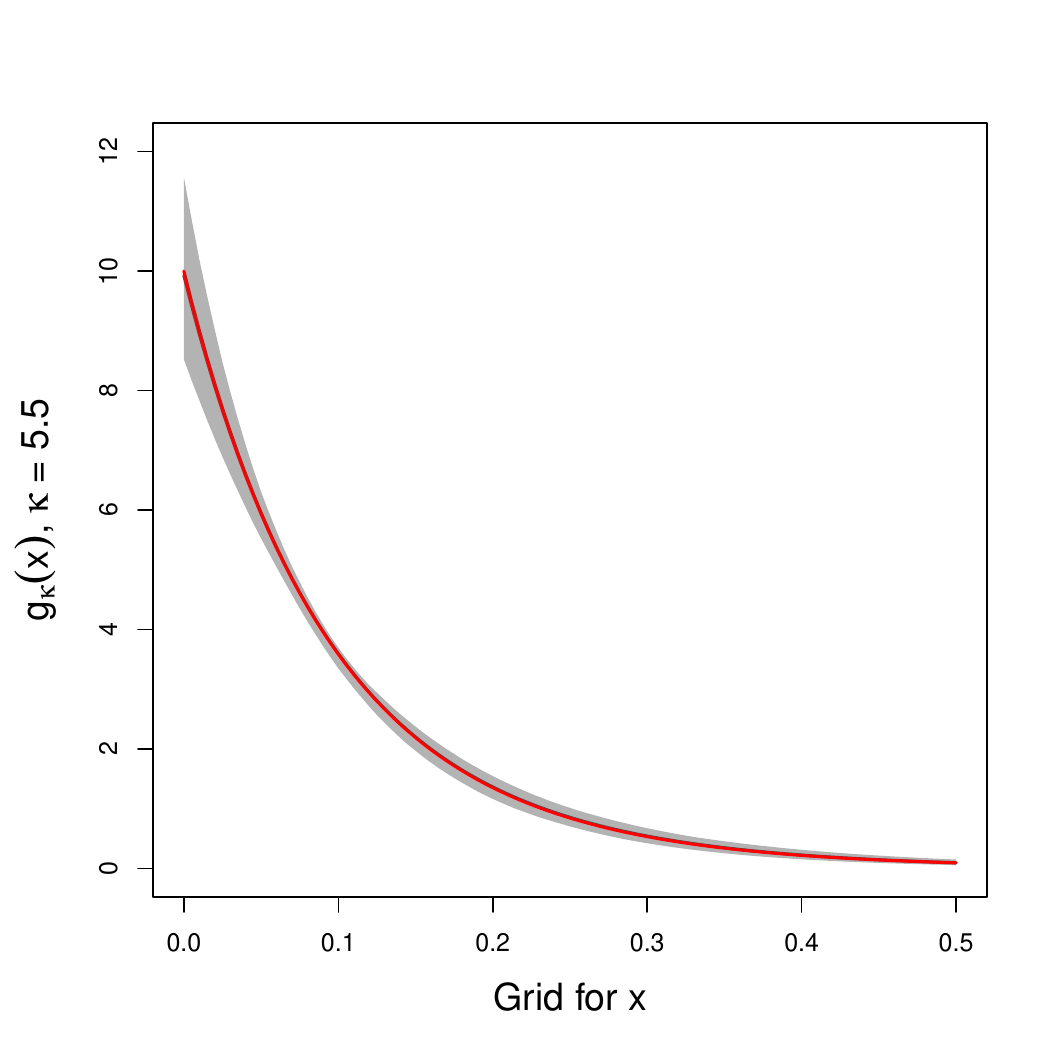}
\includegraphics[width=0.19\textwidth]{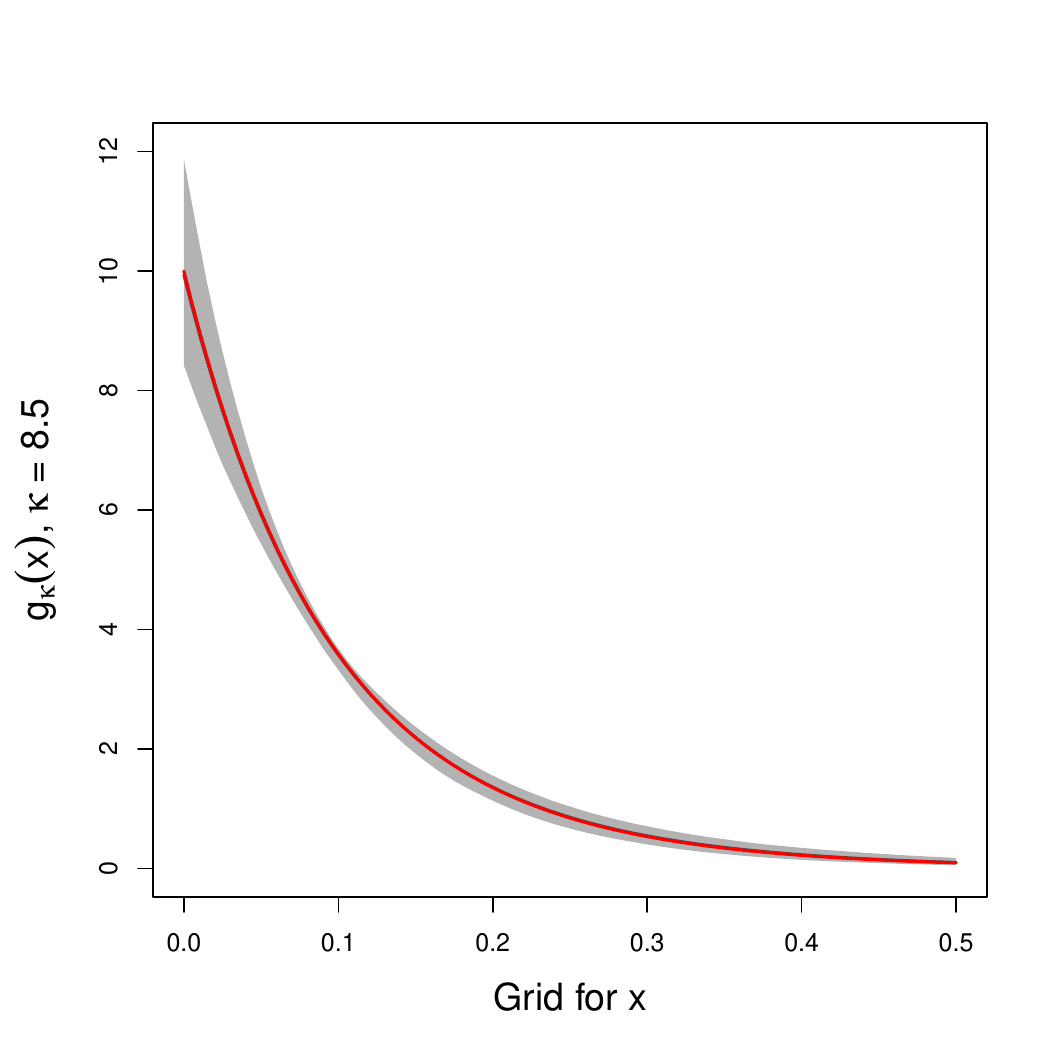}
\includegraphics[width=0.19\textwidth]{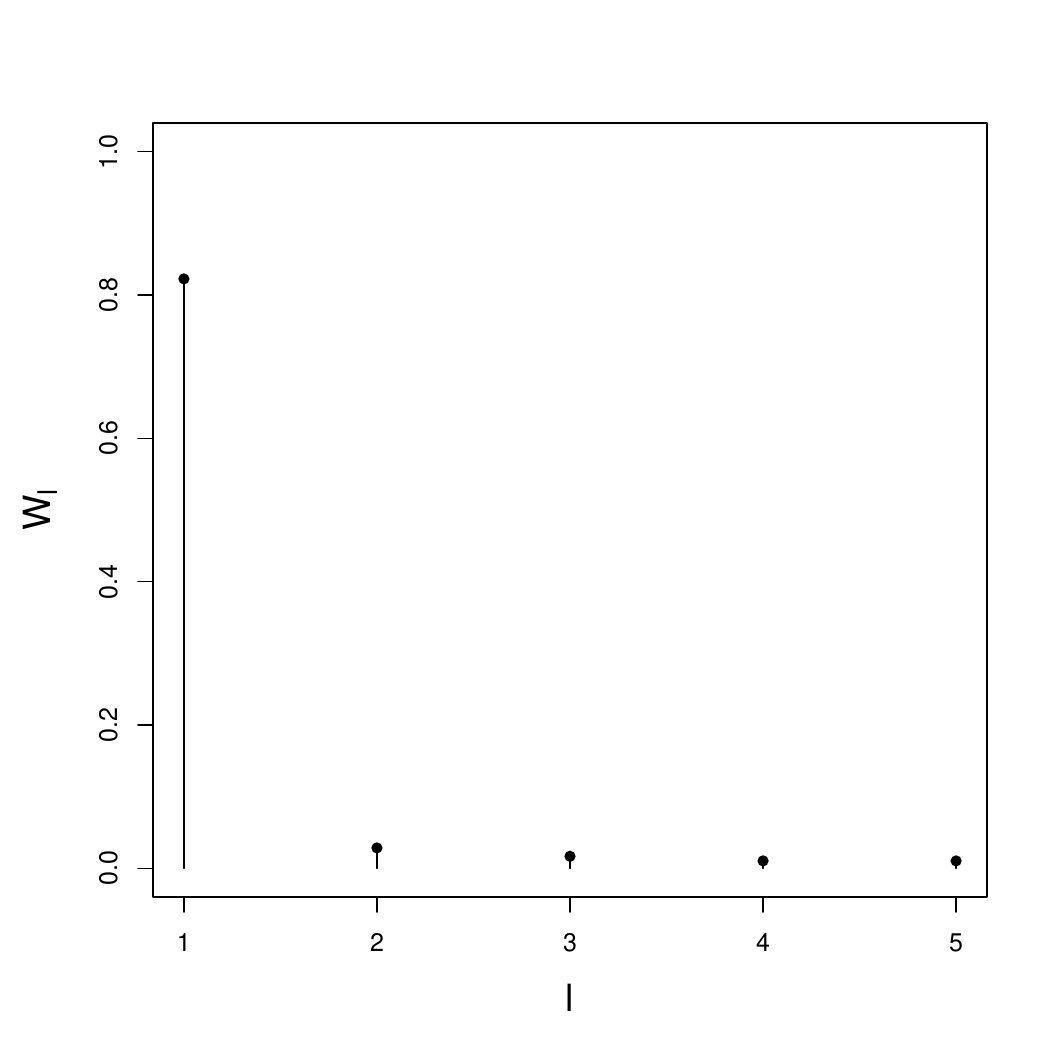}
\includegraphics[width=0.19\textwidth]{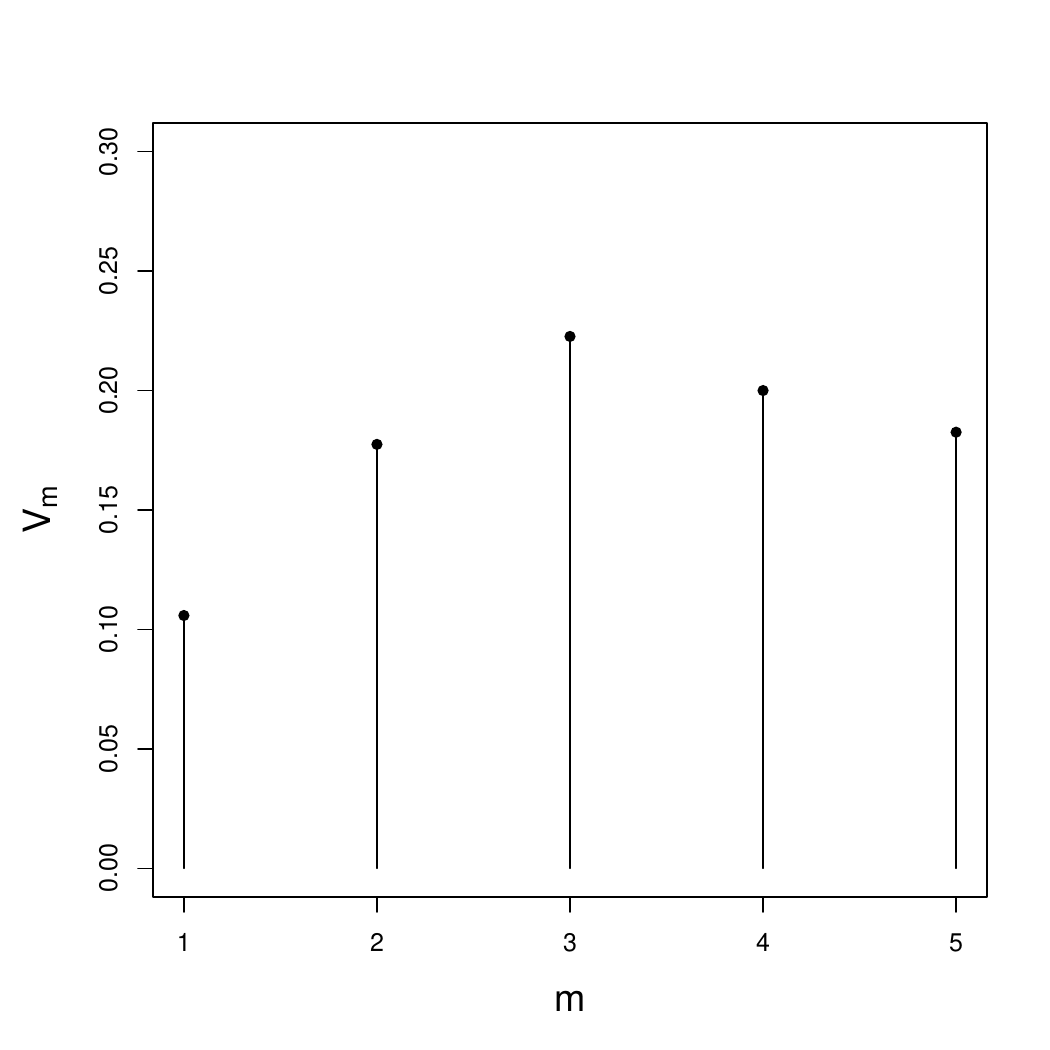} \\
\includegraphics[width=0.19\textwidth]{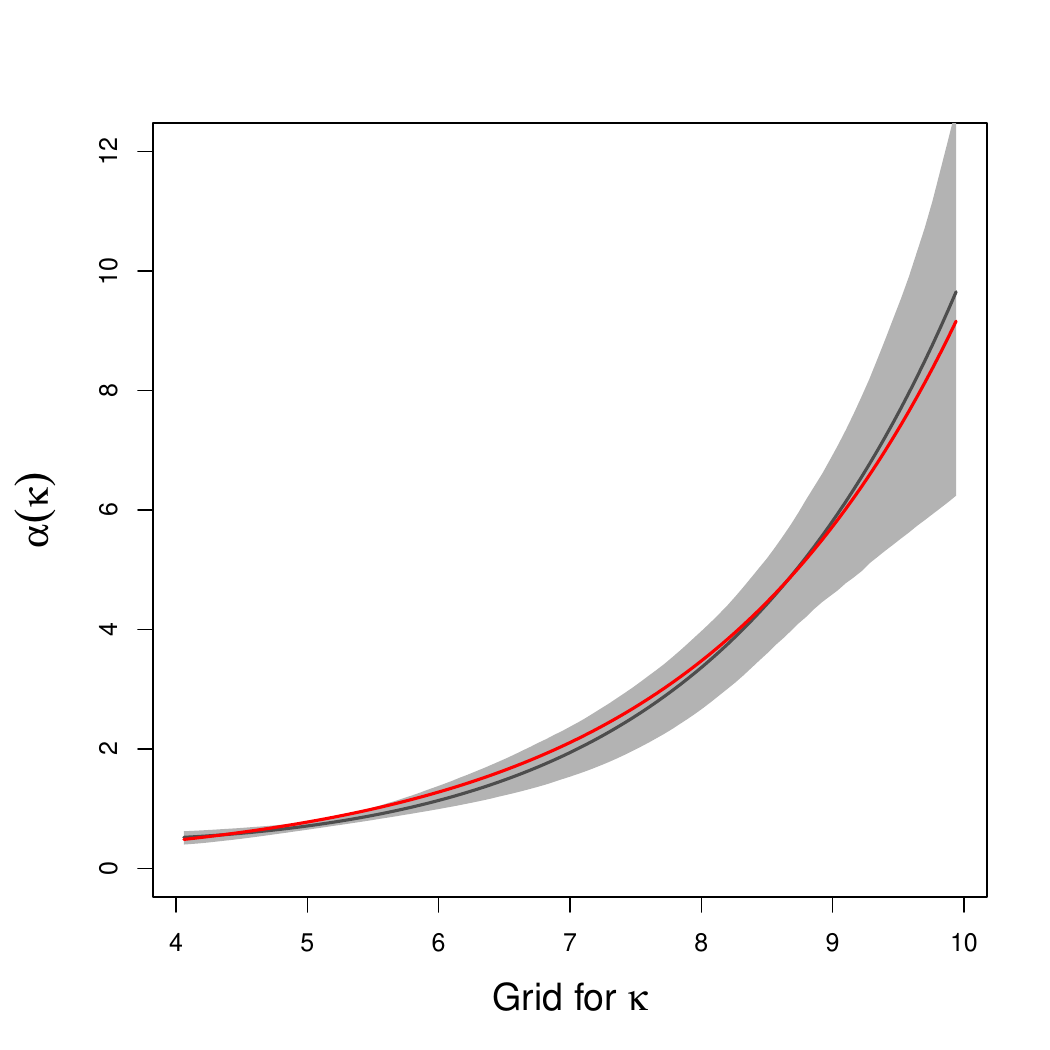}
\includegraphics[width=0.19\textwidth]{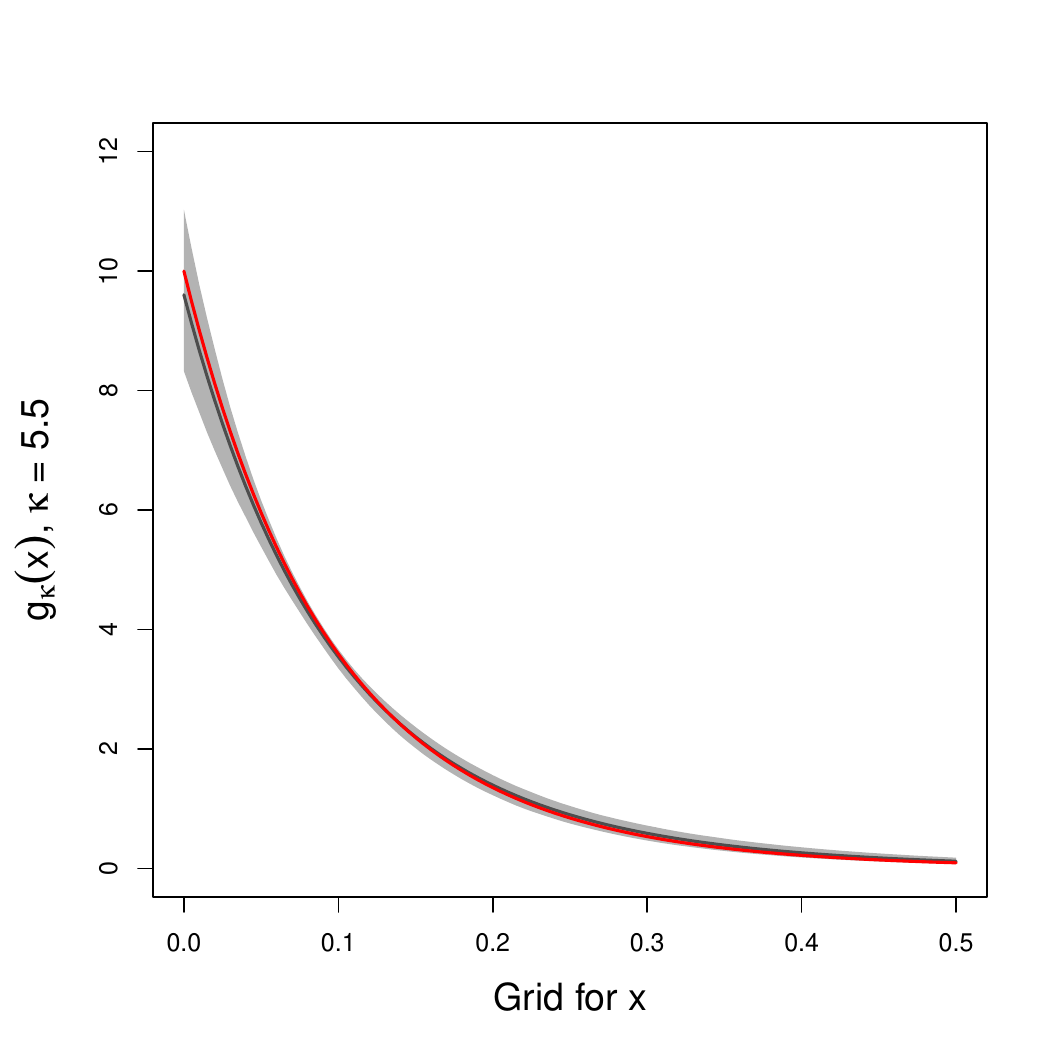}
\includegraphics[width=0.19\textwidth]{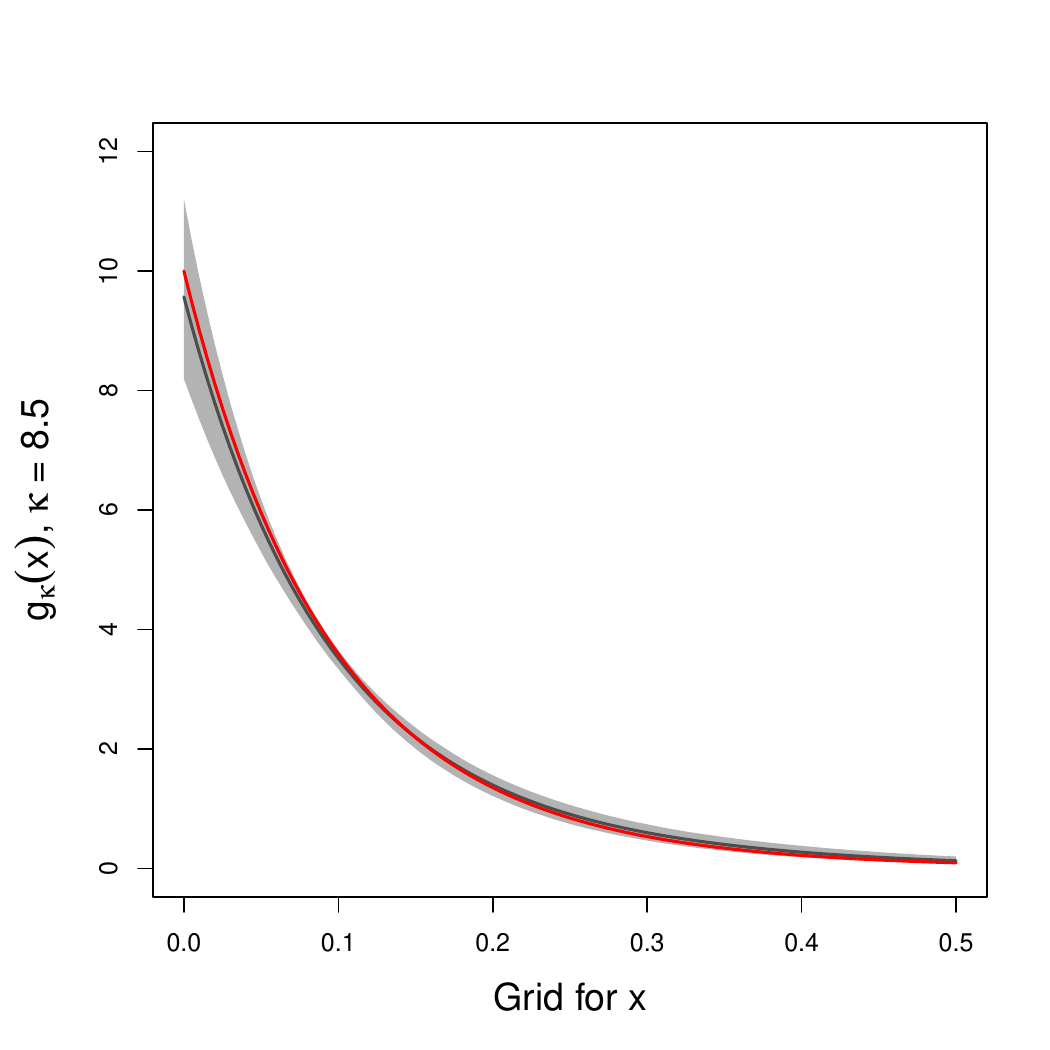}
\includegraphics[width=0.19\textwidth]{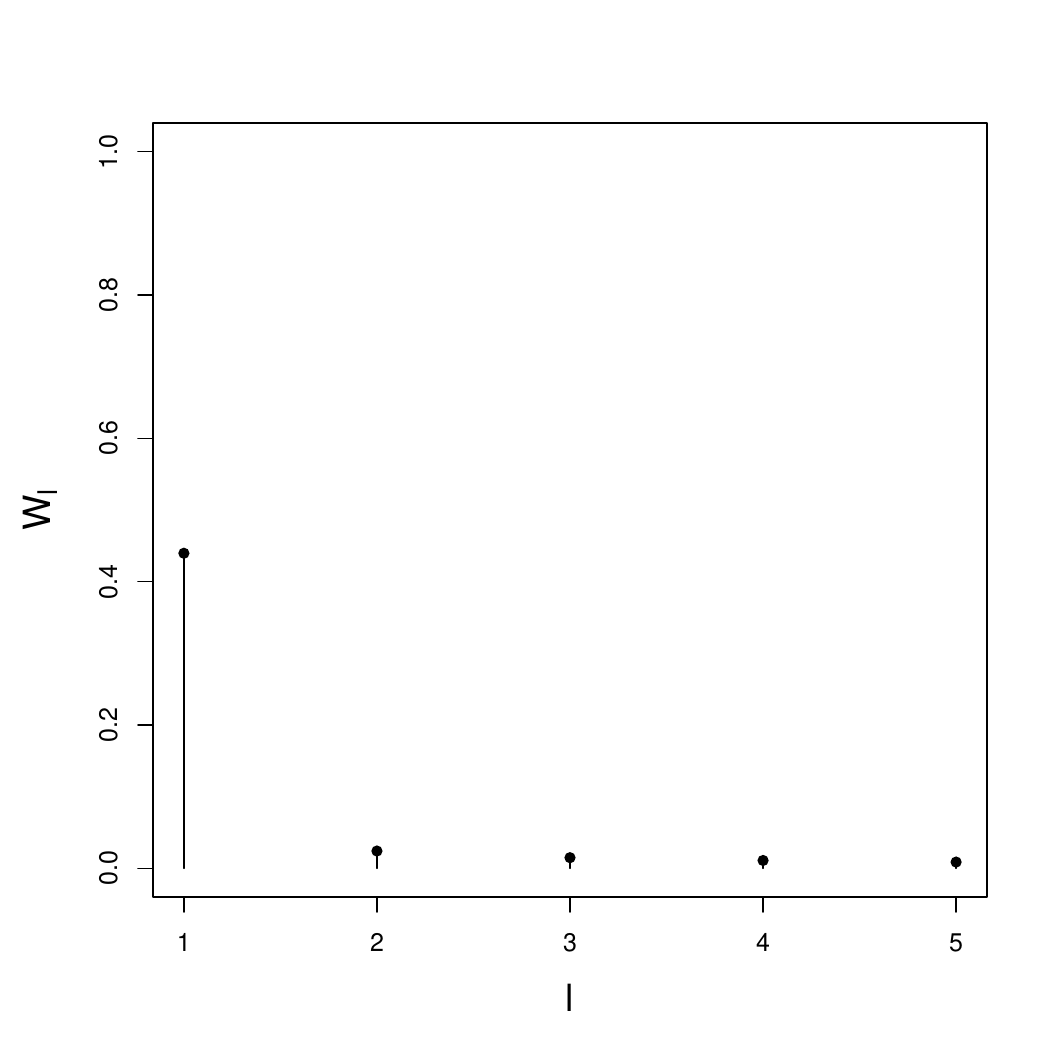}
\includegraphics[width=0.19\textwidth]{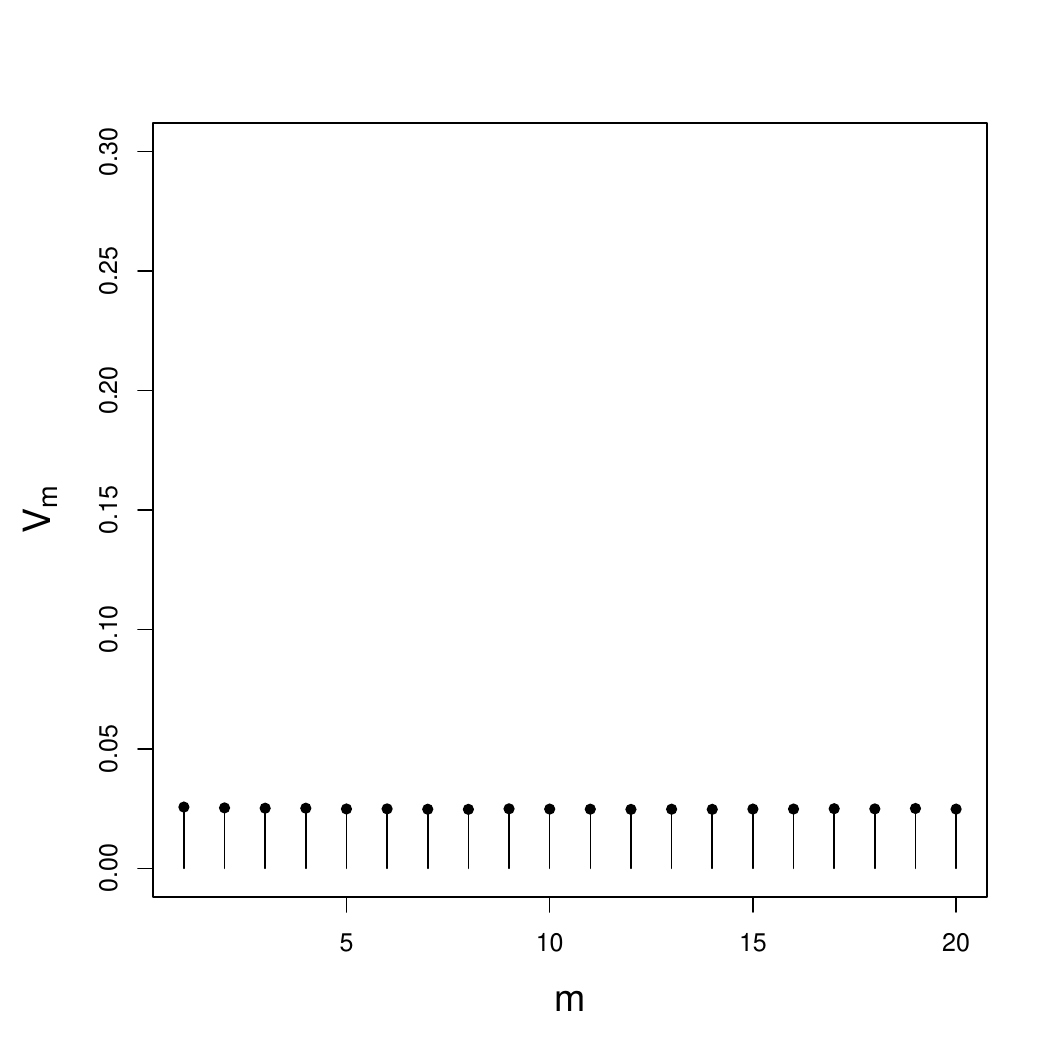} \\
\includegraphics[width=0.19\textwidth]{./figs/ak_nonpara_powlaw_seed275_L15M20_2.pdf}
\includegraphics[width=0.19\textwidth]{./figs/gkx_nonpara_powlaw_seed275_L15M20_k1_2.pdf}
\includegraphics[width=0.19\textwidth]{./figs/gkx_nonpara_powlaw_seed275_L15M20_k2_2.pdf}
\includegraphics[width=0.19\textwidth]{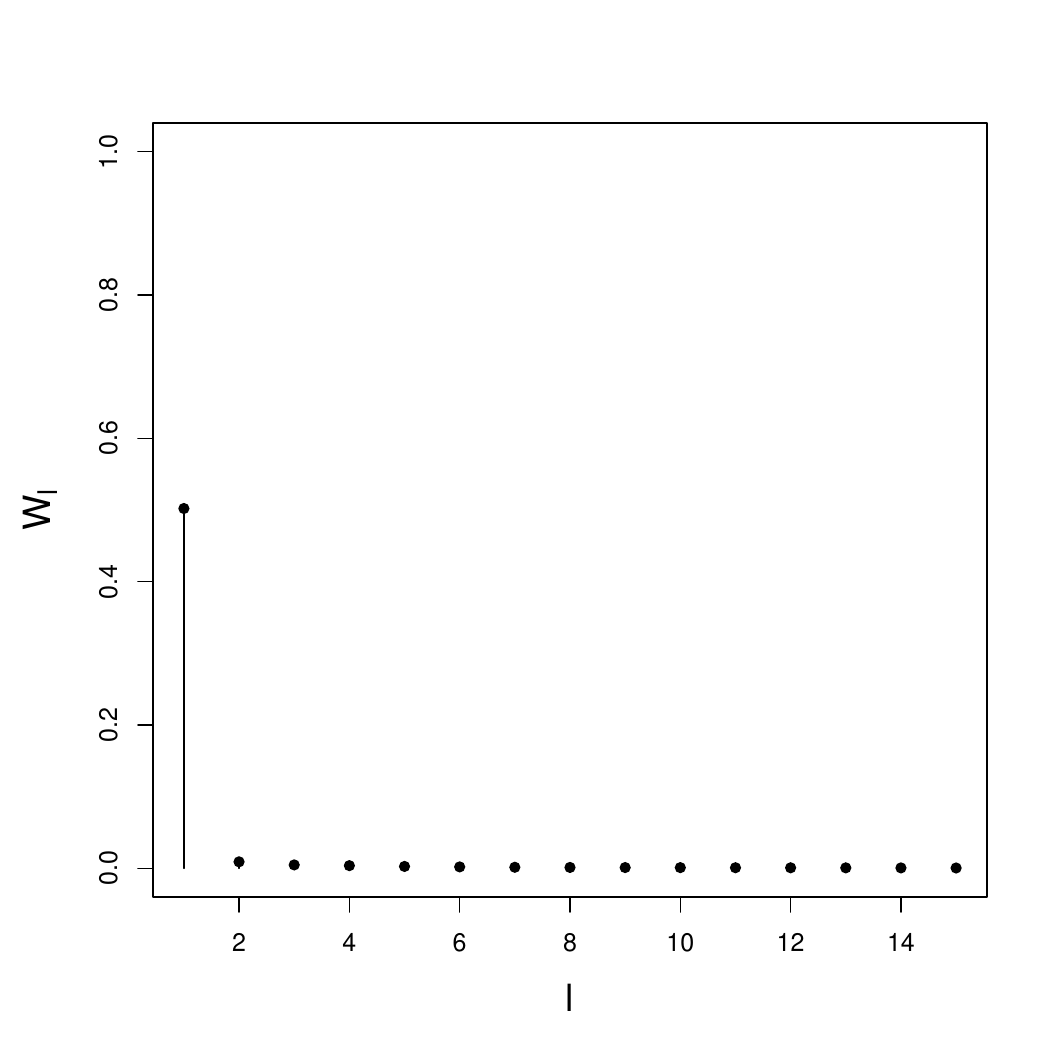}
\includegraphics[width=0.19\textwidth]{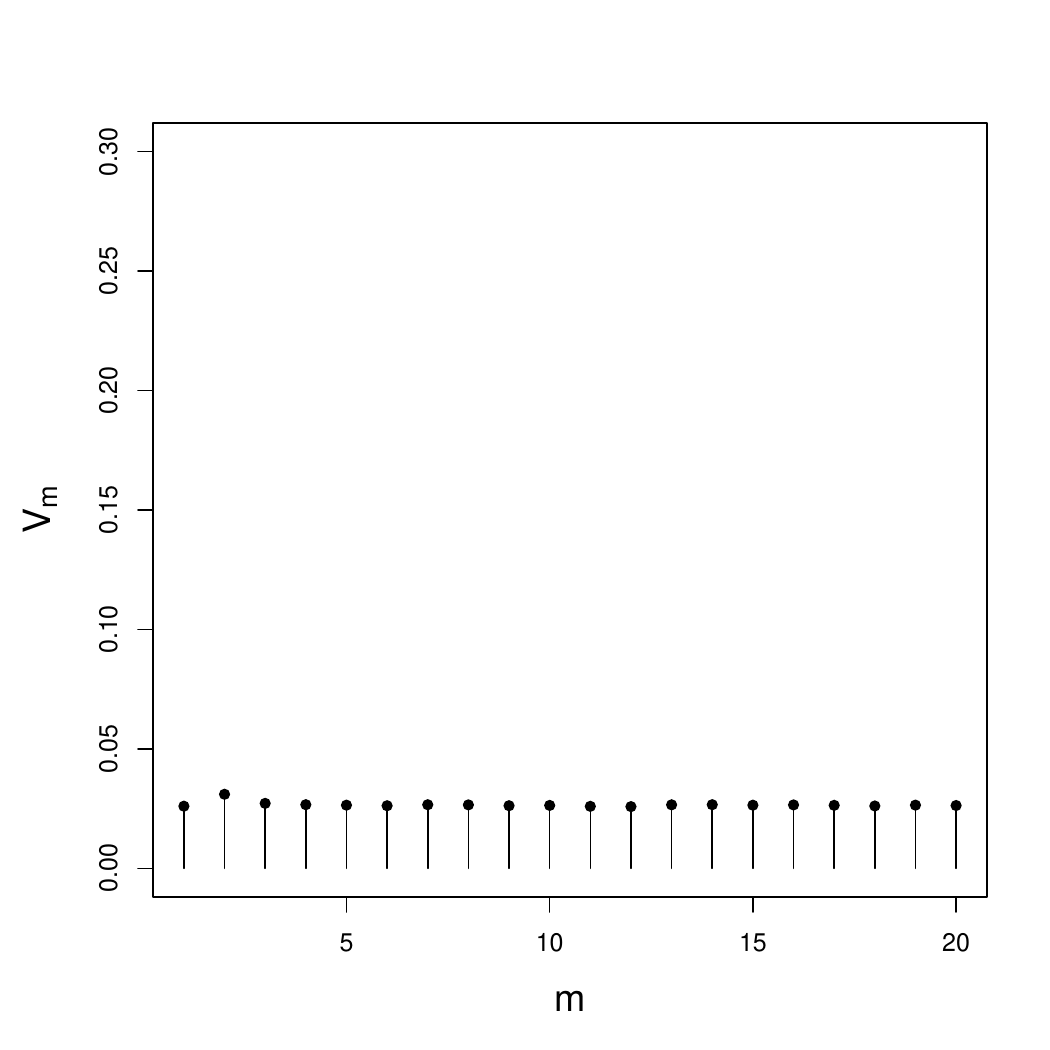} \\
\caption{
{\small Lomax density simulation example (Section \ref{SM:mhp_simulation_powlaw}). 
Nonparametric model results under different values for the number of basis components: 
$(L,M)=(5,5)$ (top row), $(5,20)$ (middle row), and $(15,20)$ (bottom row).
Plotted in each row are: the posterior mean (black line) and 95\% uncertainty bands 
for the total oﬀspring intensity (first column) and for the offspring density for two
mark values, $\kappa = 5.5, 8.5$ (second and third columns), and the posterior means of 
the weights $W_l =$ $\sum_{m=1}^M \nu_{lm}$ and $V_m =$ $\sum_{l=1}^L \nu_{lm}$ 
(last two columns). The red line denotes the function used to generate the data.}}
    \label{fig:sensitivity_sim}
\end{figure}

Figure \ref{fig:sensitivity_sim} provides some results from sensitivity analysis 
regarding the values for the numbers of basis components, $L$ and $M$, working with 
the synthetic data example presented in Section \ref{SM:mhp_simulation_powlaw}.
It is interesting to note that there is little sensitivity with respect to the 
inference results (both point and interval estimates) for the offspring density, with 
accurate estimates obtained even when $L=5$. Inference quality for the total offspring 
intensity is affected when $M$ (and $L$) are set to small values. The estimation of 
the $\alpha(\kappa)$ function improves noticeably as $M$ increases from $5$ to $20$.

\begin{figure}[!t]
\centering
\includegraphics[width=0.118\textwidth]{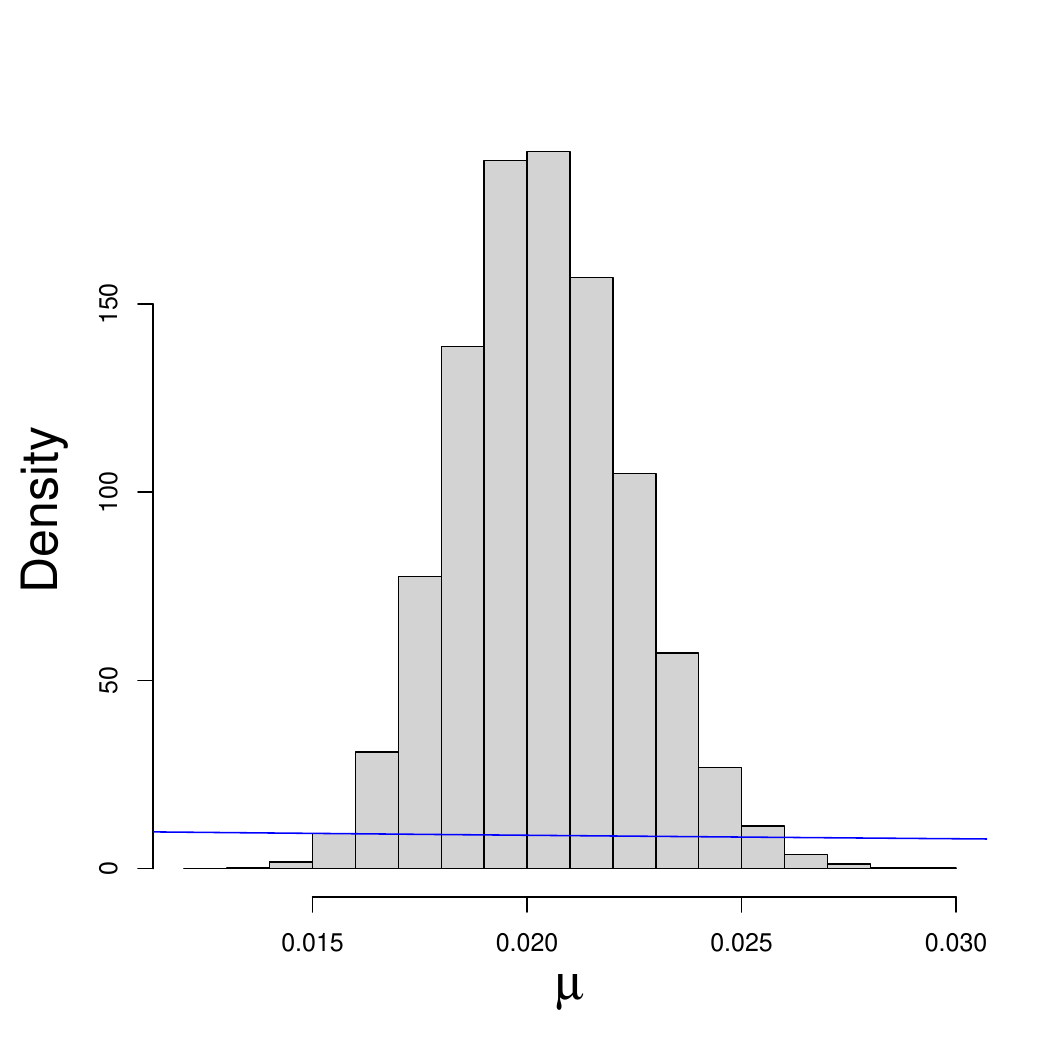}
\includegraphics[width=0.118\textwidth]{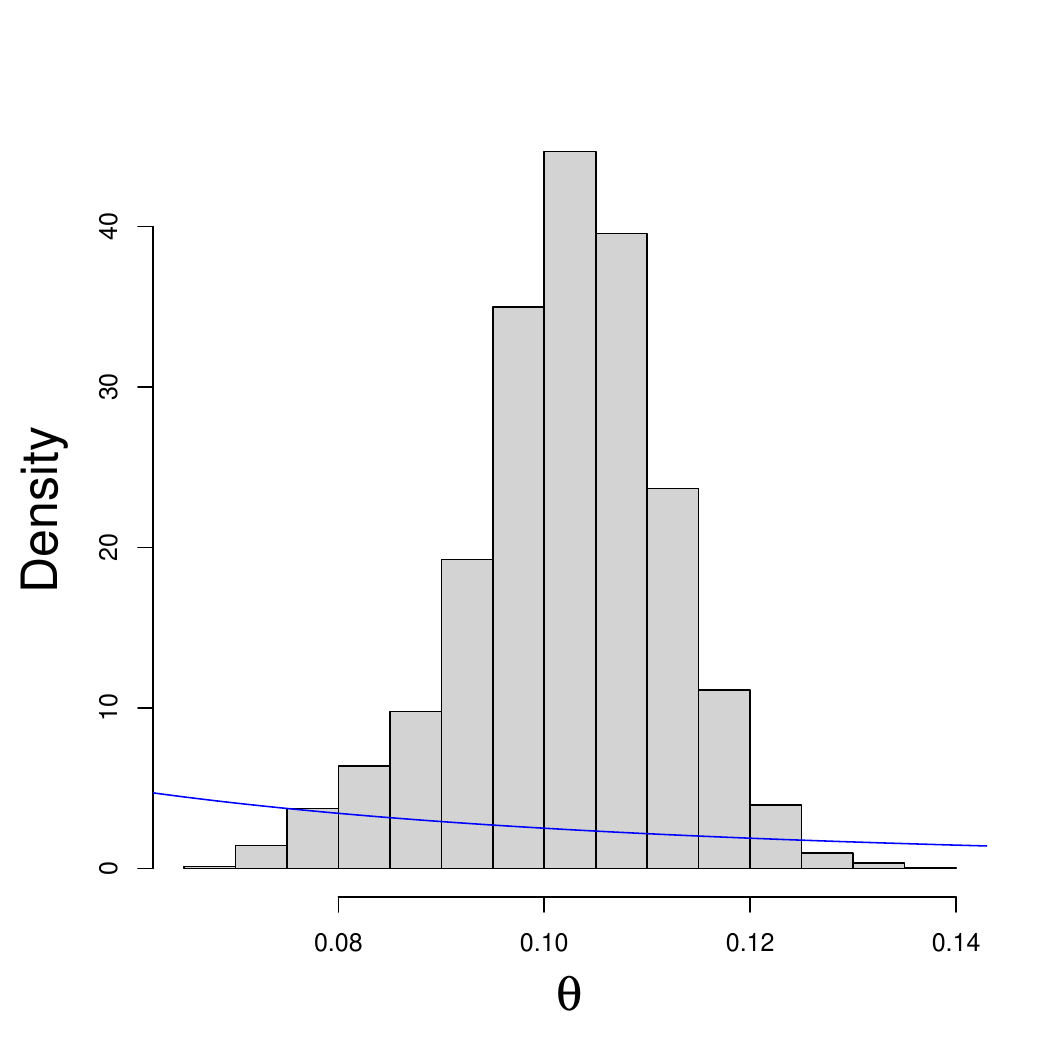}
\includegraphics[width=0.118\textwidth]{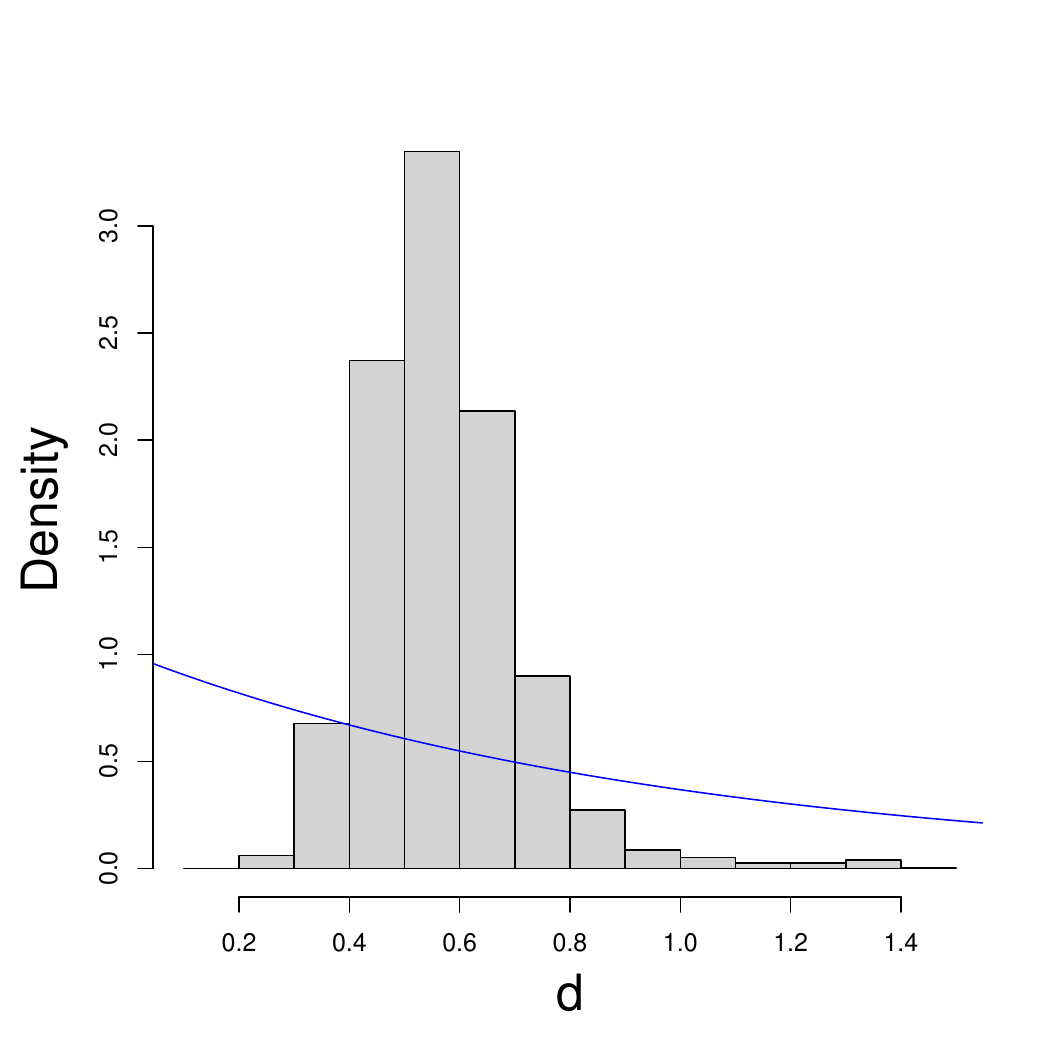}
\includegraphics[width=0.118\textwidth]{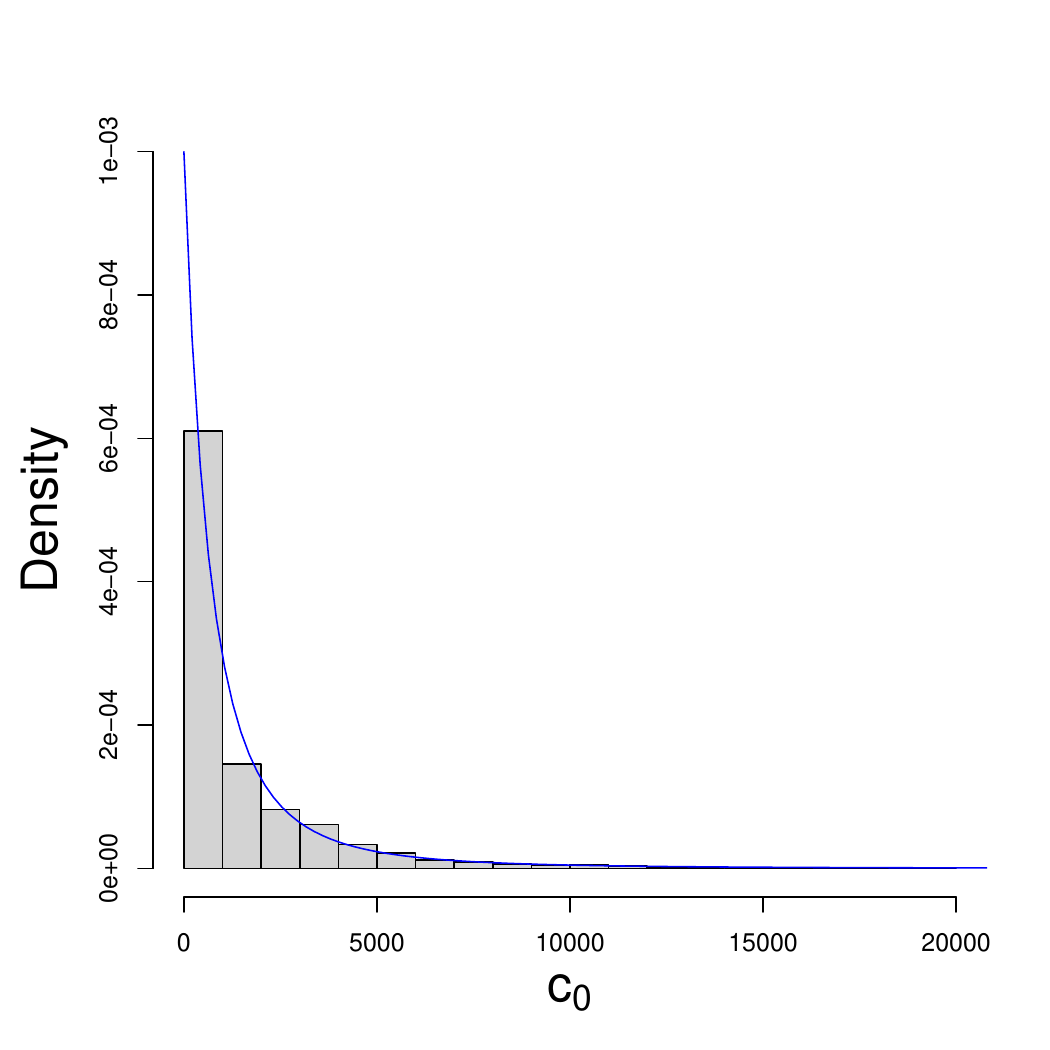}
\includegraphics[width=0.118\textwidth]{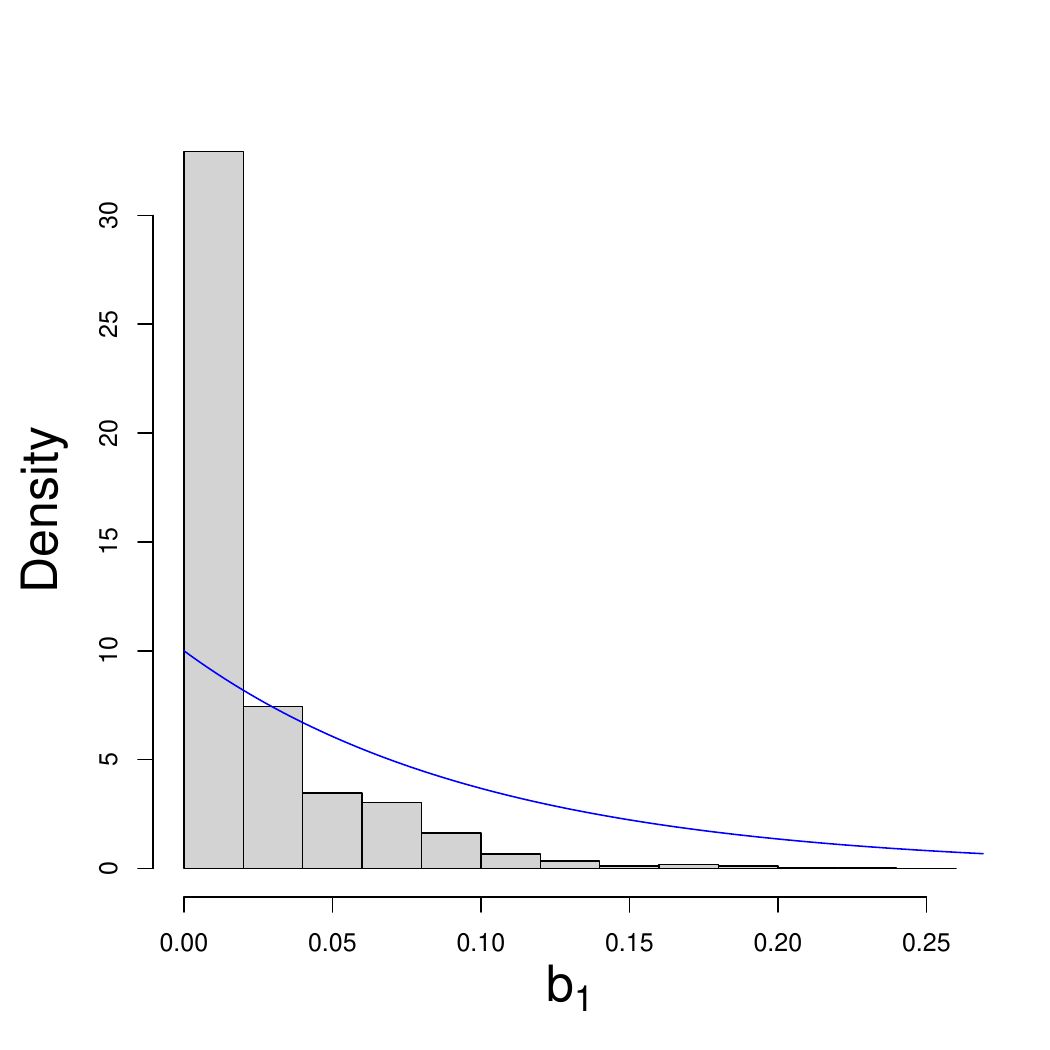}
\includegraphics[width=0.118\textwidth]{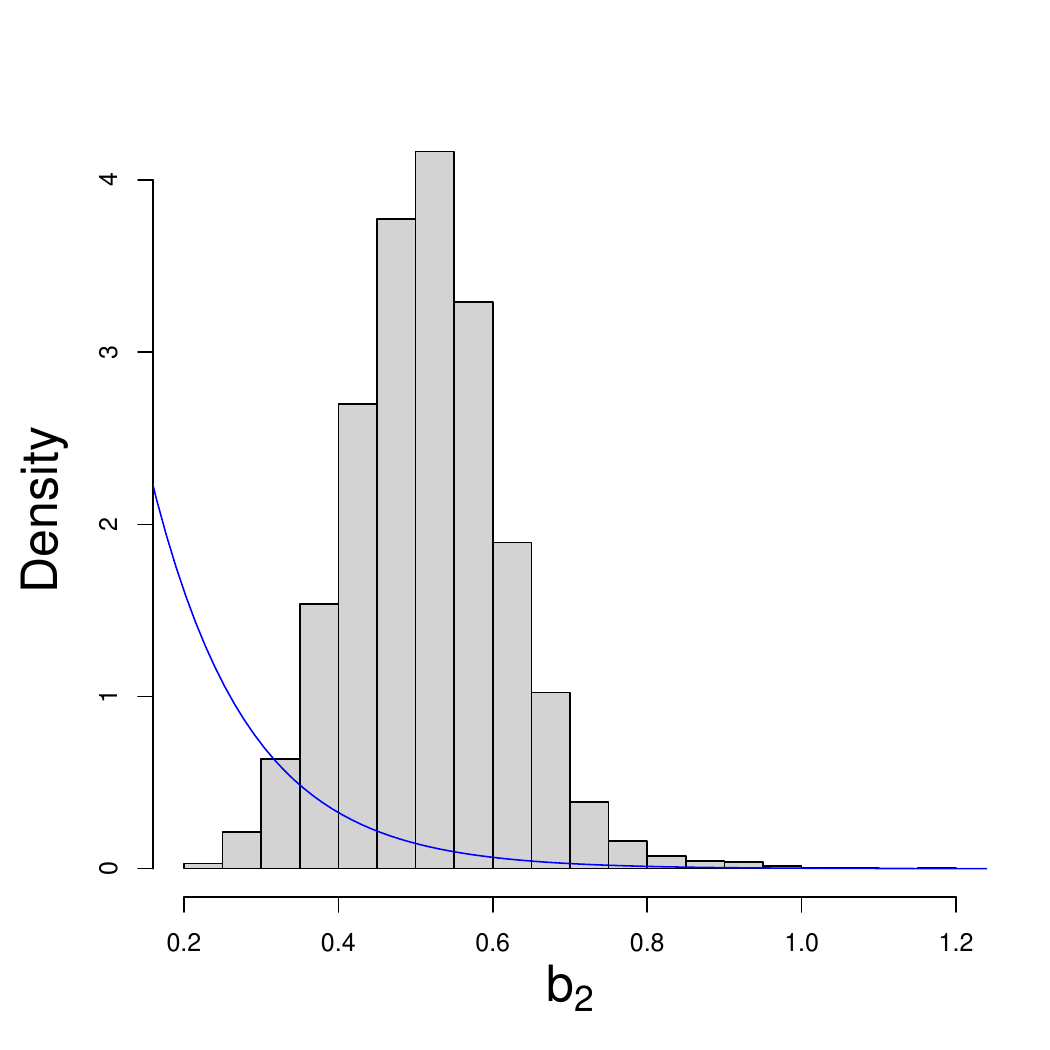}
\includegraphics[width=0.118\textwidth]{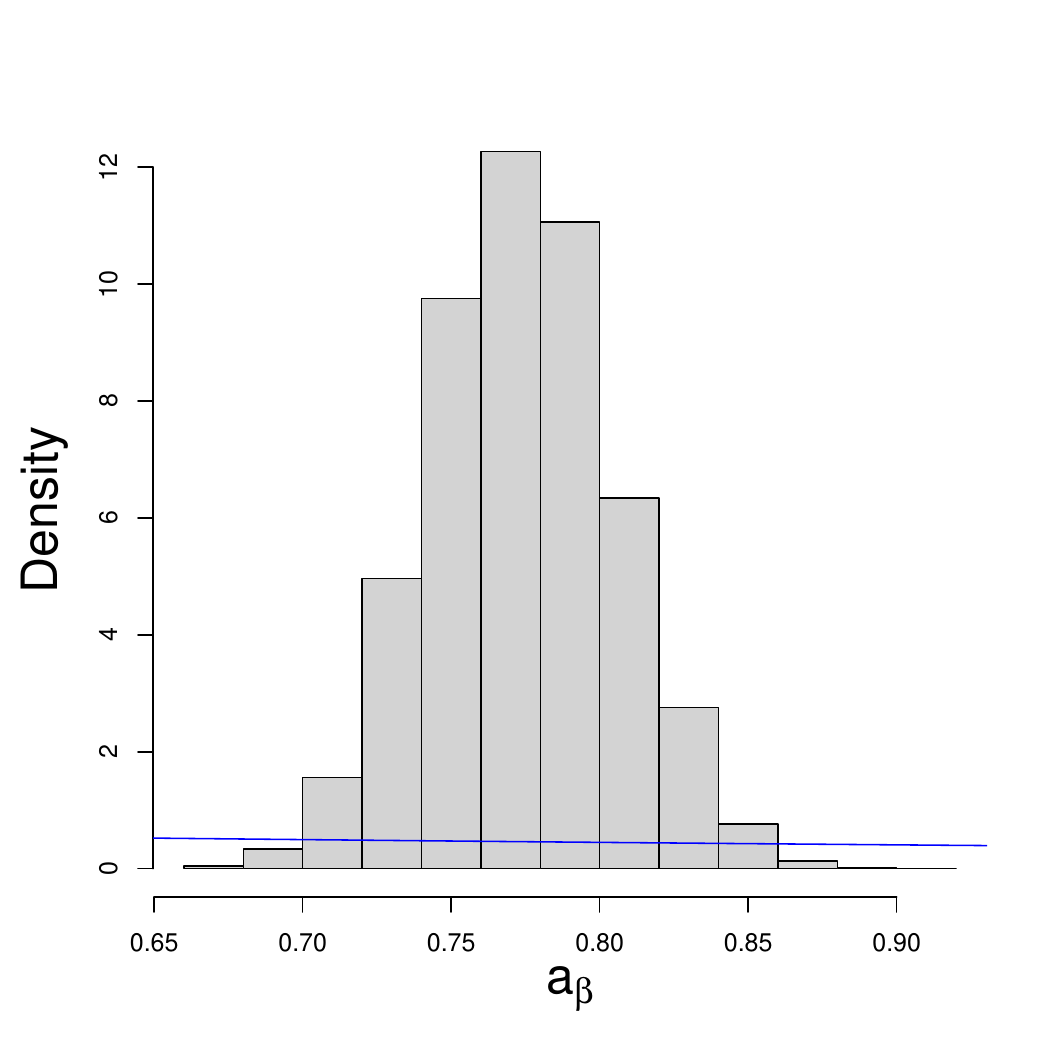}
\includegraphics[width=0.118\textwidth]{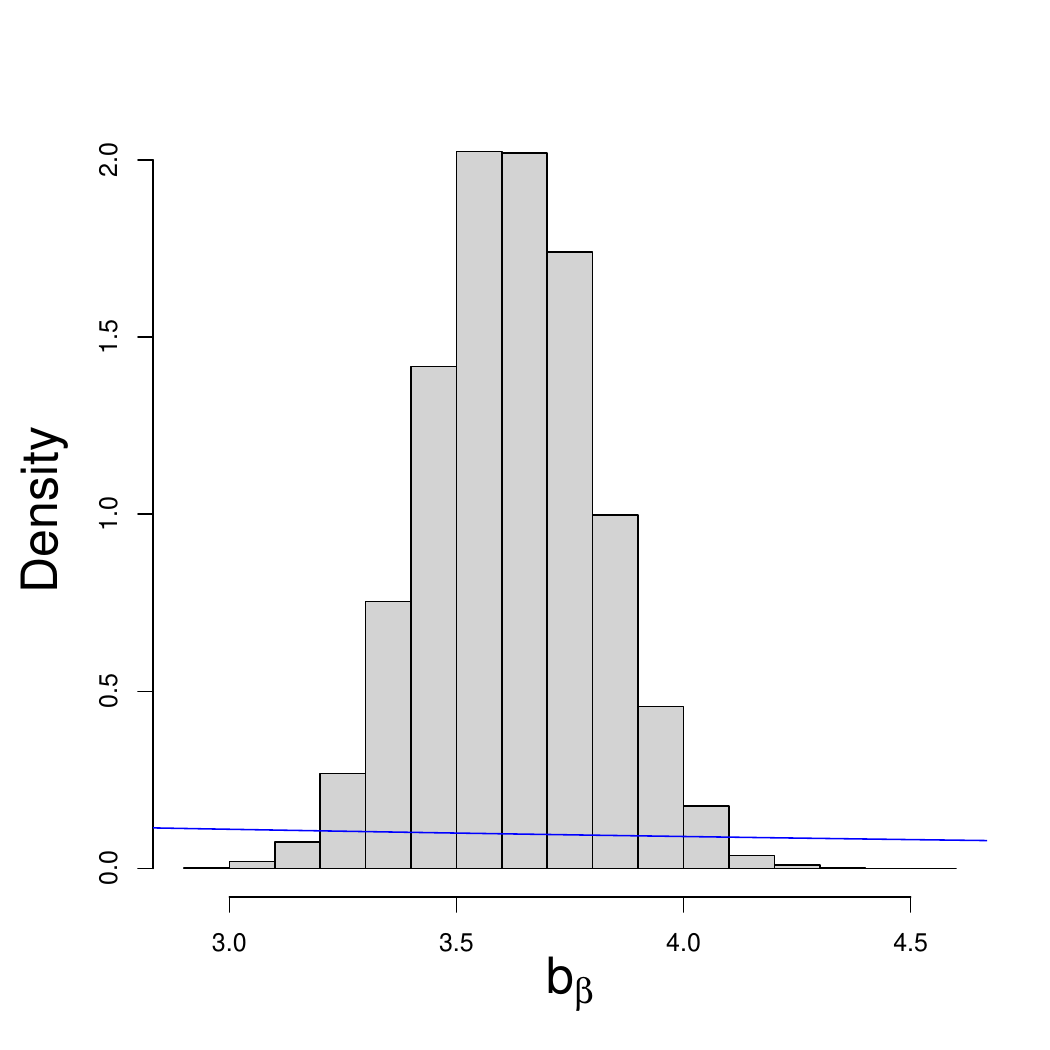} 
\\
\includegraphics[width=0.118\textwidth]{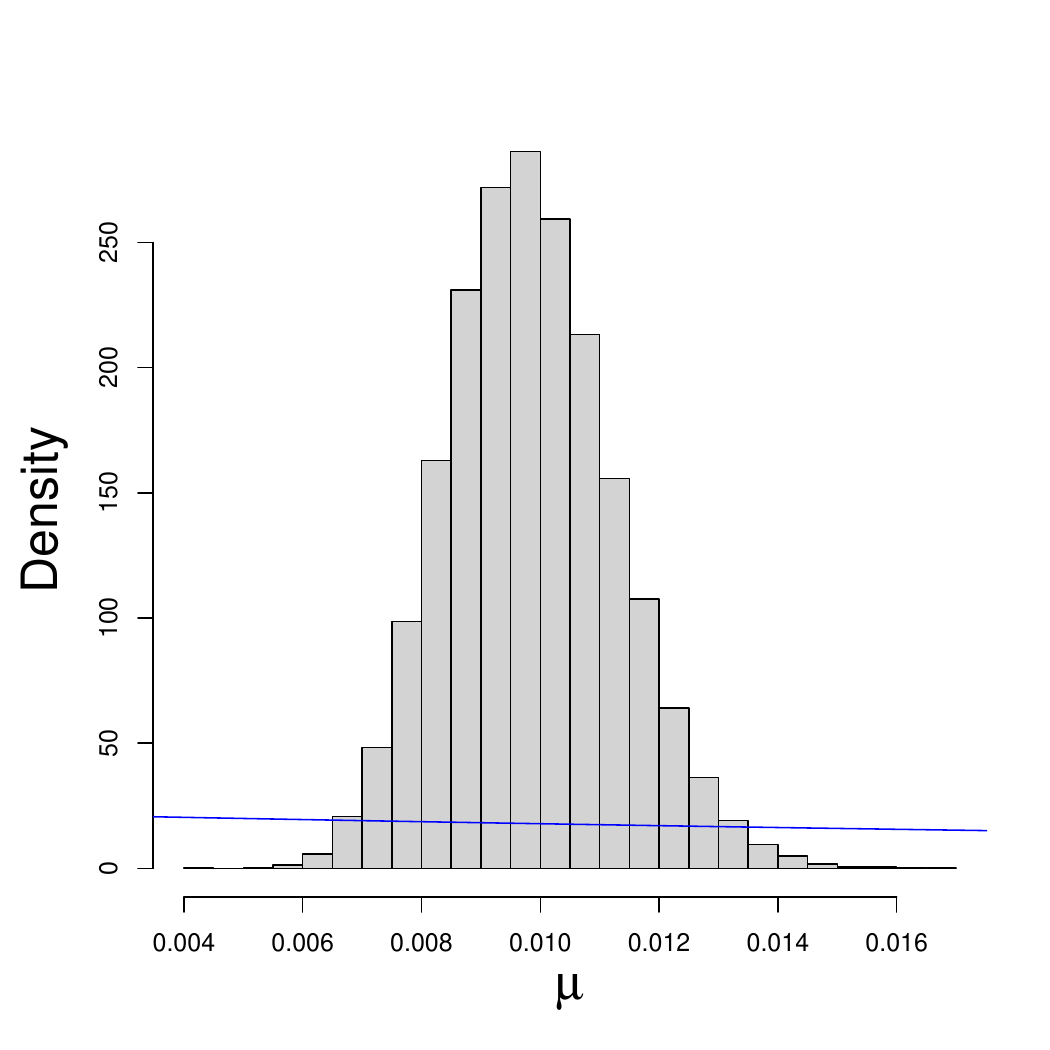}
\includegraphics[width=0.118\textwidth]{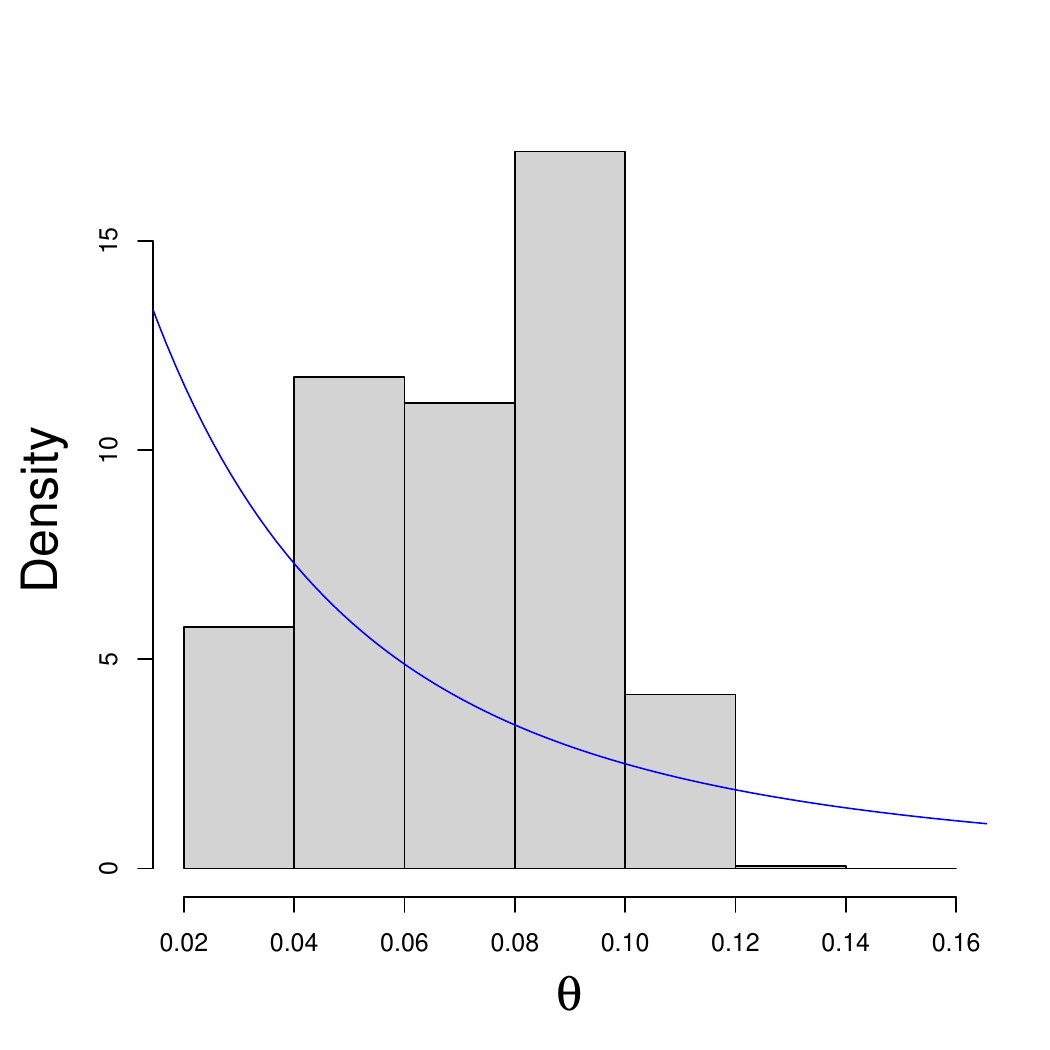}
\includegraphics[width=0.118\textwidth]{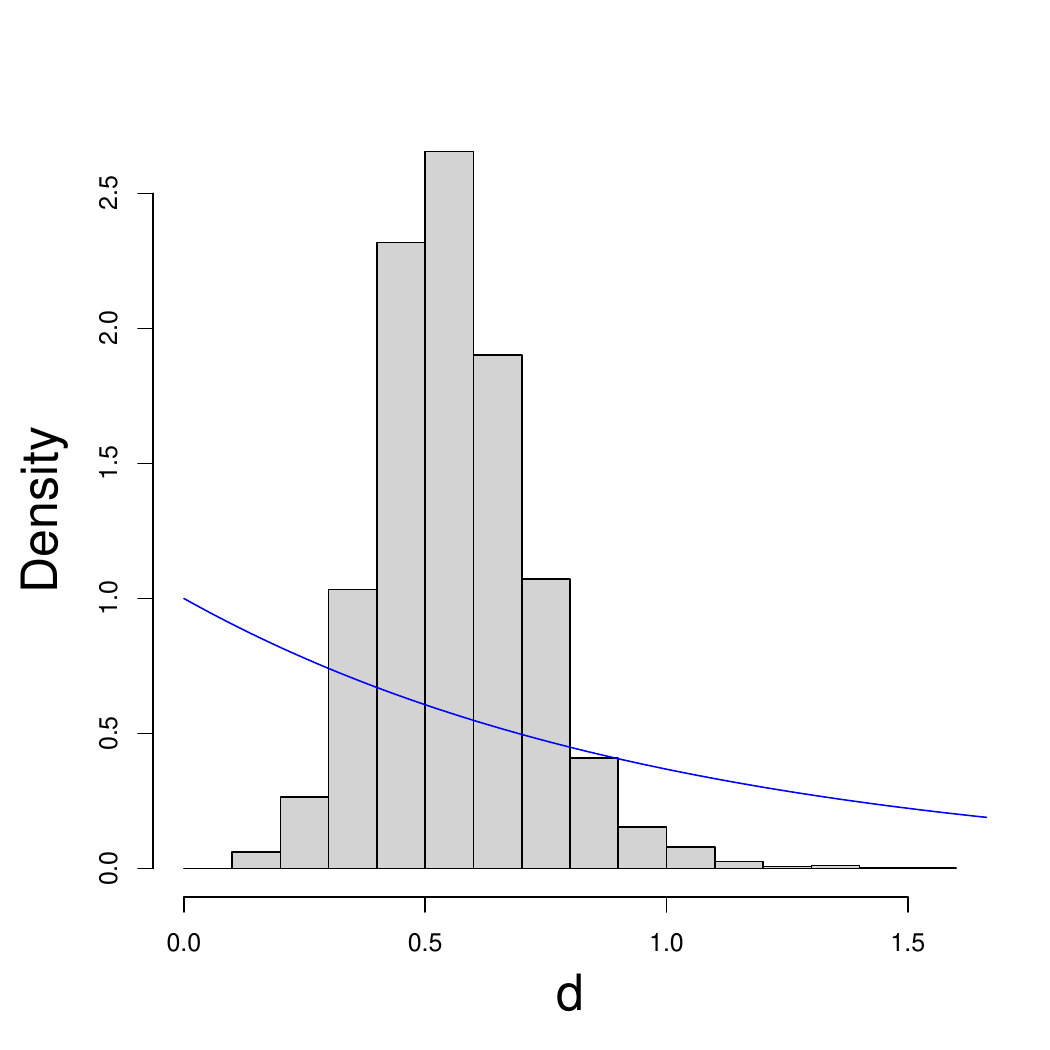}
\includegraphics[width=0.118\textwidth]{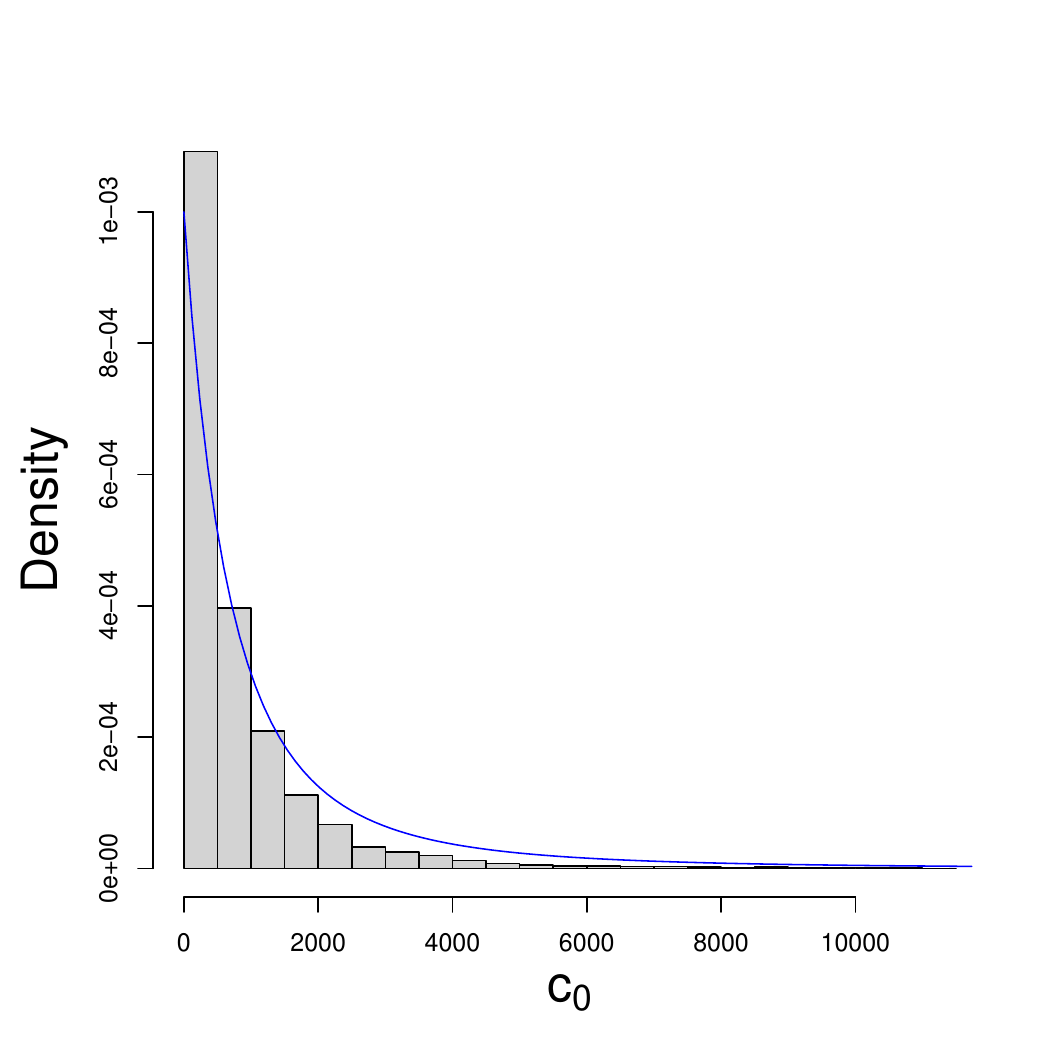}
\includegraphics[width=0.118\textwidth]{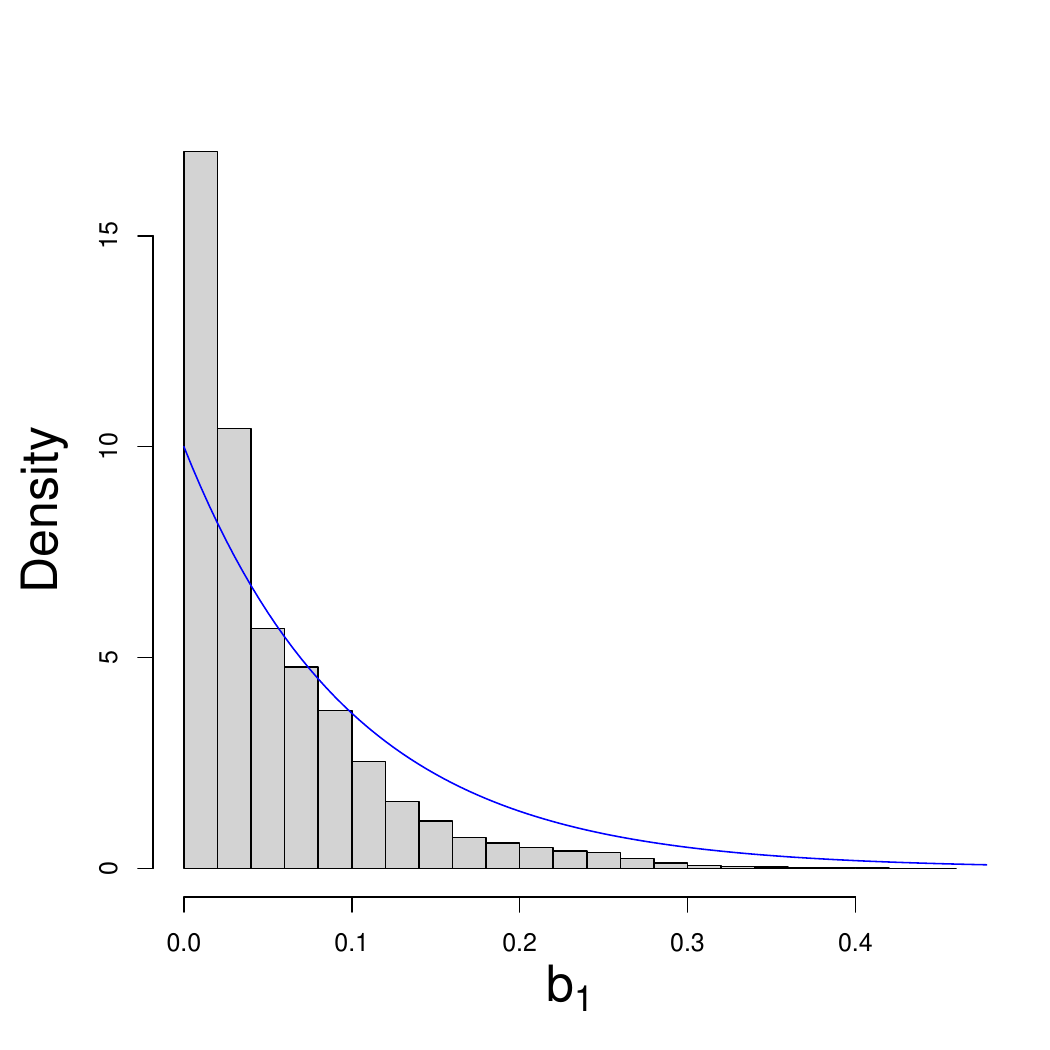}
\includegraphics[width=0.118\textwidth]{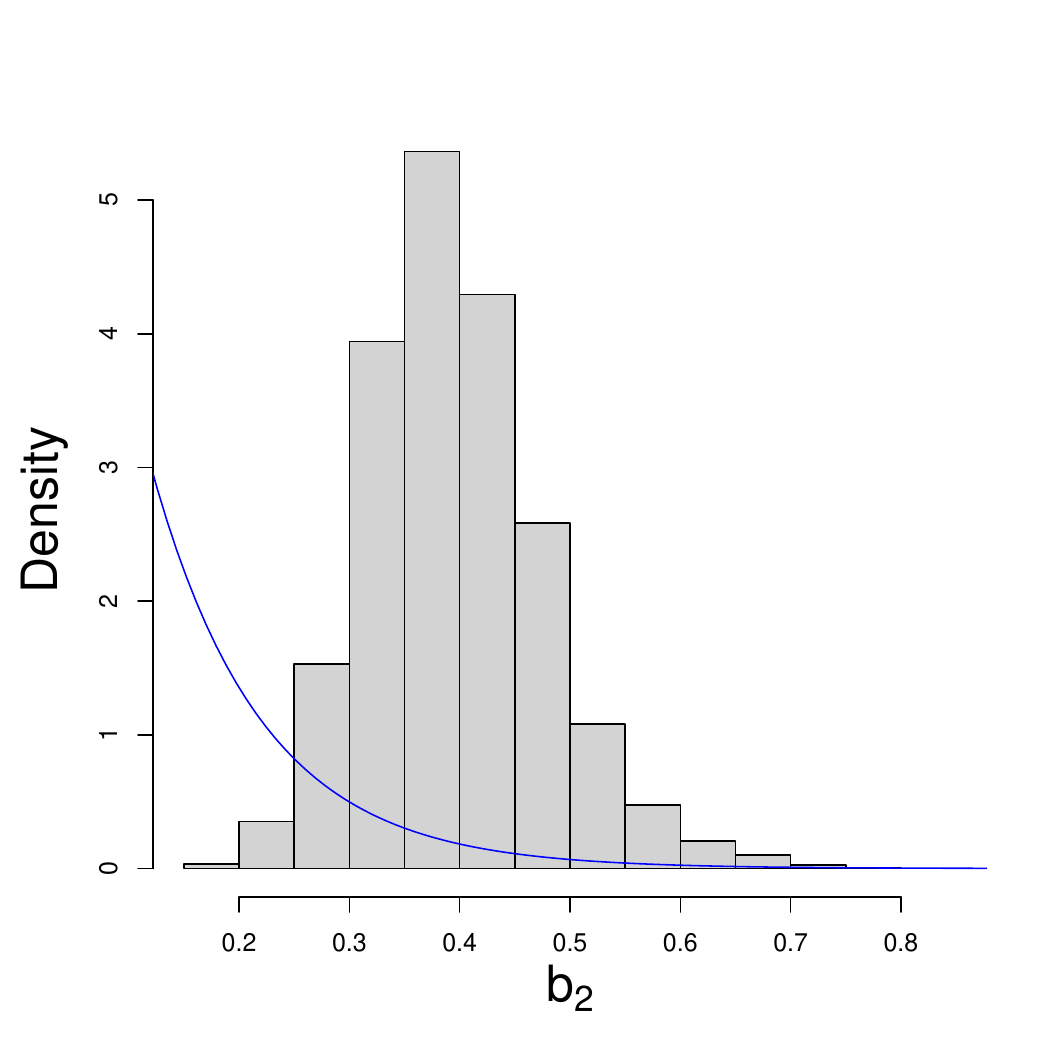}
\includegraphics[width=0.118\textwidth]{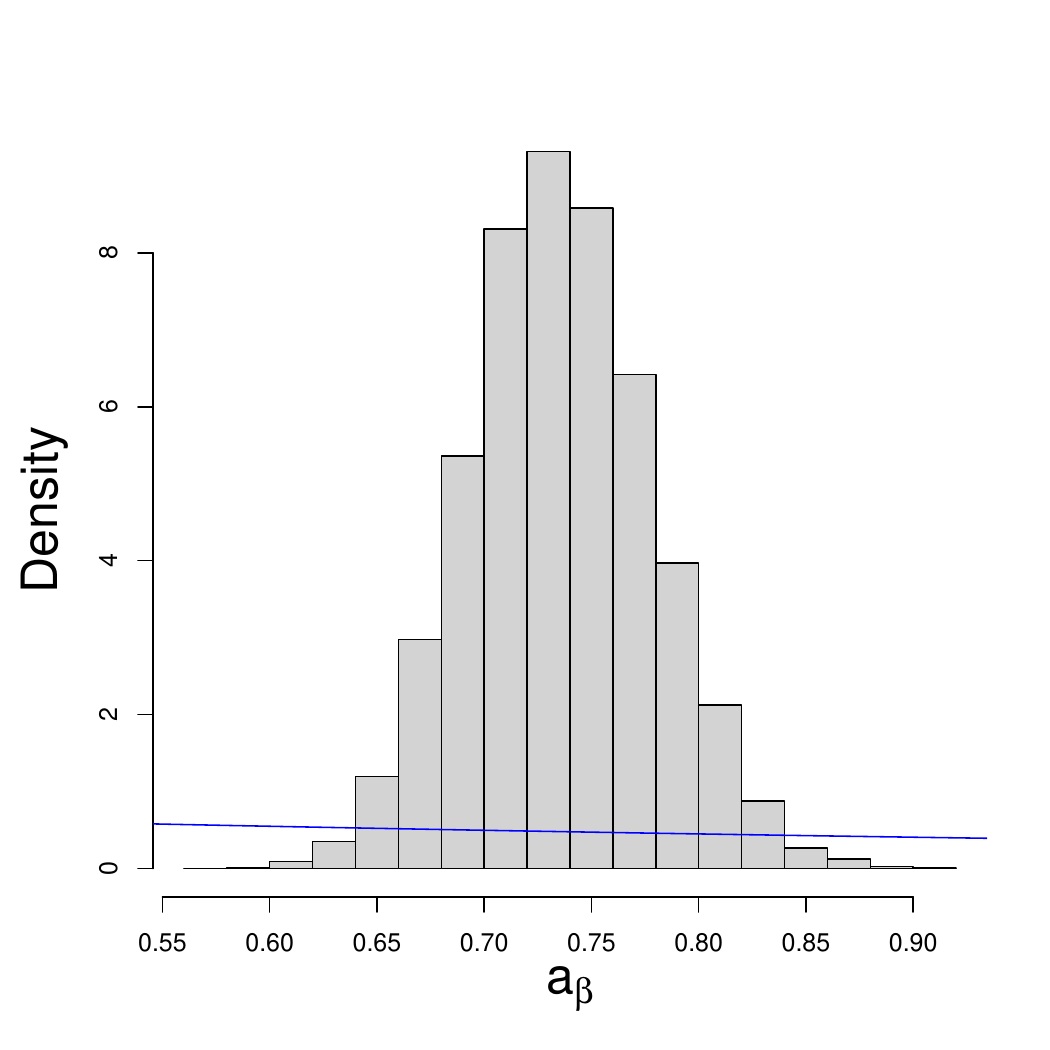}
\includegraphics[width=0.118\textwidth]{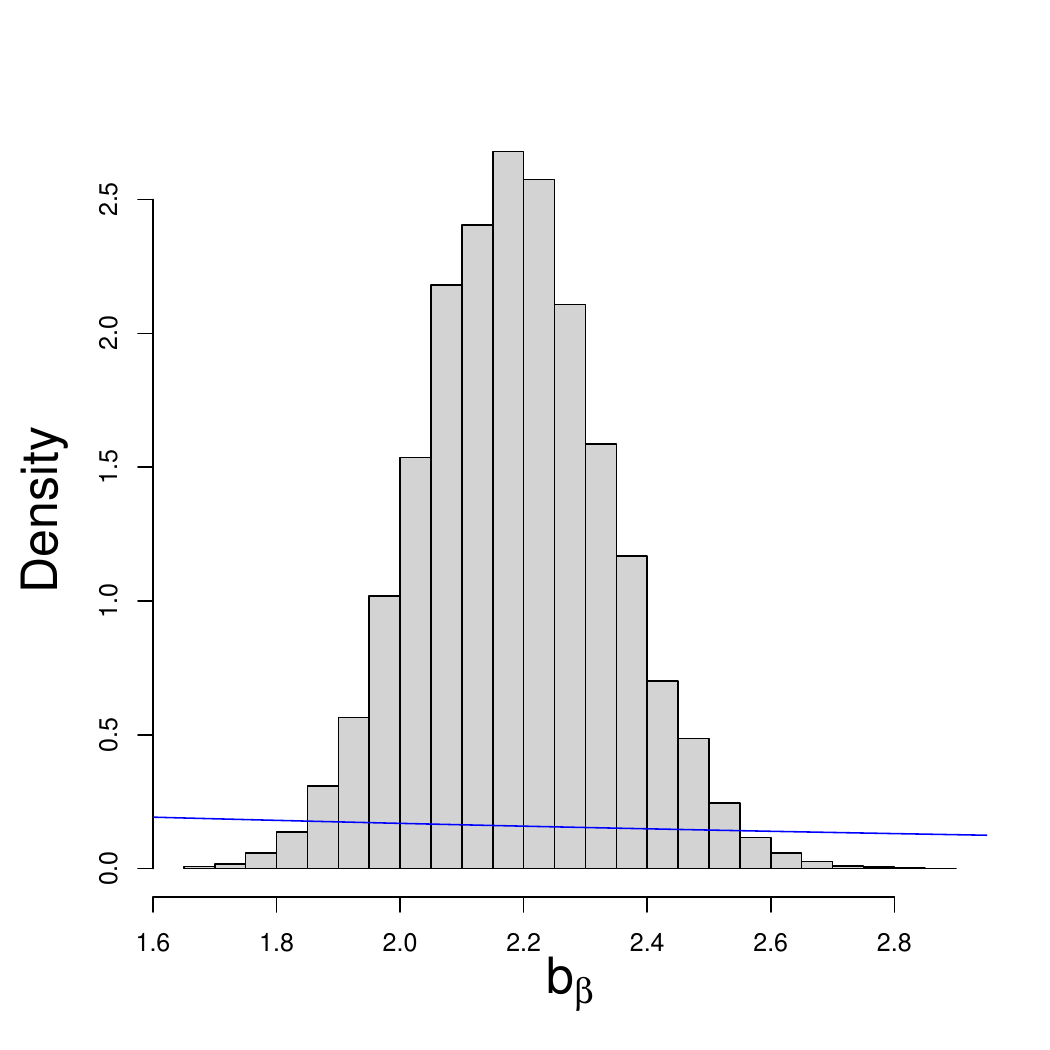}
\\
\includegraphics[width=0.118\textwidth]{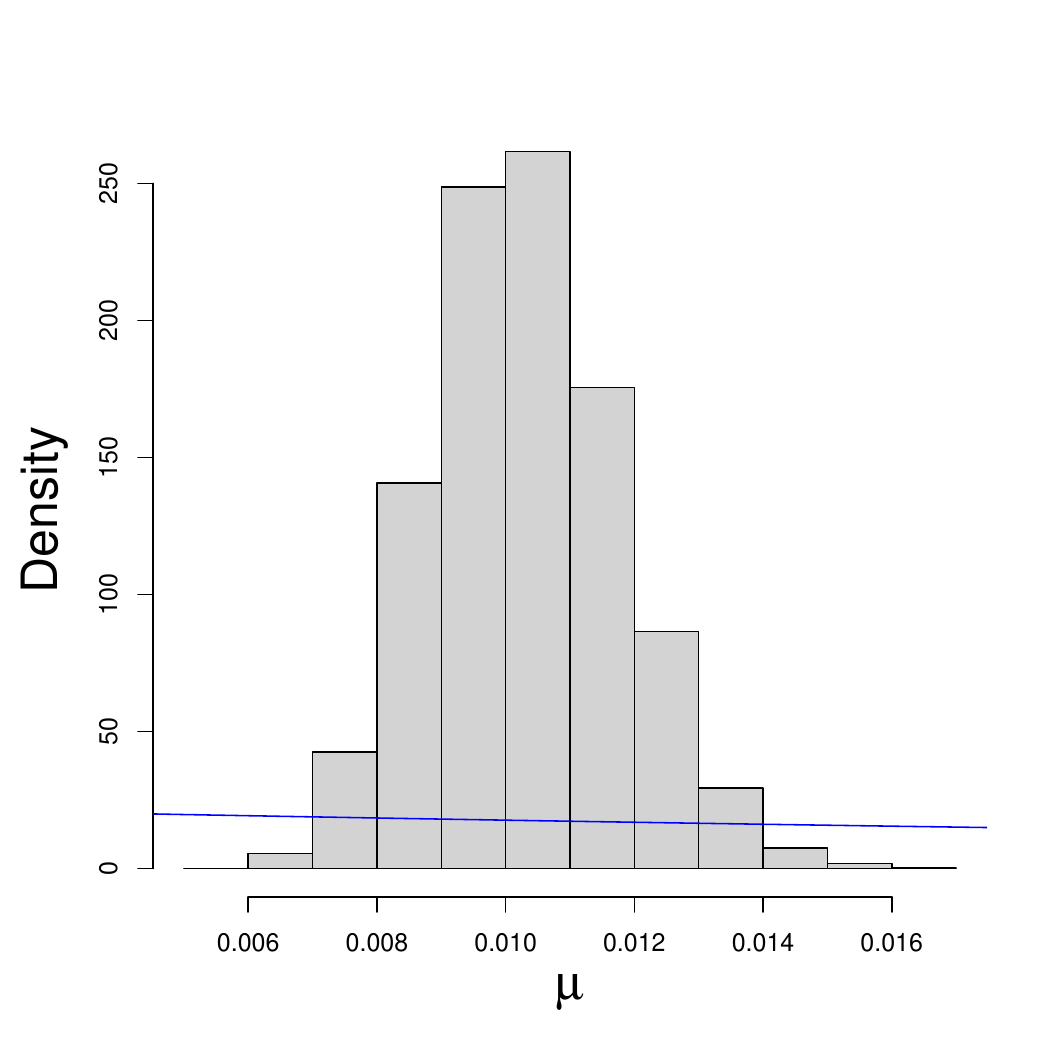}
\includegraphics[width=0.118\textwidth]{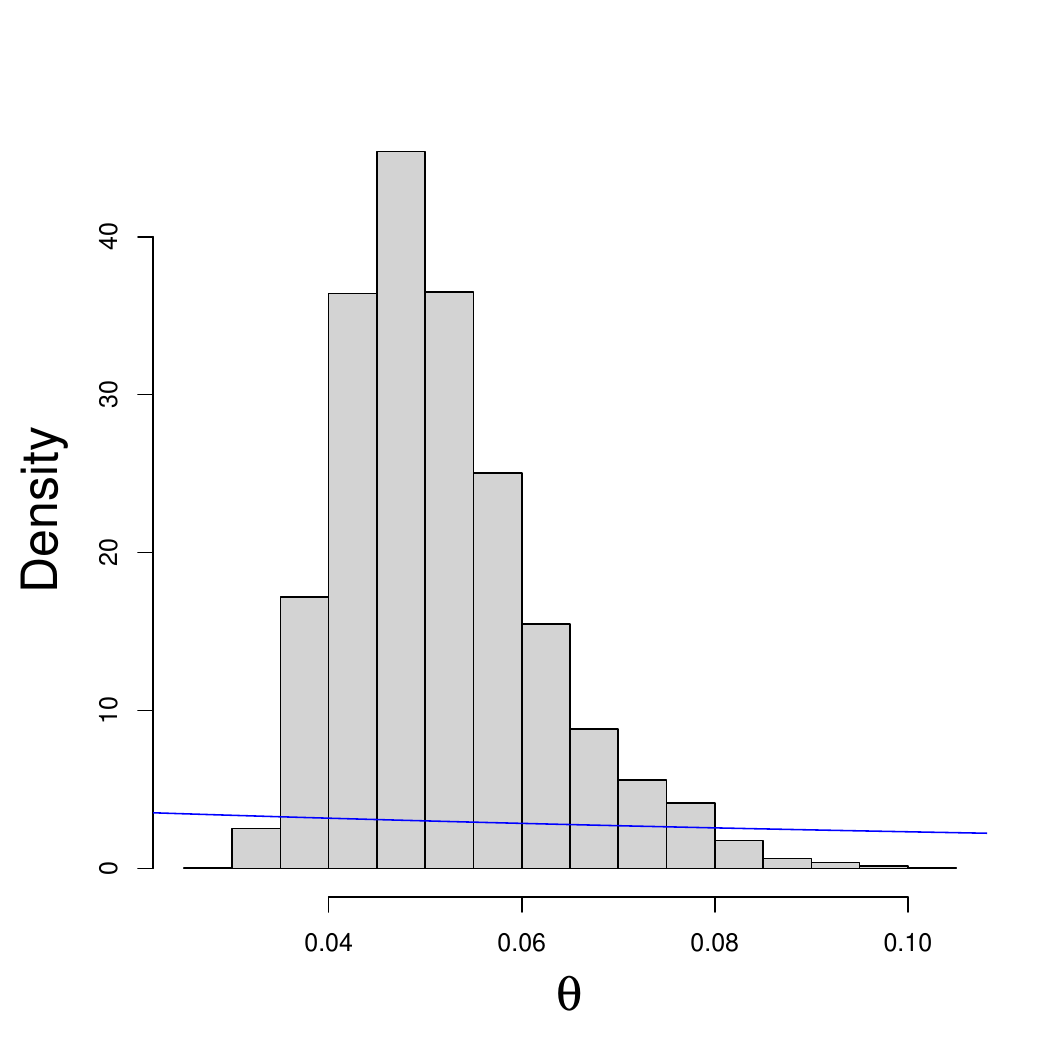}
\includegraphics[width=0.118\textwidth]{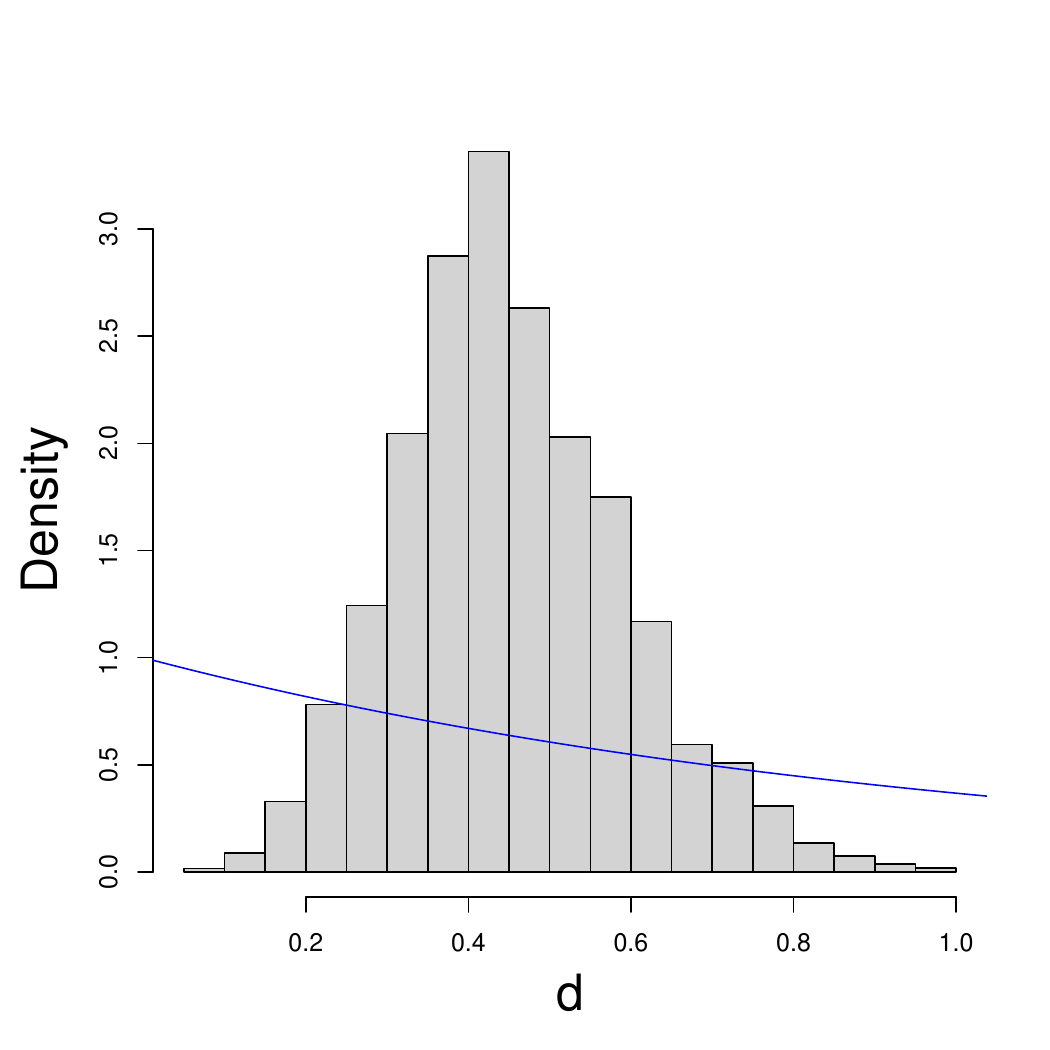}
\includegraphics[width=0.118\textwidth]{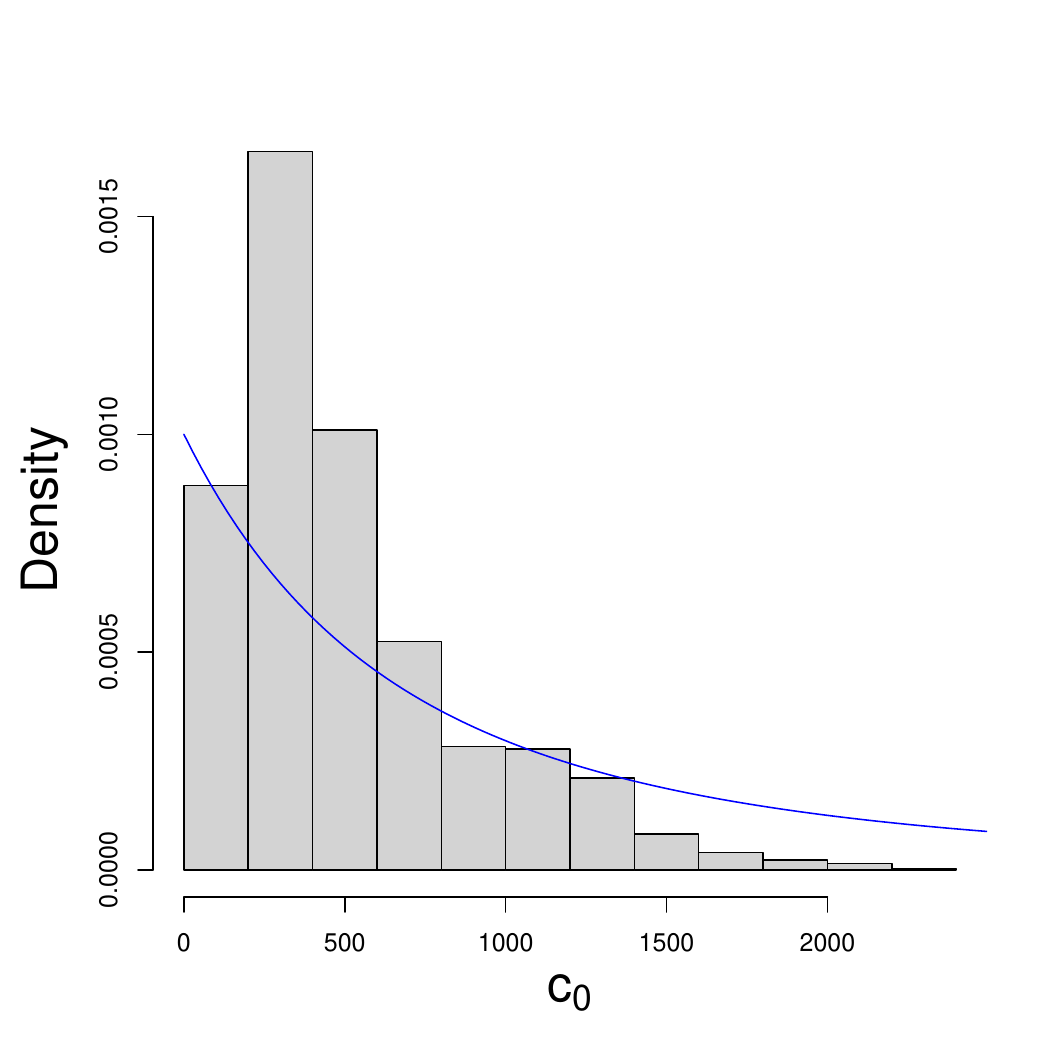}
\includegraphics[width=0.118\textwidth]{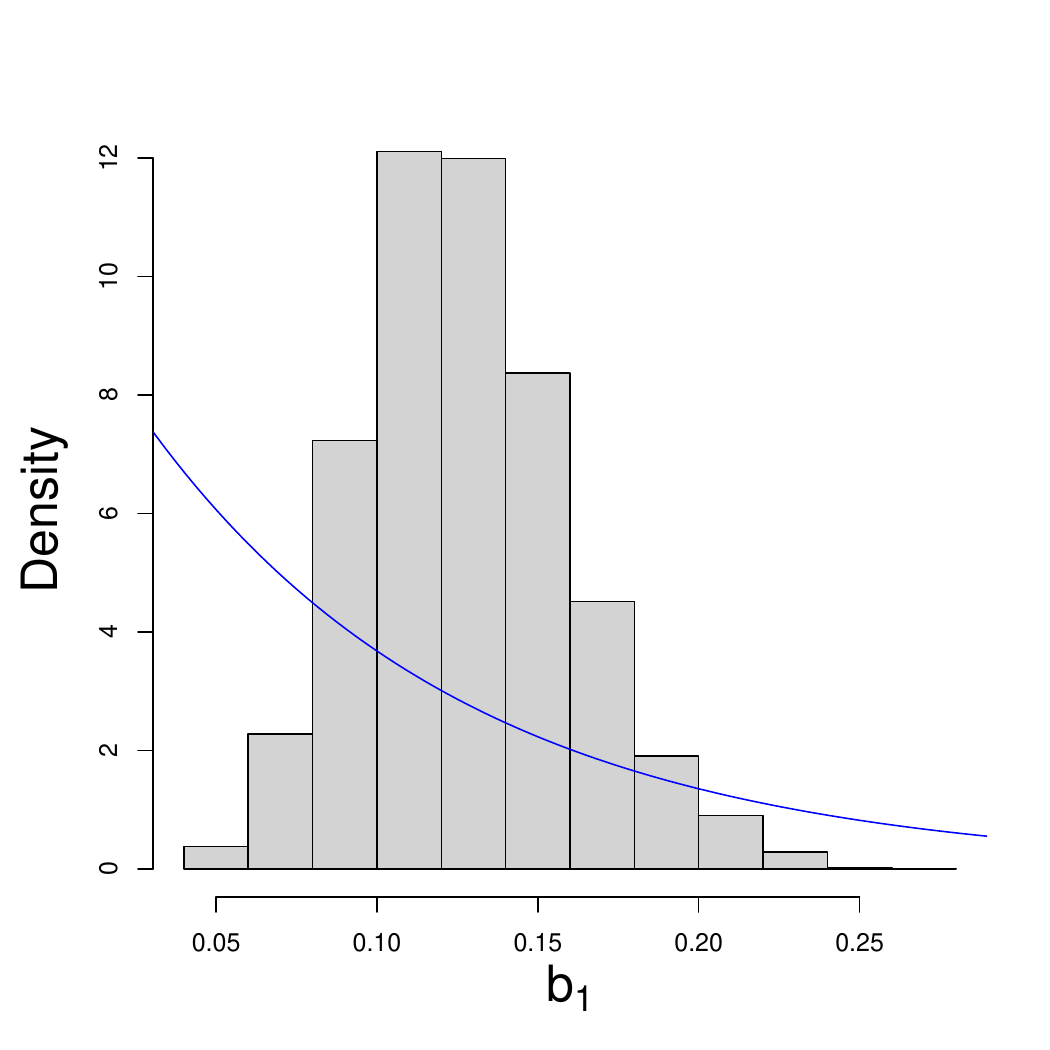}
\includegraphics[width=0.118\textwidth]{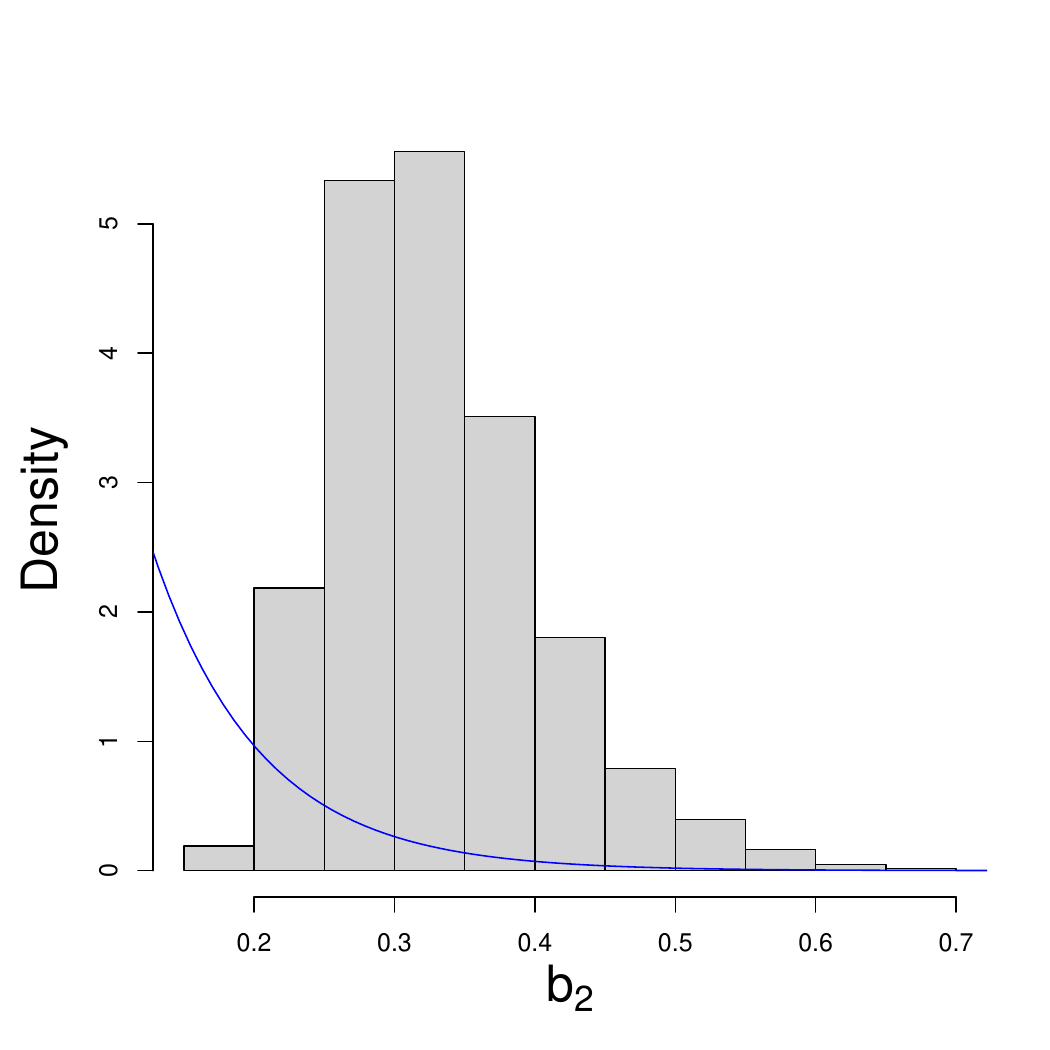}
\includegraphics[width=0.118\textwidth]{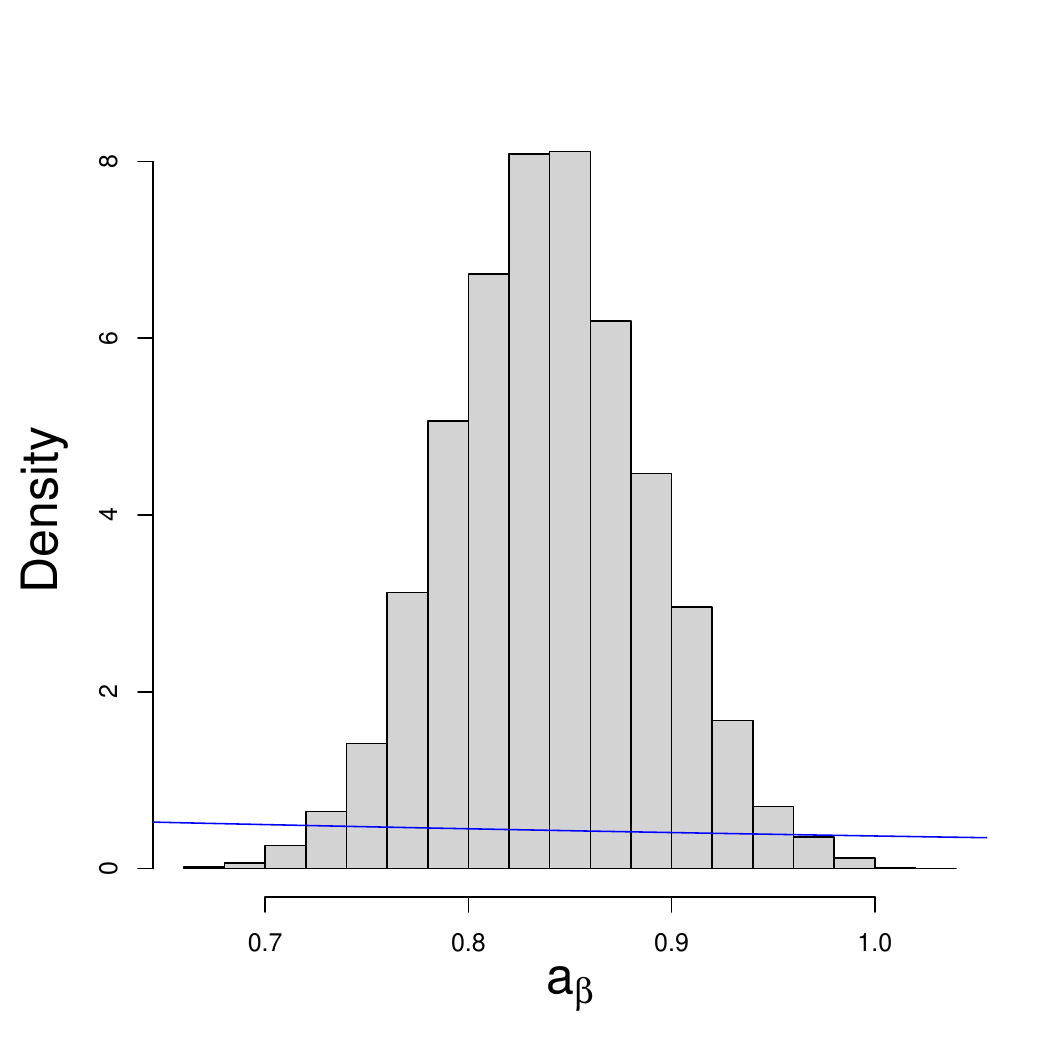}
\includegraphics[width=0.118\textwidth]{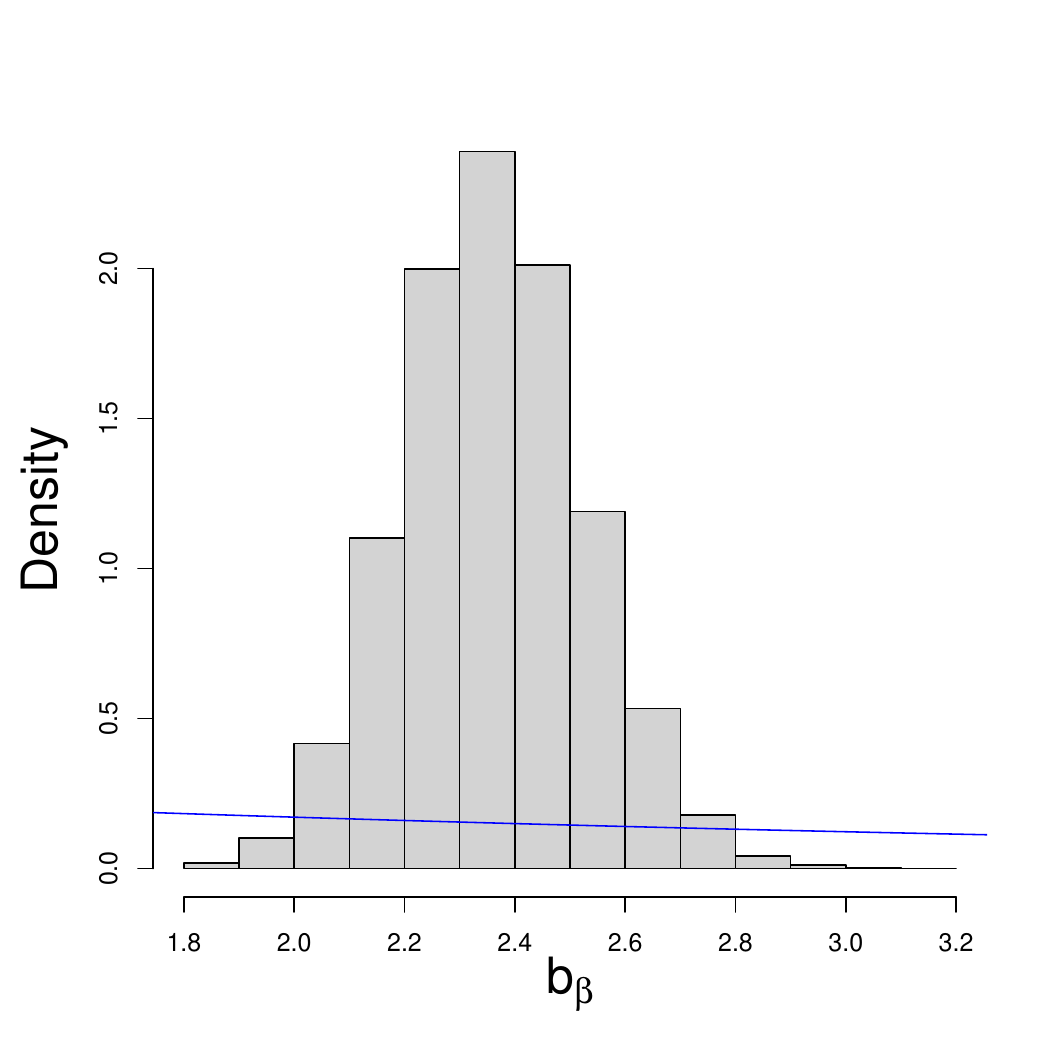}
\\
\caption{
{\small Simulation study (Section 4 of the main paper, and 
Section \ref{SM:mhp_simulation_powlaw}): Lomax density example (top row), 
mark-dependent Lomax density example (middle row), mixture of Lomax densities example 
(bottom row). Histograms of posterior samples for the model parameters 
$(\mu, \theta, d, c_0, b_1, b_2, a_\beta, b_\beta)$, along with the corresponding 
prior densities (blue lines).}}
\label{fig:simulation_examples_prior_posterior}
\end{figure}

For each of the simulation examples, Figure \ref{fig:simulation_examples_prior_posterior}
presents the prior and posterior densities for the parameters 
$(\mu, \theta, d, c_0, b_1, b_2, a_\beta, b_\beta)$ of the nonparametric model.

\section{Earthquake data analysis: additional results}
\label{sec:additional_results_realdata}

\subsection{Prior specification}
\label{SM_real_data_priors}

The priors for the three models with immigrant intensity constant in time 
(Section 5.1 of the main paper) were as follows. 
\begin{itemize}
\item 
ETAS model: $\text{Exp}(139)$ prior for $\mu$; $\text{Exp}(2)$, $\text{Exp}(0.2)$, 
and $\text{Exp}(0.1)$ priors for $a$, $b$, and $\psi$, respectively, with truncation
such that the MHP stability condition $0 < a\psi/(\psi - b) < 1$ is satisfied (with probability
$1$ in the prior); $\text{Exp}(0.005)$ and $\text{Exp}(0.1)$ priors for $p$ and $c$. 
\item 
Semiparametric model: $\text{Ga}(5, 0.25)$, $\text{Exp}(1)$, and $\text{Exp}(1.5)$ priors
for $\alpha_0$, $a_0$, and $b_0$, respectively; the priors for $\mu$, $a$, $b$, and $\psi$ 
are the same as in the ETAS model. 
\item Nonparametric model: $\text{Exp}(139)$ prior for $\mu$; $\text{Lomax}(2, 9)$ prior for $\theta$; 
$\text{Exp}(10)$ and $\text{Exp}(60)$ priors for $b_1$ and $b_2$; $\text{Exp}(1)$ prior for $d$; 
$\text{Lomax}(2, 2000)$ prior for $c_0$; $\text{Exp}(1)$ and $\text{Exp}(0.241)$ priors for 
$a_\beta$ and $b_\beta$. The values of $L = 20$ and $M = 160$ were calibrated using 
sensitivity analysis (see Section \ref{subsec:LM}). 
\end{itemize}

For the analysis under the general MHP model (Section 5.2 of the main paper), we also need 
to specify the hyperpriors corresponding to the nonparametric prior for the immigrant intensity 
function (Section 3.1.3 of the main paper). A $\text{Lomax}(2,600)$ prior is assigned to 
$\phi$, an $\text{Exp}(0.1)$ prior to $e_0$, and an $\text{Exp}(0.007)$ prior to $b_{G_0}$. 
The number of basis densities is set to $J=100$.

\subsection{Inference for the mark density and for model parameters}
\label{subsec:mark_density}

\begin{figure}[!t]
\centering
\includegraphics[width=0.54\textwidth]
{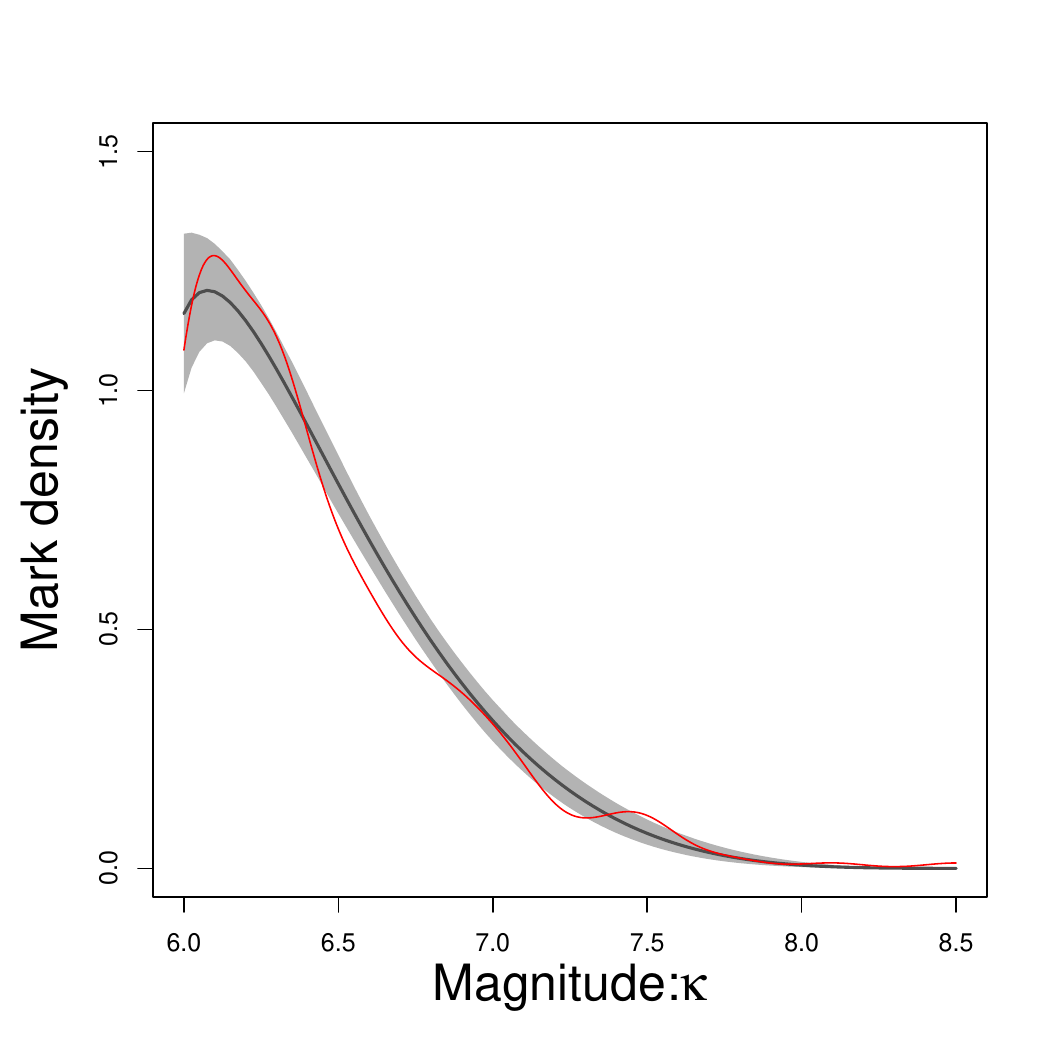} \\
\caption{{\small 
Earthquake data analysis (Section 5 of the main paper).
Posterior mean (black solid line) and 95\% posterior uncertainty bands (gray shaded area) 
for the mark density of earthquake magnitudes, under the nonparametric model.
The red line shows the kernel density estimate (\texttt{R function density()}) of 
the observed earthquake magnitudes.}}
\label{fig:markDensity}
\end{figure}

As discussed in Section 3 of the main paper, the mark density for the nonparametric model 
is a (rescaled) beta density with support over the earthquake magnitude interval relevant 
to the particular application. Figure \ref{fig:markDensity} plots the mark density estimates 
for the earthquake data from Japan.

\begin{figure}[t]
\centering
\includegraphics[width=0.118\linewidth]{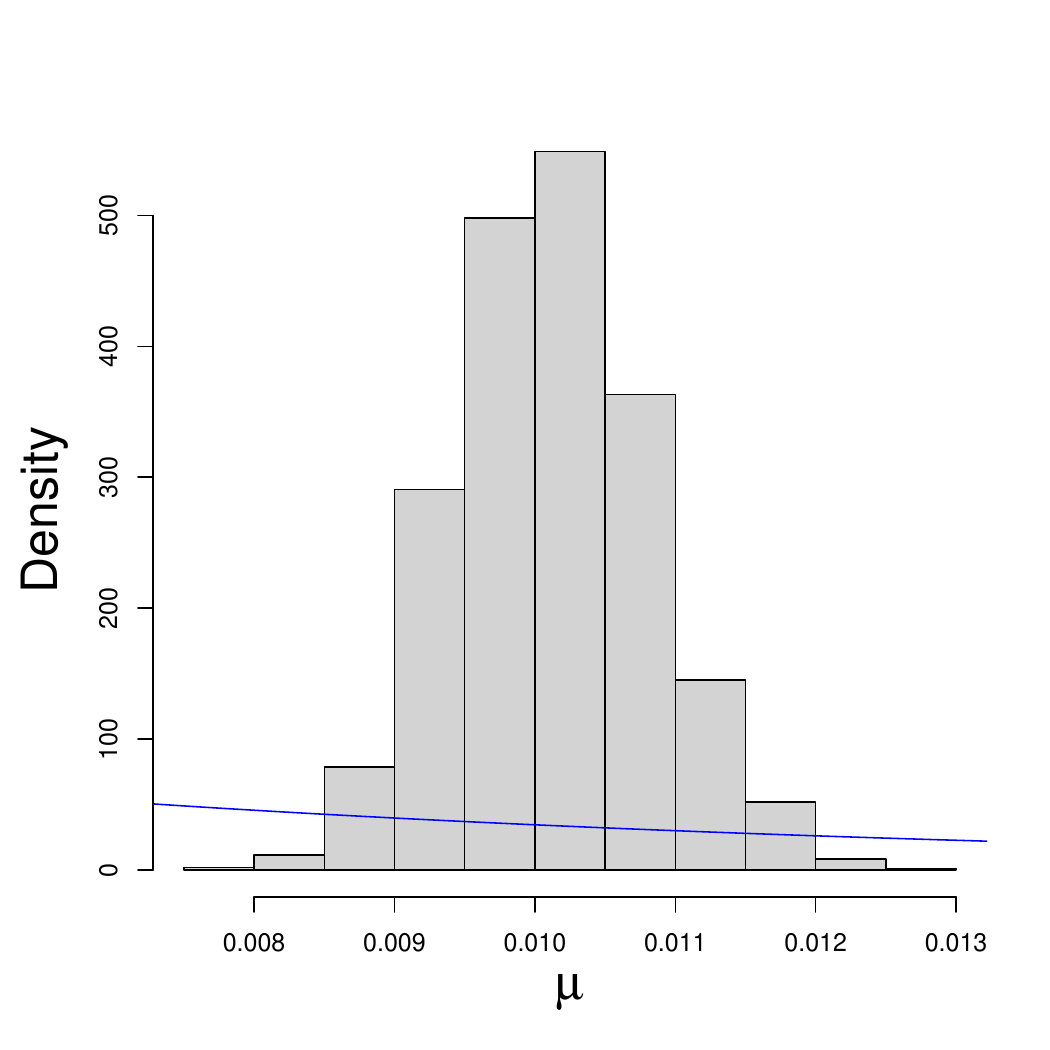}
\includegraphics[width=0.118\linewidth]{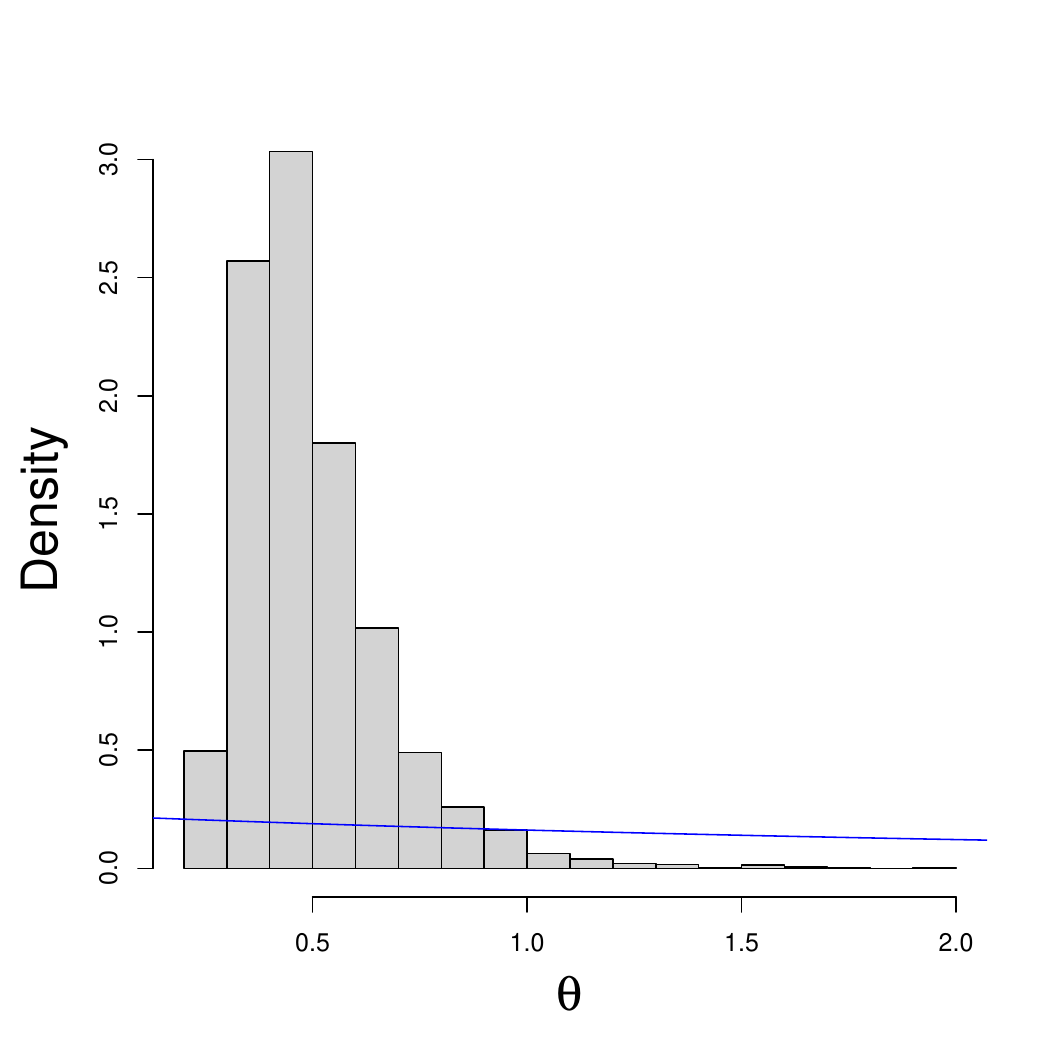}
\includegraphics[width=0.118\linewidth]{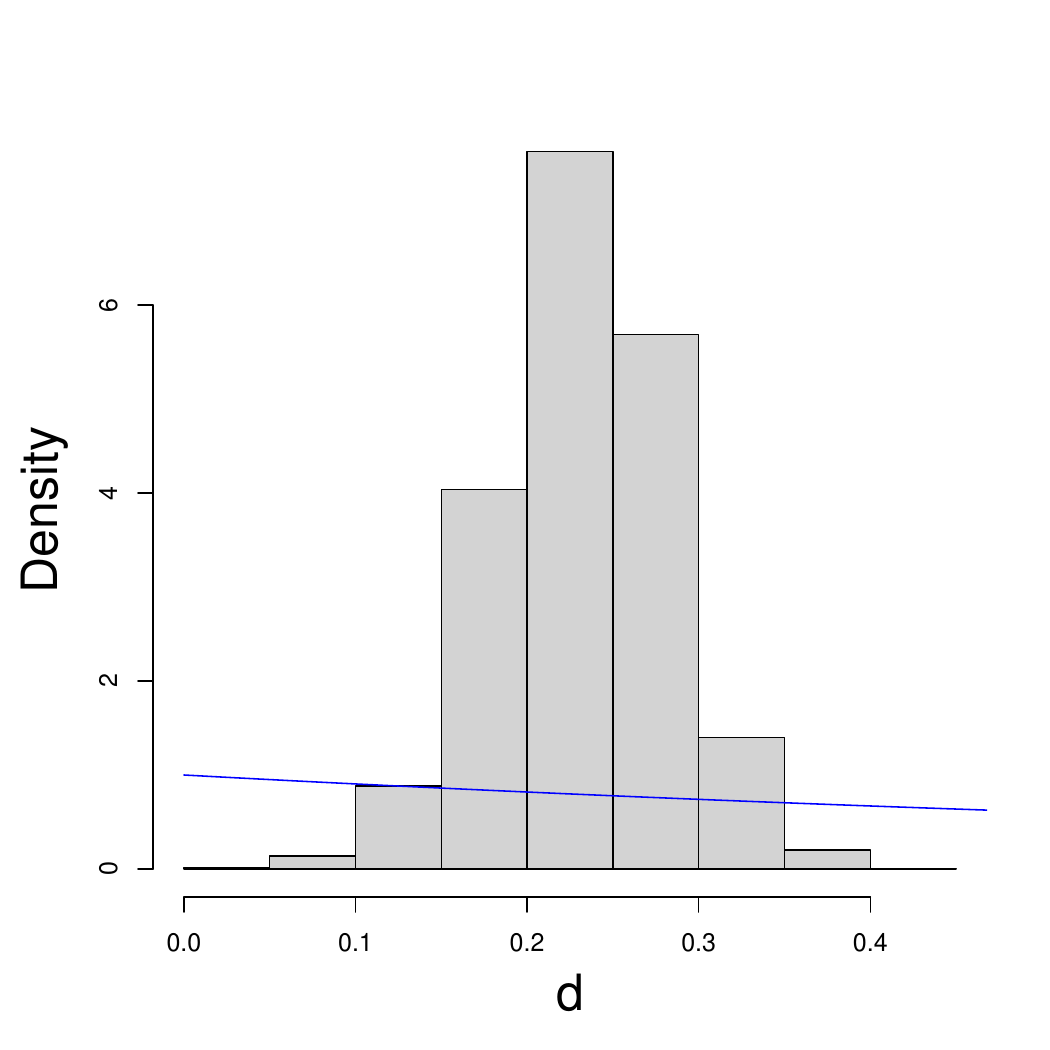}
\includegraphics[width=0.118\linewidth]{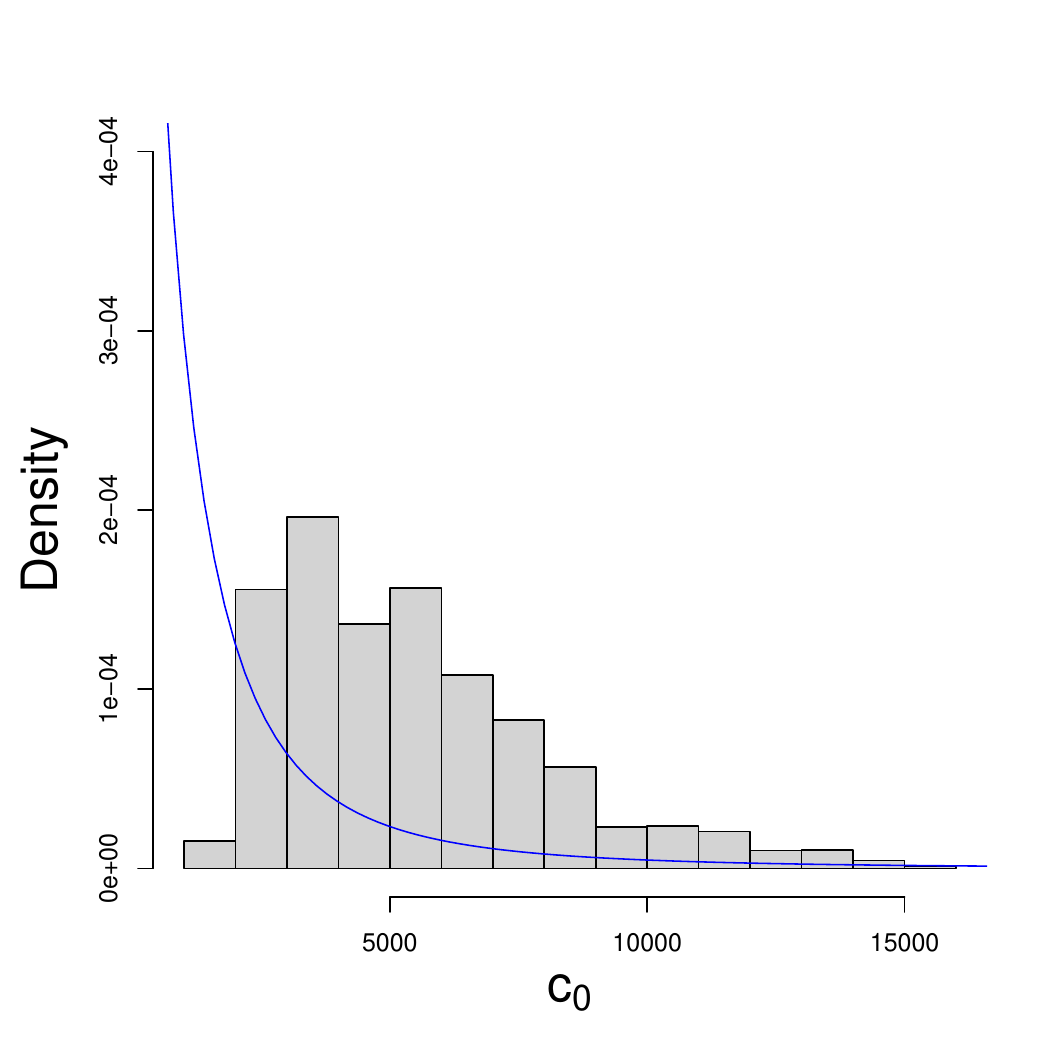}
\includegraphics[width=0.118\linewidth]{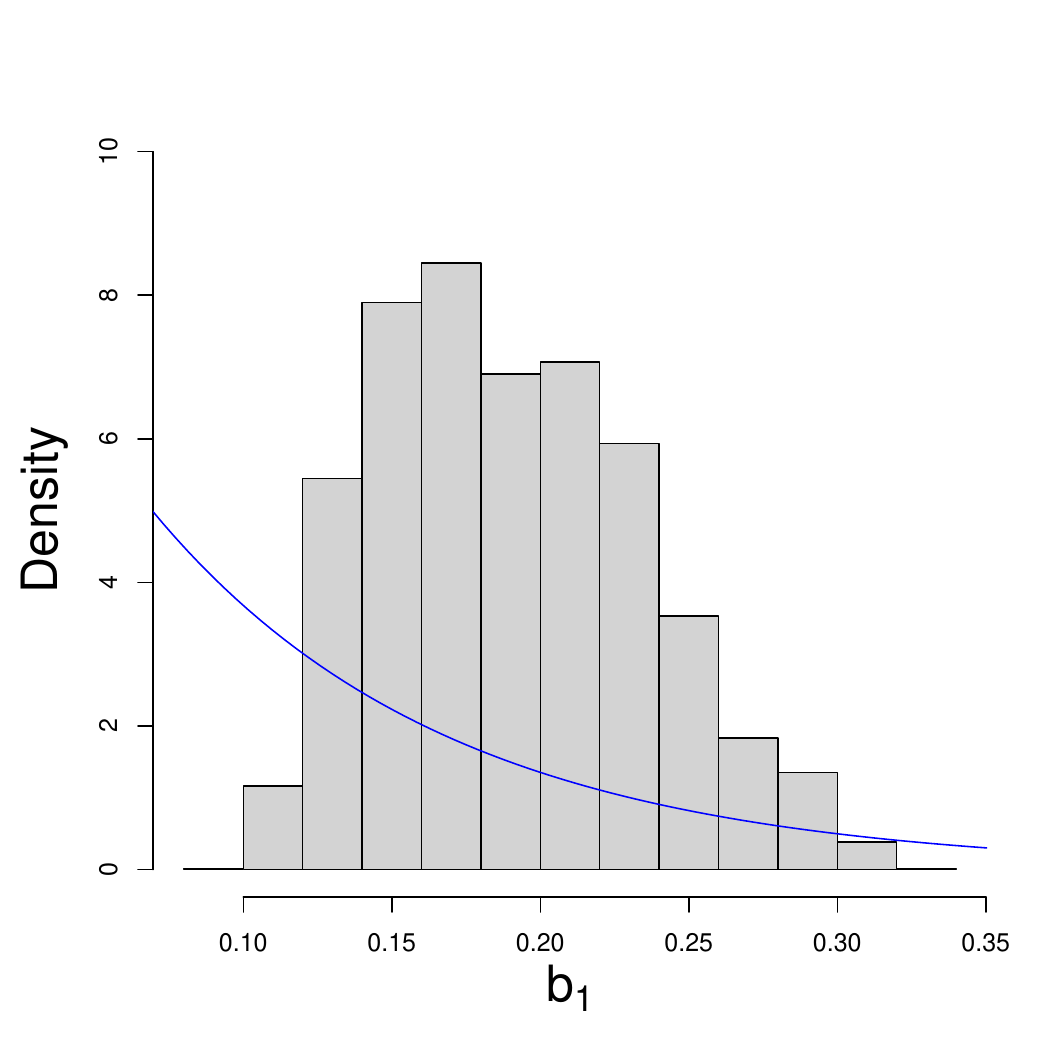}
\includegraphics[width=0.118\linewidth]{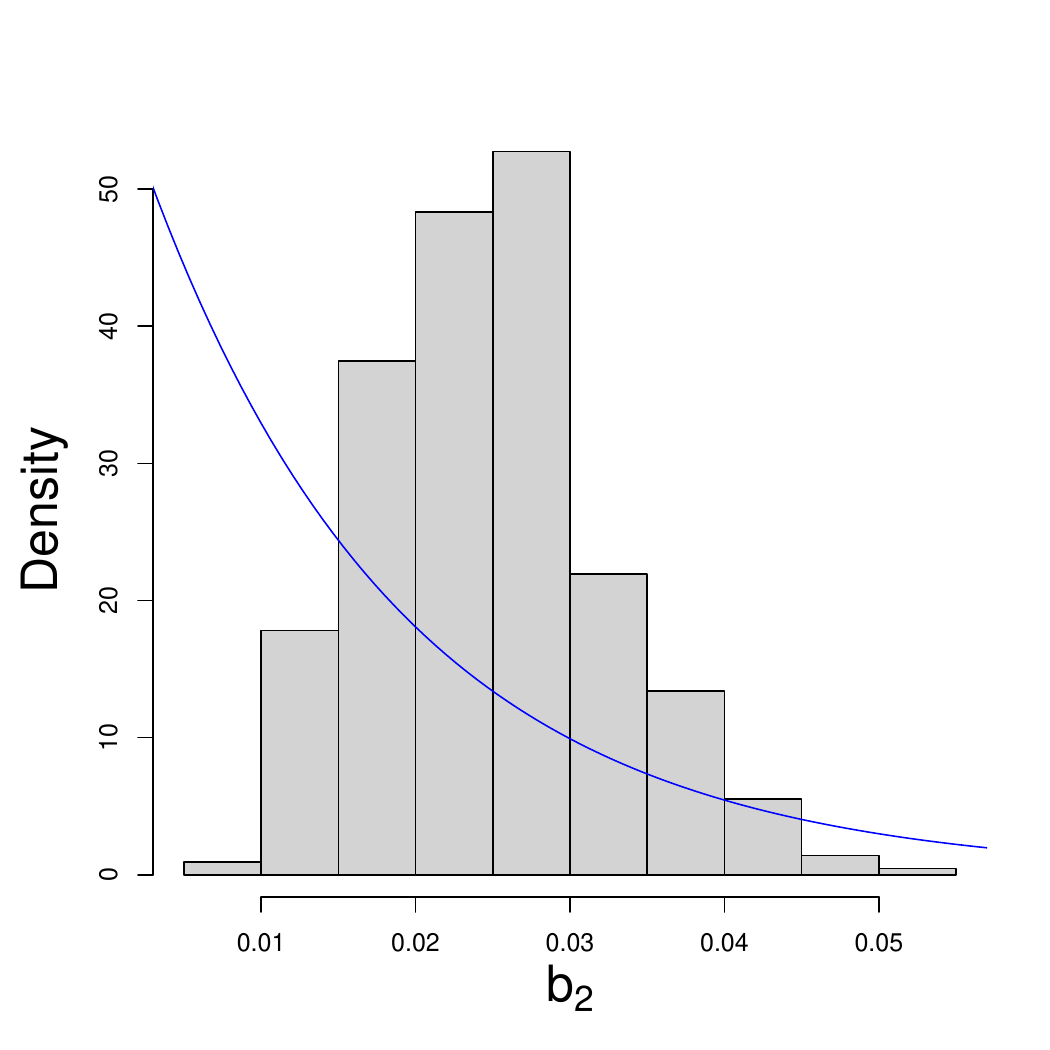}
\includegraphics[width=0.118\linewidth]{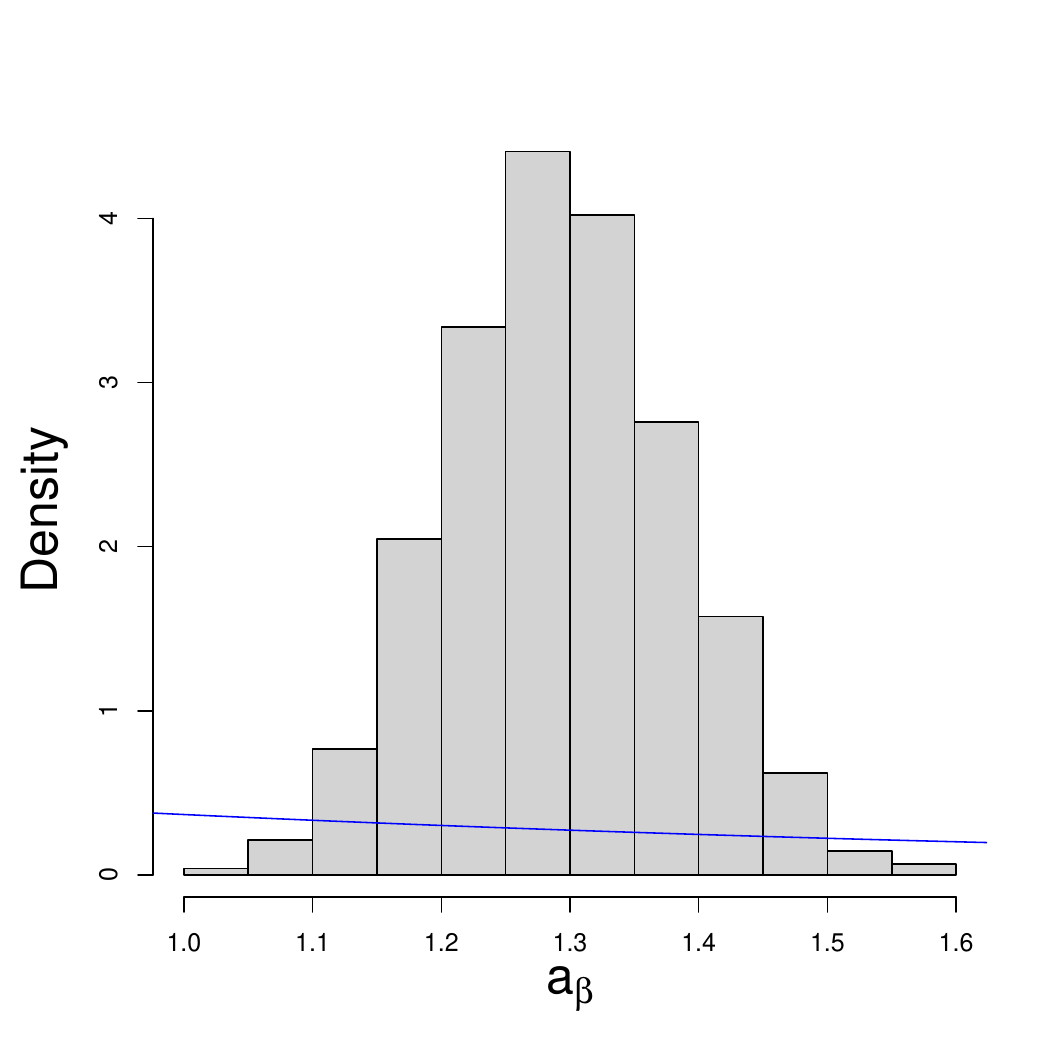}
\includegraphics[width=0.118\linewidth]{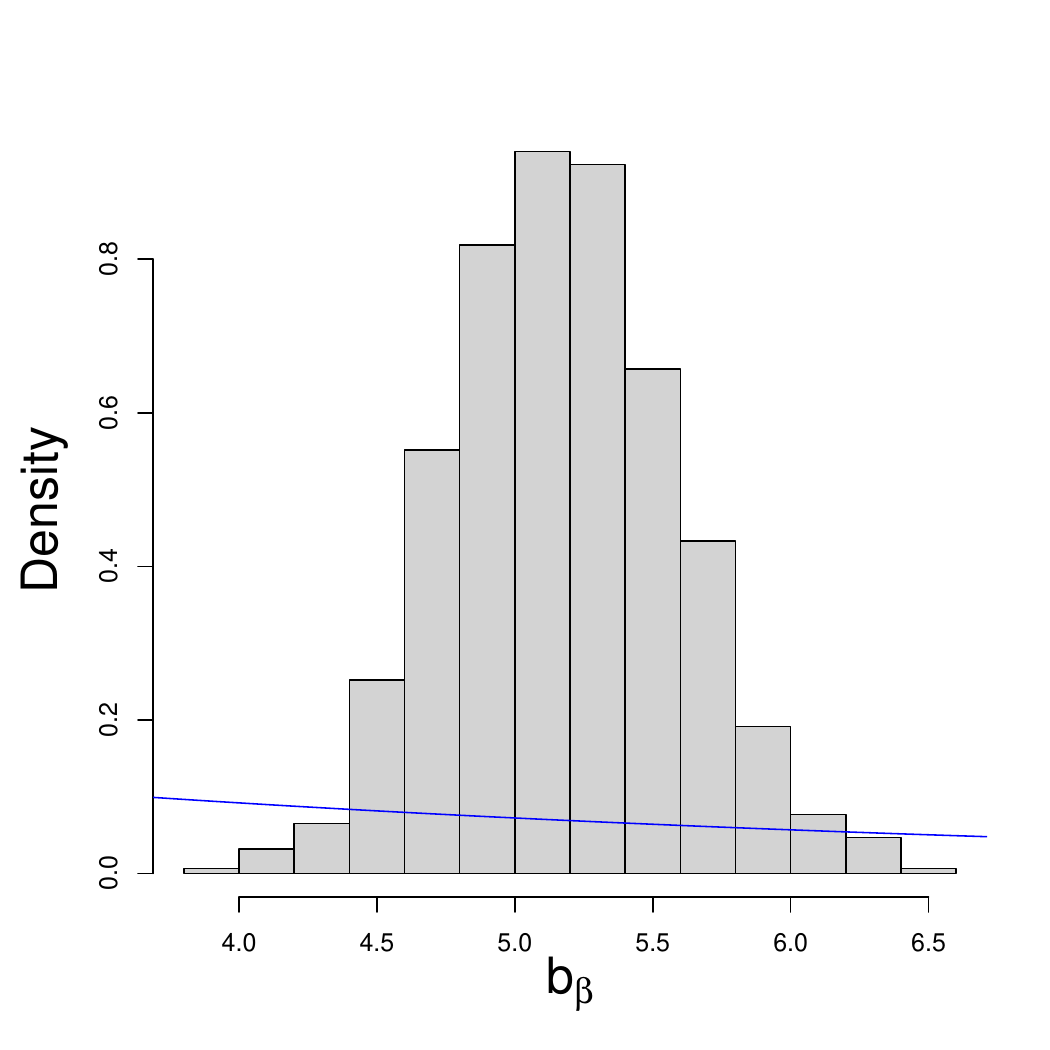}
\caption{{\small 
Earthquake data analysis (Section 5 of the main paper).
Histograms of posterior samples and prior densities (blue lines) 
for the parameters $(\mu, \theta, d, c_0, b_1, b_2, a_\beta, b_\beta)$ of the 
nonparametric model with immigrant intensity constant in time.}}
\label{fig:realdata_prior_posterior}
\end{figure}

For the version of the nonparametric model with immigrant intensity constant in time, 
Figure \ref{fig:realdata_prior_posterior} plots the prior and posterior densities for 
the model parameters. We note prior-to-posterior learning for all parameters.

\subsection{Prior sensitivity analysis}
\label{subsec:sensitivity}

\begin{figure}[!t]
\centering
\includegraphics[width=0.19\textwidth]{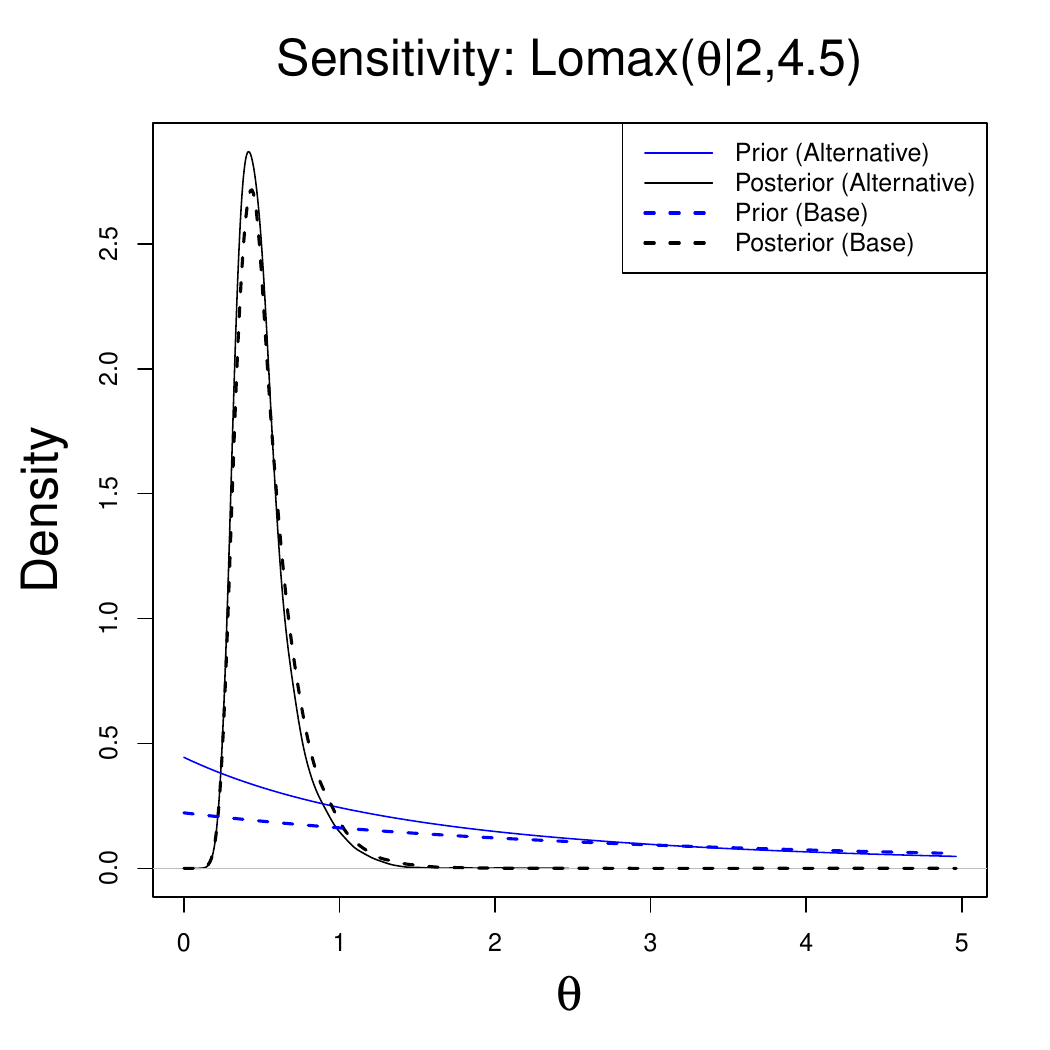}
\includegraphics[width=0.19\textwidth]{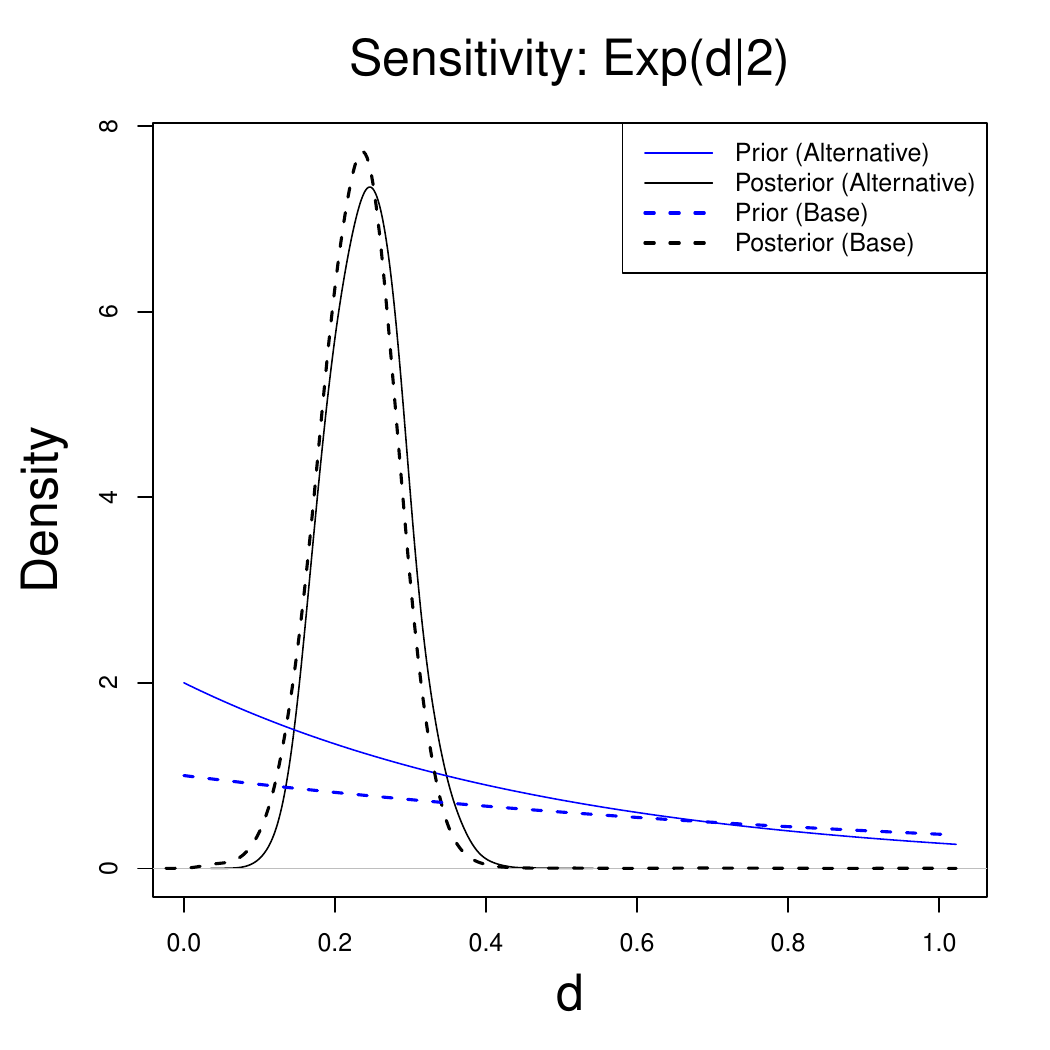}
\includegraphics[width=0.19\textwidth]{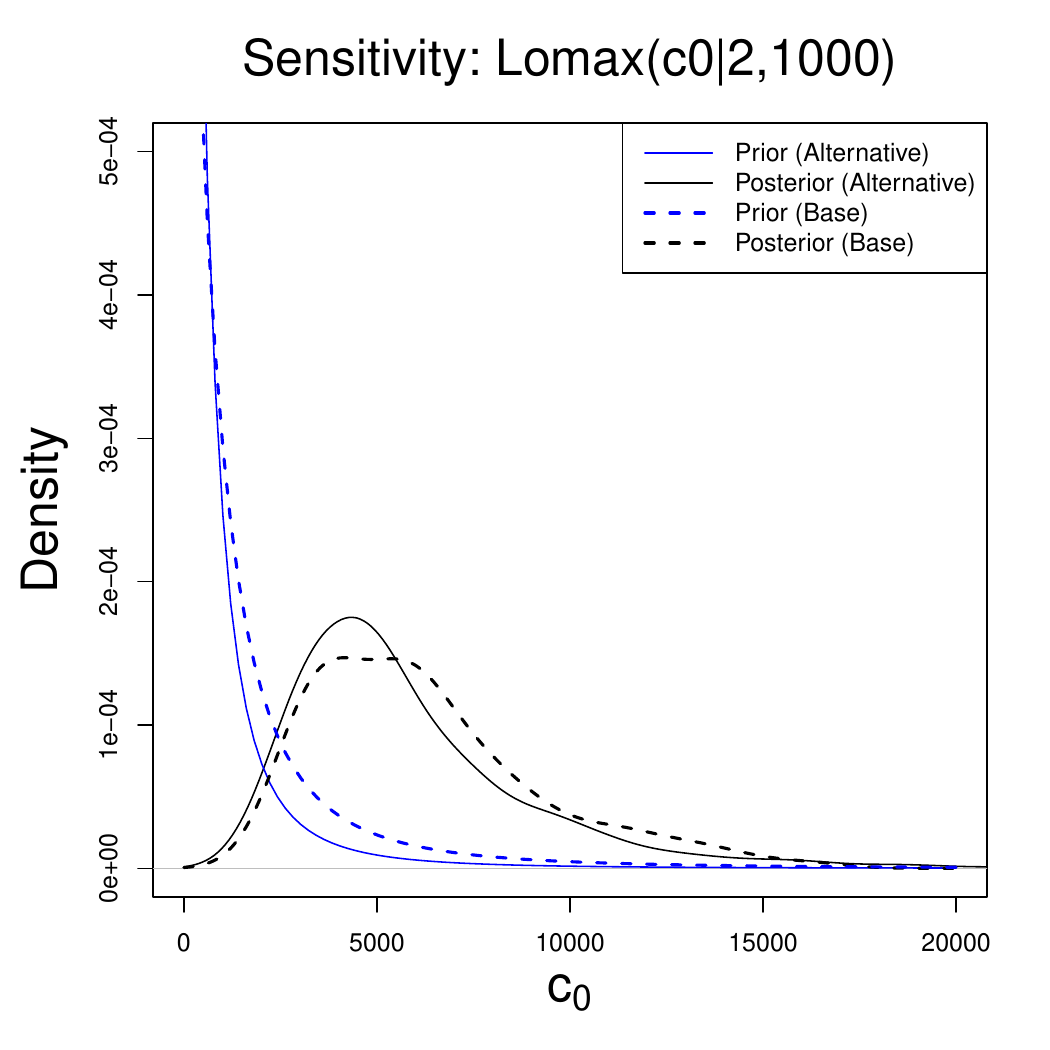}
\includegraphics[width=0.19\textwidth]{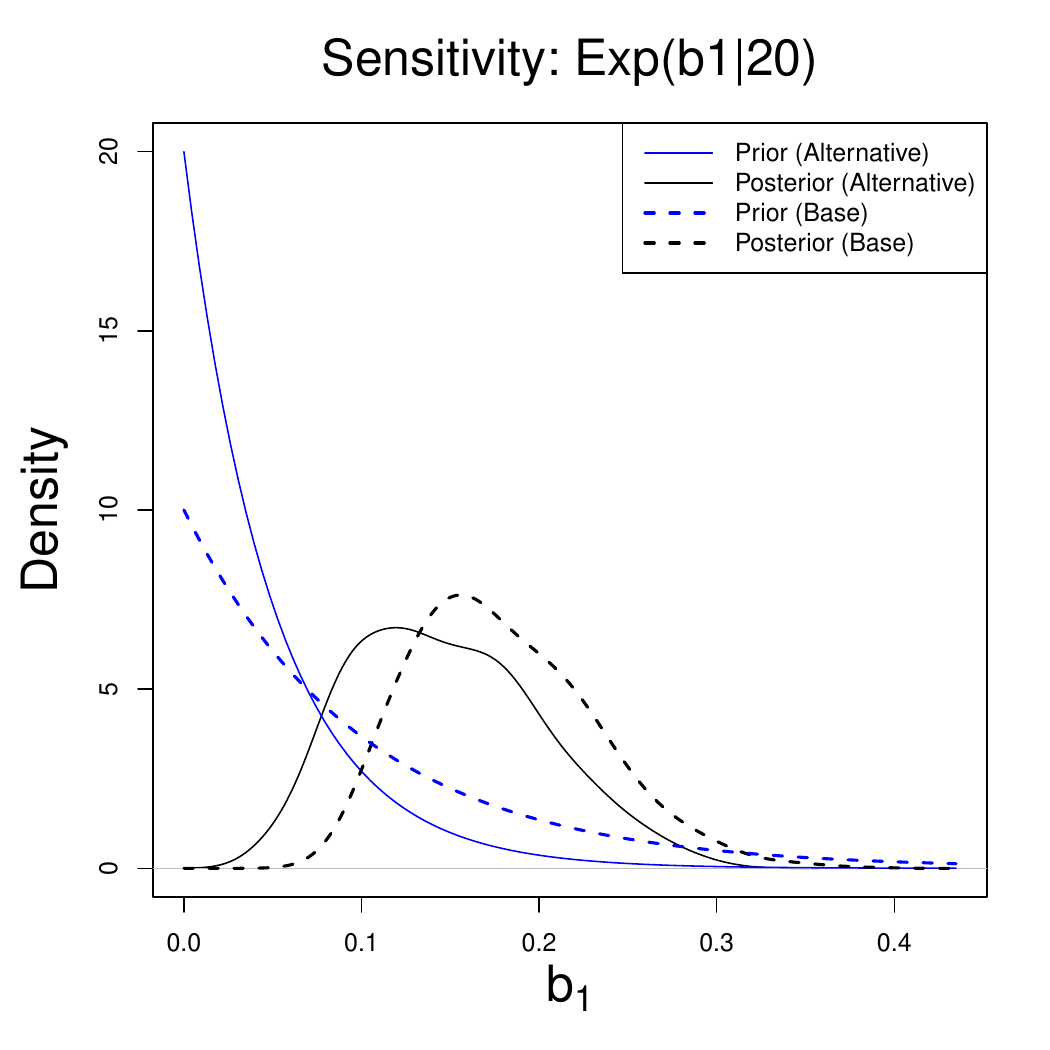}
\includegraphics[width=0.19\textwidth]{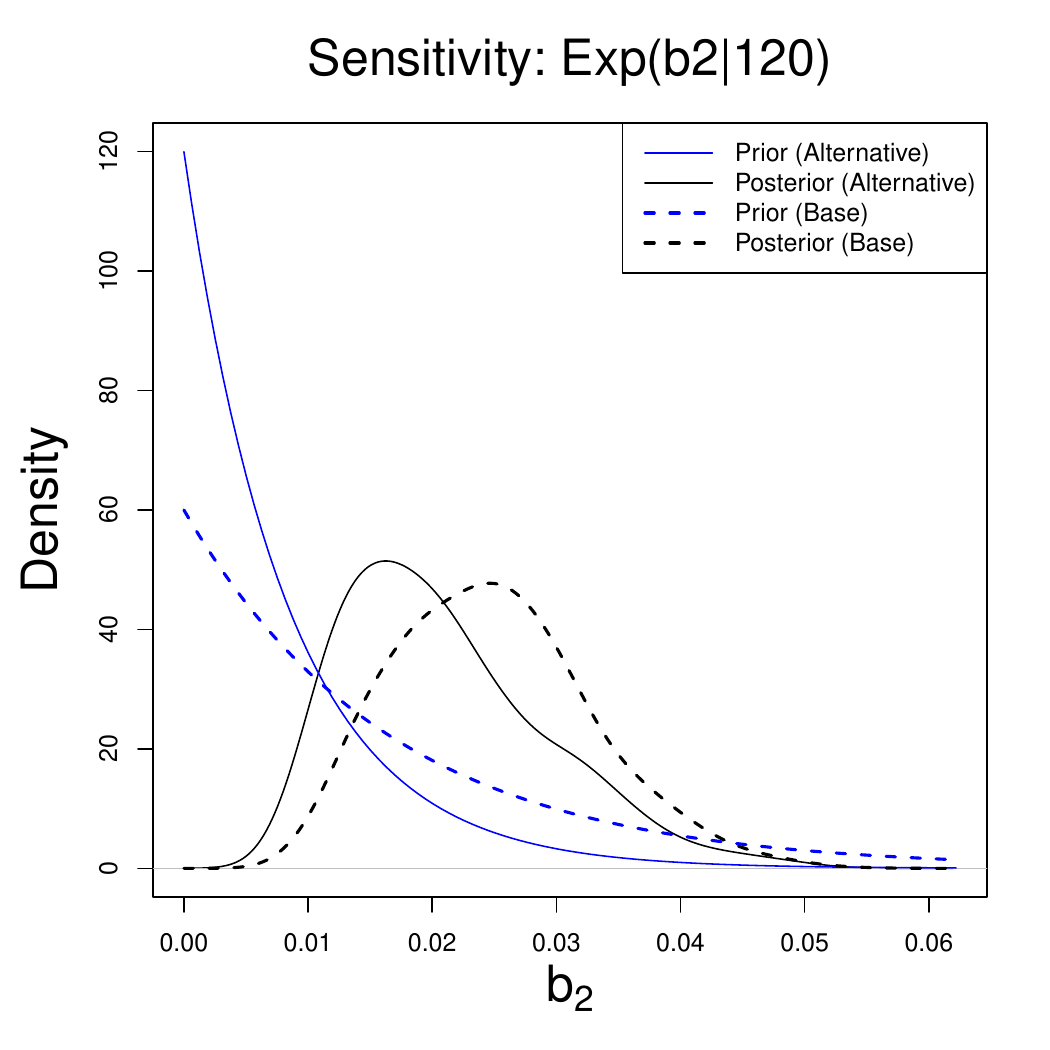} \\
\includegraphics[width=0.19\textwidth]{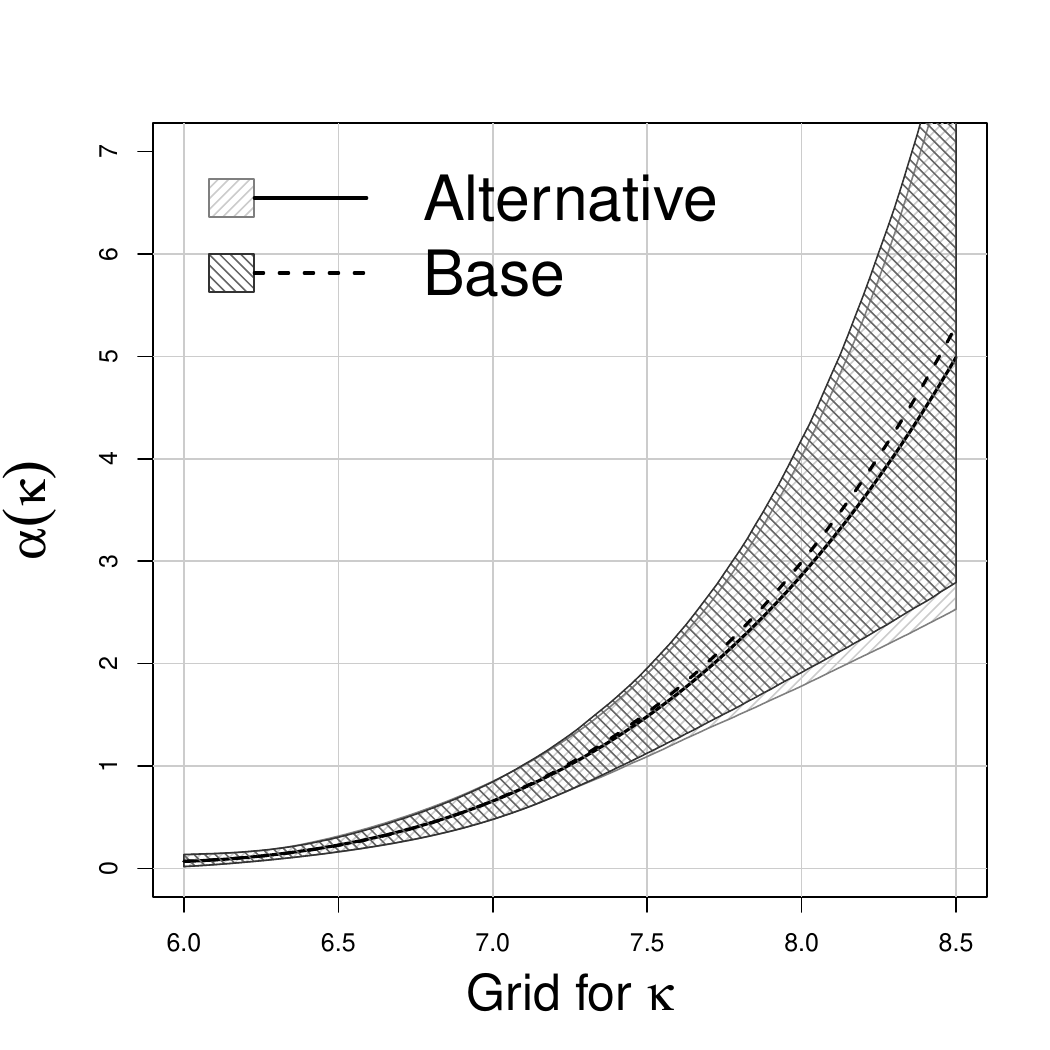}
\includegraphics[width=0.19\textwidth]{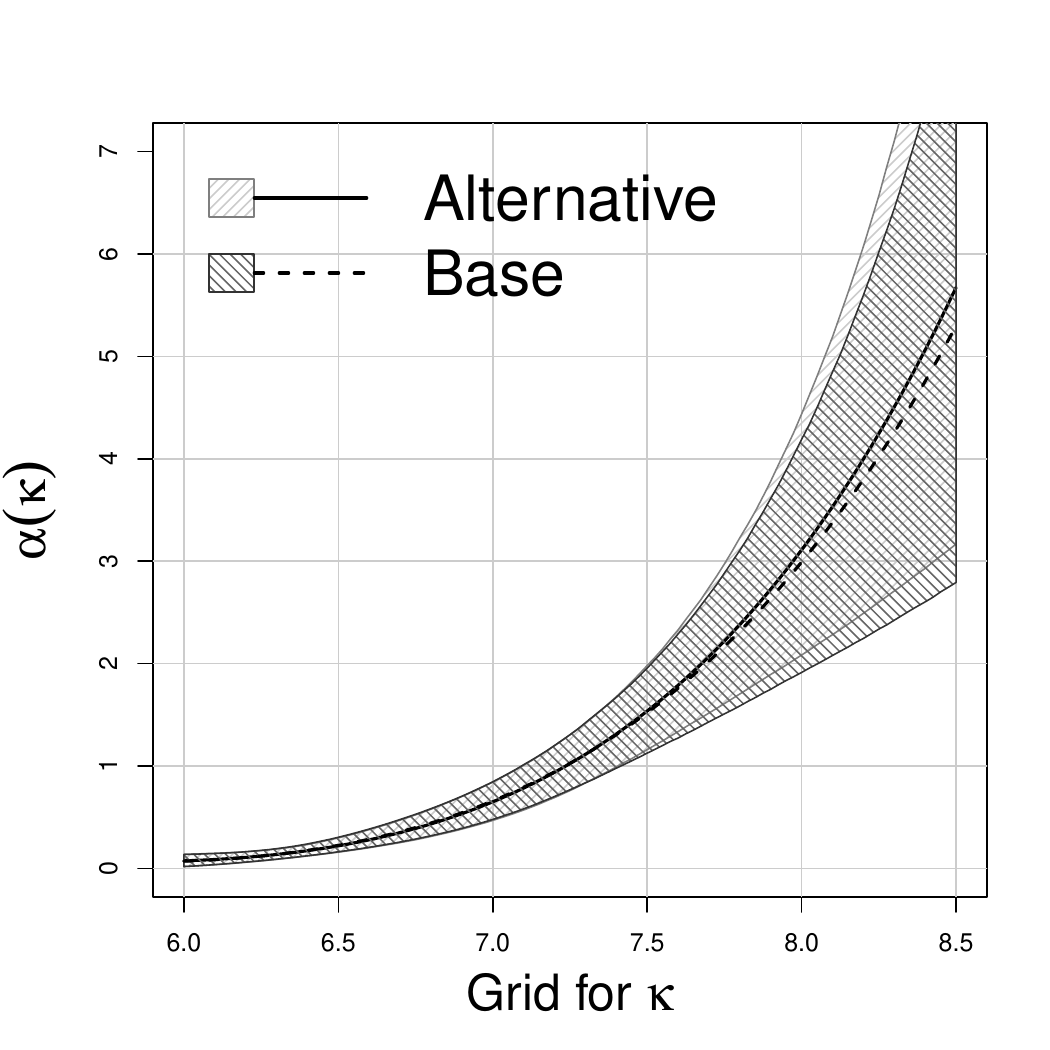}
\includegraphics[width=0.19\textwidth]{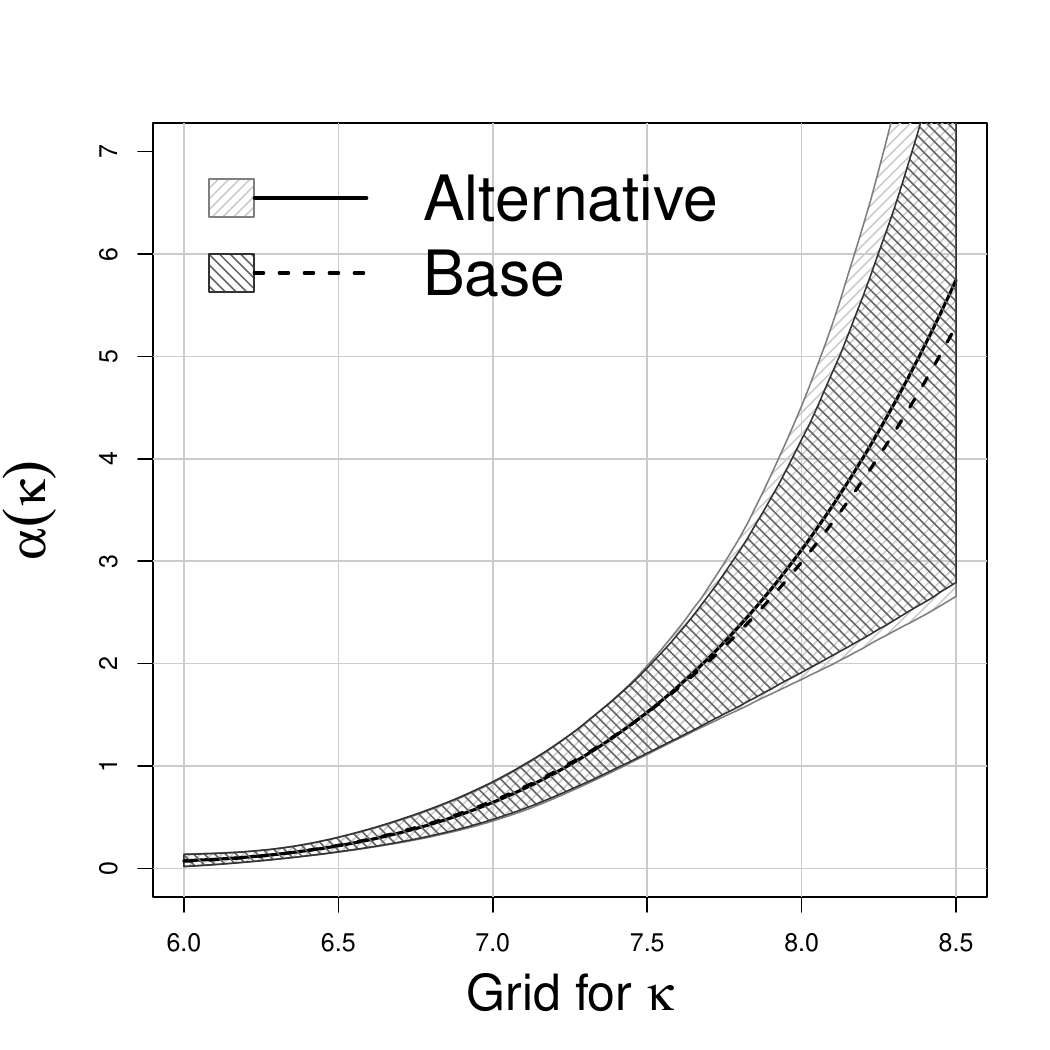}
\includegraphics[width=0.19\textwidth]{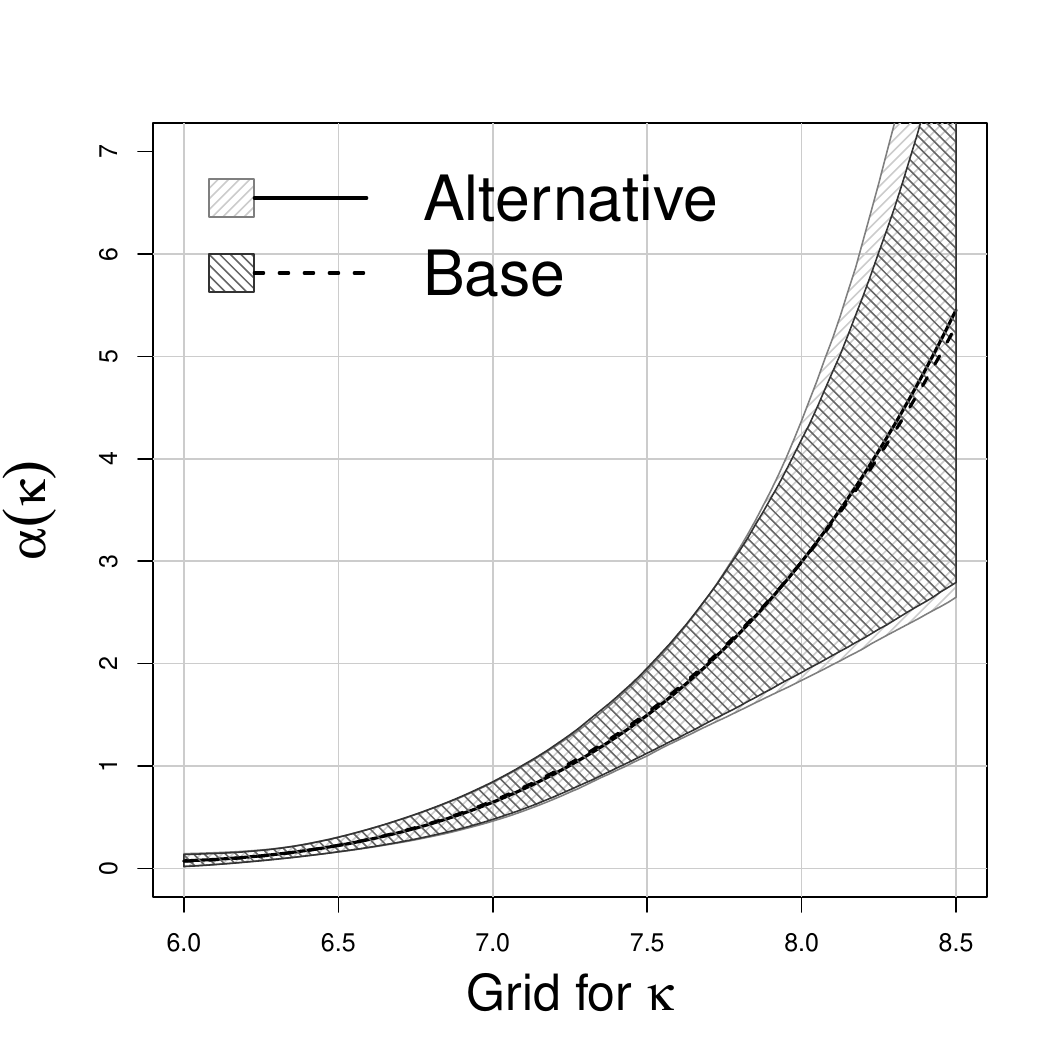}
\includegraphics[width=0.19\textwidth]{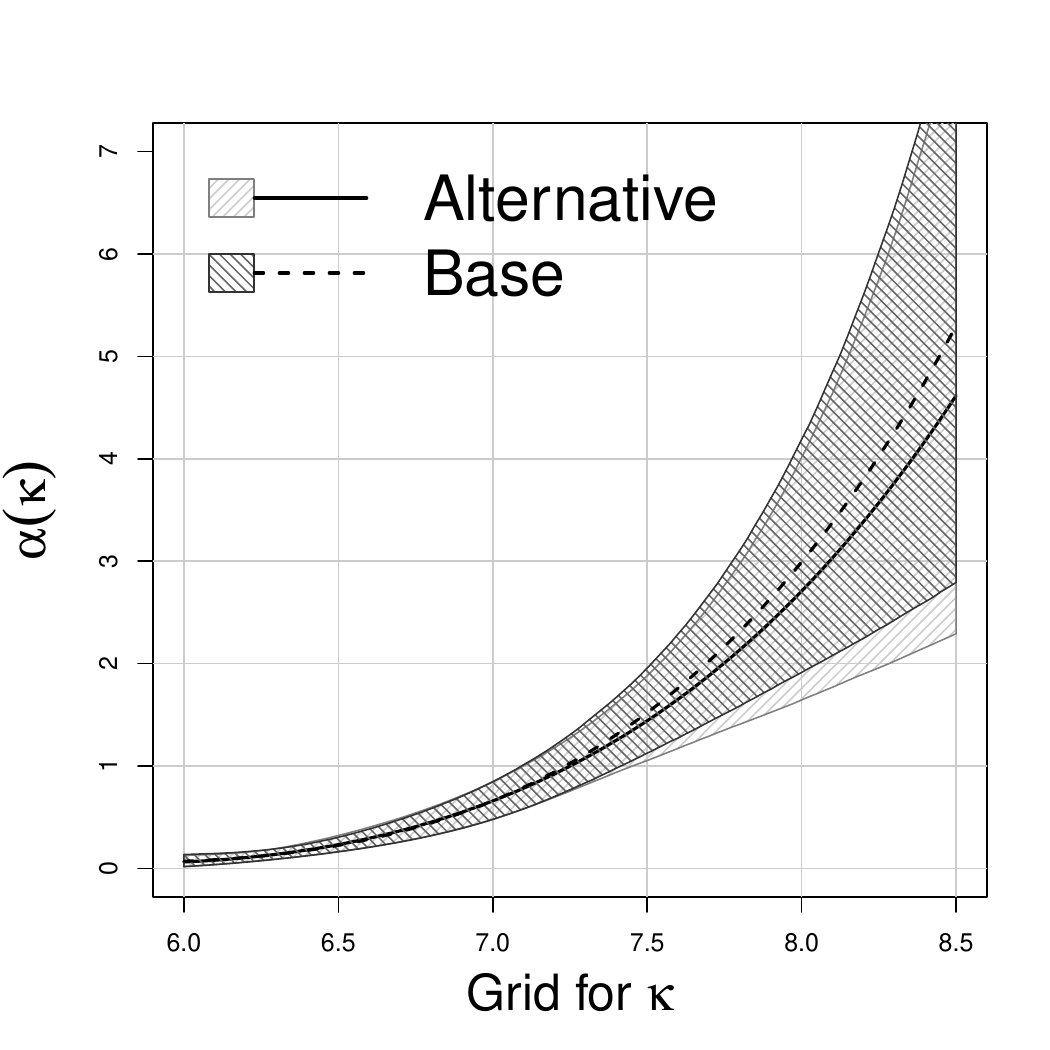} \\
\includegraphics[width=0.19\textwidth]{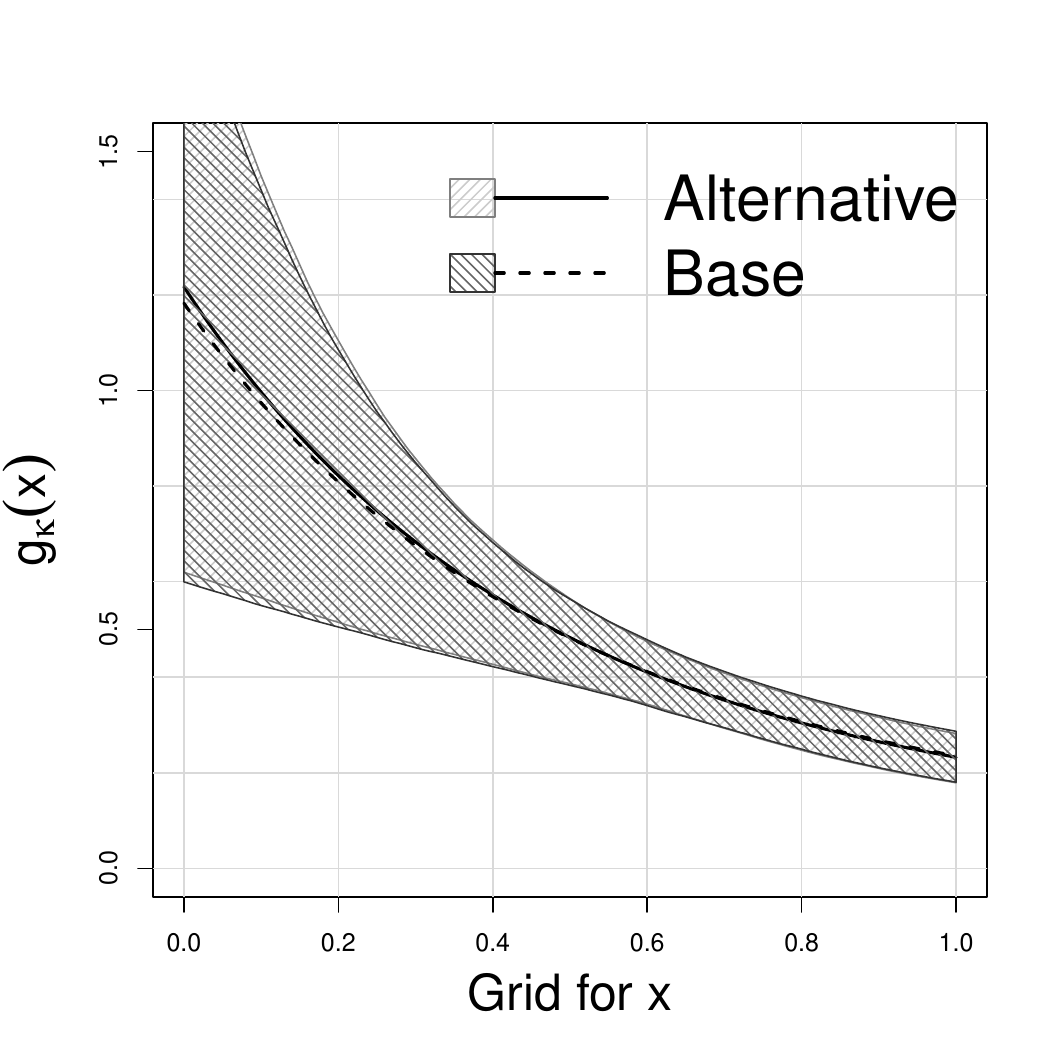}
\includegraphics[width=0.19\textwidth]{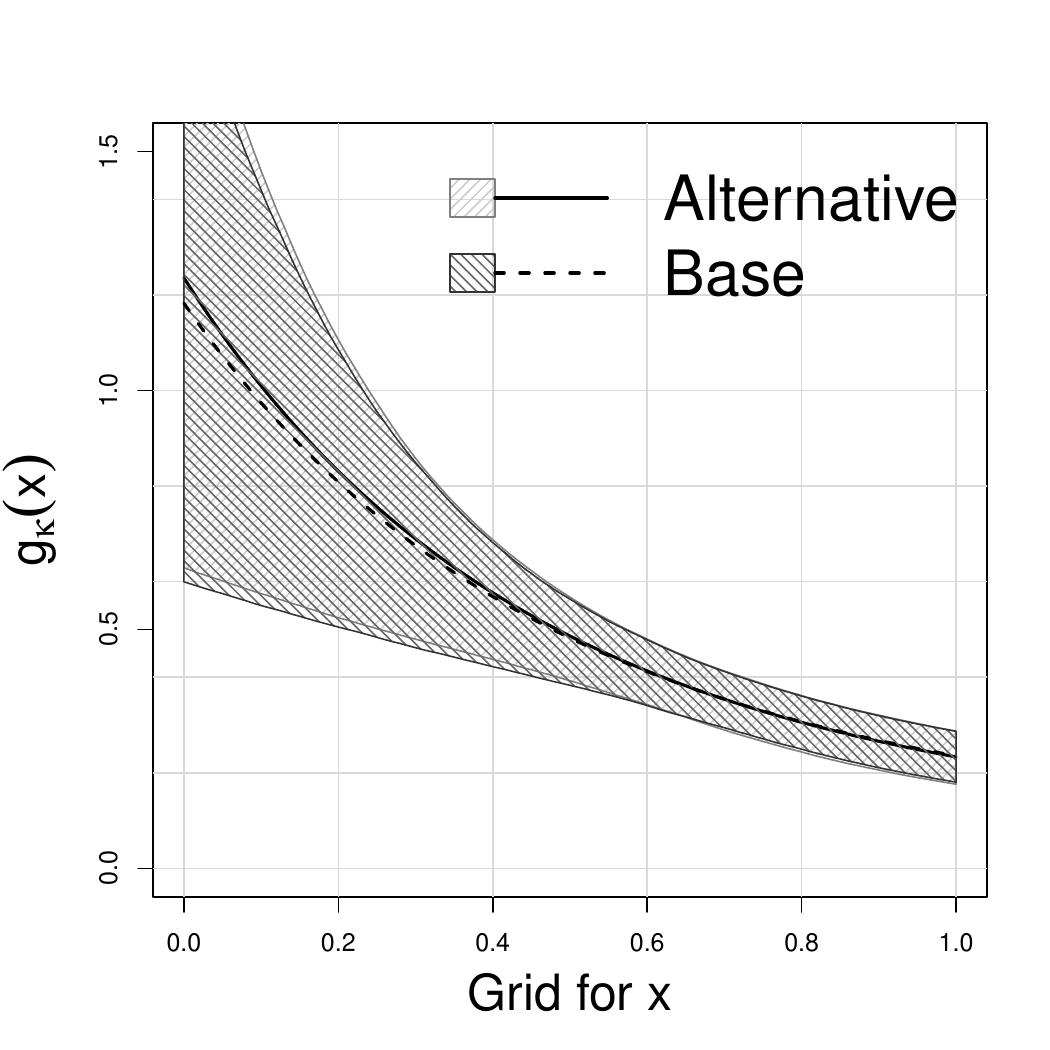}
\includegraphics[width=0.19\textwidth]{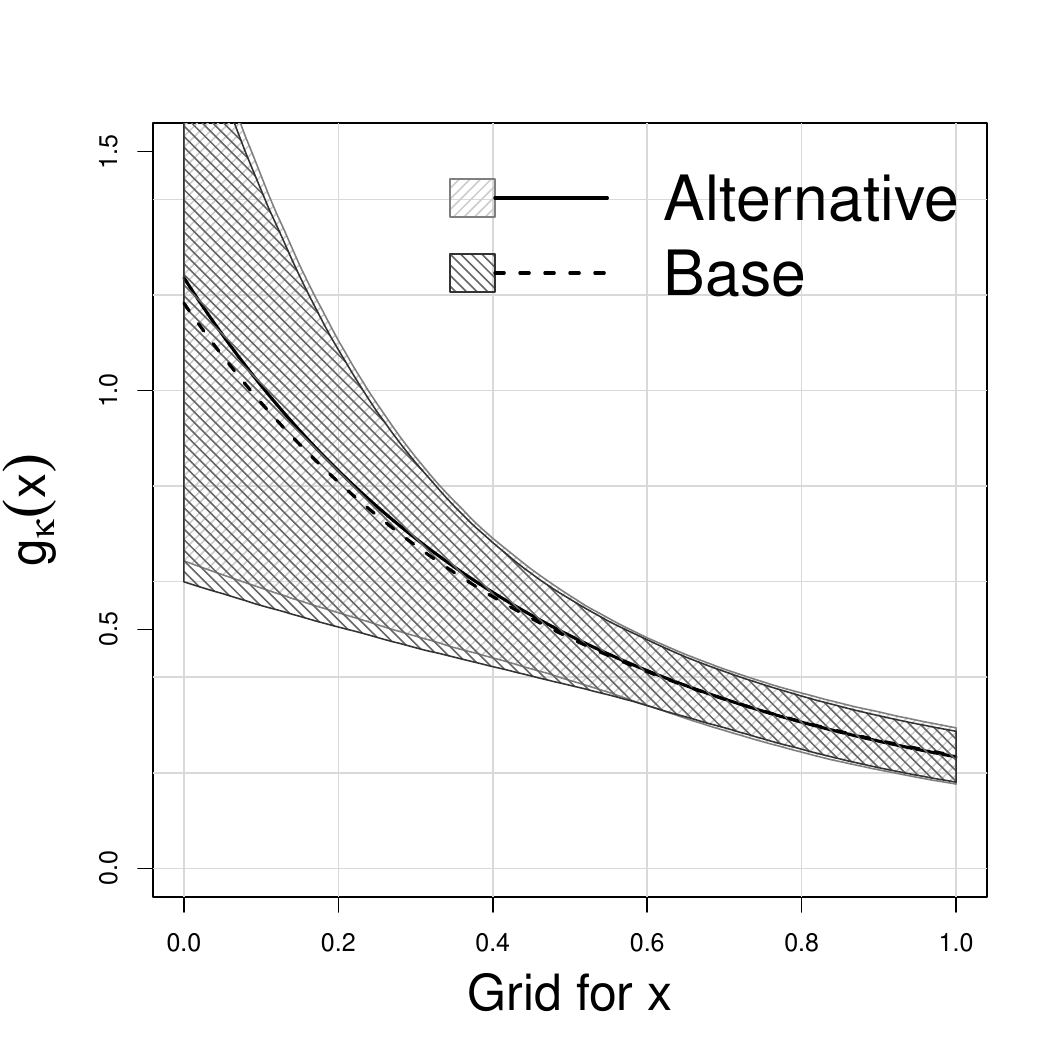}
\includegraphics[width=0.19\textwidth]{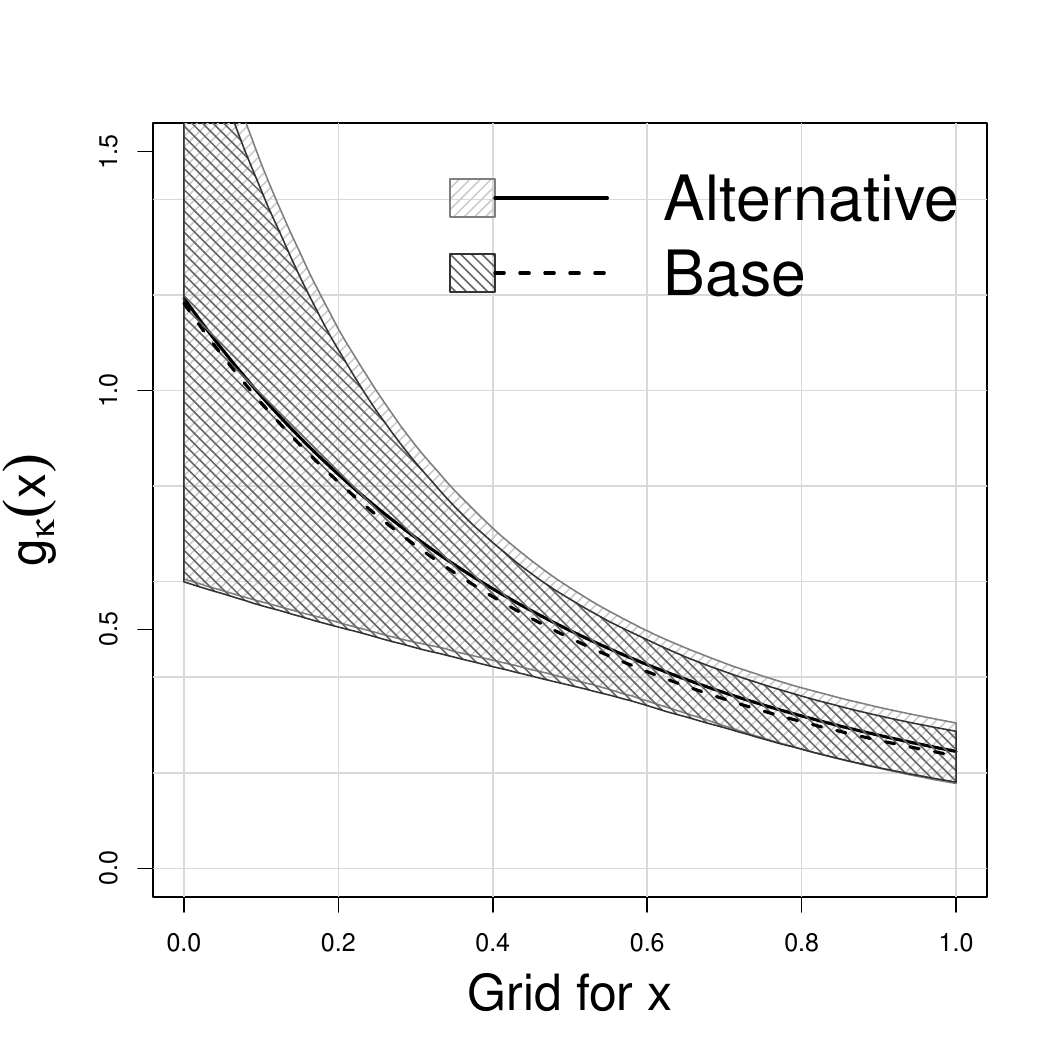}
\includegraphics[width=0.19\textwidth]{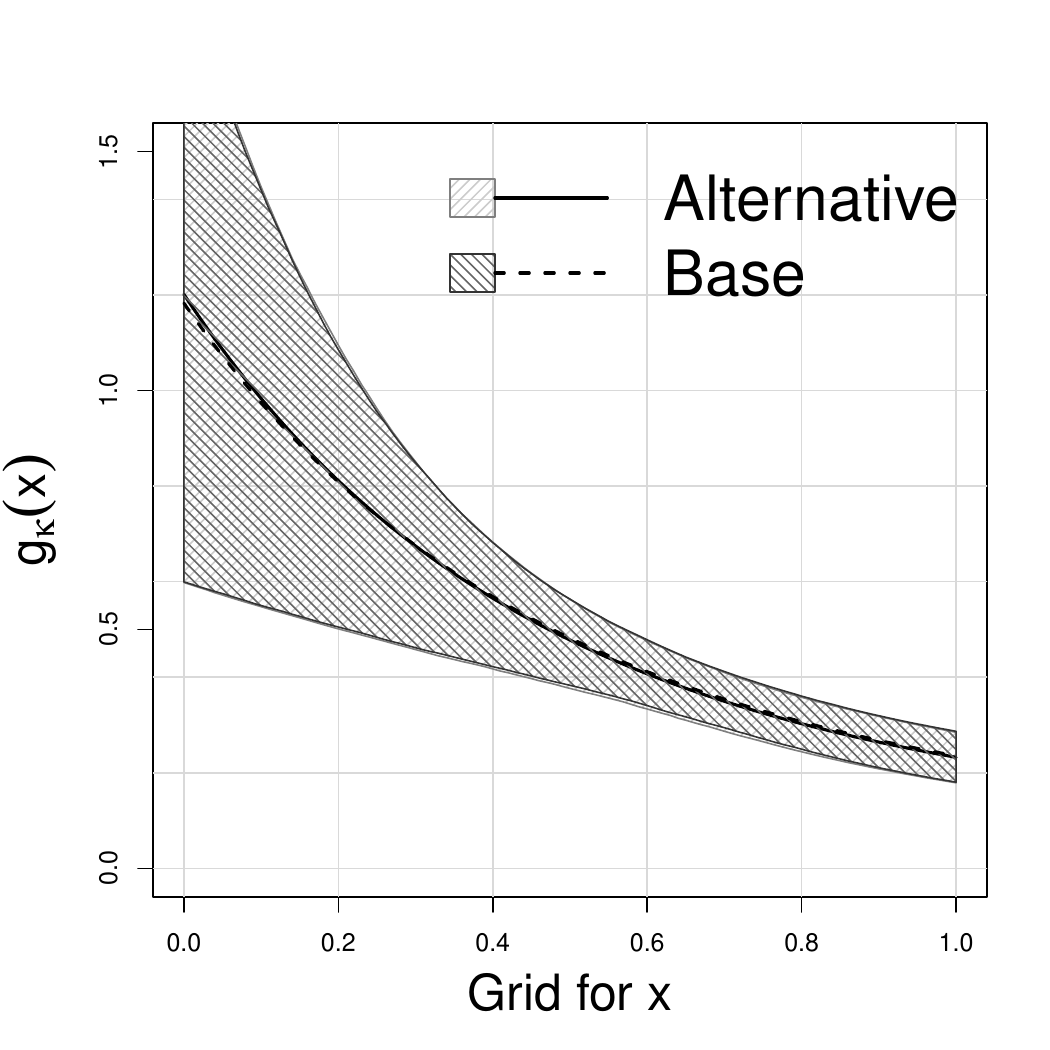} \\
\caption{{\small 
Earthquake data analysis (Section 5 of the main paper): prior sensitivity analysis 
for the nonparametric model with immigrant intensity constant in time. 
Results are shown for the original (``Base'') prior and the modified (``Alternative'') prior. 
Each column corresponds to a distinct prior setting, where the prior for the associated 
parameter (shown in the top row) is changed one-at-a-time. 
Top row: prior and posterior densities for parameters $\theta$, $d$, $c_0$, $b_1$, and $b_2$.
Middle row: posterior mean and 95\% interval estimates for the total offspring intensity. 
Bottom row: posterior mean and 95\% interval estimates for the offspring density at 
magnitude $\kappa = 7.25$.}}
\label{fig:sensitivity}
\end{figure}

Here, we present some results from prior sensitivity analysis, working with the version of 
the nonparametric model with immigrant intensity constant in time. We focus on the parameters 
involved in the prior model for the MHP excitation function, that is, parameters $\theta$, 
$d$, $c_0$, $b_1$, and $b_2$. 
Using two prior choices for each of these parameters, Figure \ref{fig:sensitivity} compares 
the posterior density for the parameter, as well as the estimates for the total offspring 
intensity and for the offspring density at magnitude $\kappa = 7.25$. The prior specification 
used for the data analysis (given in Section \ref{SM_real_data_priors}) is referred to as 
the ``Base'' prior in Figure \ref{fig:sensitivity}. The modified prior specification used 
for these sensitivity analysis results (referred to as ``Alternative'' prior in 
Figure \ref{fig:sensitivity}) changes one-at-a-time the prior for each one of the 
five parameters above. In particular, the modified prior for $d$, $b_1$, and $b_2$ is 
exponential with mean given by one-half of the original prior mean. The modified prior for 
$\theta$ and $c_0$ is Lomax with shape parameter kept equal to $2$, but with mean set to 
one-half of the original prior mean.

We note from the top row of Figure \ref{fig:sensitivity} that the posterior densities for 
parameters $\theta$ and $d$ are practically the same under the two different prior choices. 
The other three parameters are the hyperparameters of the gamma process prior that defines 
the weights in the basis representation for the MHP excitation function. As to be expected, 
the posterior densities for $c_0$, $b_1$, and $b_2$ are (somewhat) more sensitive to the prior 
change, but there is still evident prior-to-posterior learning for these parameters.
Importantly, estimation and uncertainty quantification for the total offspring intensity 
function and, especially, for the offspring density is robust to the different prior settings.

\subsection{Sensitivity analysis for $L$ and $M$}
\label{subsec:LM}

\begin{figure}[!t]
\centering
\includegraphics[width=0.24\textwidth]{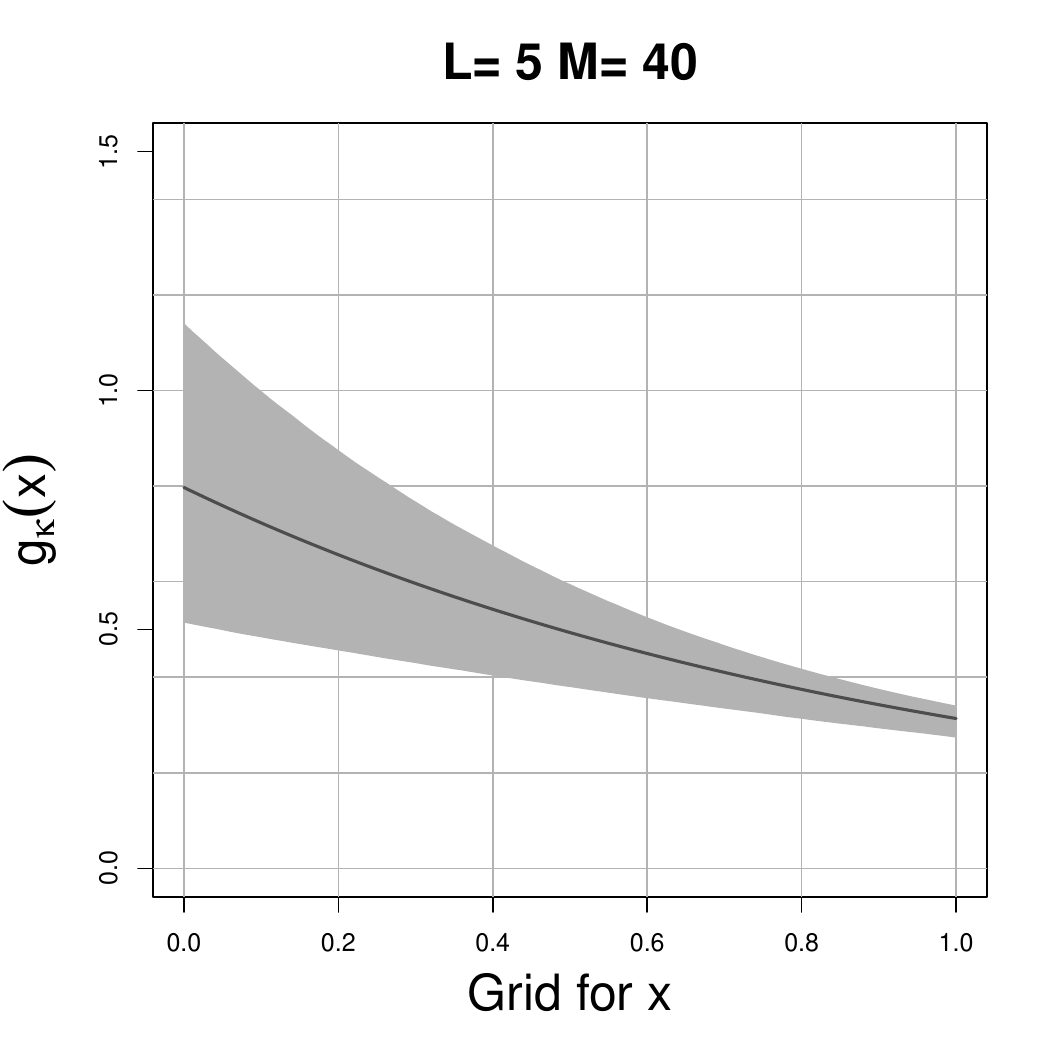}
\includegraphics[width=0.24\textwidth]{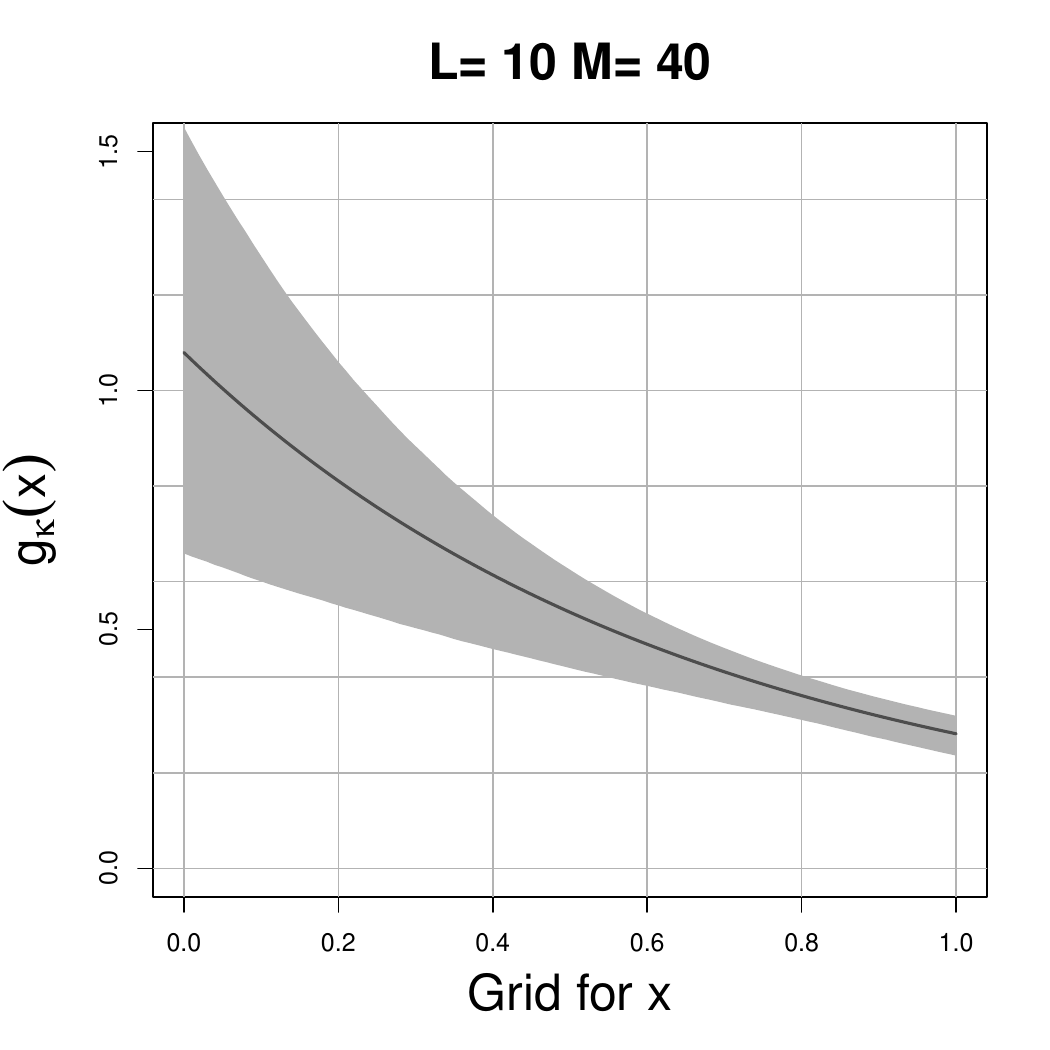}
\includegraphics[width=0.24\textwidth]{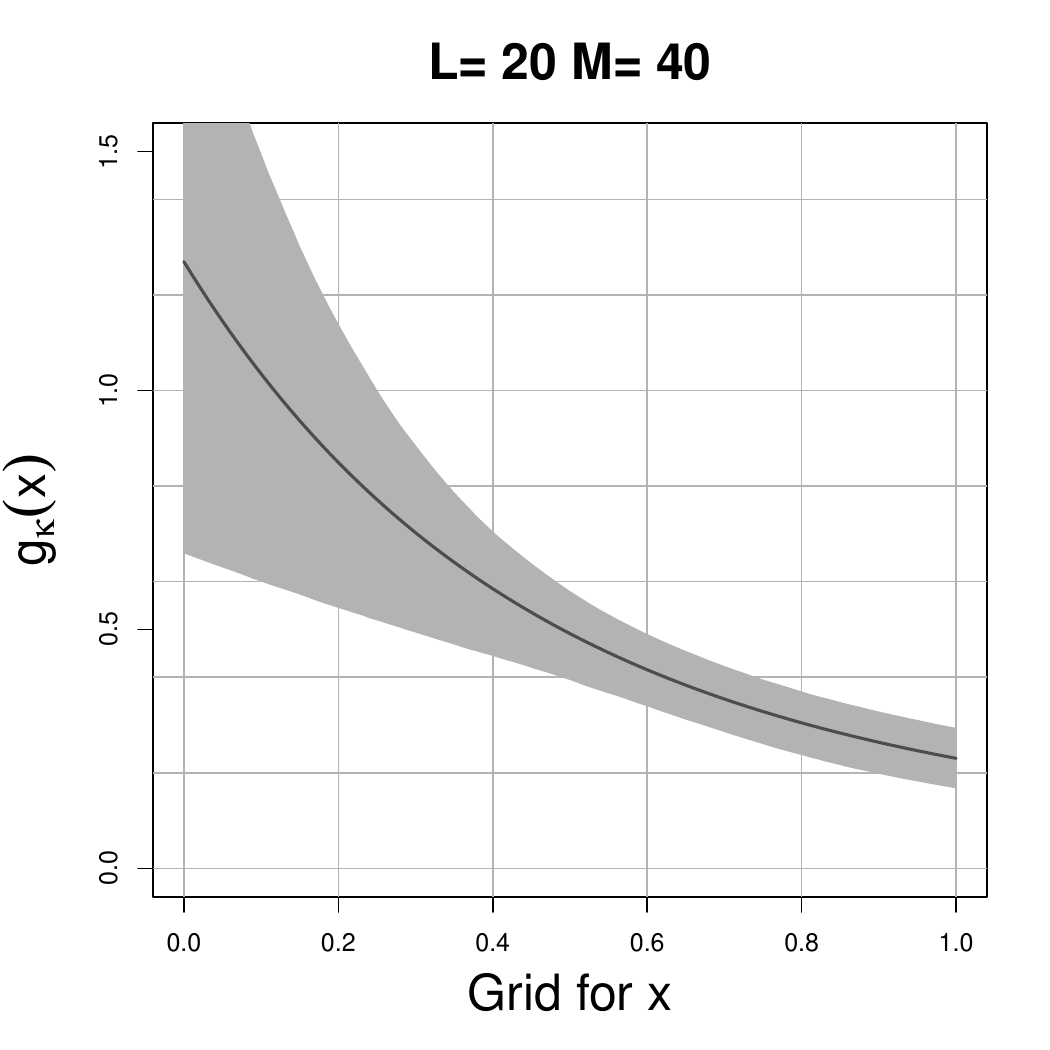}
\includegraphics[width=0.24\textwidth]{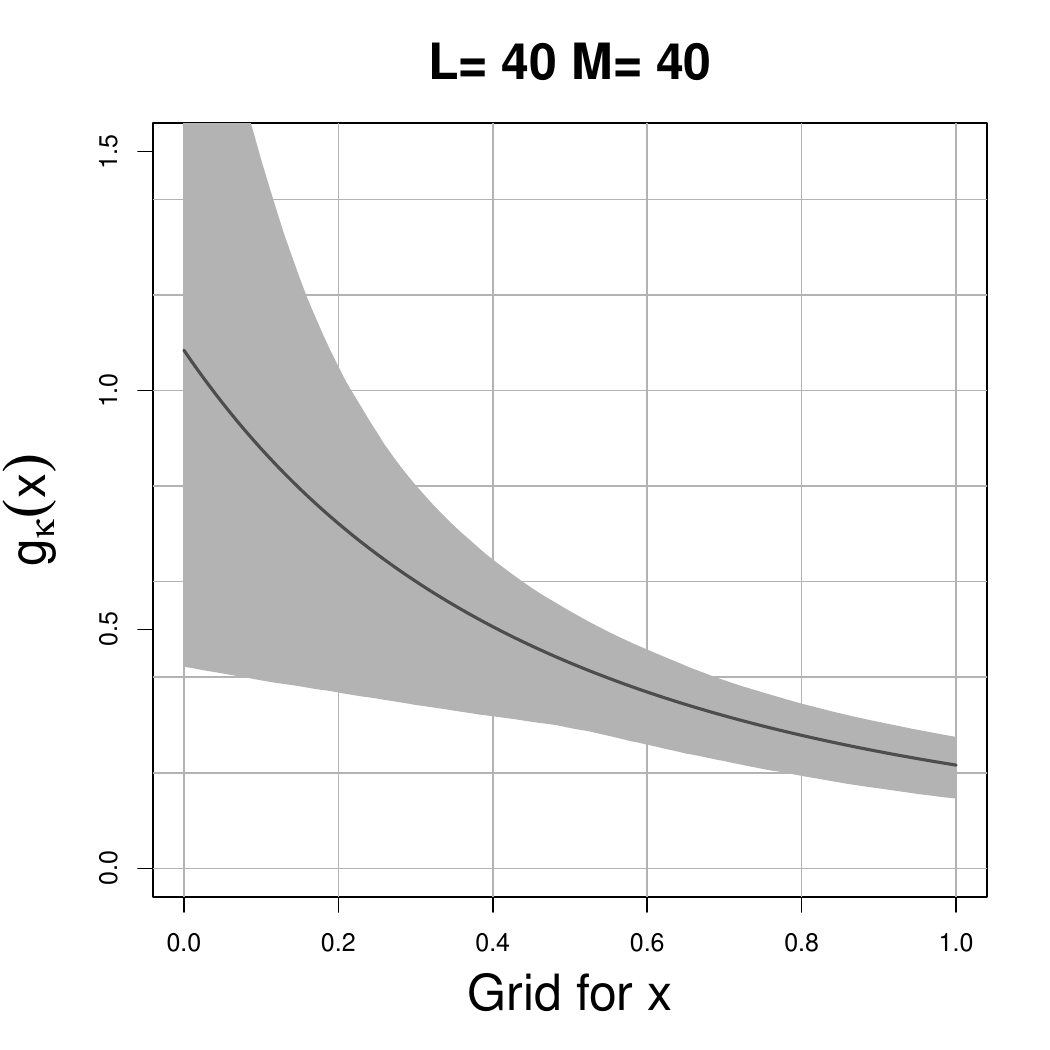} \\
\caption{{\small 
Earthquake data analysis (Section 5 of the main paper).
Posterior mean and 95\% interval estimates for the offspring density at magnitude 
$\kappa = 7.25$, for $L = {5, 10, 20, 40}$ (with $M = 40$ in all cases).}}
\label{fig:sensitivity_L}
\end{figure}

\begin{figure}[!t]
\centering
\includegraphics[width=0.24\textwidth]{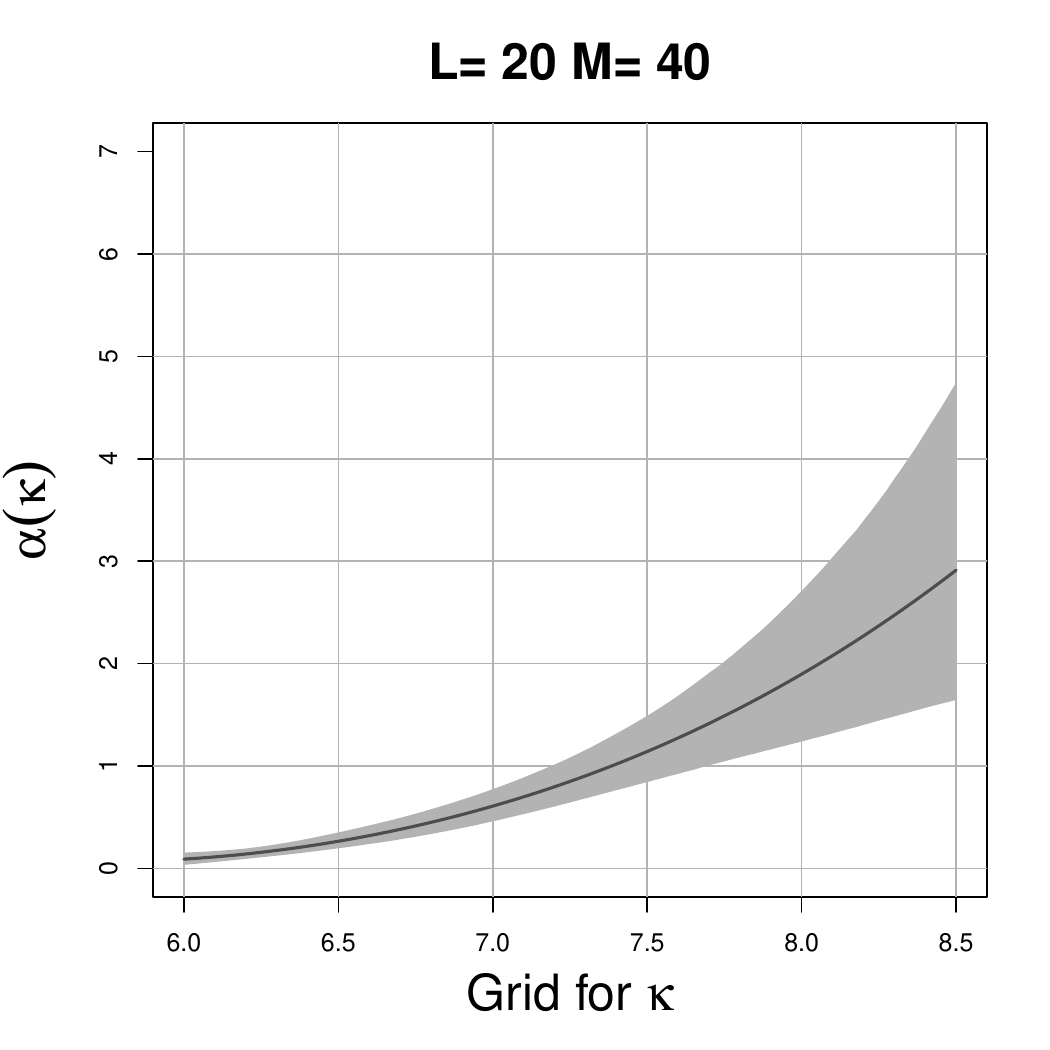}
\includegraphics[width=0.24\textwidth]{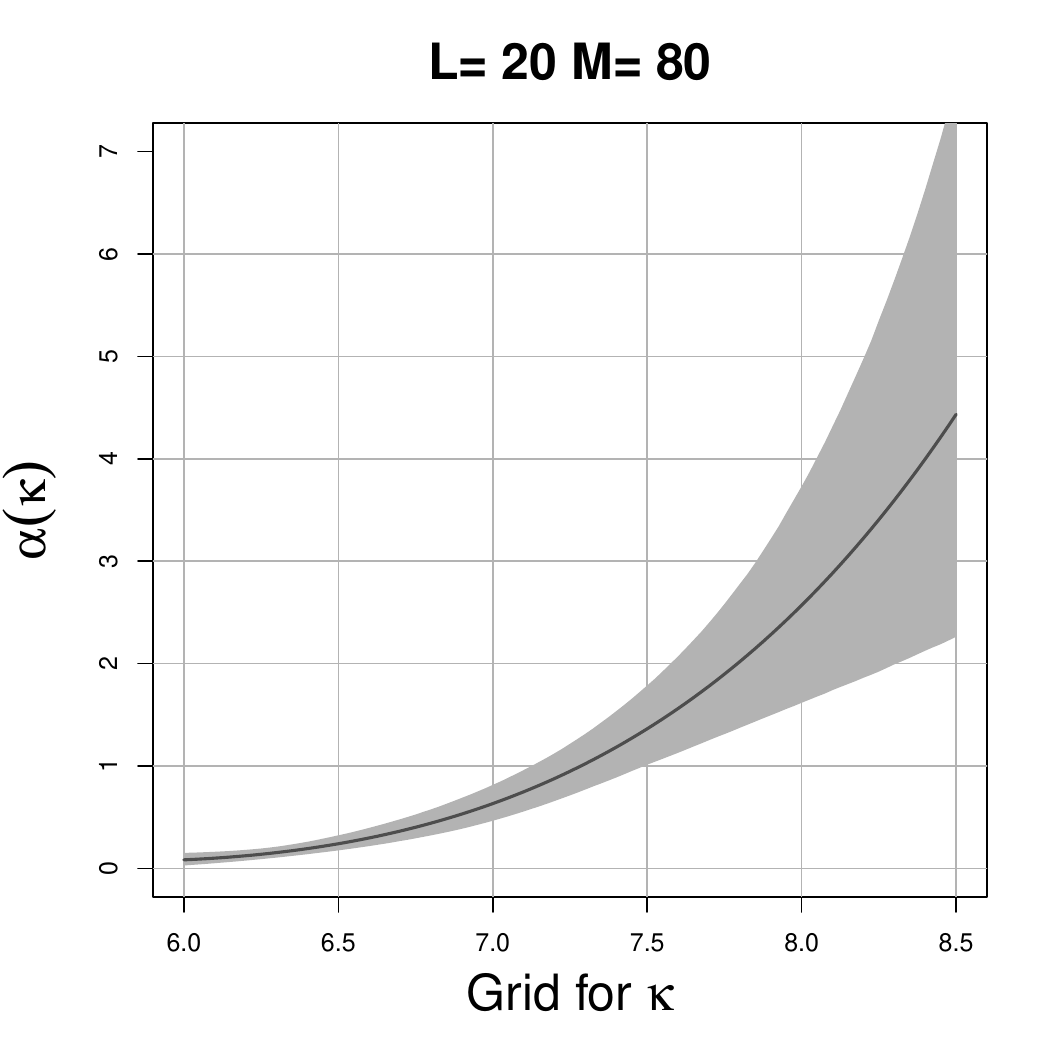}
\includegraphics[width=0.24\textwidth]{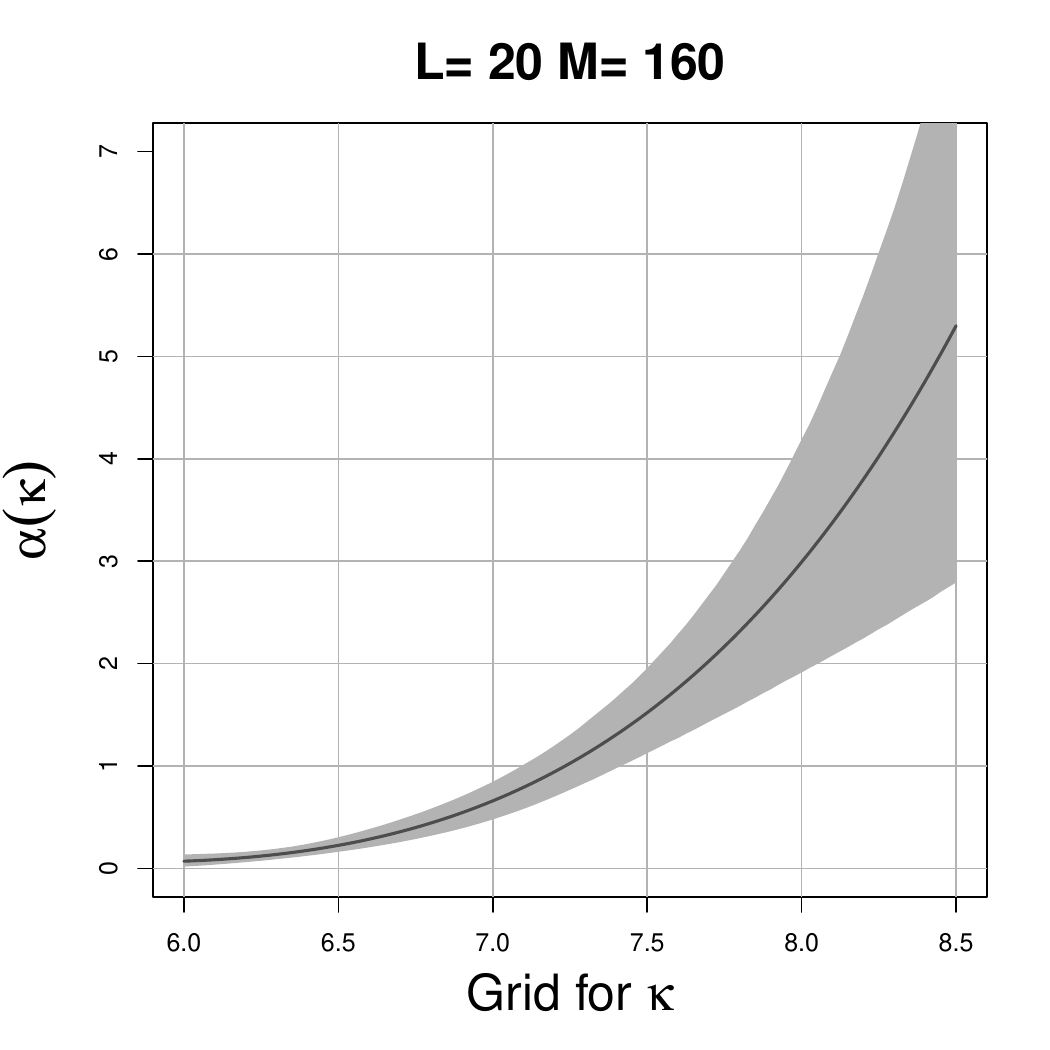}
\includegraphics[width=0.24\textwidth]{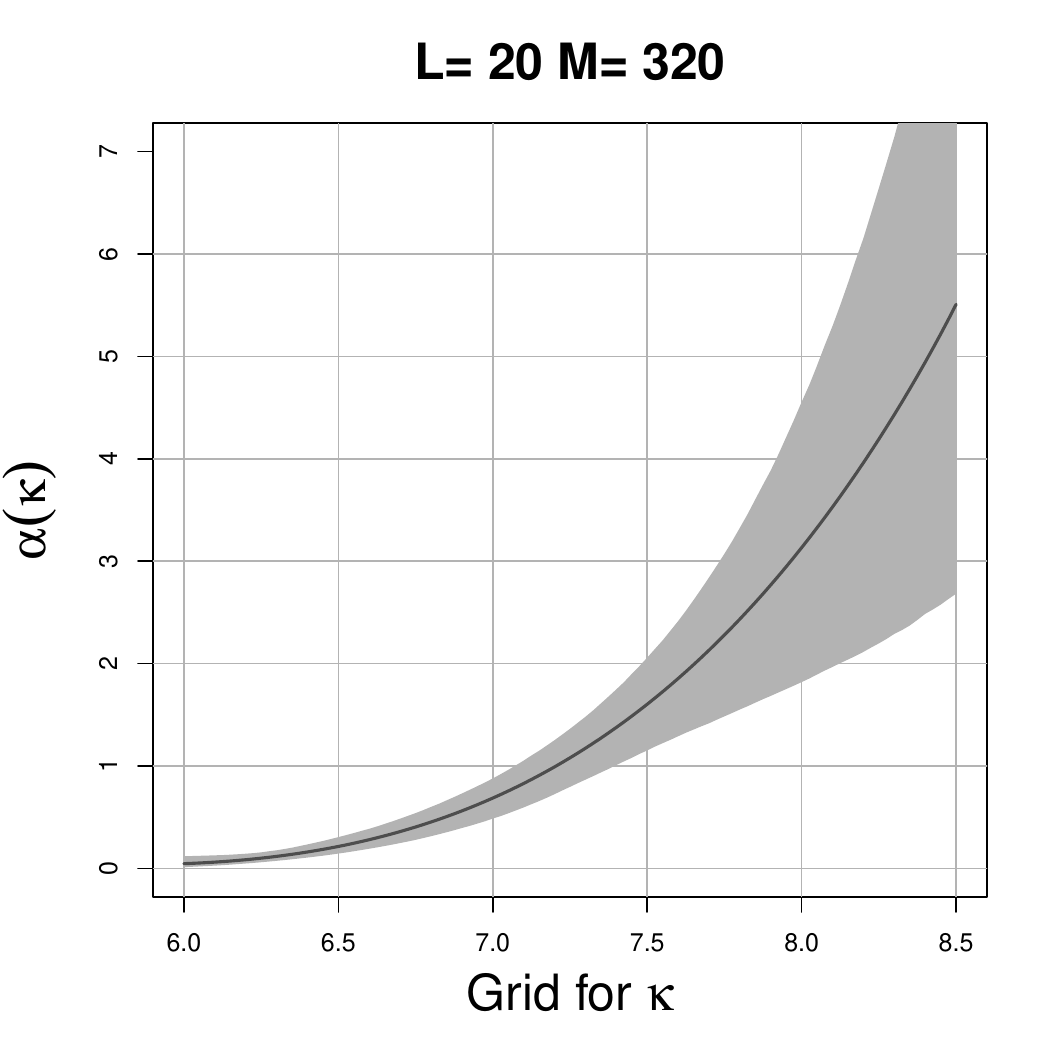} \\
\caption{{\small 
Earthquake data analysis (Section 5 of the main paper).
Posterior mean and 95\% interval estimates for the total offspring intensity function, 
for $M = \{40,80,160,320\}$ (with $L = 20$ in all cases).}}
\label{fig:sensitivity_M}
\end{figure}

\begin{figure}[!t]
\centering
\includegraphics[width=0.44\textwidth]{./figs/ak_L20M160_realdata.pdf}
\includegraphics[width=0.44\textwidth]{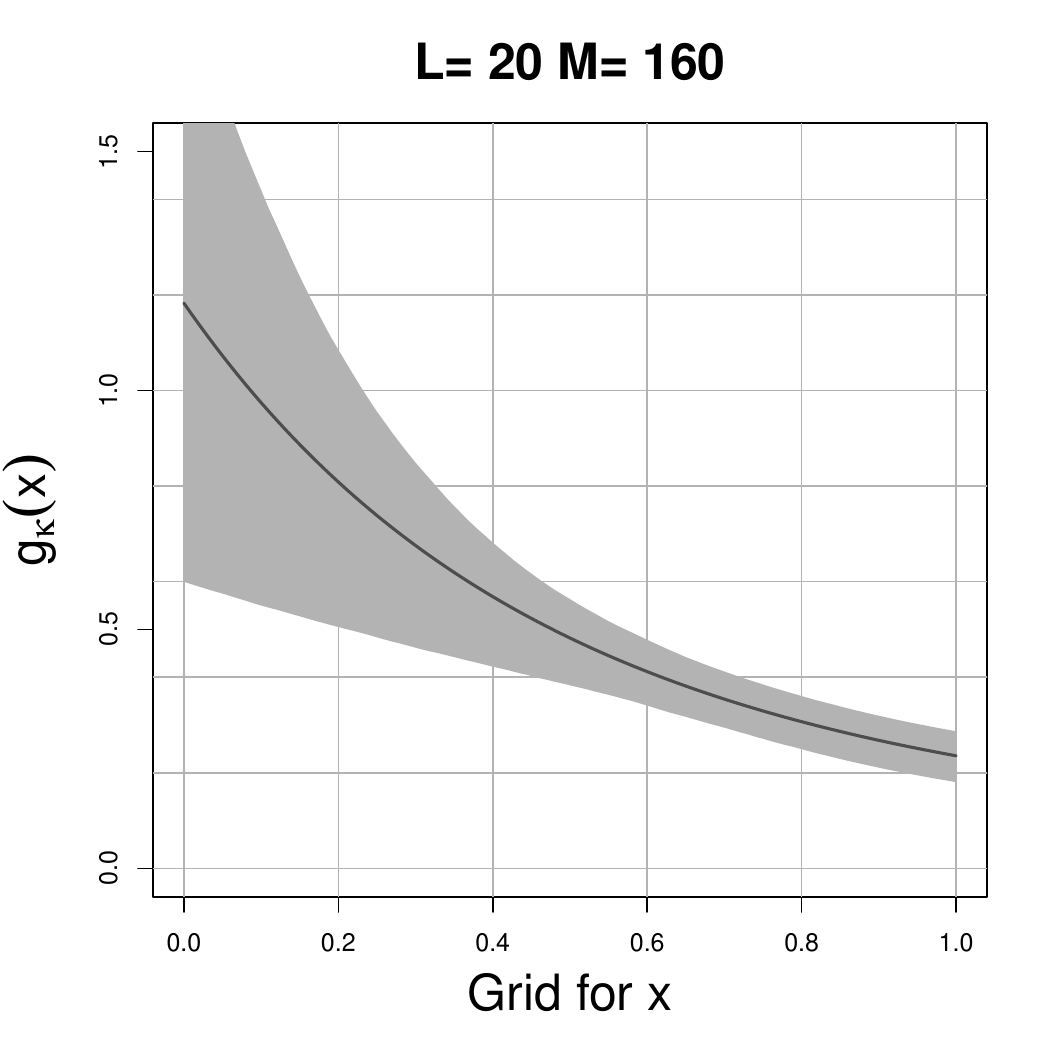} \\
\caption{{\small 
Earthquake data analysis (Section 5 of the main paper).
Posterior mean and 95\% interval estimates for the total offspring intensity function
(left panel) and for the offspring density at magnitude $\kappa = 7.25$ (right panel), 
under $L = 20$ and $M=160$.}}
\label{fig:selectedLM}
\end{figure}

Figures \ref{fig:sensitivity_L} and \ref{fig:sensitivity_M} explore the effect of the 
values for the number of basis components $L$ and $M$ on the estimation of the offspring 
density and the total offspring intensity, respectively. 
In Figure \ref{fig:sensitivity_L} (Figure \ref{fig:sensitivity_M}), from left to right, 
$L$ ($M$) is doubled from 5 (40), while $M$ ($L$) is held fixed. As the number of components 
increases, the slopes of the functions change before stabilizing at particular values, given 
by $L=20$ and $M=160$ for this point pattern. The results under these values for the number 
of basis components are shown in Figure \ref{fig:selectedLM}.
%
%

\end{document}